\def\PrintCover{0}
\def\PrintTitle{1}
\def\PrintISBN{1}
\def\PrintTOC{1}
\def\PrintAcronyms{0}
\def\PrintPreface{0}
\def\PrintIntroduction{1}
\def\PrintChTwo{1}
\def\PrintChThree{1}
\def\PrintChFour{1}
\def\PrintChFive{1}
\def\PrintConclusion{0}
\def\PrintAppendices{0}
\def\PrintSamenvatting{1}
\def\PrintPublications{1}
\def\PrintCV{1}
\def\PrintAcknowledgements{1}
\def\PrintBackCover{0}

\def\PrintAll{1}
\if\PrintAll1
\def\PrintIntroduction{1}
\def\PrintChTwo{1}
\def\PrintChThree{1}
\def\PrintChFour{1}
\def\PrintChFive{1}
\def\PrintConclusion{1}
\def\PrintAppendices{1}
\else 
\def\PrintIntroduction{0}
\def\PrintChTwo{0}
\def\PrintChThree{0}
\def\PrintChFour{0}
\def\PrintChFive{0}
\def\PrintConclusion{0}
\def\PrintAppendices{0}
\fi

\newcommand{\isAppendix}{0}
\newcommand{\isBackmatter}{0}

\documentclass[]{Resources/Template_Files/dissertation}

\title[]{Differential Reductions and Cosmological Correlations}

\author{Arno}{Hoefnagels}

\usepackage{lipsum}

\usepackage{marvosym}

\usepackage{datetime}
\newdateformat{monthyeardate}{\monthname[\THEMONTH] \THEYEAR}

\usepackage{acro}
\acsetup{make-links=true}

\usepackage{layout}

\usepackage{tocloft}

\usepackage[parfill]{parskip}

\usepackage[T1]{fontenc}

\usepackage{setspace}

\usepackage{import}

\usepackage{lipsum}
\usepackage{mdframed}

\usepackage{subcaption}

\usepackage{afterpage}

\usepackage{bm}

\usepackage[many]{tcolorbox}

\usepackage{xltabular}
\usepackage{booktabs}

\usepackage{csquotes}

\usepackage[style=ieee,citestyle=numeric-comp,backend=biber,doi=false,isbn= false,maxbibnames=30]{biblatex}
\AtEveryBibitem{
    \clearfield{urlyear}
    \clearfield{urlmonth}
    \clearfield{urlday}
}

\usepackage{tikzscale}
\usepackage{tikz}

\usetikzlibrary{shapes,arrows,chains}

\colorlet{lcnorm}{black}

\renewcommand{\headrulewidth}{1.5pt}

\LetLtxMacro\latexincludegraphics\includegraphics 

\makeatletter

\let\LaTeXStandardPart\part%
\newcommand{\unstarredpart@@noopt}[1]{%
	\unstarredpart@@opt[#1]{#1}%
}%

\newcommand{\unstarredpart@@opt}[2][]{%
        \clearpage
	\begingroup%
	\let\newpage\relax%
	\LaTeXStandardPart[#1]{#2}%
	\endgroup%
}%

\newcommand{\starredpart}[1]{%
	\LaTeXStandardPart*{#1}%
}%

\newcommand{\unstarredpart}{%
	\@ifnextchar[{\unstarredpart@@opt}{\unstarredpart@@noopt}%
}%

\renewcommand{\part}{%
	\@ifstar{\starredpart}{\unstarredpart}%
}%

\makeatother

\titlecontents{part}%
[0pt]{\color{thumb\arabic{colorcounter}}\bfseries\large\protect\addvspace{10pt}\titlerule[1pt]\addvspace{1.3ex}}
{}{\partname~}
{\hfill\contentspage}%
[\addvspace{0.7ex} {\titlerule[1pt]} \addvspace{5pt}]%

\renewcommand{\thepart}{\color{thumb\arabic{colorcounter}}\Roman{part}} 

\newcounter{thumbcounter}
\newcounter{colorcounter}

\usepackage{tikz-feynman}
\usepackage{tikz-cd}
\usepackage{array}
\usepackage{mdframed}
\usepackage{empheq}
\usepackage{subcaption}
\usepackage{braket}

\usepackage{microtype}

\usetikzlibrary{decorations.pathreplacing,calc}

\usepackage[export]{adjustbox}

\newcommand{\nc}{\newcommand}
\nc{\lb}{\llbracket}
\nc{\rb}{\rrbracket}
\nc{\gl}{\llbracket}
\nc{\gr}{\rrbracket}
\nc{\bbR}{\mathbb{R}}
\nc{\bbC}{\mathbb{C}}
\nc{\bbZ}{\mathbb{Z}}
\nc{\cO}{\mathcal{O}}
\nc{\cS}{\mathcal{S}}
\nc{\cM}{\mathcal{M}}
\nc{\cT}{\mathcal{T}}
\nc{\cX}{\mathcal{X}}
\nc{\cQ}{\mathcal{Q}}
\nc{\cD}{\mathcal{D}}
\nc{\cC}{\mathcal{C}}
\nc{\cG}{\mathcal{G}}
\nc{\cF}{\mathcal{F}}
\nc{\cI}{\mathcal{I}}
\nc{\pd}{\partial}
\nc{\eps}{\epsilon}
\nc{\la}{\lambda}
\newcommand\beq{\begin{equation}}
\newcommand\eeq{\end{equation}}
\nc{\del}{\partial}
\nc{\tri}{\hspace{-31pt}\vartriangle\hspace{-31pt}}
\nc{\blacktri}{\blacktriangle}
\nc{\eq}[1]{\begin{equation}
                     \begin{split} #1 \end{split}
                     \end{equation}}
\nc{\ul}{\underline}
\nc{\ov}{\overline}
\nc{\fa}{\hat}
\nc{\fb}{\MakeUppercase}
\nc{\fc}{\tilde }
\nc{\Lie}{{\cal L}} 
\nc{\lambdabar}{{\mkern0.75mu\mathchar '26\mkern -9.75mu\lambda}}

\renewcommand{\P}{\mathbb{P}}
\newcommand{\R}{\mathbb{R}}

\renewcommand{\S}{\mathbb{S}}
\newcommand{\C}{\mathbb{C}}

\renewcommand{\L}{\mathcal{L}}
\renewcommand{\H}{\mathcal{H}}

\renewcommand{\O}{\mathcal{O}}

\newcommand{\A}{\mathcal{A}}

\newcommand{\N}{\mathbb{N}}
\newcommand{\Z}{\mathbb{Z}}
\newcommand{\M}{\mathcal{M}}
\newcommand{\D}{\mathcal{D}}
\newcommand{\E}{\mathcal{E}}
\newcommand{\I}{\mathcal{I}}
\newcommand{\F}{\mathcal{F}}
\newcommand{\im}{\mathrm{im}}
\newcommand{\Hom}{\mathrm{Hom}}
\newcommand{\av}{\alpha_v}
\newcommand{\nuv}{\nu_v}
\newcommand{\Nv}{N_{\rm v}}

\tikzset{connect/.style={rounded corners=#1,
        to path= ($(\tikztostart)!-#1!(\tikztotarget)!-#1!-90:(\tikztotarget)$) -- ($(\tikztotarget)!-#1!(\tikztostart)!-#1!90:(\tikztostart)$) --
        ($(\tikztotarget)!-#1!(\tikztostart)!#1!90:(\tikztostart)$) -- ($(\tikztostart)!-#1!(\tikztotarget)!-#1!90:(\tikztotarget)$) -- cycle (\tikztotarget)
}}
\tikzset{connect/.default=4mm}

\tikzset{connectup/.style={rounded corners=#1,
        to path= ($(\tikztostart)!-#1!(\tikztotarget)!-#1!-90:(\tikztotarget)$) -- ($(\tikztotarget)!-#1!(\tikztostart)!-#1!90:(\tikztostart)$) --
        ($(\tikztotarget)!-#1!(\tikztostart)!#1!90:(\tikztostart)$) -- ($(\tikztostart)!-#1!(\tikztotarget)!-#1!90:(\tikztotarget)$)  (\tikztotarget)
}}
\tikzset{connectup/.default=4mm}

\tikzset{connectsmall/.style={rounded corners=#1,
        to path= ($(\tikztostart)!-#1!(\tikztotarget)!-#1!-90:(\tikztotarget)$) -- ($(\tikztotarget)!-#1!(\tikztostart)!-#1!90:(\tikztostart)$) --
        ($(\tikztotarget)!-#1!(\tikztostart)!#1!90:(\tikztostart)$) -- ($(\tikztostart)!-#1!(\tikztotarget)!-#1!90:(\tikztotarget)$) -- cycle (\tikztotarget)
}}
\tikzset{connectsmall/.default=1.5mm}
\tikzfeynmanset{
bulkblob/.style={
/tikz/shape=circle,
/tikz/draw=green,
/tikz/pattern=north west lines,
/tikz/pattern color=green,
/tikz/minimum size = 1.2 cm,}}
\tikzfeynmanset{
boundarydot/.style={
/tikzfeynman/dot,
/tikz/color=blue,}}

\tikzfeynmanset{
bulkdot/.style={
/tikzfeynman/dot,
/tikz/color=green,}}

\def\BoolDrawFeyn{1} 

\if\BoolDrawFeyn1
    \newcommand{\drawFM}[1]{ \begin{tikzpicture}[baseline=-0.5ex] \begin{feynman} #1 \end{feynman}\end{tikzpicture}}
\else
    \newcommand{\drawFM}[1]{}
\fi

\begin{document}


\pagenumbering{roman}

\if\PrintCover1
\fi

\if\PrintTitle1

\begin{titlepage}


\begin{center}

        \vspace*{2\bigskipamount}

        {\makeatletter
                \titlestyle\bfseries\Huge\@title
                \makeatother}

        {\makeatletter
                \ifx\@subtitle\undefined\else
                \bigskip
                \titlefont\titleshape\LARGE\@subtitle
                \fi
                \makeatother}

        \vfill

        {\makeatletter
                \titlestyle\bfseries\LARGE\@firstname~{\titleshape\@lastname}
                \makeatother}

        \vspace*{4\bigskipamount}

\end{center}

\if\PrintISBN1
\newpage

\thispagestyle{empty}

\vspace*{\fill}




\noindent
\begin{tabular}{@{}p{0.25\textwidth}@{}p{0.75\textwidth}}
        \textit{ISBN:} &  978-90-393-7944-8 \\[\medskipamount]
        \textit{DOI:} & 10.33540/3172 \\[\medskipamount]
        \textit{Cover by:} & Frits Hoefnagels \\[\medskipamount]
        \textit{About the cover:} & The cover consists of two elements: an abstract, intricate landscape and a bridge that stretches across it. 

        The bridge is modeled on the Prins Clausbrug, visible from our balcony and windows, and represents an unexpected shortcut. It is a path that, once we have found it, allows us to cross this landscape much more directly and simply.

        Before each chapter, a new part of the bridge is revealed and gradually the full bridge emerges. This mirrors how the hidden pathway through the mathematics is uncovered step by step as the dissertation progresses.

        Finally, the background color reflects a more personal note: it is the color of my electric guitar, which I bought with the very first salary of my PhD. This guitar represents my other great passion, music, and its color felt like the perfect link between my personal and academic journeys.
\end{tabular}





\fi


\cleardoublepage

\thispagestyle{empty}
\dedication{\epigraph{Science is built up of facts, as a house is built of stones; but an accumulation of facts is no more a science than a heap of stones is a house.}{Henri Poincaré}}


\clearpage

\thispagestyle{empty}

\begin{center}

        \vspace*{2\bigskipamount}

        {\makeatletter
                \titlestyle\bfseries\Huge\@title
                \makeatother}

        {\makeatletter
                \ifx\@subtitle\undefined\else
                \bigskip
                \titlefont\titleshape\LARGE\@subtitle
                \fi
                \makeatother}

        \vfill

        {\LARGE\titlefont\bfseries Differenti\"ele Reducties en \\[\medskipamount]
        Kosmologische Correlaties}


        (met een samenvatting in het Nederlands)

        \bigskip
        \bigskip

        {\Large\titlefont Proefschrift}

        \bigskip
        \bigskip

        ter verkrijging van de graad van doctor aan de\\[\medskipamount]
         Universiteit Utrecht\\[\medskipamount]
        op gezag van de \\[\medskipamount]
        rector magnificus, prof. dr. ir. W. Hazeleger,\\[\medskipamount]
        ingevolge het besluit van het College voor Promoties\\[\medskipamount]
        in het openbaar te verdedigen op\\[\medskipamount]
        woensdag 8 oktober 2025 des ochtends te 10.15 uur

        \bigskip
        \bigskip

        door

        \bigskip
        \bigskip

        {\makeatletter
        \Large\titlefont\bfseries\@firstname~{\titleshape\@lastname}
        \makeatother}

        \bigskip
        \bigskip

        geboren op 21 januari 1997 \\[\medskipamount]
        te Amersfoort

        \vspace*{2\bigskipamount}

\end{center}


\clearpage

\thispagestyle{empty}

Promotor: \\[\medskipamount]    
        \textit{Prof.\,dr.\ T.W.~Grimm}  \\[\medskipamount]     

Copromotor: \\[\medskipamount] 
        \textit{Dr.\ N.E.~Chisari}  \\[\medskipamount]

\vspace*{2\bigskipamount}

Beoordelingscommissie: \\[\medskipamount]      
        \textit{Prof.\,dr.\ R.A.~Britto-Pacumio}\\[\medskipamount]   
        \textit{Prof.\,dr.\ E.L.M.P.~Laenen}\\[\medskipamount]   
        \textit{Dr.\ G.L.~Pimentel}\\[\medskipamount]   
        \textit{Dr.\ D.~Schuricht}\\[\medskipamount]   
        \textit{Prof.\,dr.\ S.J.G.~Vandoren}\\[\medskipamount]

\vfill

\vspace*{2\bigskipamount}

\noindent This research is supported by the Dutch Research Council (NWO) via
a Vici grant.

\end{titlepage}

\fi

\if\PrintTOC1
\setcounter{tocdepth}{1}
\tableofcontents
\fi

\startcontents

\if\PrintAcronyms1
\printacronyms
\addcontentsline{toc}{chapter}{Acronyms}
\setheader{Acronyms}
\fi

\if\PrintPreface1

\cleardoublepage

\fi

\if\PrintPublications1

\chapter*{List of Scientific Publications}
\label{ch:publications}

\addcontentsline{toc}{chapter}{List of Scientific Publications}

\setheader{List of Scientific Publications}

\textbf{Chapter~\ref{ch:generalGKZ}} contains the general results on Gelfand-Kapranov-Zelevinsky (GKZ) systems. It starts with a review of their generic construction, including their construction in terms of D-modules and Euler-Koszul homologies. Then it describes the use and interpretation of reducibility in terms of D-modules, and shows how subsystems be obtained using what we call reduction operators. This chapter contains the results of
\begin{itemize}
    \item[\cite{grimm_reductions_2025}] \fullcite{grimm_reductions_2025}
\end{itemize}
that concern general GKZ systems. It also contains some previously unpublished results regarding solution counting for GKZ systems and reduction operators.

\textbf{Chapter~\ref{ch:cosmology}} begins our study of cosmological correlator. Here, we will introduce the toy model we will consider throughout the remaining chapters, as well as study one particular cosmological correlator in detail using the reduction operators. This chapter contains the results of~\cite{grimm_reductions_2025} that directly concern cosmological correlators.

\textbf{Chapter~\ref{ch:reductionalgorithm}} is concerned with a general study of cosmological correlators within the toy model. Here, we use reduction operators to construct differential systems for general tree-level correlators. Furthermore, these operators give rise to many algebraic relations and symmetries. We explain how to leverage these in the recursive reduction algorithm, which expresses cosmological correlators in terms of only a small number of simple functions. This chapter contains the results of
\begin{itemize}
    \item[\cite{grimm_reduction_2025}] \AtNextCite{\defcounter{maxnames}{99}} \fullcite{grimm_reduction_2025}
\end{itemize}
that concern the recursive reduction algorithm and the general study of cosmological correlators.

\textbf{Chapter~\ref{ch:complexity}} contains our study of complexity for cosmological correlators. Here we compute the complexity of cosmological correlators using the Pfaffian framework, resulting in explicit bounds on topological and computational quantities satisfied by cosmological correlators. Furthermore, we study complexity reduction for the recursive reduction algorithm. This chapter contains the results of
\begin{itemize}
    \item[\cite{grimm_structure_2024}] \AtNextCite{\defcounter{maxnames}{99}} \fullcite{grimm_structure_2024}
\end{itemize}
as well as the results of~\cite{grimm_reduction_2025} that pertain directly to complexity.
\fi





\pagenumbering{arabic}
\thumbtrue

\if\PrintIntroduction1
\setcounter{thumbcounter}{1}
\setcounter{colorcounter}{1}

%
%
%


\chapter{Introduction}\label{ch:Introduction}

Throughout physics, studying a theory and obtaining its experimental predictions ultimately reduces to evaluating integrals or solving systems of differential equations. 
Often, this evaluation step introduces a large amount of intermediate complexity yet, in many cases, the solutions obtained after this procedure are surprisingly simple.
Of course, simple could mean a variety of things, depending on the observer. 
However, let us continue with a somewhat nebulous, intuitive understanding of simplicity, leaving more precise definitions for later on.

To illustrate such unexpected simplicity, consider the tree-level amplitudes of quantum field theories (QFT) containing gauged degrees of freedom. 
Obtaining these amplitudes naively involves summing large amounts of Feynman diagrams. One finds, however, that the resulting amplitude is astoundingly simple, in the sense that it can be expressed quite compactly in terms of momenta in spinor space.
In fact, the simplicity of these amplitudes is a consequence of various recursive relations, which are not immediately manifest when considering only single Feynman diagrams~\cite{berends_recursive_1988,britto_new_2005,britto_direct_2005}.
In general, one would also expect that such simplifications signify the existence of additional structure not yet taken advantage of. This naturally leads us to question how such structure can be obtained in generality.

In this thesis we take steps towards this general study of simplifications, with a focus on applications in perturbative quantum field theories.
In this setting, physical amplitudes can be computed using sums of Feynman integrals. 
The evaluation of such integrals is a challenging task, and considerable research is dedicated to developing new techniques to address it.
Frequently, the evaluation can be greatly simplified by interpreting the Feynman integrals as solutions to a particular set of differential equations \cite{anastasiou_higgs_2002,anastasiou_dilepton_2003,anastasiou_automatic_2004,argeri_feynman_2007,cachazo_sharpening_2008,arkani-hamed_allloop_2011,henn_multiloop_2013,gehrmann_qcd_2014,henn_lectures_2015,vanhove_differential_2021,armadillo_evaluation_2023,giroux_loopbyloop_2023,kreimer_bananas_2023,henn_first_2023,gorges_procedure_2023,hidding_feynman_2023,dlapa_algorithmic_2023,britto_generalized_2023,ananthanarayan_method_2023,he_symbology_2023,marzucca_recent_2024,he_symbology_2024,fevola_landau_2024,frellesvig_epsilonfactorised_2024,henn_dmodule_2024,calisto_learning_2024,jiang_symbol_2025}.
The choice of differential equations varies significantly, as do the methods used to derive them \cite{tkachov_theorem_1981,chetyrkin_integration_1981,kotikov_differential_1991,remiddi_differential_1997,gehrmann_differential_2000,artico_integrationbyparts_2024,bloch_feynman_2015,bloch_local_2017,bourjaily_traintracks_2018,bourjaily_bounded_2019,bourjaily_embedding_2020,klemm_lloop_2020,bonisch_analytic_2021,bonisch_feynman_2022,gasparotto_cohomology_2023,lairez_algorithms_2023,delacruz_algorithm_2024}.

The challenge of evaluating such observables becomes even more pronounced in non-flat space-times, such as in cosmological settings that describe an expanding  universe. As we will describe in more detail in section~\ref{sec:cosmology}, the natural observables in this setting are cosmological correlators, which are evaluated on a fixed time-slice \cite{weinberg_quantum_2005,benincasa_amplitudes_2022,lee_records_2024}. 
Understanding the general mathematical structure of these correlators is an essential aspect of describing the history of our universe and forms an active topic of research \cite{baumann_bootstrapping_2020,sleight_bootstrapping_2020,arkani-hamed_cosmological_2020,baumann_cosmological_2021,baumann_cosmological_2021,pajer_boostless_2020,goodhew_cosmological_2021,kuhne_faces_2022,pimentel_boostless_2022,hogervorst_nonperturbative_2023,wang_bootstrapping_2023,arkani-hamed_differential_2023,jazayeri_parity_2023,jazayeri_shapes_2023,benincasa_geometry_2025,chowdhury_subtle_2025,choudhury_primordial_2025,fan_cosmological_2024,aoki_cosmological_2024,kuhne_faces_2024,albayrak_perturbative_2024,liu_dispersive_2025,arkani-hamed_differential_2023,baumann_new_2025,lee_records_2024,werth_cosmoflow_2024,anninos_sitter_2024,melville_sitter_2024,goodhew_cosmological_2024,werth_spectral_2024,fevola_algebraic_2025,de_physical_2025,qin_cosmological_2025,cespedes_massive_2025}. 
In this thesis, we take a significant step forward by introducing a new approach that systematically explores the space of tree-level correlators with a focus on maximizing the use of algebraic and permutative identities. 

To study the simplifications of cosmological correlators, we will use a general system of differential equations originally due to Gelfand, Kapranov and Zelevinsky \cite{gelfand_generalized_1990,gelfand_hypergeometric_1991,gelfand_discriminants_1994}. 
Known as the \textit{GKZ system}, this differential system  can be readily obtained for large classes of integrals and has been extensively studied \cite{saito_grobner_2000,nasrollahpoursamami_periods_2016,delacruz_feynman_2019,feng_gkzhypergeometric_2020,klausen_hypergeometric_2020,klausen_kinematic_2022,klausen_hypergeometric_2023a,ananthanarayan_feyngkz_2023,blumlein_hypergeometric_2021,blumlein_hypergeometric_2023,chestnov_restrictions_2023,caloro_ahypergeometric_2023}.
The class of integrals for which GKZ systems can be found is quite broad, including examples from all over physics. This makes it a natural framework for a general study of simplifications, as any results obtained for GKZ systems immediately inherit a wide domain of applicability.

Studying integrals through the lens of differential equations brings with it some notions of complexity, which will be key in our study of simplifications. 
One immediate example of such a notion is the number of linearly independent solutions. 
For example, one would intuitively expect that the solution to a first-order differential equation is less complex than a higher-order one.
Thus, we could define the complexity of a function as the minimal number of linearly independent solutions of the differential system it satisfies.
Note here that the complexity depends on the specific system of differential equations. This dependence is a crucial ingredient in our search for unexpected simplifications. 
In particular, let us consider some complicated system of differential equations. Then, if one of the solutions is also a solution of another, simpler, system of differential equations, we could conclude that this solution is unexpectedly simple.\footnote{For generic systems of differential equations, such solutions do not exist. Thus we could reasonably conclude that the existence of such a function is unexpected.}

Deferring to section~\ref{sec:introreducibility} for more detail, let us mention here that the above situation is an example of a mathematical property known as reducibility. Briefly, reducibility can be understood as the existence of smaller, simpler subsystems within some larger system.
The existence of such subsystems lets us use a ``divide and conquer'' approach, where we solve the various subsystems separately and combining the solutions in order to solve the full system.
Crucially, the solutions of the subsystems will be simpler than one would expect from the larger system. This provides us with a general framework in which solutions can be unexpectedly simple.

Crucially for our purposes, reducibility of GKZ systems has been completely classified using the notion of resonance~\cite{walther_duality_2005,beukers_irreducibility_2011,schulze_resonance_2012}. In chapter~\ref{ch:generalGKZ}, we use the proofs in these papers to obtain subsystems of GKZ systems explicitly, in terms of certain differential operators that we call \textit{reduction operators}. These additional differential operators allow us to obtain partial solution bases, realizing the divide and conquer approach laid out above. Alternatively, we show that, in many cases, the reduction operators lead to inhomogeneous differential equations relating different GKZ systems. These inhomogeneous equations naturally lend themselves to iterative solutions, something that will be crucial in our general study of cosmological correlators.
We note that the results of chapter~\ref{ch:generalGKZ} hold for general GKZ systems, and can thus be applied to any integral for which a GKZ system can be obtained.

Applied to cosmological correlators, we show that the above perspective is especially fruitful. Here, we broadly followed two approaches. One approach, which we follow in chapter~\ref{ch:cosmology}, is to obtain and solve subsystems directly using the reduction operators. Each subsystem corresponds to some system of differential equations and, as this system will be simpler than the original system, sometimes its solutions can be obtained directly. We apply this approach to a particular correlator, and show that it can be obtained in this manner.

For a general study of cosmological correlators, we show that the inhomogeneous differential equations are particularly useful, and use these to uncover a rich algebraic and symmetric structure in these correlators in chapter~\ref{ch:reductionalgorithm}.
Interestingly, this structure includes diagrammatical identities, where cutting or contracting an edge within a diagram can be realized using certain differential operators.
Such identities share some curious similarities with the tree-level amplitudes for gauged quantum field theories we discussed before, as here, too, we find that these relations can be expressed recursively.
One difference between the two settings is that, for cosmological correlators, the recursive identities involve differential operators, while for ordinary tree-level QFT amplitudes the recursive relations are of algebraic nature. 
Interestingly, one can view the differential relations as arising from corresponding algebraic relations of the integrand. The process of integrating over some domain then turns these algebraic relations into differential ones.
We expect that this is a general phenomenon, and want to extend the above analysis to loop-level Feynman integrals in future work.

There is also a fascinating interplay between the reducibility of the underlying differential equations and the appearance of algebraic identities between various correlators. 
Specifically, reducibility first allows us to obtain a convenient closed set of basis functions for the differential equations. 
Afterwards, we can leverage reducibility again and obtain additional differential relations between various basis functions. 
As the basis functions are already closed when taking of differentials, any additional differential relation turns into an algebraic one. In this way we obtain a large number of algebraic relations between different basis functions, greatly simplifying the process of obtaining the actual solutions.

In addition to the algebraic relations, there also appears a permutation symmetry for cosmological correlators, where only a handful of functions appear many times but with permuted inputs. 
From this observation, we put forward the recursive reduction algorithm in chapter~\ref{ch:reductionalgorithm}, where we solve for cosmological correlators in two parts. 
First, we exploit algebraic identities and permutation symmetries to express any cosmological in terms of these minimal functions. Secondly, we solve for these functions, that we call the minimal representation functions, using the differential equations they satisfy. Finally, we substitute the minimal representation functions back to obtain the correlator itself.
The advantage of the above algorithm stems from the fact that managing many algebraic or symmetry identities, though tedious, is conceptually simple.
By contrast, solving large systems of differential equations directly is often much more complex. 
Therefore, it is useful to push complexity into the algebraic and symmetric manipulations wherever possible, so as to minimize the difficult differential steps.

Reducing the complexity of a system using such relations naturally leads us to study complexity reductions in a more formal manner. Specifically, we have used the framework of Pfaffian functions for this, which assigns a numerical complexity to functions through the differential equations they satisfy. 
In contrast to the coarse measure of complexity we have been discussing so far, the resulting Pfaffian complexity also allows for concrete bounds on a number of topological and computational quantities.
For example, it is possible to bound the number of poles of zeros of a function using Pfaffian complexity \cite{vorobjov_complexity_2004}.
Interestingly, the Pfaffian complexity is obtained using the specific differential equations one considers, which we call the representation. This implies that one finds different bounds for the same function depending on which representation is considered.
Hence, an over-estimate of the Pfaffian complexity is an indication that there could exist a simpler representation, or that there are symmetries or relations not exploited.

We apply this analysis to cosmological correlators in chapter~\ref{ch:complexity}, first using the kinematic flow algorithm of~\cite{arkani-hamed_differential_2023}.
Here, we find that using the kinematic flow algorithm does lead to an overestimation of the complexity of cosmological correlators. This observation was one of the original motivations for the recursive reduction algorithm of chapter~\ref{ch:reductionalgorithm}, which aims to incorporate symmetries and algebraic relations directly leading to a lower complexity. 
While the recursive reduction algorithm succeeded in its goal of incorporating the relations described above, analyzing it with the Pfaffian framework led to an unexpected complication: certain operations that seem to reduce complexity on an intuitive level do not appear to do so in the Pfaffian sense.
For example, when obtaining cosmological correlators permuting a functions input is a trivial operation. 
However, from a Pfaffian perspective the permuted function is completely independent, leading to a large increase in complexity. We expect that there is a framework which does incorporate such symmetries but leave that for future work.

In summary, in this thesis we have focused on reducing complexity via the reducibility of differential systems.
For cosmological correlators we have shown that this approach is quite fruitful, leading to the novel recursive reduction algorithm, which tries to incorporate all present symmetries and relations in a systematic manner. We expect that the same approach also yields results in similar settings, such as Feynman integrals or string theory compactifications, something that we will explore in future work.

\paragraph{Structure of the thesis.}

The remaining sections of this chapter are devoted to a general discussion of reducibility in section~\ref{sec:introreducibility}, and cosmological correlators in section~\ref{sec:cosmology}. In chapter~\ref{ch:generalGKZ} we discuss the most technical aspects of this thesis. Here, we will discuss GKZ systems and explain how their reducibility is classified. Then, we apply these tools and provide techniques to exploit reducibility when solving general integrals.
Following this, we turn our eye towards some application of these techniques in chapter~\ref{ch:cosmology}, where we start discussing cosmological correlators.
We introduce the model we will study throughout the rest of this thesis there, a conformally coupled scaler in a deSitter space-time.
Furthermore, we focus on a particular cosmological correlator and investigate it in detail using the techniques of the previous chapter.
Afterwards, in chapter~\ref{ch:reductionalgorithm}, we turn our attention to a vast generalization of these techniques for cosmological correlators. 
Here, we demonstrate how reducibility can be leveraged for general tree-level cosmological correlators within the model. 
We find that there is an immense amount of symmetry within these correlators, and leverage this symmetry in the recursive reduction algorithm. In chapter~\ref{ch:complexity}, we discuss the Pfaffian perspective on complexity for cosmological correlators, yielding explicit bounds and also highlighting some of its limitations. Finally, in chapter~\ref{ch:conclusion} we summarize our results and provide an outlook towards the future.

\section{Reducibility and differential equations}\label{sec:introreducibility}

In this section, we will give a high level overview of the concept of reducibility for differential equations and why it is useful when solving them. 
Here, we will be somewhat hands-on, focusing on the implications of reducibility for the solutions of a system of differential equations, referring to chapter~\ref{ch:generalGKZ} for a more formal study.
Crucially, reducibility implies that some of the solution are ``simpler'' in a precise sense. 
This will be a coarse but useful measure of complexity that we will consider throughout this thesis. In order to obtain a more fine-grained measure, we will also give a high-level description of Pfaffian complexity.

To study reducibility, it is useful to start with an example. Therefore, we will now consider Euler's hypergeometric function.
This function solves a particular differential equation which, besides the variable $z$ the function depends on, involves three parameters. Crucially, the differential equation factorizes in certain limits of these parameters. Correspondingly, its solutions are surprisingly simple in these limits. This is emblematic of general behavior for GKZ systems, using which Euler's hypergeometric function can also be obtained. In fact, the dependence on the parameters corresponds exactly to the resonance of GKZ systems discussed in chapter~\ref{ch:generalGKZ}.

\paragraph{The hypergeometric function.}

The example we will discuss is that of the hypergeometric function ${}_2F_1$. While usually defined from a series expansion, it is a well known fact that this expansion is equal to the integral
\begin{equation}\label{eq:2f1int}
    {}_2F_1(a,b,c;z) = \frac{1}{B(b,c-b)}\int_0^1 x^{b-1}(1-x)^{c-b-1}(1-zx)^{-a} dx\, ,
\end{equation}
provided the real parts of $b$ and $c$ satisfy $\mathrm{Re}(c)>\mathrm{Re}(b)>0$. Here, $B$ denotes the Euler beta function. Furthermore, it is also well known that the hypergeometric function satisfies the differential equation
\begin{equation}\label{eq:2f1difeq}
\left(z(1-z)\frac{d^2}{dz^2} + \big(c-(a+b+1)z\big)\frac{d}{dz}- ab\right)\, {}_2F_1(a,b,c;z) =0 \, ,
\end{equation}
which can be confirmed using the integral representation of~\ref{eq:2f1int}. In fact, the differential equation above has two linearly independent solutions, one of which is the hypergeometric function as above. We will study the solutions to this differential equation for various values of the parameters $a$, $b$ and $c$.

For generic values of the parameters, both of the solutions to the differential equation above are of hypergeometric type, in the sense that both consist of a hypergeometric function, possibly with a relatively simple pre-factor and up to a change of variables. For example, provided that $c$ is not an integer, the second solution is of the form
\begin{equation}
    z^{1-c} \, {}_2F_1(1+a-c,1+b-c,2-c;z)\, .
\end{equation}
This leads us to the following question: are there any values of $a$, $b$ and $c$ such that one (or both) of the solutions simplify?
One obvious candidate is restricting to $a=0$. Then, from the integral representation, it follows that ${}_2F_1(0,b,c;z)$ is a constant. From the differential equation, too, something interesting happens. Namely, one finds that it factorizes as 
\begin{equation}
    \left( c-(b+1)z + z(1-z)\frac{d}{dz}\right) \frac{d}{dz} \,{}_2F_1(0,b,c;z) = 0\,.
\end{equation}
From the form of this equation, it also becomes clear that the constant function is a solution, providing us with an alternate perspective on the simplification we observed. 
Furthermore, from the differential equation one finds a similar factorization for $b=0$ from which we find that the differential equation again has a constant solution. In particular, one can check that in this case, again, ${}_2F_1(a,0,c;z)$ is constant.
Crucially, $b=0$ lays outside of the domain of the integral representation and therefore could not have been observed from this perspective.

While the above simplifications are rather trivial, it turns out that there are other special values in which the differential equation factorizes. For example, if one considers the case where $c=b$, one finds that the differential equation~\eqref{eq:2f1difeq} can be written as
\begin{equation}\label{eq:hypergeomfac}
    \left(z \frac{d}{dz}+b\right)\left( (1-z)\frac{d}{dz}-a\right){}_2F_1(a,b,b;z) = 0\, ,
\end{equation}
and again we find that the differential equation factorizes. Similarly, this factorization results in a simplification of the solutions. 
In particular, in this limit the hypergeometric function simplifies to ${}_2F_1(a,b,b;z)= (1-z)^{-a}$. 

\subsection{Reducibility}

From the previous discussion, some natural questions emerge: is it possible to systematically classify when such simplifications arise? And if so, can one use this help in solving the differential equations? In order to answer these questions, let us highlight two aspects of the observations above.
Firstly, although it is clear that on an intuitive level that a function such as $(1-z)^{-a}$ is simpler than a generic hypergeometric function, it is worthwhile to make this intuition a bit more precise.
In particular, from the perspective of the differential equations, the function $(1-z)^{-a}$ satisfies first-order differential equation
\begin{equation}
    \left( (1-z)\frac{d}{dz}-a\right)(1-z)^{-a} = 0 \,
\end{equation}
while a generic hypergeometric minimally satisfies a differential equation of second order. Note that this differential operator is also the right-most factor of equation~\eqref{eq:hypergeomfac}. This leads us to posit that, for a given ordinary differential equation, a solution is ``simpler'' than expected if it can also be obtained by solving an ordinary differential equation of strictly lower order.

Secondly, we have found these special solutions by studying the \textit{factorizations} of the original hypergeometric differential equation. Furthermore, we note that the existence of these factorizations directly implies the existence of ``simpler" solutions as defined above. To see this, note that if a differential operator $P$ factorizes as $P= QR$, where $Q$ and $R$ are differential operators of strictly lower order, then the solutions to $R f =0$ will automatically be solutions to $P f =0$. Because the equation $R f =0$ is of lower order than the original differential equation, its solutions are more restricted and therefore can be considered to be more simple.

\paragraph{Simplicity and reduciblity.}

To extend this notion of simplicity to systems of differential equations involving multiple variables, we note that in all of these cases there existed an additional differential operators annihilating some, but not all of the solutions. For example, for the general factorization $P = QR$ only some of the solutions of $Pf=0$ can be annihilated by $R$. If all of the solutions would be annihilated by $R$, then both operators would have the same number of solutions. However, this is impossible as $R$ is of strictly lower order than $P$.

The existence of such differential operators is deeply linked with the mathematical notion of reducibility. In fact, for a system of 
differential equations reducibility can be described as follows:
\begin{center} \textit{A set of differential equations is reducible if there exists a strict subset of solutions that are annihilated by additional differential operators.} 
\end{center}
These additional differential operators will be what we call the reduction operators. 

The usefulness of reducibility lies in the fact that, due to these extra constraints, finding solutions to this more constricted set of differential equations is almost always easier than solving the entire system at once. Moreover, there may be multiple subsystems of a given set of differential equations, each associated with a partial solution basis. If enough of such subsystems exist, one can build up the full solution space by solving each subsystem separately, resulting in a full solution basis while never having to solve the full system of differential equations at once. This approach will be highlighted in chapter~\ref{ch:cosmology}, where this formalism is applied to a specific cosmological correlator.

\paragraph{The language of D-modules.}

The concept of reducibility as defined above is somewhat different than the usual definition of reducibility for differential systems, which is usually phrased in terms of the underlying differential modules. We will describe the precise relation between the two in chapter~\ref{ch:generalGKZ}. However, it is useful to introduce some of the basic concepts here already in a more lightweight form. Both to familiarize ourselves with some of the more technical aspects of this thesis, but also because it can bring some insights into how a more formal and mathematical perspective can bring with it great simplifications when applied in the right setting.

In particular, various notions of reducibility appear all throughout mathematics, and almost always for the same reason. Instead of trying to describe a complicated object at once, it is often much simpler to break it down into its simplest parts, its irreducible components, and study these parts separately. Afterwards, one can then try to relate these parts to the whole object if desired. 

To make the study of such subsystems precise, it is first necessary to define exactly what we mean with a differential system. Initially, we might define this simply as the system of differential equations we are given. However, it turns out that it is useful to instead consider linear differential equations as \textit{equivalence relations} on a ring of differential operators. For example, instead of the differential equation
\begin{equation}\label{eq:ddxmin3f}
    \left(\frac{d}{dx}-3\right)f(x) =0\, ,
\end{equation} 
one could instead study the ring of polynomial differential operators in the variable $x$, modulo the equivalence relation
\begin{equation}\label{eq:reduciblitymathex}
    \frac{d}{dx}-3 \sim 0\,.
\end{equation}
This equivalence relation encodes the fact that, when acting on solutions to equation~\eqref{eq:ddxmin3f}, $\frac{d}{dx}-3$ will act equivalently to multiplication by zero.
Note that as this perspective contains the same information as the original differential equation, it also contains all information regarding the solutions $f$. Technically, the ring of differential operators modulo equivalence relations is an example of a so-called D-module. The technical results of this thesis thus consist of results on various such D-modules.

In this language, reducibility turns out to be equivalent to the possibility of adding non-trivial additional equivalence relations. 
Here, non-trivial means that, after adding these additional relations, the new differential system has a number of linearly independent solutions which is strictly smaller than the original system but non-zero.
The existence of such additional relations is far from guaranteed. For example, for the system defined by the relation~\eqref{eq:reduciblitymathex}, adding a relation of the form $\frac{d}{dx}-2 \sim 0$ implies that its solutions satisfy
\begin{equation}
\left(\frac{d}{dx}-2\right)f(x) -  \left(\frac{d}{dx}-3\right)f(x)   = 0 \, \implies \, f(x) = 0\, .
\end{equation}
In other words, this system has no linearly independent solutions and therefore does not consist of a subsystem. In fact, no such additional non-trivial relations exist for this system. Returning to the language of D-modules, the existence of such additional relations is equivalent to the existence of \textit{sub-modules}. In turn, this is equivalent to the original module being reducible. This motivates the terminology used above, where reducibility was purely framed from the perspective of solutions to the differential equations.

\paragraph{Reduciblity and maps between modules.}

An interesting observation is that the study of subsystems can be related to the study of maps between different systems. One example of this that comes to mind is from the study of vector spaces. Here, whenever we are provided with a linear map $f$ between two vector spaces, we immediately obtain two different subsystems, the kernel and image of $f$. A similar story holds for general modules over rings, where the existence of a module homomorphism similarly implies the existence of the kernel and image sub-modules as well. 
Relating this observation to the D-modules discussed above, we find that if a D-module is irreducible, any map from this module to another (arbitrary) D-module must be the zero map or an isomorphism. On the contrary, the existence of \textit{any} non-trivial map between two D-modules implies that at least one of the two is reducible.

As we have discussed, we will focus mostly on GKZ systems. In this setting, it turns out that there is a natural map between different (but similar) GKZ systems. This map simply sends a differential operator $P$ modulo the equivalence relations of the first GKZ system, to the operator $P \partial_I$ modulo the equivalence relations of the second GKZ system.
Here, $\partial_I$ is the partial derivative with respect to any of the variables $z_I$ that the solutions of the GKZ system can depend on.
As it turns out, reducibility of a GKZ system can be completely understood in terms of these maps, as follows from the proof in~\cite{schulze_resonance_2012}. 
Furthermore, we find that often the reduction operators themselves also act as maps between subsystems. These maps have very useful properties, allowing us to perform the recursive reduction algorithm of chapter~\ref{ch:reductionalgorithm}.

The beautiful consequence of introducing all of these technical constructions is that finding and constructing various subsystems becomes almost trivial once the correct formalism is laid out. In fact, many of the results on the reduction operators are based on just a single exact sequence. Amazingly, it is then possible to translate these abstract statements into practical applications, resulting in concrete algorithms that can be used when actually solving integrals or their differential equations. 

\subsection{Quantifying complexity.}

So far, we have been using a relatively coarse measure of complexity for functions, namely the number of linearly independent solutions of the differential equations it satisfies. Now, we will proceed with some more precise notions of complexity with origins in logic~\cite{dries_tame_1998}.
In particular, we will work in a framework of o-minimality, which, in recent years, has seen increasing application in physics \cite{grimm_tameness_2022,grimm_taming_2022,douglas_tameness_2022,douglas_tameness_2023,lanza_machine_2024,lanza_neural_2024,grimm_complexity_2024,grimm_complexity_2024a,grimm_structure_2024,grimm_taming_2025,grimm_reduction_2025,grimm_tame_2025}.

The main use of o-minimality is that it allows us to define precise notions of complexity for classes of functions, which imply concrete bounds on various topological and computational properties associated to these functions. In particular, we will focus on the framework of Pfaffian functions, for which the complexity of a function is encoded in the differential equations it satisfies. As the rest of this thesis is devoted to investigating the differential equations satisfied by various functions, this framework is a natural choice for us.

\paragraph{Different representations of a function.}

We will leave a general discussion of Pfaffian functions and complexities to chapter~\ref{ch:complexity}. However, we would like to discuss some interesting conceptual aspects of this framework here already. For example, one aspect of Pfaffian functions, as well as many other frameworks of o-minimality, is that the complexity of a function can depend on how it is represented. 

In order to do that, let us describe schematically how the Pfaffian framework is used. In this setting, one is interested in some function that satisfies a known system of differential equations with some special properties, usually we refer to this system as being a \textit{representation} of the function.
As an example of the required properties, for linear differential equations of the form 
\begin{equation}\label{eq:pfaffianmatrix}
    \partial_i f = A_i f\, ,
\end{equation}
with a matrix $A_i$ for each partial derivative $\partial_i$, then all of the matrices $A_i$ must be upper-triangular. We refer to chapter~\ref{ch:complexity} for more details.

Given this system of differential equations, one can use this system to obtain the topological and computational bounds. However, a function can have many different systems of differential equations it satisfies, generically resulting in different bounds.
In order to avoid this problem, one can try to take the minimal over all representations of a function, which leads to the strictest bounds. 
Unfortunately though, it is usually impossible to consider all representations.\footnote{See for example Maxwell's conjecture regarding the number of equilibrium points of the electrostatic field, for which, even after well over a hundred years, the observed bounds have not been generally proven \cite{maxwell_treatise_1873,morse_critical_1978,gabrielov_mystery_2007}.}
Therefore, the true complexity of a function is often out of reach and one has to be satisfied with bounds on the complexity that arise from accessible representations.

\paragraph{Overestimating complexity and simpler algorithms.}

The above may seem to be problematic, especially if the topological and computational bounds are which what one is interested in. However, a slight change in perspective here can turn the above into an advantage. In particular, let us consider a (Pfaffian) set of differential equations for some functions. For example, certain classes of Feynman diagrams or, as we will study in chapters~\ref{ch:cosmology} and~\ref{ch:reductionalgorithm}, cosmological correlators. Then, as we have seen, this set of differential equations brings with it some bounds related to the functions we are considering.

However, now let us imagine we have solved the differential equations in some specific examples. Then, it is possible to check the bounds explicitly and get an idea of how strict the bounds are. 
Interestingly, one often finds that these bounds largely overestimate their true values. Again, we see an example of unexpected simplifications, but now in a more precise form. 

The overestimation of the true bounds could be due to two reasons, firstly it could be the case that the framework used to obtain the bounds is not sufficient, either because it is too general, because its bounds can be improved within the framework, or some other reason. The second reason is much more interesting though, because this overestimation could also hint towards the existence of another, simpler representation of our original function. Returning to the example of Pfaffian differential equations, there could exist a system of differential equations that our function satisfies which is much simpler than the system we are currently studying. And in turn, this simpler representation should result in stricter bounds.

\paragraph{Origins of simplifications.}

Given the existence of a simpler representation for some function, it is natural to wonder hwo such simplifications can arise. 
One possibility is to try to construct completely new representations of the functions we are interested. Inspired by the hint at the existence of such systems from the knowledge that the current bounds are not optimal, and thus such a representation could exist.

A second approach is somewhat more direct and consists of identifying possible simplifications within an existing representation. This approach turns out to be quite fruitful when applied to the cosmological correlators we consider, as we will see in chapter~\ref{ch:reductionalgorithm}. In particular, we construct a Pfaffian system of differential equations for such correlators.
Note that, in practice, one can think of this as some large matrix set of matrix differential equations of the type found in equation~\ref{eq:pfaffianmatrix}.

For cosmological correlators, the construction of this system is already a non-trivial fact and, in our framework, follows using various reduction operators for a more complicated system of differential equations. Surprisingly, it turns out that there are many more reduction operators that are not used in this construction. Practically, these lead to large number of algebraic relations between basis functions, providing us with another avenue of simplifications.

\paragraph{Limitations of the Pfaffian framework.}

Interestingly, even the above algebraic relations do not make use of all available reduction operators. In particular, there is a large number of relations between different diagrams as well in the form of factorization and permutation identities. Unfortunately though, the simplifications due to these additional relations are not easily visible from the Pfaffian perspective.

For Pfaffian complexity, the only input is the particular system of differential equations. Relations between the different solutions to these differential equations imply a clear redundancy which allows us to directly simplify these systems.
However, relating solutions to this system to solutions of a different system, even a simpler one, are hard to leverage within this framework.

Similarly, one can consider the same function with permuted inputs. From a practical standpoint, this represents a significant simplification as, when the function has been obtained once, all permutations are freely available. From the Pfaffian perspective, however, this does not immediately lead to a reduced complexity, as the permuted function satisfies a completely new and unrelated system of differential equations. Combining the two systems then results in a large increase in complexity
Note that it could be possible that there exists a different system of differential equations that contains both permuted functions in a clever way, decreasing the overall complexity. If such a system exists, than it would be possible to incorporate the symmetries directly from the Pfaffian perspective. However, we were unable to construct such a system and leave such possibilities to future research.

Thus, we have seen simplicity and complexity reductions from two different perspectives. For any system of differential equations one can consider the number of solutions and its reducibility, from which we obtain a variety of tools that can be aided in finding the solutions to this system. However, in order to make this precise we are still aided by the Pfaffian framework. If this framework can be applied the notions of complexity become significantly more fine-grained, leading to explicit bounds in various topological and computational quantities. These can then make precise the notion of a function being ``simpler'' than expected and inspire the search for simpler representations or additional relations, a procedure that we have applied to cosmological correlators in this thesis.

\section{Background in cosmology}\label{sec:cosmology}

One part of this thesis is spend on the formal and theoretical aspects of integrals and differential equations. The other part concerns the physical application of these techniques that we will consider. 
In this section, we will turn our attention to this application: cosmological correlators. 
For these in particular, the general formalism we develop turns out to be quite powerful, as we will see in chapters~\ref{ch:cosmology} and~\ref{ch:reductionalgorithm}.

We will discuss the physical origin of these correlators in this section, which leave physical imprints today on experimental data such as the large scale structure (LSS) and the cosmic microwave background (CMB). Furthermore, we will discuss some of their structure, from which we will find that they can be accurately described using equal time correlators on an almost deSitter background. 
Then, we will focus on some of the more theoretical aspects of cosmological correlators. In particular, we will describe how they can be described equivalently as in-in correlators or as the so-called wave function of the universe.

\subsection{Initial conditions of the universe}

In many areas of physics, some of the most important observational data available comes in the form of correlators, and in this cosmology is no different. Using available data on, for example, the cosmic microwave background \cite{dicke_cosmic_1965,penzias_measurement_1965,komatsu_fiveyear_2009,durrer_cosmic_2015,collaboration_planck_2020} and the large scale structure \cite{skrutskie_two_2006,collaboration_dark_2016,blanton_sloan_2017} of the universe, it is possible to fit our best models of cosmology with astounding accuracy. With these models, it is possible to evolve backwards in time in order to obtain the initial conditions giving rise to our current universe.

It turns out that these conditions, called the \textit{initial conditions of the universe}, exhibit a large amount of structure.
In what follows we will describe some of this structure, and see that it seems to suggest that there was a period of accelerated expansion giving rise to these initial conditions, called inflation \cite{guth_inflationary_1981,linde_new_1982,albrecht_cosmology_1982}.

\begin{figure}
    \centering
    \includegraphics[width = 0.6\linewidth]{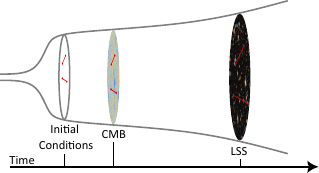}
    \caption{Imprints of the initial conditions of the universe are still visible in the CMB and LSS. Pictures of the CMB and LSS are adapted from \cite{esa_planck_2013} and \cite{nasa_hubble_2014} respectively.}
\end{figure}

\paragraph{Structure in the initial conditions of the universe.}

Now, we will focus on some specific aspects of these correlations, as laid out in~\cite{lee_records_2024}. From these we will conclude that these fluctuations can be accurately described using equal time correlations arising from a scalar quantum field theory on an almost deSitter background. Let us begin by discussing some evidence for the need of inflationary models in general. Famously, part of the original motivations for inflation came from the horizon and flatness problems in big bang cosmology \cite{dodelson_modern_2024}. Another interesting piece of evidence comes from the anisotropies in the CMB. Here, the largest scales exhibit a clear anti-correlation, which is also naturally explained using inflation \cite{spergel_cmb_1997,dodelson_coherent_2003}.

There is a lot of interesting structure in the primordial perturbations. For example, they are adiabatic to an astounding accuracy \cite{akrami_planck_2020}, meaning that they are sourced by only a single scalar degree of freedom. This implies that one only needs to study effective theories of this single degree of freedom \cite{piazza_effective_2013}. Additionally, the primordial perturbations are almost scale invariant \cite{collaboration_planck_2020}. Assuming this invariance emerged from space-time isometries during inflation, this naturally leads one to consider an inflationary theory on deSitter space with weakly broken boost symmetry \cite{piazza_effective_2013,lee_amplitudes_2024}.

In fact, these two facts combined lead us to consider a quantum field theory on (almost) deSitter, consisting of a single scalar field. In the next section, we will describe what particular kinds of correlators we will consider for such a theory.

\subsection{Cosmological Correlators}

Although there are clear indications from experiment on what kind of theories should be considered to describe primordial perturbations, it should be noted that the correlators in this setting are somewhat different from those usually considered in quantum field theories. Therefore, we will now spend some time describing these differences. In the process, we will introduce one of the main object of interests in this thesis: the wave function of the universe. 

\paragraph{The in-in formalism.}

In quantum field theory settings one is often interested in so-called in-out correlators, where an initial state $\ket{\rm in}$ is prepared and the transition amplitude $\braket{\rm out|in \rm }$ to an out state $\ket{\rm out}$ is calculated. In cosmology the situation is somewhat different. Here, we want to consider some initial state of the universe, evolve it to a late time and obtain the expectation value of some operators at this time. Therefore, we must consider an initial state $\ket{\Omega}$, evolve it to some later time $t$, insert the operators of interest and finally evolve backwards. Such correlators are called \textit{in-in correlators} and, in the interaction picture, take the form \cite{weinberg_quantum_2005,baumann_snowmass_2022,lee_amplitudes_2024}
\begin{equation}
    \langle\;  O(t) \;\rangle = \bra{\Omega}\bigg(\overline{T}e^{i\int_{-\infty}^t H_{\rm int}(t')dt'}\bigg)O_I(t)\bigg(Te^{-i\int_{-\infty}^t H_{\rm int}(t')dt'}\bigg)\ket{\Omega}\ ,
\end{equation}
where $H_{\rm int}$ is the interaction Lagrangian and $O_I$ is the operator $O$ in the interaction picture, $T$ is the time-ordering operator and $\overline{T}$ is the reverse-time-ordering operator.

As we will consider inflationary correlators, the time-slice on which we will consider our correlators is somewhat distinguished. In particular, we will consider a Friedmann-Robertson-Walker (FRW) metric in conformal coordinates given by
\begin{equation}
    ds^2 = a(\eta)^2(-d \eta^2 + d\vec{x}^2)\,,
\end{equation}
where $\eta$ is the conformal time and $a$ is the scale factor. In these coordinates, the end of inflation corresponds to $\eta = 0$ and thus we will consider equal time correlators on this particular time-slice.

In the definition of in-in correlators, as well as for cosmological correlators in general, a choice of initial state is necessary.
For our purposes, a natural choice of such an initial choice comes in the form of the \textit{Bunch-Davies vacuum} \cite{bunch_quantum_1978}. This choice of vacuum is uniquely defined by requiring that it satisfies the deSitter isometries and that on very small scales the Green's functions behave of a free field asymptote to those on a Minkowski space-time.
Recall that deSitter appears here due to the near scale-invariance of the primordial perturbations.

\paragraph{The wavefunction of the universe.}

Although the in-in formalism is useful and used throughout the literature, we will take another common point of view. Instead of calculating the correlators directly, we will focus on an intermediate object, the so-called \textit{wavefunction of the universe} \cite{weinberg_quantum_2005,anninos_latetime_2015} and references therein.
This object is defined by considering an eigenstate $\ket{\varphi}$ of the field operator $\hat{\phi}(\Vec{x},\eta)$ satisfying
\begin{equation}
    \hat{\phi}(\Vec{x},0)\ket{\varphi}=\varphi(\Vec{x})\ket{\varphi} \,,
\end{equation}
where $\varphi(\Vec{x})$ is some field configuration depending only on the spatial coordinates. The wavefunction of the universe is then the functional of $\varphi$ defined as $\Psi[\varphi]=\braket{\varphi|\Omega}$ with $\ket{\Omega}$ the Bunch-Davies vacuum.

The use of this wave-form becomes clear when one tries to calculate equal time correlation functions of the fields $\varphi$, since these can be expressed as
\begin{equation}
    \langle \;\prod_i \varphi(\Vec{x}_i)\;\rangle = \frac{\int \D \varphi\;  \vert \Psi[\varphi]\vert^2  \prod_i \varphi(\Vec{x}_i)}{\int \D \varphi\;  \vert \Psi[\varphi]\vert^2} \, .
\end{equation}
Therefore, finding the wavefunction of the universe leads to an alternative way of calculating correlators.

As is often the case, it is useful to perform an expansion of this wavefunction. To be precise, we will expand its logarithm as
\begin{equation}
    \log(\Psi[\varphi])=\sum_{n=0}^\infty \frac{1}{n!}\int \prod_{i=1}^n \left( d^3\Vec{x}_i \;\varphi(\Vec{x}_i)\right)\; \Psi_n(\Vec{x}_1,\cdots,\Vec{x}_n)
\end{equation}
and we are interested in finding the functions $\Psi_n(\Vec{x}_1,\cdots,\Vec{x}_n)$. It turns out to be useful to consider their Fourier transforms $\Psi_n(\Vec{k})$ and factor out the delta function to obtain
\begin{equation}
    \Psi_n(\Vec{k}_1,\cdots,\Vec{k}_n)=\delta^3\left( \sum_{i=1}^n \Vec{k}_i \right) \psi_n(\Vec{k}_1,\cdots,\Vec{k}_n) \, .
\end{equation}
The functions $\psi_n(\Vec{k}_1,\cdots,\Vec{k}_n)$ are called the \textit{wavefunction coefficients}. Note that, from now on, we will refer to wavefunction coefficients as cosmological correlators, since they are directly related \cite{maldacena_nongaussian_2003,goodhew_cosmological_2021,benincasa_amplitudes_2022,stefanyszyn_there_2024}.

These wavefunction coefficients will be the main physical objects that we will apply our results on GKZ systems and reducibility to. In particular, we will study these wavefunction coefficients for a particular toy model, which we will discuss in more detail in chapter~\ref{ch:cosmology}.

\newpage
\fi

\if\PrintChTwo1



\stepcounter{thumbcounter}
\setcounter{colorcounter}{2}
\chapter{General GKZ Systems}
\label{ch:generalGKZ}

In this chapter, we will consider GKZ systems, which will be of importance throughout this thesis. In section~\ref{sec:genGKZ}, we explain for which kind of integrals GKZ systems be obtained, as well as the differential equations that define the system.
Then, we will consider a more mathematical perspective on these systems in section~\ref{sec:Dmodules} by considering GKZ systems as D-modules.
Afterwards, we use this perspective to obtain reduction operators in~\ref{sec:reductionoperators}, based on results obtained in~\cite{grimm_reductions_2025}.
Finally, we will recap the results from section~\ref{sec:reductionoperators} in slightly less technical terms, as well as provide explicit algorithms for their construction in section~\ref{sec:reductionconstruction}.
These reduction operators will be applied extensively throughout this thesis and their construction and properties constitute one of its main results.

\section{Differential equations from integrals}\label{sec:genGKZ}

In this section we will review some of the general properties of GKZ systems. 
These properties are well-known, with many of results already obtained in from the original papers by Gelfand, Kapranov, and Zelevinsky~\cite{gelfand_generalized_1990,gelfand_hypergeometric_1991,gelfand_discriminants_1994} or being from the excellent book~\cite{saito_grobner_2000}.
The presentation and results described here will follow this last reference, as well as some more modern references that focus specifically on Feynman integrals~\cite{nasrollahpoursamami_periods_2016,delacruz_feynman_2019,blumlein_hypergeometric_2021,klausen_hypergeometric_2020}.

\subsection{Constructing the GKZ data}

A GKZ system can be obtained for any integral of the form
\begin{equation}\label{eq:gengkzint}
   I(z;\alpha,\beta)= \int_\Gamma d^n \omega\; \frac{\prod_{i=1}^n \omega_i^{\alpha_i-1}}{\prod_{j=1}^k p_j(z,\omega)^{\beta_j}}\, ,
\end{equation}
where $\Gamma$ is an arbitrary integration cycle, $\alpha_i$ and $\beta_j$ are complex numbers and the $p_j$ are polynomials in the $\omega_i$ with coefficients $z_{j,n}$. To be specific we will take $p_j$ to be written as
\begin{equation}\label{p-form}
    p_j(z,\omega)= \sum_m z_{j,m} \prod_{i=1}^n \omega_i^{(a_{j,m})_i}\, ,
\end{equation}
where the $a_{j,m}$ are vectors describing the powers of $x$ in each term of $p_j$. For example, the polynomial
\begin{equation}\label{eq:gkzexamp}
    p_j(z,\omega)=z_{j,1} \;\omega_2+z_{j,2}\;\omega_1^3 \omega_2^2
\end{equation}
results in the vectors $a_{j,1}=(0,1)^T$ and $a_{j,2}=(3,2)^T$, with $T$ being the transpose. The coefficients $z_{j,m}$ will function as the variables that the integral depends on and the differential equations solved by the integral will be differential equations in these variables. 

\paragraph{The matrix defining a GKZ system.}

The differential equations defining a GKZ system are defined from a certain matrix $\A$. These matrices follow from the polynomials $p_j$ as follows. One begins by defining a set of matrices $\A_j$ for each polynomial $p_j$, simply by interpreting the vectors $a_{j,m}$ as column vectors of these matrix. Continuing with the example of equation~\eqref{eq:gkzexamp}, this matrix takes the form
\begin{equation}
    \A_j=\begin{pmatrix}
    0 & 3\\
    1 & 2
    \end{pmatrix}\, .
\end{equation}
One combines the matrices $\A_j$ in a new matrix $\A$ as 
\begin{equation}\label{eq:Adef}
    \A\coloneqq \begin{pmatrix} \mathbf{1} & \mathbf{0} &\cdots\\
    \mathbf{0} & \mathbf{1} & \cdots \\
    \vdots&\ddots&\ddots\\
    \A_1 & \A_2 &\cdots 
    \end{pmatrix}\ ,
\end{equation}
where the $\mathbf{1}$ are row vectors with $1$ at every entry and the $\mathbf{0}$ are row vectors of zeroes to fill in the gaps. Adding these zeroes and ones is called homogenization and the full matrix $\A$ is called the homogenized matrix. Combined with the parameters $\alpha_i$ and $\beta_j$, this determines the GKZ system completely. Again, it is much easier to see this by example where combining the two matrices
\begin{equation}
    \begin{array}{lr}
        \A_1=\begin{pmatrix}
    0 & 3\\
    1 & 2
    \end{pmatrix}\, ,    &     \A_2=\begin{pmatrix}
    2 & 1\\
    0 & 0
    \end{pmatrix}\, ,
    \end{array}
\end{equation}
results in the homogenized matrix
\begin{equation}\label{eq:Aexample}
    \A= \begin{pmatrix}
    1 & 1 & 0 & 0\\
    0 & 0 & 1 & 1\\
    0 & 3 & 2 & 1\\
    1 & 2 & 0 & 0
    \end{pmatrix}\, ,
\end{equation}
and we can identify $\A_1$ and $\A_2$ as being contained in the bottom two rows of $\A$.

\subsection{Differential equations from the GKZ data}

The differential equations which the integral~\eqref{eq:gengkzint} is a solution to can be obtained from the matrix $\A$ and the parameter $\nu$. 
But, before we explain this, it is useful to introduce some simplifying notation. 
First of all, we will write $N$ for the number of columns of $\A$, and $M$ for the number of rows. 
Secondly, we will write $a_I$ for the $I$-th column vector of $\A$. Since we can associate a variable $z_{j,m}$ with each column of $\A$, we will rename these to $z_I$ with $I$ being the corresponding column. 
We will also use the shorthand
\begin{equation}
    \partial_I\coloneqq \frac{\partial}{\partial z_I}
\end{equation}
for derivatives with respect to $z_I$ and
\begin{equation}\label{eq:thetadef}
    \theta_I\coloneqq z_I \partial_I
\end{equation}
for the associated homogeneous derivative. Finally, for each $l$-dimensional vector and list of $l$ objects $c_i$, we will write
\begin{equation}
    c^v \coloneqq \prod_{i=1}^l c_i^{v_i}\, .
\end{equation}
For example, the integral~\eqref{eq:gengkzint} can be written as
\begin{equation}
     f_\Gamma(z;\nu)= \int_\Gamma d^nx\; \frac{x^{\alpha-1}}{p(z,x)^\beta} \, .
\end{equation}
using this last notation. Note that here, we have also combined the complex exponents $\alpha$ and $\beta$ into the single vector $\nu$.

Furthermore, recall that the matrix $\A$ was constructed from a set of partial matrices $\A_j$ for each polynomial $p_j$, with a homogenization step that fills the top $k$ rows with ones and zeroes. These components of 
$\A$ will exhibit different behaviors when considering the differential equations of this GKZ system. Therefore, we will refer to the top $k$ rows as the homogeneous part and the bottom $n$ rows as the polynomial part. We will denote these by $\A_\beta$ and $\A_\alpha$, respectively. Schematically, this split takes the form
\begin{equation}\label{eq:Aalphabetadef}
    \A\coloneqq \begin{pmatrix}
    \mathbf{1} & \mathbf{0} &\cdots\\
    \mathbf{0} & \mathbf{1} & \cdots \\
    \vdots&\ddots&\ddots\\
    \A_1 & \A_2 &\cdots 
    \end{pmatrix}
    \begin{array}{l}
         \left\}\begin{array}{l}
              \, \\
            \A_\beta    \\
              \,
         \end{array} \right.\\
          \left\}\begin{array}{l}
           \; \A_\alpha
         \end{array} \right.
    \end{array}
\end{equation}
where $\mathbf{1}$ and $\mathbf{0}$ denote row vectors consisting of either ones or zeroes. 

Similarly, we can separate each vector $a_I$ into its first $k$ components $a_{I,\alpha}$ and its last $n$ components $a_{I,\beta}$, such that $a_I=(a_{I,\beta},a_{I,\alpha})$, similar to how we split $\nu=(\beta,\alpha)$. Note that $a_{I,\alpha}$ is the exponent of the term
\begin{equation}\label{eq:aialpha}
  p_j(z,x)=  \cdots+z_I x^{a_{I,\alpha}}+\cdots\ 
\end{equation}
of the polynomial $p_j$, while $a_{I,\beta}$ is the unit vector in the $j$-th direction. Therefore, one can interpret $a_{I,\beta}$ as indicating the polynomial in which $z_I$ appears, while $a_{I,\alpha}$ specifies the exponent of $x$ with which it is associated.

We are now ready to obtain the differential equations from the GKZ data. 
We will find that these come in two parts, which we will treat separately. There are the toric equations, depending only on the matrix $\A$, as well as the Euler equations, depending on both $\A$ and $\nu$. We will begin our treatment with the toric equation.

\paragraph{Toric equations.}

To obtain the toric operators, let us begin by observing that, regardless of the integration cycle, acting with a partial derivative $\partial_I$ on $p^{-\beta}$ results in
\begin{equation}\label{eq:dzjm}
    \partial_I p(z;x)^{-\beta} = -(\beta\cdot a_{I,\beta})\, x^{a_{I,\alpha}}\, p(z,x)^{-\beta-a_{I,\beta}}\, ,
\end{equation}
where we recall that $a_I=(a_{I,\beta},a_{I,\alpha})$ and $\cdot$ denotes the vector dot product.
Similarly, for an $N$-dimensional vector $u$ with positive integer entries, one can act with $\partial^u$ on $p^{-\beta}$ yielding
\begin{equation}\label{eq:gengkzpdu}
    \partial^u p(z,x)^{-\beta}=c_u(\beta)\,x^{\A_\alpha u} \,p(z,x)^{-\beta-\A_\beta u}\, ,
\end{equation}
where $c_u(\beta)$ arises from an iterative application of equation~\eqref{eq:dzjm}.
Furthermore, careful evaluating of the pre-factor $c_u(\beta)$ reveals that it depends only on $\A_\beta u$.

Now, let us consider two vectors $u$ and $v$, both in $\N^N$, such that
\begin{equation}
    \A u= \A v\,.
\end{equation}
Then, clearly this implies that both $\A_\alpha u_\alpha=\A_\alpha v_\alpha$ and $\A_\beta u_\beta = \A_\beta v_\beta$, where we have used the decompositions from~\eqref{eq:Aalphabetadef} and the corresponding decomposition for the vectors $u$ and $v$.
Comparing to equation~\eqref{eq:gengkzpdu}, 
 if $\A u=\A v$ for two vectors with positive integer entries, it follows that
\begin{equation}
    (\partial^u-\partial^v) p(z,x)^{-\beta} =0\, .
\end{equation}
However, the partial derivatives commute with the integration of the GKZ integral~\eqref{eq:gengkzint}, thus we find that it satisfies
\begin{equation}
    (\partial^u-\partial^v)f_\Gamma(z;\nu)=0\,.
\end{equation}
These operators of the form $\partial^u-\partial^u$ for $\A u= \A v$ are the toric operators of the GKZ system. From now on, we will denote these as
\begin{equation}\label{eq:toricopdef}
   \L_{u,v}=\partial^u -\partial^v\, , 
\end{equation}
where it is implicit that we must impose $\A u=\A v$. For each such operator, we obtain a toric equation of the form
\begin{equation}\label{eq:gentoriceq}
   \boxed{\rule[-.1cm]{0cm}{.55cm}\quad \L_{u,v}f_\Gamma(z;\nu)=0\ .\quad }
\end{equation}
satisfied by all solutions $f_\Gamma(z;\nu)$ of the GKZ system.

Note that for any positive integer vectors $u_1$, $u_2$, $v_1$ and $v_2$ satisfying $\A u_1 = \A v_1$ and $\A u_2 =\A v_2$. Clearly, the sum $u=u_1+u_2$ and $v=v_1+v_2$ also satisfies $\A u= \A v$. However, this does not provide us with an independent toric operator as, for any function $f(z;\nu)$ satisfying 
\begin{equation}
    \L_{u_1,v_1}f(z;\nu)=\L_{u_2,v_2}f(z;\nu)=0 \, .
\end{equation}
The function $f$ will automatically satisfy
\begin{equation}
    \prod_{I=1}^N\partial_I^{(v_1)_I+(v_2)_I}f(z;\nu)=\prod_{I=1}^N \partial_I^{(u_1)_I+(u_2)_I}f(z;\nu)\, .
\end{equation}
Thus, $\L_{u,v}f=0$ is automatically satisfied. This means that one can find a basis for the integer kernel $\ker_\Z(\A)$ and use this to obtain a generating set for the possible combinations of $u$ and $v$.

\paragraph{Euler operators.}

The second set of differential equations satisfied by our integral are called the Euler equations. These are defined by taking the homogenous differentials $\theta_I$ from equation~\eqref{eq:thetadef} and combining them in a vector
\begin{equation}\label{eq:bigThetadef}
    \Theta\coloneqq (\theta_I)_{1\leq I \leq N}\, .
\end{equation}
One can apply the matrix $\A$ to this vector to obtain the vector
\begin{equation}\label{eq:Edef}
    \E \coloneqq \A \Theta\, .
\end{equation}
The components $\E_J$ of this vector are called Euler operators, where $1\leq J \leq M$ and $M$ is the number of rows of $\A$. Combined with the parameter vector $\nu$ these give the Euler equations, written as
\begin{equation}\label{eq:eulergen}
    \boxed{\rule[-.1cm]{0cm}{.55cm} \quad(\E_J+\nu_J)f(z;\nu)=0\ , \quad}
\end{equation}
where $1\leq J\leq M$. Note that these are the only equations where $\nu$ plays a role and when considering a single GKZ system one usually considers $\nu$ fixed. Although there are ways of relating solutions at different $\nu$, as we will discuss in sections~\ref{sec:reductionoperators} and~\ref{sec:reductionconstruction}.

To see that the integral~\eqref{eq:gengkzint} actually satisfies the Euler equations we will separate them into two parts, in accordance with the split of $\A$ into $\A_\alpha$ and $\A_\beta$ given in~\eqref{eq:Aalphabetadef}. We will begin by considering Euler equations arising from the homogenization of $\A$, which can be written as
\begin{equation}
    (\A_\beta\Theta +\beta)f(z;\nu) =0\,.
\end{equation}
Note that these equations correspond to the equations~\eqref{eq:eulergen} with $J\leq k$, where $k$ is the number of polynomials $p_j$. Now, we observe that the relevant Euler operators will only involve derivatives with respect to the variables contained in a single polynomial. Let us fix a polynomial $p_j$ and consider the action of the Euler operator $\E_j$ with the same index. Then, we find that
\begin{equation}\label{eq:eulerderivhom}
    \E_j\, p_j^{-\beta_j}=\left(\sum_{I\in p_j} z_I \partial_I \right)\left(\sum_{I\in p_j} z_I x^{a_I} \right)^{-\beta}=-\beta_j \,p_j^{-\beta_j}\, ,
\end{equation}
where the sum is only over the terms contained in $p_j$. Since this Euler operator acts trivially on all other $p_l$ for $l\neq j$, we obtain
\begin{equation}
    (\E_j+\nu_j)I(z;\alpha,\beta)=0
\end{equation}
for $1\leq j \leq k$.

The other Euler operators will be indexed by an index $\E_{k+i}$, with $1\leq i \leq n$ and $n$ the number of integration variables. 
To show that the Euler equations~\eqref{eq:eulergen} hold for these indices, let us perform a coordinate transformation $x_i\rightarrow s\, x_i$, with $s\in \C$. 
We assume that, for $s$ sufficiently close to $1$, the contour $\Gamma$ is invariant under this transformation. 
Considering the image of $p(z,x)$ under this transformation, one finds
\begin{equation}
    p(z,x)\rightarrow p\left(z_1 s^{(a_{\alpha,1})_i},z_2  s^{(a_{\alpha,2})_i},\cdots,z_N s^{(a_{\alpha,N})_i},x \right)\, .
\end{equation}
In other words, the transformation can be negated by a suitable inverse transformation of the $z_I$. It follows that, in total, applying this coordinate transformation to the integral~\eqref{eq:gengkzint} has the effect
\begin{equation}
     I(z;\alpha,\beta)= \int_\Gamma d^nx\; \frac{x^{\alpha-1}}{p(z,x)^\beta}=s^{\alpha_i} I\left(s^{(a_{\alpha,I})_i}z_I;\alpha,\beta\right)\, ,
\end{equation}
where the pre-factor is due to the transformation of $d^nx\, x^{\alpha-1}$. Differentiating both sides with respect to $s$ and taking the limit $s\rightarrow 1$ results in
\begin{equation}
   0= \alpha_i I(z;\alpha,\beta)+\left(\sum_{I=1}^N (a_{\alpha,I})_i z_I \partial_I\right)I(z;\alpha,\beta)\, .
\end{equation}
Recognizing that 
\begin{equation}
    \sum_{I=1}^N (a_{\alpha,I})_i\, z_I \partial_I=\E_{i+k}\, ,
\end{equation}
with $k$ the number of polynomials $p_j$, we recover the Euler equations
\begin{equation}\label{eq:eulerderiv1}
    (\E_{i+k}+\nu_{i+k})I(z;\alpha,\beta)=0
\end{equation}
for $1\leq i\leq n$. Combined with equation~\eqref{eq:eulerderivhom} this shows that the GKZ integral satisfies the Euler equations for suitable $\Gamma$. 

\subsection{General solutions}
\label{ap:series}
The original integral~\eqref{eq:gengkzint} will be a particular solution to the above differential equations but usually not the only one. To obtain the integral of interest one first determines a complete basis of solutions $f_{d}(z;\nu)$ to the  differential equations  and afterwards fix the particular coefficients either numerically or by evaluating the integral in specific limits for the $z_I$. Afterwards the integral can then be written as
\begin{equation}
    I(z;\alpha,\beta)=\sum_{d=1}^D c_d(\Gamma;\nu)f_{d}(z;\nu)\, ,
\end{equation}
where $D$ is the dimension of the solution space associated to the GKZ system. Note that $D$ is often also called the rank of the GKZ system.

To find the solutions $f_d$ one can take a few different approaches, of which we want to discuss two in more detail.
The first approach is due to the remarkable aspect of GKZ systems that, provided certain technical conditions are met, their solutions can be obtained in an entirely geometrical manner as explicit series expansions, see \cite{saito_grobner_2000} and references therein.  
An alternative approach is more common in the physics literature, and consists of making a convenient ansatz that automatically solves the Euler equations \cite{hosono_mirror_1995,hosono_gkzCY_1996,hosono_gkzapp_1996}.
This then results in a smaller set of differential equations that can be solved using a variety of methods. In this chapter we will mostly use the second method as this requires the least amount of additional theory. 

\paragraph{Convex polytopes and number of solutions.}

The main geometrical structure underlying GKZ systems is a convex polytope that we associate to the matrix $\A$. To construct it, recall that we labeled the columns of $\A$ as $a_I$. Since this is a collection of vectors, one can take their convex hull and denote it as
\begin{equation}
    \mathrm{Conv}(\A)\coloneqq \mathrm{Conv}(a_1,a_2,\cdots,a_N)\, .
\end{equation}
Note that the resulting polytope can also be obtained from the Newton polytopes of the polynomials $p_j$.

Interestingly, important information about the number of solutions to the GKZ system can be derived from the volume of this polytope. In particular, only assuming that $\A$ is homogenized, one finds that \cite[Thm 3.5.1]{saito_grobner_2000}
\begin{equation}\label{eq:solsandvol}
     \text{number of solutions }\geq \text{Vol}(\A)\, ,
\end{equation}
where $\text{Vol}(\A)$ is volume of the polytope $\mathrm{Conv}(\A)$, normalized such that the standard simplex has volume one. If we assume that $\A$ is normal, an assumption satisfied by most Feynman integrals \cite{tellander_cohenmacaulay_2023} as well as cosmological correlators, this inequality turns into an equality. Similarly, if the parameter $\nu$ is generic, meaning it is outside of the so-called exceptional hyperplane arrangement,\footnote{See \cite[Sec 4.5]{saito_grobner_2000} for precise definitions.} equation~\eqref{eq:solsandvol} also becomes an equality.

\paragraph{Series expressions from triangulations.}

Besides the number of solutions, the polytope $\mathrm{Conv}(\A)$ also gives rise to a way of obtaining these solutions, again assuming that $\nu$ is generic. These solutions are associated to a triangulation of this polytope, and specifically the simplices of this triangulation.

If we let $\mathcal{T}$ to be a triangulation of $\mathrm{Conv}(\A)$, and assume that it is both unimodular, in the sense that each simplex has a normalized volume of one, and regular, in the sense of \cite[Ch 8]{sturmfels_grobner_1997} each simplex will correspond to a single solution. 

The explicit series expansion of these solutions are known as the canonical series, or the $\Gamma$-series of the GKZ system. To describe these functions, it is first necessary to introduce some notation. Recall that for any subset $F\subset \{1,\cdots,N\}$, we have defined the matrix $\A_F$ as the matrix with column vectors $a_I$ for $I\in F$. Furthermore, we have denoted $\bar{F}$ as the complement of $F$ in $\{1,\cdots, N\}$. We will also denote the set of coordinates indexed by $I\in F$ as $z_F$. Finally, we will need the multivariate Pochhammer symbol
\begin{equation}
(a)_n = \prod_k \frac{\Gamma(a_k+n_k)}{\Gamma(a_k)}
\end{equation}
with $\Gamma$ the $\Gamma$-function
\footnote{Not to be confused with the $\Gamma$-\textit{series} we are defining here. This $\Gamma(z)$ is the familiar generalization of the factorial.} 
as well as the multivariate factorial $n!\coloneqq \prod_k n_k!$.

With these definitions at hand, the canonical series solution of a simplex $\sigma$ of the unimodular regular triangulation $\mathcal{T}$ is defined as \cite{klausen_hypergeometric_2020}
\begin{equation}\label{eq:canonicalseries}
    \phi_\sigma(z;\nu) \coloneqq z_\sigma^{-\A_\sigma^{-1}\nu} \sum_{n\in \N^r} \frac{(\A_\sigma^{-1} \nu)_{\A^{-1}_\sigma \A_{\bar{\sigma}}n}}{n!} \frac{z_{\bar{\sigma}}^n}{(-z_\sigma)^{\A_\sigma^{-1} \A_{\bar{\sigma}}n}}\, ,
\end{equation}
with $r=\mathrm{dim}(\ker(\A))$. Note that the simplices will correspond to square invertible matrices $\A_\sigma$, making their inverses well-defined. Furthermore, the number of summation parameters is exactly the number of independent parameter $s_i$.

We want to emphasize that, even if these series expressions can be obtained and result in a basis of solutions to the GKZ system, complications can arise as the number of summation parameters grows. Since then these types of series become increasingly difficult to evaluate. Furthermore, analytical continuations of such series expansions can also pose problems in such situations.

\paragraph{Solving the GKZ system using an ansatz.}

Aside from the more geometric approach above, one can also solve the system directly.
Starting with a useful ansatz that automatically solves the Euler equations and then inserting this into the toric equations. This idea, first introduced in \cite{hosono_mirror_1995,hosono_gkzCY_1996,hosono_gkzapp_1996} stems from the fact that 
Euler equations mostly determine the scaling of the solutions, while the toric equations determine their actual form. To see this, we let $u_i$, $v_i$ be $\mathrm{dim}(\ker(\A))$ independent vectors satisfying
\begin{equation}
    \A u_i=\A v_i
\end{equation}
and define associated variables
\begin{equation}\label{eq:sidef}
    s_i\coloneqq \frac{\prod_{I=1}^N z_I^{(u_i)_I}}{\prod_{I=1}^N z_I^{(v_i)_I}}\, ,
\end{equation}
which are known as the \textit{homogeneous variables}. Then, any function of the form
\begin{equation}\label{eq:fpgansatz}
    f(z;\nu)=P(z;\nu)g(s;\nu)
\end{equation}
will have
\begin{equation}
    (\E_J+\nu_J)f(z;\nu)=g(s;\nu)(\E_J+\nu_J)P(z;\nu)\, ,
\end{equation}
where $g(s;\nu)$ denotes that $g(s;\nu)$ can only depend on the variables $s_i$, and not other combinations of the $z_I$.\footnote{Note that $g(s;\nu)$ is also dependent on $\nu$, since the toric equations will mix $P$ and $g$.} Therefore, given any particular solution $P$ to 
\begin{equation}
    (\E_J+\nu_J)P(z;\nu)=0\, ,
\end{equation}
it is possible to make an ansatz of the form~\eqref{eq:fpgansatz} which is guaranteed to  satisfy the Euler equations. This is especially useful since the Euler equations can be easily solved by considering their scaling properties. Inserting the ansatz into the toric equations results in a system of partial differential equations for $g(s;\nu)$. Often this system is too difficult to solve directly. However, in this paper we show that if the underlying GKZ is reducible it is possible to find other differential operators that annihilate particular solutions. In these cases, if a convenient ansatz for $P(z;\nu)$ is chosen it is actually possible to solve the resulting differential equations for $g(s;\nu)$.

\paragraph{Relating solutions at different parameters.} 
One further useful property of GKZ systems is that it, given a GKZ system defined by a matrix $\A$ and a parameter $\nu$, it is possible to obtain solutions at different parameters $\tilde{\nu}$ by applying suitable differential operators to the solutions.
 Namely, consider a differential operator $\mathcal{O}$ that simultaneously satisfies
\begin{equation}\label{eq:comlo=0}
    [\L_{u,v},\mathcal{O}]=0\ , \qquad [\E_J,\mathcal{O}]=c_J\; \mathcal{O}\ , 
\end{equation}
for all toric operators $\L_{u,v}$, all Euler operators $\E_J$ and some  complex numbers $c_J$. For any such differential operator, one can apply it to a solution $f(z;\nu)$ of the GKZ system at $\nu$, in order to obtain a new solution at parameter $\nu-c$,
\begin{equation}
    f(z;\nu-c)\coloneqq \mathcal{O} f(z;\nu)\, ,
\end{equation}
where $c$ is the vector with components $c_J$. Of course, if $\mathcal{O}$ annihilates all solutions to the GKZ system at $\nu$, this will lead only to the trivial solutions $f(z;\nu-c)=0$.\footnote{Note that $f=0$ is always a solution to a GKZ system, since all the differential equations are homogenous.}

One example of this arises by considering a partial derivatives $\partial_I$. This clearly commutes with the toric operators while it also satisfies
\begin{equation}
    [\E_J,\partial_I]=-a_{J,I}\partial_I\ , 
\end{equation}
with $a_{J,I}$ being the element of $\A$ at the $J$-th row and the $I$-th column. This implies that it maps a solution $f$ to a solution
\begin{equation}
    f (z;\nu+a_I)\coloneqq \partial_I f(z;\nu)
\end{equation}
at $\nu+a_I$.  In fact, unless there are solutions at $\nu$ with $\partial_I f(z;\nu) =0$, all of the solutions at $\nu+a_I$ can be obtained in this way, which matches our discussion around~\eqref{eq:relating_sol1}. Crucially, it can happen that in order to reduce a GKZ system one first has to apply integer parameter shifts. The above discussion implies that these can be realized by applying partial derivatives. Or, for inverse shifts, by applying a suitable inverse operator, as described in \cite{beukers_irreducibility_2011,bitoun_feynman_2018,caloro_ahypergeometric_2023}.

\section{D-modules and reducibility}\label{sec:Dmodules}

Having discussed GKZ systems as systems of differential equations, we will now consider a more formal framework of studying these systems. In particular we will use the language of D-modules and construct a so-called Euler-Koszul complex. The original differential equations above will then correspond to the zeroth homology of this complex. 

In order to keep our exposition concise, we will not present 
all details about these mathematical concepts and instead refer to interested reader to the existing literature on these subjects. In particular, references \cite{andres_constructive_2010,brodmann_notes_2018,sattelberger_dmodules_2019} give a general discussion of $\D$-modules, while the works  \cite{saito_grobner_2000,cattani_three_2006,stienstra_gkz_2005,klausen_hypergeometric_2020,feng_gkzhypergeometric_2020,henn_dmodule_2024} provide an overview of how GKZ systems are related with $\D$-modules. Introductions to Euler-Koszul homologies can be found in  \cite{matusevich_homological_2004,reichelt_algebraic_2021}.

This section is structured as follows. We start in section~\ref{ssec:gkzdmods} with a brief introduction to $\D$-modules and describe the $\D$-module associated to a GKZ system. Afterwards we will introduce the framework of Euler-Koszul homologies in section~\ref{ssec:eulerkoszul} and discuss their relation to GKZ systems. In section~\ref{ssec:reducibility} we discuss how reducibility of a $\D$-module gives rise to `simpler' solutions associated to a submodule and relate this observation to known results about the reducibility of GKZ systems. In  section~\ref{ssec:reductionoperators} we build upon these results to obtain additional submodules of GKZ system, where the associated solutions are annihilated by special operators -- the reduction operators.

\subsection{GKZ systems as D-modules}\label{ssec:gkzdmods}

We start by defining the Weyl algebra $\D_N$ in $N$ coordinates $z_I$. This is obtained by considering the free algebra $A_{2N}$ in the $2N$ variables $z_I$ and $\partial_I$, modulo the commutation relations
\begin{equation}
    [z_I,z_J]=0,\quad [\partial_I,\partial_J]=0,\quad [\partial_I,z_J]=\delta_{I,J}\ ,
\end{equation}
where $\delta_{I,J}$ is the Kronecker delta. In other words, the Weyl algebra in $N$ variables is defined as the quotient
\begin{equation}
  \D_N \coloneqq A_{2N}/\big( [z_I,z_J]\sim 0,[\partial_I,\partial_J]\sim 0, [\partial_I,z_J]-\delta_{I,J}\sim 0 \big)\, .
\end{equation}
Note that if it is clear from the context we will drop the subscript $N$ and write $\D=\D_N$.

We can use the Weyl algebra to study differential equations in the following way. Consider a set of differential operators $P_i$, it is possible to define the left $\D$-ideal $\I$ generated by these operators as
\begin{equation}
    \I=\langle P_i\rangle_\D \coloneqq \sum_i \D \cdot P_i\ ,
\end{equation}
which allows us to define the $\D$-module
\begin{equation}
    \M=\D/\I\, .
\end{equation}
The $\D$-module homomorphisms of this module are then related to the solutions of
\begin{equation}\label{eq:pif=0}
    P_i f=0
\end{equation}
as follows. Consider the space of holomorphic functions $\O$ in $N$ variables on an open subset of $\C^N$, where this open subset lies outside of the singular locus of the differential equations. We then consider the $\D$-homomorphisms $\Hom_\D(\M,\O)$ and claim that these are in one to one correspondence with solutions of equation~\eqref{eq:pif=0}. This correspondence is as follows, for any such homomorphism $\phi \in \Hom_\D(\M,\O)$ we have that
\begin{equation}
    0=\phi(0)=\phi(P_i)=P_i\phi(1)\, .
\end{equation}
since $P_i=0$ as an element of $\M$. Therefore, $\phi(1)$ is an element of $\O$ that satisfies the equations~\eqref{eq:pif=0}. Conversely, for any $f$ satisfying equation~\eqref{eq:pif=0} we obtain a homomorphism by defining $\phi(1)=f$. Since such a homomorphism is completely determined by its action on the identity we obtain the required result. Because of this relation, we will move between the different perspectives if one is more useful than the other. We will call such a $\phi$ or such an $f$ a \textit{solution of $\M$}, and write the vector space of such solutions as $\mathrm{Sol}(\M)$ where we have suppressed the dependence on $\O$.

In general, one often studies the sheafified versions of $\O$ and $\D$ over some algebraic variety. One then considers the solution complex $R \mathrm{Hom}_{\D_X}(\M,\O_X)$ in the derived category \cite{hotta_dmodules_2008}. Since we are only interested in obtaining the solutions around some point, we are free to take $X$ such that $\M$ is non-singular. In this case the homology of $R \mathrm{Hom}_{\D_X}(\M,\O_X)$ becomes concentrated in the zero-th degree and we obtain the solutions as described above. In particular, this implies that, when acting on such non-singular $\M$, the solution functor is exact.\footnote{We are grateful to Andreas Hohl and Anna-Laura Sattelberger for insightful discussions regarding this matter.}

It is natural to ask how many independent solutions there are for a given $\D$-module $\M$, or equivalently, what the dimension is of $\mathrm{Sol}(\M)$. This is known as the \textit{rank} of $\M$. For us, we will consider only so-called holonomic $\D$-modules, which implies that the rank is finite.

\paragraph{The GKZ module.}

We are now ready to rephrase the set of differential equations considered in section~\ref{sec:genGKZ} in terms of $\D$-modules. We consider an $M$ by $N$ matrix $\A$ and an $M$-dimensional complex vector $\nu$. The first step is to use the differential equations to define an ideal. Recall that these come in two parts, we first consider the toric equations from equation~\eqref{eq:toricopdef}, leading to the ideal
\begin{equation}
    \I_\A \coloneqq \langle \L_{u,v} \; \vert \; \A u=\A v, \rangle_\D
\end{equation}
where $u,v$ are elements of $\N^M$. Similarly we can define an ideal by considering the Euler operators
\begin{equation}
    \langle \E_J+\nu_J\; \vert \;1\leq J \leq M\rangle_\D
\end{equation}
defined in equation~\eqref{eq:eulergen}. Combining these two ideals we obtain the GKZ ideal
\begin{equation}
    \H_\A(\nu)\coloneqq \I_\A+\langle \E_J+\nu_J\rangle_\D\, ,
\end{equation}
and with it, the GKZ module
\begin{equation}\label{eq:gkzmod}
\M_\A(\nu)\coloneq \D/\H_\A(\nu)\, ,
\end{equation}
which will be the $\D$-module whose solutions we want to obtain.

Similarly, for any subset $F\subset A\coloneqq \{1,\cdots,N\}$, we will write $z_F$, $\partial_F$ for the coordinates indexed by $F$. With this it is then possible to proceed along the same lines as before and define the Weyl algebra $\D_F$ for these coordinates. If we then define the matrix $\A_F$ by combining the column vectors of $\A$ for $i\in F$, we can consider the GKZ system this matrix defines. This GKZ system will then be a $\D_F$-module and we will write it as $\M_{\A_F}(\nu)$ for a parameter $\nu$. Note that by \cite[lemma 4.9]{matusevich_homological_2004}, $\cM_F(\nu)$ will have rank zero if $\nu$ is not in the $\C$-span of $F$

\subsection{Euler-Koszul homologies}\label{ssec:eulerkoszul}

While the $\D$-module $\M_\A(\nu)$ describes the GKZ system, it will be useful for us to consider a slightly different perspective as well. Instead of defining the GKZ system as above, we will consider it as the zero-th homology of a so-called Euler-Koszul complex \cite{matusevich_homological_2004,walther_duality_2005,matusevich_combinatorics_2004,berkesch_agraded_2009,schulze_hypergeometric_2009,berkesch_rank_2011,schulze_resonance_2012,reichelt_bfunctions_2018,steiner_ahypergeometric_2019,steiner_dualizing_2019,reichelt_algebraic_2021}. In this section we will briefly review its construction, referring to \cite{matusevich_homological_2004} for more details and proofs.

In order to define the Euler-Koszul complex, we first consider the commutative subring $R$ of $\D$, consisting only of the partial differentials:
\begin{equation}\label{eq:partialdifring}
   R\coloneqq \C[\partial_I]\simeq\D/\langle z_I\rangle_\D \ .
\end{equation}
 Since the toric operators involve only these partial differentials, it is also possible to consider the $R$-ideal generated by the toric operator $I_\A$ and use it to define the ring
\begin{equation}\label{eq:SAdef}
    S_\A\coloneqq R/I_\A\, .
\end{equation}
We will often switch between considering this quotient as a $\D$-module, as an $R$-module or as a ring itself, depending on the application. Furthermore, ideals of this ring will be written as $\langle \cdots \rangle$ without a subscript.

Note that $S_\A$ is naturally $\Z^N$ graded by defining the degree operator $\mathrm{deg}$ as
\begin{equation}
    \mathrm{deg}(\partial_I)=-a_I,\quad \mathrm{deg}(P_1 P_2)=\mathrm{deg}(P_1)+\mathrm{deg}(P_2)\, ,
\end{equation}
where $a_I$ is the $I$-th column of $\A$.\footnote{Note that the degree of an operator is a vector since the $a_I$ are vectors.} It is also possible to extend this grading to the full Weyl algebra $\cD$ by defining $\mathrm{deg}(z_I)=a_I$. We will call any module compatible with this grading to be \textit{$\A$ graded}. 

\paragraph{The Euler-Koszul complex.}

For any homogeneous operator $P_\alpha$ of degree $\alpha$, we can define the maps
\begin{equation}\label{eq:sjdef}
  s_J: P_\alpha \mapsto\ (\E_J+\nu_J-\alpha_J)P_\alpha\, ,
\end{equation}
where $\E_J$ is the $J$-th Euler operator defined in equation~\eqref{eq:eulergen} and $1\leq J\leq M$. These maps can then be linearly extended to non-homogeneous operators.  Furthermore, for a vector of operators $P \in (S_\A)^M$, we define
\begin{equation}
    s(P)= \sum_{J=1}^M s_J(P_J)
\end{equation}
where $P_J$ is the $J$-th element of $P$. With the map $s$, it is possible to define the Koszul Complex
\begin{equation}
   K_\bullet(\E+\nu):\quad 0\rightarrow \bigwedge^M \D^M\xrightarrow{d_M} \cdots\xrightarrow{d_1} \bigwedge^1 \D^M \xrightarrow{d_0} \D\rightarrow 0\, ,
\end{equation}
where the differentials are given by
\begin{equation}\label{eq:differentialdef}
    d_k(P_1\wedge\cdots\wedge P_k)=\sum_{i=1}^k (-1)^{i+1}s(P_i) \;P_1\wedge \cdots \wedge \widehat{P}_i\wedge \cdots \wedge P_k
\end{equation}
and $\widehat{P}_i$ means that this term is omitted. For any $\A$ graded $R$-module $S$ we consider it as a $\D$-module by taking $\mathcal{S}\coloneqq\D \otimes_{R} S$ and define the \textit{Euler-Koszul complex}
\begin{equation}
    K_\bullet(\E+\nu,S)\coloneqq K_\bullet(\E+\nu) \otimes_{\D} \mathcal{S}\, ,
\end{equation}
where the differentials are induced by the differential~\eqref{eq:differentialdef}. The \textit{Euler-Koszul homology} $H_i(\E+\nu,S)$ is then the $i$-th homology of this complex. Note that the zero-th homology $H_0(\E+\nu,S_\A)$ recovers the GKZ system from equation~\eqref{eq:gkzmod}.

\paragraph{Induced exact sequences.}

One useful property of the Euler-Koszul homologies is that a short exact sequence of $\A$ graded $R$-modules
\begin{equation}
    0\rightarrow S_1\rightarrow S_2\rightarrow S_3 \rightarrow 0
\end{equation}
with homogeneous maps will induce a long exact sequence of the form
\begin{equation}
\begin{array}{rll}
    \cdots &\rightarrow H_{i+1}(\E+\nu,S_3)&\rightarrow H_i(\E+\nu,S_1)\rightarrow H_i(\E+\nu,S_2)\\
    &\rightarrow H_i(\E+\nu,S_3) &\rightarrow\quad\cdots
\end{array}
\end{equation}
on the Euler-Koszul homologies. One particularly useful example of this is the exact sequence
\begin{equation}\label{eq:dIseq}
    0\rightarrow S_\A(a_I)\xrightarrow{\cdot \partial_I} S_\A \rightarrow S_\A/\langle \partial_I\rangle \rightarrow 0\ ,
\end{equation}
where $S_\A(a_I)$ is the module $S_\A$ with the degrees shifted by $a_I$ and $\cdot \partial_I$ denotes right multiplication with $\partial _I$. This particular sequence was used extensively in \cite{walther_duality_2005} and will play a major role in the results obtained in this paper. 

The exact sequence~\eqref{eq:dIseq} becomes especially useful when one considers that, for an $\A$ graded $R$-module $S$ and a vector $\alpha\in \C^M$, we can define its twist $S(\alpha)$ obtained by shifting the degrees of each operator with $\alpha$. Since the maps $s_J$ from equation~\eqref{eq:sjdef} are sensitive to this shift, we find
\begin{equation}
H_i(\E+\nu,S(\alpha))=H_i(\E+\nu+\alpha,S)\, .
\end{equation}
Therefore the sequence in equation~\eqref{eq:dIseq} results in the long exact sequence
\begin{equation}
\begin{array}{rll}
    \cdots&\rightarrow H_1(\E+\nu,S_\A/\langle \partial_I\rangle)&\rightarrow H_0(\E+\nu+a_I,S_\A)\xrightarrow{\cdot \partial_I} H_0(\E+\nu,S_\A)\\
    & \rightarrow H_0(\E+\nu,S_\A/\langle \partial_I\rangle)&\rightarrow 0
\end{array}
\end{equation}
which will be one of the key sequences used in this paper. Note that if $S_\A$ is a Cohen-Macaulay ring, all terms to the left of $H_1(\E+\nu,S_\A/\langle \partial_I\rangle)$ will be zero \cite[Remark 6.4]{matusevich_homological_2004}. This happens in many examples of interest, see for example the recent discussion in \cite{tellander_cohenmacaulay_2023}.\footnote{For a Cohen-Macaulay subring of $S_\A$, it is not guaranteed that the higher homology groups are zero since it may not be of maximal dimension.}

\subsection{Reducibility of GKZ systems and solution spaces}\label{ssec:reducibility}
In this section we want to briefly explain how the existence of submodules leads to the ability to obtain a partial basis of the solution space. Afterwards we will apply this observation to a specific subsystem found in \cite{schulze_resonance_2012} and show how this subsystem manifests itself in terms of the solutions to the GKZ system.

\paragraph{Submodules and solution spaces.}

The crucial observation is that a submodule gives rise to a subset of the solutions satisfying additional of differential equations. This can be seen most clearly when considering solutions as $\D$-module homomorphisms. In this case,  any surjective $\D$-module homomorphism
\begin{equation}
    \Phi:\M\twoheadrightarrow\mathcal{N}
\end{equation}
induces an injective map on the solution spaces
\begin{equation}\label{eq:inducedsolmap}
    \Phi^*:\mathrm{Sol}(\mathcal{N})\hookrightarrow \mathrm{Sol}(\M)
\end{equation}
given simply by precomposition with $\Phi$. 

The solutions of $\M$ in the image of this map have some interesting properties. For any differential operator $P\in \ker(\Phi)$ and $\phi\in \mathrm{Sol}(\mathcal{N})$, we have that
\begin{equation}
   P\;\Phi^*(\phi)(1)= \Phi^*(\phi)(P)=0\, .
\end{equation}
In other words, the solutions in the image of $\Phi^*$ are exactly those solutions of $\M$ that are also annihilated by every $P\in \ker(\Phi)$. Therefore, if this kernel is non-trivial, there will be a part of the solution space that satisfies additional differential equations with respect to a general solution of $\M$. We now want to apply this perspective to GKZ systems.

\paragraph{Resonance and submodules.}
 
 In \cite{schulze_resonance_2012} the reducibility of a GKZ system was characterized in terms of its resonance. Furthermore, in the proof of \cite[Thm. 4.1]{schulze_resonance_2012} an explicit submodule was found, on the condition that there exists a non-trivial face $F$ such that $\nu$ is in the $\C$-span of $F$ and $\A$ is not a pyramid over $F$. Defining
\begin{equation}
    \bar{F}\coloneqq A \setminus F\,
\end{equation}
the natural surjection
\begin{equation}
    S_\A\twoheadrightarrow S_{\A_F}\simeq S_\A/ \langle \partial_{\bar{F}}\rangle\, ,
\end{equation}
induces a surjection 
\begin{equation}\label{eq:Fsurjection}
    H_0(\E+\nu,S_\A)\twoheadrightarrow H_0(\E+\nu,S_{\A_F})
\end{equation}
on the Euler-Koszul homologies, where we recall that $A\coloneqq \{1,\cdots,N\}$ and $\langle \partial_{\bar{F}}\rangle$ is the ideal generated by the partial derivatives not in $F$. Thus we find a surjective $\D$-module morphism as required to obtain relate the solution spaces as in~\eqref{eq:inducedsolmap}. 

Now let us consider some properties of this map and the solutions in its image. Clearly, the kernel of~\eqref{eq:Fsurjection} is generated by $\partial_I$ with $I\in A\setminus F$. Therefore, the associated solutions are exactly those with
\begin{equation}\label{eq:pif=02}
    \partial_I f=0
\end{equation}
for $I\in A\setminus F$. This provides us with the first set of subsystems described in section~\ref{ssec:reductionoperators}. 

There is an alternative characterization of the solutions associated to this submodule as solutions to a different GKZ system. This is due to the fact that, if $F$ is a face of $\A$ there is an isomorphism \cite[Lemma 4.8]{matusevich_homological_2004}
\begin{equation}\label{eq:subfiso}
    H_0(\E+\nu,S_{\A_F})\simeq \C[z_{A\setminus F}]\otimes_\C \M_{\A_F}(\nu)\, ,
\end{equation}
where we recall that $\M_{\A_F}(\nu)$ is the GKZ system defined by the matrix $\A_F$. From the perspective of the solutions of $\M_{\A_F}(\nu)$, this isomorphism simply maps
\begin{equation}
    \phi(1)\mapsto \phi(1)\, .
\end{equation}
Therefore, it is possible to lift the solutions of $\M_{\A_F}(\nu)$ to obtain solutions of $\M_\A(\nu)$. Note that these solutions will automatically satisfy equation~\eqref{eq:pif=02}.

We will now show that a similar story holds even if $F$ is not a face. 
For any subset $F$ of $A$, recall that its toric ring of $\A_F$ is given by
\begin{equation}\label{eq:saf}
    S_{\A_F}\simeq R_\A / \left( I_{\A_F}+  \langle \partial_{\bar{F}}\rangle \right)\,.
\end{equation}
with $R_\A$ the ring of partial derivatives $\partial_I$. Let us consider an $\vert F \vert$-dimensional integer vector $u_F$. It is possible to lift this to an $N$-dimensional vector $u$ by defining $u_I=(u_F)_I$ if $I\in F$ and zero otherwise. Then, any two vectors $u_F$ and $v_F$ satisfying $\A_F u_F=\A_F v_F$ lift to vectors $u$ and $v$ satisfying $\A u= \A v$, since
\begin{equation}
    \A u= \sum_I u_I a_I =\sum_{I\in F}u_I a_I = \A_F u_F\,.
\end{equation}
This implies that 
\begin{equation}
    I_{\A_F}+  \langle \partial_{\bar{F}}\rangle \subseteq I_{\A}+ \langle \partial_{\bar{F}}\rangle\,,
\end{equation}
and applying the third isomorphism theorem results in
\begin{equation}
    R_\A/\left(I_{\A}+  \langle \partial_{\bar{F}}\rangle \right) \simeq \frac{R_\A /\left(I_{\A_F}+  \langle \partial_{\bar{F}}\rangle \right)}{\left(I_{\A}+ \langle \partial_{\bar{F}}\rangle \right)/\left(I_{\A_F}+ \langle \partial_{\bar{F}}\rangle \right)}\, .
\end{equation}
Identifying the left hand side as $S_\A/\langle \partial_{\bar{F}}\rangle$ and relating the right hand side to~\eqref{eq:saf}, this provides us with a surjection
\begin{equation}\label{eq:SAfsurjection}
    S_{\A_F} \twoheadrightarrow S_\A /\langle \partial_{\bar{F}}\rangle\,.
\end{equation}
By the same arguments as before, this surjection implies that all solutions of $H_0(\E+\nu,S_{\A}/\langle \partial_{\bar{F}}\rangle)$ lift to solutions of $H_0(\E+\nu,S_{\A_F})$. To relate this to the solutions of the actual GKZ system defined by $\A_F$, we note that the Euler operators of $\A_F$ are simply the Euler operators of $\A$ under the map that sends $\partial_I$ to zero for $I$ not in $F$. Therefore, we find that $H_0(\E+\nu, S_{\A_F})$ is isomorphic to the GKZ system defined by $\A_F$ and solutions to $H_0(\E+\nu,S_\A / \langle \partial_{\bar{F}}\rangle)$ lift to solutions to the GKZ system defined by $\A_F$. 

\paragraph{Example of a subsystem.}  Before continue with the task of obtaining reduction operators, we will briefly provide an example of the subsystems obtained in the previous paragraph. We focus on the case where we want to find a subsystem with solutions satisfying
\begin{equation}\label{eq:pifnu=0}
    \partial_I f(z;\nu)=0  \text{ for } I \in F
\end{equation}
for some set $F\subseteq \{1,\cdots,N\}$, with $N$ the number of columns of the matrix $\A$. One natural thing to do is to consider the matrix $\A$ defining the original GKZ system, and pick out the column vectors $a_J$ for $J in F$. Defining $\A_F$ to be this matrix, we can consider the GKZ system that it defines. For example, considering $\A$ as in equation~\eqref{eq:Aexample} and $F=\{1,2,4\}$, one finds the submatrix
\begin{equation}
    \A_F= \begin{pmatrix}
    1 & 1  & 0\\
    0 & 0  & 1\\
    0 & 3  & 1\\
    1 & 2  & 0
    \end{pmatrix}
\end{equation}
of $\A$. Its solutions will be functions of $z_J$ with $J\in F$ and therefore automatically satisfy~\eqref{eq:pifnu=0} for $I$ not in $F$. 

In fact, from the discussion around equation~\eqref{eq:SAfsurjection} we find that all solutions of the full GKZ system satisfying equation~\eqref{eq:pifnu=0} will also be solutions of the GKZ system defined by $\A_F$. However, the converse is not guaranteed, as the matrix $\A_F$ can have less toric operators than $\A$. This implies that it is not guaranteed that all solutions of $\A_F$ will lift to solutions of $\A$. However, if $F$ is a face of $\A$, as we will define below, then all solutions of $\A_F$ will lift to solutions of $\A$.

\section{Reduction operators from Euler-Koszul homologies}\label{sec:reductionoperators}

In the previous section, we have seen that, if a GKZ system is reducible, it is possible to obtain differential operators that annihilate a subspace of the full solution space. In this section, we will we use combine this observation with general results obtained for the Euler-Koszul homologies of GKZ systems. We begin with a study on how the reduction operators follow from the reducibility of a GKZ system, and then show various of their properties in section~\ref{ssec:reductionoperators}. 
Finally, we will provide some hitherto unpublished results regarding the number of solutions annihilated by combinations of reduction operators in section~\ref{ssec:reductionoperators}.

\subsection{Reduction operators from Euler-Koszul homologies}\label{ssec:reductionoperators}

From the discussion in section~\ref{sec:Dmodules} we have seen that, if the parameter $\nu$ is in the $\C$-span for some face $F$, it is possible to obtain a subsystem associated to this face. In general, any surjection of the type
\begin{equation}\label{eq:disurjection}
    S_\A\twoheadrightarrow S_\A/\langle \partial_I\rangle
\end{equation}
can result in a similar subsystem, composed of the solutions of $H_0(\E+\nu,S_\A)$ with $\partial_I f=0$. Note that the existence of such solutions is implied by the existence of a resonant face $F\subseteq A\setminus \{I\}$. In this section, we will show that there exists a different type of submodules for GKZ systems, whose solutions are annihilated by a set of operators we will call reduction operators.

\paragraph{Reduction operators from long exact sequences.}

Equation~\eqref{eq:disurjection} is the end of the exact sequence from equation~\eqref{eq:dIseq}, induced by multiplication with $\partial_I$. Therefore, it is natural to study the induced long exact sequence
\begin{equation}\label{eq:deltaexactseq}
\begin{array}{rll}
    &H_1(\E+\nu,S_\A/\langle \partial_I\rangle)&\xrightarrow{\delta} H_0(\E+\nu+a_I,S_\A)\xrightarrow{\cdot \partial_I} H_0(\E+\nu,S_\A)\\
     \rightarrow &H_0(\E+\nu,S_\A/\langle \partial_I\rangle)&\rightarrow 0\, ,
\end{array}
\end{equation}
where we define $\delta$ as the boundary map arising from the zig-zag lemma. Splicing this long exact sequence in two short exact sequences results in
\begin{equation}\label{eq:imd1ses}
    0\rightarrow \mathrm{im}(\cdot \partial_I)\rightarrow H_0(\E+\nu,S_\A) \rightarrow H_0(\E+\nu,S_\A/\langle \partial_I\rangle)\rightarrow 0\, ,
\end{equation}
and
\begin{equation}\label{eq:imdelses}
    0\rightarrow \mathrm{im}(\delta)\rightarrow H_0(\E+\nu+a_I,S_\A) \xrightarrow{\cdot \partial_I} \mathrm{im}(\cdot \partial_I)\rightarrow 0\, ,
\end{equation}
where $\im(\delta)$ and $\im(\cdot \partial_I)$ are the images of their respective maps. 

The second exact sequence provides us with a submodule of $H_0(\E+\nu+a_I,S_\A)$ by considering its quotient by $\im(\delta)$. The associated solutions will therefore be annihilated by all $Q\in \im(\delta)$. We will call such operators $Q$ reduction operators at the parameter $\nu$ in the direction $I$, and we will study their properties in the remaining part of this section.

\paragraph{Number of solutions.}

From combining the exact sequences~\eqref{eq:imd1ses} and~\eqref{eq:imdelses} one can obtain the short exact sequence 
\begin{equation}\label{eq:boundarysequence}
\begin{tikzcd}
    0\arrow[r]& H_0(\E+\nu+a_I,S_\A)/\mathrm{im}(\delta) \arrow[r,"\cdot \partial_I"]& H_0(\E+\nu,S_\A)\arrow[dl]  \\
    & H_0(\E+\nu,S_\A/\langle \partial_I\rangle)\arrow[r] & 0\, .
\end{tikzcd}
\end{equation}
Since this is a short exact sequence of holonomic $\D$-modules, it implies that the rank satisfies
\begin{equation}\label{eq:h0imderanks}
\begin{array}{rll}
   \mathrm{rank}\left(H_0(\E+\nu+a_I,S_\A)/\mathrm{im}(\delta)\right)=& &\mathrm{rank}\left(H_0(\E+\nu,S_\A)\right)\\
   &-&\mathrm{rank}\left(H_0(\E+\nu,S_\A/\langle \partial_I\rangle)\right)\, .
\end{array}
\end{equation}
Therefore, the number of solutions annihilated by the reduction operators at $\nu+a_I$ is determined by the number of solutions annihilated by the partial derivative $\partial_I$ at $\nu$. Correspondingly, if the GKZ system is not rank-jumping the quotient by $\im(\delta)$  is non-trivial if and only if the rank of $H_0(\E+\nu,S_\A/\langle \partial_I\rangle)$ is non-zero.

\paragraph{Partial derivatives map solution spaces.}

Recall from the discussion in section~\ref{ssec:gkzdmods} that, as we consider the solutions of these modules around some generic point, we can take the modules we consider to be non-singular. Furthermore, this implies that, when acting on these modules, the solution functor is exact. Applying this observation to the exact sequence~\eqref{eq:boundarysequence} results in
\begin{equation}\label{eq:solimdiseq}
\begin{tikzcd}
        0\arrow[r]& \mathrm{Sol}(H_0(\E+\nu,S_\A/\langle \partial_I \rangle ))\arrow[r] &\mathrm{Sol}(H_0(\E+\nu,S_\A)) \arrow[dl,"\cdot \partial_I"]\\
        &\mathrm{Sol}(H_0(\E+\nu+a_I,S_\A)/\mathrm{im}(\delta))\arrow[r]&0\, ,
\end{tikzcd}
\end{equation}
which provides us with a surjective mapping between the solutions of $H_0(\E+\nu,S_\A)$ and the solutions of $H_0(\E+\nu+a_I,S_\A)/\mathrm{im}(\delta)$ which simply sends $f\mapsto \partial_I f$.

We will now explain how to obtain the image of $\delta$ explicitly, explaining the basis behind the algorithm of section~\ref{ssec:reductionoperators}.

\paragraph{Reduction operators from the zig-zag lemma.}

The map $\delta$ in the sequence~\eqref{eq:dIseq} arises due to the zig-zag lemma, therefore its image can be obtained simply by explicitly performing the diagram chasing.

First, we consider an operator vector $P\in K_1(\E+\nu,S_\A/\langle \partial_I\rangle)$ and require that it satisfies
\begin{equation}\label{eq:tild1p=0}
    \tilde{d}_1(P)=0\, ,
\end{equation}
where $\tilde{d}_1$ is the differential between $K_1(\E+\nu,S_\A/\langle \partial_I\rangle)\rightarrow K_0(\E+\nu,S_\A/\langle \partial_I\rangle)$ . Note that, since this particular $S_\A$ module is not shifted, the differential can be written as
\begin{equation}
    \tilde{d}_1(P)=P\cdot(\A\Theta+\nu)\mod \partial_I\sim0 \in K_0(\E+\nu,S_\A/\langle \partial_I\rangle)\, ,
\end{equation}
where $\cdot$ denotes the vector dot product and we note that working in $K_0(\E+\nu,S_\A/\langle \partial_I\rangle)$ implies that we must impose $\partial_I\sim 0$.

Secondly, we lift $P$ to an element of $K_1(\E+\nu,S_\A)$ and apply the differential $d_1:K_0(\E+\nu,S_\A)\rightarrow K_0(\E+\nu,S_\A)$ to $P$. In order to do this, we again calculate $P\cdot (\A\Theta+\nu)$, however, since we are now considering elements of $ K_0(\E+\nu,S_\A)$ we must no longer set $\partial_I ~0$. This recovers the second step of the algorithm.

Finally, this procedure guarantees that 
\begin{equation}
    P\cdot (\A \Theta+\nu)=Q\partial_I \in K_0(\E+\nu,S_\A)
\end{equation}
for some operator $Q$. The action of $\delta$ on $P$ is then defined as
\begin{equation}
    \delta(P)=Q
\end{equation}
resulting in the reduction operator $Q$. In practice, it is enough to find the generators of $\im(\delta)$, since these will be the relevant operators when discussing solutions. 

In order to obtain these generators, one can start by considering the different solutions to equation~\eqref{eq:tild1p=0}. Noting that since we want to consider elements of
\begin{equation}
    H_1(\E+\nu,S_\A/\langle \partial_I\rangle)=\ker(\tilde{d}_1)/\im(\tilde{d}_2)
\end{equation}
we should ignore relations due to operators in $\im(\tilde{d}_2)$. In practice, this requires us to use the toric operators of $\A$ that have a term proportional to $\partial_I$. From these toric operators it is often clear how many such solutions are possible and from these solutions one can then obtain a full basis of generators of $\im(\delta)$.

\paragraph{Homogeneity and $\partial_I Q=0$.}

Although the algorithm above works for any $P \in K_1(\E+\nu,S_\A/\langle \partial_I\rangle)$, there are two particular classes of $P$ satisfying some additional useful properties. Firstly, it is often useful to consider only those reduction operators that are homogeneous with respect to the $\A$ grading. This is possible since $K_\bullet (\E+\nu,S_\A)$ and $K_\bullet (\E+\nu,S_\A/\langle \partial_I \rangle)$, as well as their differentials, inherit the $\A$ grading \cite[Lemma 4.3]{matusevich_homological_2004}. From this, one finds that by decomposing 
\begin{equation}
    P=\sum_{\alpha} P_\alpha
\end{equation}
into its homogeneous elements $P_\alpha$, each $P_\alpha$ must satisfy
\begin{equation}
    \tilde{d}_1(P_\alpha)=0
\end{equation}
separately. Therefore it is possible to obtain a reduction operator
\begin{equation}
    Q_\alpha\coloneqq \delta(P_\alpha)\, ,
\end{equation}
for each $P_\alpha$. Furthermore, since $\delta$ is compatible with the grading, $Q_\alpha$ is homogeneous. This allows us to always find a homogeneous set of generators for $\im(\delta)$. Note that this homogeneity implies that
\begin{equation}
    [\A \theta,Q_\alpha]=-\alpha Q
\end{equation}
since $\E_J$ act as the grading operators.

Secondly, we will show that it is always possible to obtain generators $Q$ that satisfy
\begin{equation}
    \partial_I Q=0 \in H_0(\E+\nu+a_I,S_\A)\, ,
\end{equation}
where the derivative $\partial_I$ acts on everything to the right. To show this we need that it is possible to obtain operator vectors $P$ independent of $z_I$. To see this, note that
\begin{equation}
    \D\otimes_{R} S_\A/\langle\partial_I\rangle \simeq \C[z_I]\otimes_\C( \D_{A\setminus \{I\}}\otimes_{R_{A\setminus \{I\}}}S_\A/\langle \partial_I \rangle)\, ,
\end{equation}
where we recall that $\D_{A\setminus\{I\}}$ and $R_{A\setminus\{I\}}$ denote the Weyl algebra and polynomial ring in the variables indexed by $A\setminus \{I\}$. Furthermore, writing $\E^{A\setminus\{I]}$ for the Euler operators with $\partial_I$ set to zero, it is possible to adapt the proof of \cite[Lemma 4.8]{matusevich_homological_2004} to arbitrary subsets $F\subset A$. This allows us to decompose
\begin{equation}
    K_\bullet (\E+\nu,S_\A/\langle \partial_I \rangle)\simeq \C[z_I]\otimes_{\C} K^{A\setminus\{I\}}_\bullet (\E^{A\setminus\{I\}}+\nu,S_\A/\langle \partial_I \rangle)\, ,
\end{equation}
where we have defined $K^{A\setminus\{I\}}_\bullet$ as the Euler-Koszul complex over the ring $R_{A\setminus \{I\}}$ and $\E^{A\setminus\{I\}}$ as the associated reduction operators.\footnote{Note that, while the complex $K^{A\setminus\{I\}}(\E^{A\setminus\{I\}}+\nu)$ is exactly the complex obtained from the matrix $\A_{A\setminus\{I\}}$, the rings $S_\A/\langle \partial_I\rangle$ and $S_{\A_{A\setminus\{I\}}}$ are not isomorphic in general. Therefore, this construction does not immediately result in a map between ordinary GKZ systems.} The differential of this complex acts trivially on $\C[z_I]$, therefore the generators $P$ can be chosen to be independent of $z_I$.

A consequence of $(\partial_I P)=0$ is that its reduction operator $Q$ satisfies
\begin{equation}\label{eq:partialiqi}
    (\partial_I Q)=P\cdot a_I\, ,
\end{equation}
which follows from the observation that
\begin{equation}
    P\cdot \A\Theta=P_1+(P\cdot a_I)z_I\partial_I
\end{equation}
with $P_1$ independent of $z_I$. Since $Q\partial_I = P\cdot( \A \Theta+\nu)$, equation~\eqref{eq:partialiqi} follows. In order to use this to show that $\partial_I Q=0$, simply commute
\begin{equation}
    \partial_I Q=Q\partial_I+(\partial_I Q) = P\cdot (\A\Theta+\nu)+P\cdot a_I\, ,
\end{equation}
where in the second equality we have applied the definition of the reduction operator. Since all components of the vector $A\Theta+\nu+a_I$ are zero in $H_0(\E+\nu+a_I,S_\A)$, we find that $\partial_I Q$ vanishes.

\subsection{Counting formulae}\label{ssec:reductioncounting}

In this section we will provide ways of counting the number of solution annihilated by various reduction operators. In particular, we will consider the reduction operators as generators of $\ker \partial_I$ and obtain the number of solutions annihilated by combinations of such kernels.
In order to obtain these results we will assume that the ring $S_\A$ is Cohen-Macaulay and furthermore that the various partial derivatives form a regular sequence of $S_\A$, the reasons for which will become clear in what follows.
This section consists fully of previously unpublished work.

\paragraph{Intersections of kernels for two partial derivatives.}

Before we tackle the general case, let us first consider a GKZ system $H_0(\E+\nu,S_\A)$ and consider the kernels of $\partial_I$ and $\partial_J$ when considered as maps of from this GKZ system to the same system at parameters $\nu-a_I$ and $\nu-a_J$ respectively. In this case, the problem we are trying to solve is that we want to know the rank of $H_0(\E+\nu,S_\A)/(\ker \partial_I +\ker \partial_J)$ where $\ker$ again denote the respective kernels.
Assuming the rank of of the full GKZ module is known, this is equivalent to obtaining the rank of $\ker \partial_I +\ker \partial_J$ which by the inclusion exclusion principle is given by
\begin{equation}\label{eq:twoderivkernelintersec}
    \mathrm{rank}(\ker \partial_I + \ker \partial_J)=  \mathrm{rank}(\ker \partial_I)+\mathrm{rank}(\ker \partial_J) - \mathrm{rank}(\ker \partial_I \cap \ker \partial_J)\, .
\end{equation}
Now, we note that the Intersection of these kernels can be obtained as the kernel of the map
\begin{equation}
    \partial_I\oplus\partial_J\,: H_0(\E+\nu,S_\A) \longrightarrow H_0(\E+\nu-a_I,S_\A)\oplus H_0(\E+\nu-a_J,S_\A)
\end{equation}
induced from a similar map on the level of toric rings. 
From the pair $(\partial_I,\partial_J)$ one can construct the the ordinary Koszul complex, which is given by
\begin{equation}\label{eq:partialIJexactseq}
   T_\bullet :\; S_\A \xrightarrow{(\partial_I ,\partial_J)^{\rm T}} S_\A(-a_I) \oplus S_\A(-a_J) \xrightarrow{(\partial_J,-\partial_I)}S_\A(-a_I-a_J)\rightarrow 0,
\end{equation}
where $\rm T$ denotes the transpose. If $(\partial_I,\partial_J)$ is a regular sequence on $S_\mathcal{A}$, the homology of this complex is concentrated in the zeroth degree and is equal to $H_0(T_\bullet)= S_\A(-a_I-a_J)/\langle \partial_I,\partial_J\rangle$.
To relate the complex $T_\bullet$ to the GKZ systems, we will construct the double complex\footnote{Note that due to the degree shifts in equation~\eqref{eq:partialIJexactseq}, the differentials on $T_\bullet$ commute with those of the Euler-Koszul complex.}
\begin{equation}
    E^{\,0}_{p,q}\coloneqq K_q(\E+\nu, T_p) \, .
\end{equation}
and study its spectral sequences.

\paragraph{Spectral sequence for the two derivative case.}

Let us first consider the spectral sequence obtained by taking the horizontal differential first on $E^0$. Note that from the definition of $T_\bullet$, we find that
\begin{equation}
    \begin{array}{rl}
        E^0_{0,q} &= K_q(\E+\nu-a_I-a_J,S_\A) \, , \\
        E^0_{2,q} &= K_q(\E+\nu,S_\A)\, .
    \end{array}
\end{equation}
Furthermore, as the tensor product commutes with taking direct sums, the remaining term splits as
\begin{equation}
    E^0_{1,q} = K_q(\E+\nu-a_I,S_\A) \oplus K_q(\E+\nu-a_J,S_\A)\,.
\end{equation}
The vertical differential on this spectral sequence is simply the Euler-Koszul differential and, as $S_/A$ is assumed to be Cohen-Macaulay, we find that $E^1_{p,q}=0$ for all $q\neq 0$. The remaining non-zero terms are then given by
\begin{equation}
    \begin{array}{rl}
        {}_v E^1_{0,0} & = H_0(\E+\nu-a_I-a_J,S_\A)\, , \\
        {}_v E^1_{1,0} &= H_0(\E+\nu-a_I,S_\A) \oplus H_q(\E+\nu-a_J,S_\A)\,.\\
        {}_v E^1_{2,0} &= H_0(\E+\nu,S_\A)\, ,
    \end{array}
\end{equation}
where ${}_v E^1$ denotes that we are taking the vertical differential first. As the sequence has collapsed we find that this spectral sequence abuts on the second page to
\begin{equation}\label{eq:twoderivvertabut}
    \begin{array}{rl}
    {}_v E^{\infty}_{0,0}  & = H_0(\E+\nu-a_I-a_J,S_\A)/(\im \partial_I + \im \partial_J)\, \\
    {}_v E^{\infty}_{1,0}  & = \ker \partial_J /\im\partial_I \oplus \ker \partial_I/\im \partial_J\, ,\\
    {}_v E^{\infty}_{2,0} &= \ker \partial_I \cap \ker \partial_J\, ,
    \end{array}
\end{equation}
where we have suppressed  the specific parameters $\nu$ in the various images and kernels in order to lighten the notation. All other terms in the spectral sequence abut to zero.

Conversely, it is possible to take the horizontal differential first on $E^0$, and obtain a different spectral sequence also converging to the homology of the total complex. In this case, we can use that the homology of $T_\bullet$ is concentrated in the zeroth degree and obtain
\begin{equation}
    {}_h E^1_{0,q} = K_q(\E+\nu-a_I-a_J,S_\A/\langle \partial_I,\partial_J \rangle)\, ,
\end{equation}
with all other terms in the first page being zero. Again we see that the spectral sequence collapses at the first page. In this case we find that the spectral sequence abuts to
\begin{equation}\label{eq:twoderivhorabut}
    {}_h E^{\infty}_{0,q} = H_q(\E+\nu-a_I-a_J,S_\A/\langle \partial_I,\partial_J \rangle)\, ,
\end{equation}
with all other terms abutting to zero. 

Both the vertical and horizontal spectral sequences converge to the homology of the total complex. Therefore, we find that the terms in the spectral sequence must be isomorphic to graded quotients of the homology of the total complex. In particular, denoting the homology of the total complex as $ H_k(\mathrm{Tot})=H_k\big(\mathrm{Tot}(K_\bullet(\E+\nu,T_\bullet))\big)$, we find that
\begin{equation}
    \begin{array}{rl}
        {}_v E^\infty_{p,k-p} \simeq {}_v F_p H_k(\mathrm{Tot})/ {}_v F_{p-1} H_k(\mathrm{Tot})\,, \\ 
        {}_h E^\infty_{k-q,q} \simeq {}_h F_q H_k(\mathrm{Tot})/ {}_h F_{q-1} H_k(\mathrm{Tot})\,. \\ 
    \end{array}
\end{equation}
Note that the gradings ${}_h F$ and ${}_v F$ on $H_k(\mathrm{Tot})$ will be different.
However, we note that for fixed $k$, ${}_v E^\infty_{p,k-p}$ is only non-zero for $p=k$. Therefore, $H_k$ is fully localized in the $k-th$ degree with respect to the grading ${}_v F$ and we obtain
\begin{equation}
        {}_v E^\infty_{k,0} \simeq H_k(\mathrm{Tot})\, . \\ 
\end{equation}
In a similar manner, we obtain that with respect to the grading ${}_h F$, $H_k$ is fully localized in the $k$-th degree and we have
\begin{equation}
        {}_h E^\infty_{0,k} \simeq H_k(\mathrm{Tot})\, . \\ 
\end{equation}
Comparing with the abutments of ${}_h E$ and ${}_v E$ this provides us with a number of isomorphisms. Most importantly, we obtain ${}_v E^\infty_{2,0} \simeq {}_h E^{\infty}_{0,2}$ which, comparing to equations~\eqref{eq:twoderivvertabut} and~\eqref{eq:twoderivhorabut} implies that
\begin{equation}
    \ker \partial_I \cap \ker \partial_J \simeq H_2(\E+\nu-a_I-a_J,S_\A/\langle \partial_I,\partial_J \rangle)\, .
\end{equation}
Using the methods of~\cite{berkesch_rank_2011}, the rank of the right-hand-side of this equation can be computed. Inserting the rank into equation~\eqref{eq:twoderivkernelintersec} allows us to compute the rank of the $\ker \partial_I +\ker \partial_J$ as we set out to do.

\paragraph{Generalizing to multiple partial derivatives.}

The above can be straightforwardly generalized to arbitrary regular sequences of partial derivatives. 
We will denote the index set of the partial derivatives we are interested in as $P$, and obtain the rank of $\bigcap_{I\in P} \ker\partial_I$ if the partial derivatives in $P$ define a regular sequence on $S_\A$. 
Using the inclusion-exclusion principle, this allows us to obtain the rank of $H_0(\E+\nu,S_\A)/(\sum_{I\in P} \ker \partial_I)$.

We begin again by considering the ordinary Koszul complex $T_\bullet$ for the sequence of partial derivatives. As before, its homology will be concentrated in the zeroth degree and will equal $H_0(T_\bullet) = S_\A / \langle \partial_P\rangle$.
Using this complex, we will construct the double complex
\begin{equation}
    E^0_{p,q} = K_q(\E+\nu,T_p)\,
\end{equation}
and compute its abutments. 

Let us consider the spectral sequence obtained by taking the vertical on $E^0$, resulting in ${}_v E^1_{p,q} = H_q(\E+\nu,T_p)$.
Note that for any $0 \leq p \leq \vert P\vert$, the complex at $T_p$ will be isomorphic to a direct sum of $\vert P\vert$ choose $p$ copies of $S_\A$, with its degree shifted by an appropriate combination of the column vectors of $\A$. This implies that each $H_q(\E+\nu,T_p)$ admits a similar decomposition, into terms of the form $H_q(\E+\nu-c,S_\A)$, with $c$ being the relevant degree shift.
As, again, we assume $S_\A$ is Cohen-Macaulay, we find that $H_q(\E+\nu-c,S_\A)$ is zero for any $q>0$ and shift $c$. 
Therefore, we obtain that the spectral sequence collapses to the row $q=0$, with its only non-zero entries being
\begin{equation}
{}_v E^1_{p,0} = H_0(\E_\nu,T_p) = \bigoplus_{i=1}^{\vert P \vert \choose i} H_0(\E_\nu-\gamma_{p,k}i,S_\A)\, ,
\end{equation}
where $\gamma_{p,i}$ enforces the correct degree shift and $\vert P \vert \choose i$ is the choose function.

As the sequence has collapsed, it will abut on the second page after we have applied the horizontal differential. In particular, we find that the term $E^\infty_{\vert P \vert,0}$ will be given by
\begin{equation}
    E^\infty_{\vert P \vert ,0} = \bigcap_{I \in P} \ker \partial_I\,.
\end{equation} 
Note that, as the sequence converges to the total complex and it has collapsed to a single row, we have that 
\begin{equation}\label{eq:kernelintersec=tot}
    \bigcap_{I \in P} \ker \partial_I \simeq H_k\big(\mathrm{Tot}(K_\bullet(\E+\nu),T_\bullet)\big) \, .
\end{equation} 
We will now compute $H_k\big(\mathrm{Tot}(K_\bullet(\E+\nu),T_\bullet)\big)$ by taking the horizontal differential first.

As the homology of $T_\bullet$ is concentrated in the zeroth degree, we find that the first page of the horizontal spectral sequence takes the form of
\begin{equation}
    {}_h E^{1}_{0,q} = K_q\bigg(\E+\nu-\sum_{I \in P} a_I,S_\A /\langle \partial_P \rangle\bigg)\,,
\end{equation}
where we recall that $\langle \partial_P \rangle$ is the ideal generated by all partial derivatives $\partial_I$ for $I$ in $P$. Turning our attention to the second page, we obtain the Euler-Koszul homologies of this complex and find that the spectral sequence abuts to
\begin{equation}
    {}_h E^{\infty}_{0,q} =  H_q\bigg(\E+\nu-\sum_{I \in P} a_I,S_\A /\langle \partial_P \rangle\bigg)\, ,
\end{equation}
with all other terms abutting to zero.

Now let us consider ${}_h E^{\infty}_{0,\vert P\vert}$ and note that this is isomorphic to $H_{\vert P \vert}\big(\mathrm{Tot}(K_\bullet(\E+\nu),T_\bullet)\big)$.
Comparing with equation~\eqref{eq:kernelintersec=tot}, we obtain the isomorphism
\begin{equation}
     \bigcap_{I \in P} \ker \partial_I \simeq H_{\vert P\vert}\bigg(\E+\nu-\sum_{I \in P} a_I,S_\A /\langle \partial_P \rangle\bigg)\, .
\end{equation}
As the right-hand side can be computed using the methods of~\cite{berkesch_rank_2011} this provides us with the rank of the intersected kernel. Using the inclusion-exclusion principle this allows us to obtain the number of solutions annihilated by any combination of reduction operators.

For completion, we provide the full formula for the rank of the quotient of a GKZ module by numerous reduction operators:
\begin{equation}
    \mathrm{rank}\bigg(H_0(\E+\nu,S_\A)/\bigg(\sum_{I\in F} \ker\partial_I\bigg)\bigg) = \sum_{\emptyset \subseteq P \subseteq F} (-1)^{\vert P\vert} \mathrm{rank}\left(H_{\vert P \vert}(\E+\nu_P,S_\A/\langle \partial_P\rangle)\right)\, ,
\end{equation}
where $F$ is a set such that the partial derivatives $\partial_I$ for $I \in F$ form a regular sequence and the sum is over all subsets $P$ of $F$ including the empty subset. Finally, $\nu_P$ is defined as
\begin{equation}
 \nu_P =    \nu - \sum_{I \in P} a_I\, ,
\end{equation}
with $a_I$ the column vectors of $\A$.

\section{Construction of reduction operators}\label{sec:reductionconstruction}

Having obtained numerous technical results regarding existence and properties of the reduction operators, let us now take a somewhat more pragmatic approach and rephrase the results above in a slightly less technical manner. 
We begin by describing the necessary and sufficient conditions for the existence of non-trivial reduction operators, afterwards we will outline their construction, and finally we will describe some of their properties.

\subsection{Existence of reduction operators.}

Reducibility of a GKZ system is not a generic property, as it is generally not expected that solutions to a set of differential equations also obey smaller, simpler subsystem.
We will now introduce the necessary and sufficient conditions for reducibility. These conditions are known in the mathematical literature \cite{walther_duality_2005,beukers_irreducibility_2011,schulze_resonance_2012} through a property called resonance, which occurs if there are certain relations of the parameter $\nu$ with the matrix $\mathcal{A}$. In this section, we describe these results with a view towards explicit calculations. Interestingly, we will see that resonance is a geometric property and checking the resonance of a GKZ system does not require solving any differential equations.\footnote{The procedures described in this section can also be framed in a more geometric manner, here we have chosen to focus on the differential operators description instead.} Therefore, it is possible to check if reductions are possible without having to solve any subsystems explicitly. 

In general, the reducibility of a GKZ system can be checked in three steps. Leaving the precise definitions of these properties for later, these are:
\begin{enumerate}
    \item Finding a \textit{resonant set} $F \subseteq \{1,...,N \}$ for the parameter $\nu$,
    \item Determining if $F$ is a \textit{face} of $\A$,
    \item Checking that $\A$ is not a \textit{pyramid} over $F$.
\end{enumerate}
If all of these conditions are satisfied, the GKZ system is reducible and subsystems can be determined. 

\paragraph{Definition of a face.}

Recall that a GKZ system is completely defined by its matrix $\A$ and the parameter $\nu$. We will consider index sets $F$ that are subsets of $\{1,\cdots,N\}$, where $N$ is the number of columns of $\A$. As a technical prerequisite for reducibility, we have to require that $F$ is a \textit{face} of $\A$, defined as follows. The subset $F$ is a face of $\A$ if there exists a linear functional $L_F:\bbZ^N\rightarrow \bbZ$ such that
\beq \label{eq:LF}
\begin{array}{rl}
  L_F(a_I) =0 & \text{ for } I\in F \, , \\
  L_F(a_I)>0 & \text{ for } I\not\in F \, .
\end{array}
\eeq
This property also has a geometric interpretation. Let us consider the column vectors $a_I$ of $\A$ as generating a cone. Similarly, we can consider the cone generated by the column vectors of $a_I$ with $I$ in $F$. Then, $F$ defines a face of $\A$ if the cone generated by $F$ is a face of the cone, in the geometric sense. If $F$ is a face of codimension one, we will call it a \textit{facet}. Geometrically, this will also correspond to a facet of the cone of $\A$.

\paragraph{Resonance faces.}

Given the notion of a face, we now introduce resonance. We say that $F$ is a \textit{resonance face} for $\nu$ if
there are complex numbers 
$c_I$ and integers $n_I$ such that
\begin{equation} \label{resonance_cond}
    \nu=\sum_{I\in F}c_I\; a_I+\sum_{I=1}^N n_I \;a_I\ ,
\end{equation} 
where we recall that $a_I$ are the column vectors of $\A$. In other words, the vector $\nu$ lies in the span of the face $F$, up to shifts given by integer multiples of the columns of $A$. A minimal resonance face for $\nu$ is also known as a \textit{resonance center} for $\nu$. Since \eqref{resonance_cond} contains general $c_I$, any face that contains a resonance face is also a resonance face and a resonance center picks out the smallest combination.

Finally, one needs to exclude the 
possibility that $\A$ defines a \textit{pyramid} over $F$. While the notion of a pyramid is defined geometrically, it is equivalent to the requirement that none of the toric operators contain $\partial_I$ for $I\not \in F$. This extra condition excludes the trivial factorization of solutions with an overall pre-factor depending on $z_I$ for $I$ not in $F$.

Given these definitions, we can now use Theorem 3.1 of \cite{schulze_resonance_2012} (see also \cite{grimm_reductions_2025}). It states that if one has a resonance center $F$ for $\nu$ and $\A$ is not a pyramid over $F$, then the GKZ system with data $\nu,\A$ is reducible. As we will discuss next, this ensures that one is able to construct additional differential equations of the form 
\beq
   Q f(z;\nu) = 0\ ,
\eeq
satisfied by \textit{some} of the solutions to 
the GKZ system.
These operators $Q$ will be the reduction operators.
As there will usually be multiple of such operators, we will index them either by an integer $I$ corresponding to a variable $z_I$ of the GKZ system, where $I$ is some index not in the face $F$.
Alternatively, we can emphasize the face $F$ that gives rise to the reduction operator.
This gives rise to the notation $Q^{(F)}_u$, where $u$ is some label for the different reduction operators stemming from the same face $F$. These two different notations stem from the different ways of constructing the reduction operators that we will highlight now.

\subsection{Constructing reduction operators.}\label{ssec:reductionconstructionexplicit}

If conditions outlined above are met, the GKZ system under consideration will be reducible and there will exist non-trivial reduction operators. Now, we will outline how the results from section~\ref{sec:reductionconstruction} can be used to construct these non-trivial operators explicitly.
First, we will provide the general procedure, directly adapted from the proofs in~\ref{sec:reductionconstruction}. 
Afterwards, we will consider an improvement version of this algorithm that makes the relation between the resonant face and the reduction operator more explicit.

\paragraph{Direct method.} To derive the reduction operators at a parameter $\nu$ in the direction $I$, we have to consider the toric operators~\eqref{eq:toricopdef} determined by $\A$, and interpret them as equivalence relations. To be precise, every toric operator $\L_{u,v}$ leads to an equivalence relation of the form
\begin{equation}\label{eq:gentoricrel}
    \prod_{I=1}^N (\partial_I)^{u_I}\simeq \prod_{I=1}^N (\partial_I)^{v_I}\ .
\end{equation}
This is due to the fact that, when acting on solutions of $\L_{u,v}f=0$,
both sides of~\eqref{eq:gentoricrel} will act equivalently. When considering the set of differential operators, we introduce an equivalence relation $\simeq$ that implements~\eqref{eq:gentoricrel} by saying that two operators are equivalent if they are identical when sending $\mathcal{L}_{u,v}\rightarrow 0$. 
We will also sometimes set the partial derivative $\partial_I$ to zero, which we will denote by ${\partial_I\rightarrow 0}$. One can also combine the two and impose both $\L_{u,v}\rightarrow 0$ and $\partial_I \rightarrow 0$ and the resulting equivalence relations will be denoted by $\simeq_I$. Finally, we can generalize this to an arbitrary subset $F\subset\{1,\cdots,N\}$ and define the equivalence $\simeq_F$ which sets the partial derivatives $\partial_K\rightarrow 0$ for all $K$ in $F$, as well as the toric operators. In summary, we introduce the following notation for the equivalence relations among two differential operators $\mathcal{O}_1$, $\mathcal{O}_2$:
\begin{equation}
\begin{array}{lll}
   \mathcal{O}_1 \simeq &\hspace{-5pt} \mathcal{O}_2\  \, : \quad& (\mathcal{O}_{1}-\mathcal{O}_2)|_{\mathcal{L}_{u,v}\rightarrow 0} = 0\  ,\\
   \mathcal{O}_1 \simeq_I& \hspace{-5pt}\mathcal{O}_2\ \, : \quad &(\mathcal{O}_{1}-\mathcal{O}_2)|_{\mathcal{L}_{u,v}\rightarrow 0,\partial_I \rightarrow 0} = 0 \ ,\\
   \mathcal{O}_1 \simeq_F& \hspace{-5pt}\mathcal{O}_2\ \, :\quad &(\mathcal{O}_{1}-\mathcal{O}_2)|_{\mathcal{L}_{u,v}\rightarrow 0,\partial_K \rightarrow 0, K \in F} = 0\ . 
\end{array}
\end{equation}

The algorithm for finding reduction operators is then as follows. Start with an $M$-dimensional vector of differential operators $P_I (\nu)$, where we recall that $M$ is the number of Euler operators. We then impose that, under the equivalence relations determined by $\simeq_I$, it satisfies
\begin{equation}\label{eq:vecpi=0}
    P_I (\nu)\cdot (\E+\nu-a_I)\simeq_I 0\, ,
\end{equation}
where $\cdot$ is the vector dot product and $\E$ are the Euler operators~\eqref{eq:eulergen}. 
Note that one way a vector $P_I(\nu)$ can satisfy this equation is if its components are proportional to the Euler operators themselves. When this occurs, we refer to the resulting reduction operator as trivial, meaning it only leads to a trivial subsystem.\footnote{A subsystem is considered trivial if it either consists of the same solutions as the full system or has no non-zero solutions.} To obtain non-trivial reduction operators, one must apply the toric relations~\eqref{eq:gentoricrel}. It is always possible to find $P_I(\nu)$ that are independent of $z_I$, and we will proceed under the assumption that this has been done. Note that it is possible that $P_I(\nu)$ depends on other variables $z_J$.

Equation~\eqref{eq:vecpi=0} guarantees that, if we no longer impose that the partial derivative vanishes, we have
\begin{equation}\label{eq:vecpinot0}
\boxed{\rule[-.1cm]{0cm}{.5cm}\quad 
    P_I (\nu)\cdot (\E+\nu-a_I)\simeq Q_I (\nu)\partial_I
\quad}
\end{equation}
for some differential operator $Q_I (\nu)$. Note that the difference between equations~\eqref{eq:vecpi=0} and~\eqref{eq:vecpinot0} is that, in equation~\eqref{eq:vecpi=0}, we also impose that $\partial_I$ goes to zero. The operator $Q_I (\nu)$ obtained from equation~\eqref{eq:vecpinot0} is the reduction operator we are after. If there exists multiple operator vectors solving equation~\eqref{eq:vecpi=0}, each of them will lead to a different reduction operator.

\paragraph{Reduction operators drectly from faces.} Let us now briefly explain how a reduction operator can be constructed directly from a resonant face. Our starting point is a resonance face $F$.\footnote{Note that we do not directly consider a resonance center, which is a minimal resonance face. This implies that our construction might admit further reductions.} We begin with the observation that, if a face $F$ is a resonance center, it is possible to shift the parameters $\nu$ by the column vectors of $\A$ such that the new parameter is in the span of $F$. Furthermore, denoting this new parameter as $\nu_F$, the linear functional defining $F$ will then satisfy
\begin{equation}\label{eq:LFnuF=0}
    L_F(\nu_F)= 0 \, .
\end{equation}
Recall that these shifts can be realized on the solutions by applying using partial derivatives or their inverses as in \cite{dwork_generalized_1990, beukers_irreducibility_2011,caloro_ahypergeometric_2023}.\footnote{Interestingly, the constructions of~\cite{caloro_ahypergeometric_2023} share some similarities with the algorithm for obtaining reduction operators below. Partly, this is because both are obtained from a similar construction in~\cite{dwork_generalized_1990,beukers_irreducibility_2011}. As the reduction operators were originally introduced in~\cite{grimm_reductions_2025} from a different perspective, it would be interesting to explore further how the two constructions relate.}
However, observe that shifts by the columns contained in $F$ will not change the procedure below. Because of this, and the precise form of the cosmic GKZ systems we will consider, we will not need to consider such shifts in this paper.

We will focus on the case where we have a resonant face $F$, with some fixed $I$ not in $F$, and consider the case where $\nu$ is such that $\nu-a_I$ is in the span of $F$. Equivalently, this implies that $L_F(\nu-a_I)=0$, where $L_F$ is a linear functional defining $F$. To construct a reduction operator associated to $F$ and $I$, we will first define the operator
\begin{equation}\label{eq:EFdef}
\E_F  = \sum_J L_F (a_J) \theta_J 
\end{equation}
and note that equation~\eqref{eq:LFnuF=0} implies that, at the parameter $\nu-a_I$, we have $\E_F \simeq_{\E+\nu-a_I} 0$.
Then, we will construct a vector $u$ in $\N^N$ such that 
\begin{equation}\label{eq:PEFsimQd}
     \pd_1^{u_1}\cdots \pd_N^{u_N} \E_F \simeq  Q^{(F)}_u \pd_I,
\end{equation}
and from this obtain the reduction operator $Q^{(F)}_u$.\footnote{It may happen that $\partial^u$ and $\E_F$ do not commute, in which case a small additional step is required. If any terms of the form $\partial_J^k z_J \partial_J$ appear when expanding $\partial^u \E_F$, simply replacing $\partial_J^k$ with $\prod_{j=1}^k (\theta_J-j)$ will guarantee that the expression is still proportional to $\partial_I$. For the systems we consider in this paper, we will not need this though.} Note that this reduction operator will be valid at parameter $\nu$.

To obtain the vector $u$ we can proceed in two ways. If we can immediately construct such a vector $u$ by inspecting the GKZ system, then it results in a reduction operator and we are done. This is the approach we will take in the remainder of this thesis. However, we will also provide a somewhat technical condition, proven originally in \cite{dwork_generalized_1990} but adapted from \cite[Theorem 2.1]{beukers_irreducibility_2011}, that allows us to obtain such a vector algorithmically. This condition can be stated as follows. Recall that a facet of $\A$ is a face of co-dimension one. If, for every facet $F'$ of $\A$ and $J$ not in $F$ we have that
\begin{equation}\label{eq:reducopineq}
    L_{F'}(\A u+a_J)\geq L_{F'}(a_I)\, ,
\end{equation}
then $u$ satisfies equation~\eqref{eq:PEFsimQd}. Here, $L_{F'}$ is the linear functional defining $F'$ and we note that, since $F'$ is a facet, $L_{F'}$ is unique up to a constant pre-factor. Note that, since the entries of $u$ are integers and the linear functionals $L_{F'}$ can be written in terms of matrices, the above turns into an integer linear programming problem allowing us to obtain $u$ algorithmically.

\subsection{Properties of reduction operators}

In this section we fix a GKZ system described by a matrix $\A$ and consider it at different parameters $\nu$. For simplicity, we assume that the number of solutions to this GKZ system, which we denote by $D$, does not depend on this parameter. Note that this assumption is valid for many examples of interest and was recently proved to hold for large classes of Feynman integrals \cite{tellander_cohenmacaulay_2023}.\footnote{Note that many of the mathematical results, such as the existence of the reduction operators, do not need this simplifying assumption. We refer to section~\ref{sec:reductionoperators} for more details.}

We will now state the key properties of the reduction operators, shown in section~\ref{sec:reductionoperators}. 
Let us denote a basis of linearly independent solutions to the GKZ system at parameter $\nu$ by
\beq
    f_{T}(z;\nu)\ , \qquad T=1,...,D\ . 
\eeq
In the following we assume that there are $D_I$ solutions to the GKZ system at parameter $\nu-a_I$. To indicate this dependence on the parameter we will write $D_I(\nu-a_I)$.
By assumption, these solutions satisfy 
\begin{equation}\label{eq:pift=0}
\partial_I f_t(z;\nu-a_I)=0\ , \qquad t = 1,..., D_I(\nu-a_I) \ ,
\end{equation}
where we have split the index $T=(t,\tau)$ and used $t$ as a label for those solutions annihilated by $\partial_I$. The remaining $D-D_I(\nu-a_I)$ solutions are labelled by an index $\tau$.

When assuming $D_I>0$ in~\eqref{eq:pift=0}, there are \textit{reduction operators} $Q_I (\nu)$ at $\nu$ in the direction $I$ such that:
\begin{enumerate}
    \item There are $D-D_I(\nu-a_I)$ solutions of the GKZ system at parameter $\nu$ satisfying
    \begin{equation}\label{eq:qif=0}
        Q_I (\nu)f_\tau(z;\nu)=0\ , \qquad \tau = 1,..., D-D_I(\nu-a_I)\ , 
    \end{equation}
    for all reduction operators in the direction $I$. Note that we have again labeled these solutions with $\tau$, justified by the following fact.\footnote{In general one has to be careful when considering subspaces of functions at different parameters $\nu$, as statements such as $\lim_{\nu\rightarrow \nu'} f_T(z;\nu)=f_T(z;\nu')$ are not guaranteed to hold for all $\nu$ and $\nu'$. However, we will only consider parameter shifts by the different column vectors $a_I$, which are much better behaved.}
    \item All of these solutions can be written as 
    \begin{equation}  \label{eq:relating_sol1}
    f_\tau(z;\nu)=\partial_I f_\tau(z;\nu-a_I)\ ,
    \end{equation}
    where $f_\tau(z;\nu-a_I)$ is a solution at parameter $\nu-a_I$. Note that the $f_\tau(z;\nu-a_I)$ are exactly those solutions that do not satisfy equation~\eqref{eq:pift=0}.
    \item For all solutions at $\nu$, we have 
    \begin{equation}\label{eq:relating_sol2}
        \partial_I Q_I (\nu)f_T(z;\nu)=0\, .
    \end{equation}
    Note that this is not just limited to the solutions satisfying equation~\eqref{eq:qif=0}, for which this property trivializes.
    \item The commutant of the reduction operator with the Euler operators is always of the form
        \begin{equation}\label{eq:qicommu}
        [\E_J,Q_I(\nu)]=-\left(q_I(\nu)\right)_J \,Q_I(\nu)\, 
    \end{equation}
    for some integer vector $q_I(\nu)$. If the reduction operator commutes with the toric operators~\eqref{eq:toricopdef}, this implies that there is a mapping
    \begin{equation}\label{eq:relating_sol3}
        f_T(z;\nu) \mapsto Q_I(\nu)f_T(z;\nu)
    \end{equation}
    sending solutions of the GKZ system at $\nu$ to solutions of the GKZ system at $\nu+q_I(\nu)$. Furthermore, by equation~\eqref{eq:relating_sol2}, all solutions in the image of this map will be annihilated by $\partial_I$.
\end{enumerate}
The existence of non-trivial reduction operators is determined by the value of $D_I$ and therefore depends on $\nu-a_I$. If $D_I$ is zero, the reduction operators will not reduce the solution space.
In this case, the reduction operators can still be obtained but will be proportional to the Euler operators~\eqref{eq:eulergen} and lead to trivial relations. In the rest of this thesis, we will use the properties above to study cosmological correlators.

\newpage
\fi

\if\PrintChThree1



\stepcounter{thumbcounter}
\setcounter{colorcounter}{3}
\chapter{Cosmological Correlators and Reduction Operators} \label{ch:cosmology}

In this chapter, we will begin applying the results of the previous chapter to a physical setting. 
In particular, we will consider a certain toy model used to obtain insights into cosmological correlators. 
For this particularly toy model, we zoom in on one particular correlator, associated to single particle exchanged. We will study the corresponding integral, the single-exchange integral, in detail trough the lens of GKZ systems and reduction operators. For this system, we obtain various subsystems and use these to construct partial solution bases. We note that the approach here is somewhat complementary to the one taken in chapter~\ref{ch:reductionalgorithm}.
There, we relate the differential systems for different diagrams using inhomogeneous differential equations that involve the reduction operators. 
Here, we mostly use the reduction operators to obtain homogenous differential equations, the solutions of which will be a partial solution basis for the full single-exchange system.

We will begin with some background of such correlators in section~\ref{sec:cosmology}. 
Afterwards, we will become more specific and introduce the exact type of correlators we will study in section~\ref{sec:toymodel}. 
Finally, we will proceed to section~\ref{sec:singexreducs}, where we introduce a specific example and showcase how GKZ systems and their reduction operators can aid in solving for such correlators.

\section{The toy model}\label{sec:toymodel}

In this thesis, we will work with a toy model which has the particular property that it is possible to find correlators for a variety of space-times simultaneously. We will consider a conformally coupled scalar in an FRW space-time with generic polynomial interactions. This model was introduced in \cite{arkani-hamed_cosmological_2017} and we will mostly follow the exposition  of \cite{arkani-hamed_cosmological_2017,arkani-hamed_differential_2023}. 
The action for this model can be written as 
\begin{equation}\label{model_action}
    S=\int d^3\vec{x} d\eta \; \left( \frac{1}{2}(\partial \phi)^2 -\sum_{n\geq 3} \frac{\lambda_n(\eta)}{n!} \phi^n \right)\, ,
\end{equation}
where the $\lambda_n$ are time-dependent coupling constants $
    \lambda_n(\eta)=\lambda_{n,0} \left(a(\eta) \right)^{4-n}$
and $a(\eta)$ is the pre-factor of the FRW metric. The case we will consider consist of space-times which have 
\begin{equation}
    a(\eta)= \left(\frac{\eta}{\eta_0}\right)^{-(1+\epsilon)}\, ,
\end{equation}
where $\eta_0$ and $\epsilon$ will be kept arbitrary for now. 

\paragraph{Kinematic variables.} In this work, we focus on tree-level Feynman graphs. 
Although these functions can depend on all external momenta $\vec k_1,\ldots,\vec k_n$, their dependence is actually restricted to specific combinations of these momenta. We refer to these combinations as the kinematic variables, and they are defined as follows. Every vertex with label $v$ in the diagram comes with a vertex energy 
\begin{equation}
    X_v = \sum_{i} |\vec k_i|,
\end{equation}
with the sum running over all external propagators attached to the vertex. Meanwhile, every internal propagator is associated to an internal energy variable $Y$ given by the energy flowing over that edge, which can be written in terms of the external momenta $\vec k_i$. 
An explicit example can be found in section~\ref{ssec:singexchangeint} where the single exchange integral is discussed.

\paragraph{Wavefunction coefficients from diagrams.}

One can calculate the $n$-th wavefunction coefficient $\psi_n$ using a diagrammatic approach. One draws all the possible diagrams with $n$ external particles ending on a single time-slice $\eta=0$, as is drawn in figure~\ref{fig:generalinin}. 
\begin{figure} 
\centering

\begin{tikzpicture}
    \begin{feynman}
        \vertex (topleft);
        \vertex [right=1 cm of topleft,boundarydot] (topcenter1) {};
        \vertex [right=1 cm of topcenter1,boundarydot] (topcenter2) {};
        \vertex [right=0.5 cm of topcenter2,boundarydot] (topcenter);
        \vertex [below=1 cm of topcenter,bulkblob] (diagram) {};
        \vertex [right=.2 cm of topcenter] (topcenter3) {\quad \quad\quad};
        \vertex [right=1.3 cm of topcenter3,boundarydot] (topcenter4) {};
        \vertex [right =1 cm of topcenter4] (topright) {\color{blue} $\eta=0$};

        \diagram*  {
          (topleft) --[very thick,blue] (topcenter1) --[very thick,blue] (topcenter2) --[very thick,blue] (topcenter3) --[very thick,blue] (topcenter4) --[very thick,blue] (topright);
          (diagram) --[thick,opacity=1] (topcenter1);
          (diagram) --[thick,opacity=1]  (topcenter2);
          (diagram) --[thick,opacity=1] (topcenter4);
        };
        \vertex [above=0.5 cm of topcenter2] (k2) {$\Vec{k}_2$};
        \vertex[above=0.5 cm of topcenter1] (k1) {$\Vec{k}_1$};
        \vertex[above=0.5 cm of topcenter3] (dots) {$\cdots$};
        \vertex[above=0.5 cm of topcenter4] (kn) {$\Vec{k}_n$};
    \end{feynman}
\end{tikzpicture}
\caption{The general in-in diagram corresponding to $n$ external 
particles.}\label{fig:generalinin}
\end{figure}
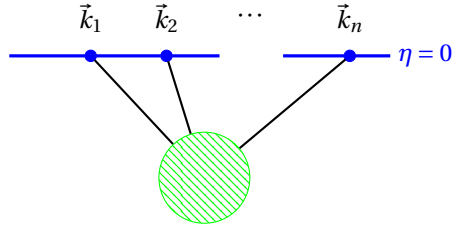

The crucial point in this evaluation is that the time of insertion is arbitrary and therefore needs to be integrated over. This means that every vertex in the Feynman diagram introduces an integral of the form
\begin{equation}
    \int_{-\infty}^0 d\eta_v\; \lambda_{n,0} \left( \frac{\eta}{\eta_0} \right)^{(n-4)(1+\epsilon)}\, .
\end{equation}
Therefore, even tree diagrams will contain integrals and are non-trivial. With this in mind one obtains the Feynman rules summarized in \cite{arkani-hamed_cosmological_2017} and can derive an integral expression for any in-in diagram of interest. 

In particular, the integrals associated to a particular diagram can be obtained schematically described in~\cite{arkani-hamed_cosmological_2017}. First, one writes down a factor of $e^{i\eta_v (X_v+x_v)}$ for each vertex $v$. Secondly, one inserts the bulk-to-bulk propagator $G_e(Y_e,\eta_{e_1},\eta_{e_2})$ for each edge $e$, where $\eta_{e_1}$ and $\eta_{e_2}$ are the variables associated to the vertices connected to the edge. The particular form of this propagator will not be important for us, but let us note that it is a solution the Green's function equation
\begin{equation}
    (\partial_{\eta_{e_1}}^2+Y_e^2)G_e=(\partial_{\eta_{e_2}}^2+Y_e^2)G_e=-i\delta(\eta_{e_1}-\eta_{e_2})\,.
\end{equation}
We will use this in section~\ref{sec:physics} to show that certain combinations of reduction operators correspond to contractions of an edge in a diagram. 

Finally, the value of the diagram is then given by the following integral
\begin{equation}\label{eq:wavefunctionpropas}
    \psi_G(X,Y;\epsilon)=\int_{\R_+^{\Nv}}d^{\Nv} x  \int_{\R_-^{\Nv}}d^{\Nv}\eta \prod_{v=1}^{\Nv} x_v^{\alpha_v-1} e^{i\eta_v(X_v+x_v)}\prod_{e=1}^{N_{\rm e}}G_e(Y_e,\eta_{e_1},\eta_{e_2})\,,
\end{equation}
where $\Nv$ is the number of vertices,  $N_{\rm e}$ the number of edges, 
and $\alpha_v$ depends on $\epsilon$ and the order of the interaction at the vertex $v$. In particular for an interaction of order $k$, $\alpha_v$ is given by 
\beq \label{alpha_nu-form}
   \alpha_v=(4-k)(1+\epsilon)
\eeq
with $\epsilon$ determining the FLRW scale factor. The variables $x_v$ effectively parameterize shifts in the kinematic variables $X_v$, and integrating over these shifts accounts for working in the FLRW spacetime. Note that these integrals also admit convenient diagrammatic interpretations as sums over so-called tubings of the diagram, a perspective which will discuss more in chapter~\ref{ch:reductionalgorithm}.

\section{A first application of reduction operators -- the single-exchange integral}
\label{sec:singexreducs}

In this section we will use the framework of GKZ systems and their reduction operators from chapter~\ref{ch:generalGKZ} to obtain a particular cosmological correlator: the single-exchange integral.
We will begin by describing the integral of interest and its physical origin in section~\ref{ssec:singexchangeint}. Afterwards, we provide the GKZ system this integral gives rise to in section~\ref{ssec:singexchGKZ}.
The resulting GKZ system is resonant, and hence reducible. From chapter~\ref{ch:generalGKZ} we know that this implies there must exist non-trivial reduction operators, and these are given in section~\ref{ssec:reduction_operators_ex}.
In sections~\ref{ssec:obtaining_solutions} and \ref{ssec:explicitsols} the reduction operators are used to construct partial solution bases, and using these to obtain the single-exchange integral. 
Afterwards, we show in section~\ref{ssec:qimoving} that the single-exchange integral also satisfies inhomogeneous differential equations, hinting towards the constructions of chapter~\ref{ch:reductionalgorithm}. This connection is then deepened in section~\ref{ssec:physics}, where we show that these inhomogeneous equations can be interpreted as local differential equations, analogously to the contraction and cut relations of section~\ref{sec:physics}.

\subsection{The single exchange integral}\label{ssec:singexchangeint}
\begin{figure}[h]
\centering
\begin{tikzpicture}
    \begin{feynman}
        \vertex (topleft);
        \vertex [right=1 cm of topleft,boundarydot] (topcenter1) {};
        \vertex [right=1 cm of topcenter1,boundarydot] (topcenter2) {};
        \vertex [right=0.5 cm of topcenter2,boundarydot] (topcenter);
        \vertex [below=1 cm of topcenter] (midcenter) ;
        \vertex[left=1cm of midcenter,bulkdot] (leftdiagram) {};
        \vertex[right=1 cm of midcenter,bulkdot] (rightdiagram){};
        \vertex [right=0.5 cm of topcenter,boundarydot] (topcenter3) {};
        \vertex [right=1 cm of topcenter3,boundarydot] (topcenter4) {};
        \vertex [right =1 cm of topcenter4] (topright) {\color{blue} $\eta=0$};

        \diagram*  {
          (topleft) --[very thick,blue] (topcenter1) --[very thick,blue] (topcenter2) --[very thick,blue] (topcenter3) --[very thick,blue] (topcenter4) --[very thick,blue] (topright);
          (leftdiagram) --[thick,opacity=1] (topcenter1);
          (leftdiagram) --[thick,opacity=1]  (topcenter2);
          (rightdiagram) --[thick,opacity=1] (topcenter3);
          (rightdiagram) --[thick,opacity=1] (topcenter4);
          (rightdiagram) --[very thick,green] (leftdiagram);
        };
        \vertex [above=0.5 cm of topcenter2] (k2) {$\Vec{k}_2$};
        \vertex[above = 0.5 cm of topcenter1] (k1) {$\Vec{k}_1$};
        \vertex[above = 0.5 cm of topcenter3] (k3) {$\Vec{k}_3$};
        \vertex[above = 0.5 cm of topcenter4] (k4) {$\Vec{k}_4$};
        \vertex[below=0.4 cm of leftdiagram] (X1) {\textbf{$X_1$}};
        \vertex[below=0.4 cm of rightdiagram] (X2) {\textbf{$X_2$}};
        \vertex[above=0cm of midcenter] (Y) {\textbf{\color{green} $Y$}};
    \end{feynman}
\end{tikzpicture}
\caption{The 4-point single-exchange diagram. Here $X_1=\vert \vec{k}_1\vert+\vert\vec{k}_2\vert$, $X_2=\vert \vec{k}_3\vert+\vert\vec{k}_4\vert$ and $Y=\vert \vec{k}_1+\vec{k}_2\vert=\vert \vec{k}_3+\vec{k}_4\vert$.  } \label{fig:singexchange}
\end{figure}

The diagram we will consider is shown in figure~\ref{fig:singexchange} and involves four external lines as well as a single propagator. Applying the Feynman rules discussed above results in the integral
\begin{equation}\label{eq:propaexchange}
\begin{split}
  -\lambda_{3,0}^2  \int_0^{\infty}\omega_1^\epsilon d\omega_1 \int_0^{\infty} \omega_2^\epsilon d\omega_2 \int_{-\infty}^0 d\eta_1 \int_{-\infty}^0 d\eta_2\; e^{i \eta_1(X_1+\omega_1)+i\eta_2(X_2+\omega_2)}G(Y,\eta_1,\eta_2)\, ,
\end{split}
\end{equation}
where $\lambda_{3,0}$ is the constant part of the three-point vertex and the exact form of the propagator is also given in~\cite{arkani-hamed_cosmological_2017}. Inserting this propagator and performing the $\eta_i$ integrals, one obtains
\begin{equation}\label{eq:singleexchange}
   I(X_1,X_2,Y,\epsilon)\coloneqq \int_0^\infty \frac{ \lambda_{3,0}^2(\omega_1 \omega_2)^\epsilon d\omega_1 d\omega_2}{(\omega_1+X_1+Y)(\omega_2+X_2+Y)(\omega_1+\omega_2+X_1+X_2)}\, .
\end{equation}
This is the integral we will study in the remainder of this chapter.

\paragraph{Comments on locality.} 

The single-exchange integral \eqref{eq:singleexchange} has special properties that are linked to the locality of the underlying theory \cite{arkani-hamed_differential_2023}. To make this more precise we replace the propagator $G(Y,\eta_1,\eta_2)$ by $-i \delta(\eta_1-\eta_2)$, which 
is equivalent to collapsing the propagator to a point.
The integrations over the $\omega_i$ can then be performed explicitly and result in 
\begin{equation}\label{eq:icont1}
   I_\mathrm{contr}= -2^{2(\epsilon +1)}\sqrt{\pi} \text{csc}(\pi \epsilon) \Gamma\big(-\epsilon -\tfrac{1}{2} \big) \Gamma(\epsilon+1) (X_2+X_1)^{2 \epsilon +1}\, ,
\end{equation}
where we have defined $I_\mathrm{contr}$ to be the contracted integral. 
Alternatively, we can use the identity
\begin{equation}
    (\partial_{\eta_1}^2+Y^2)G(Y,\eta_1,\eta_2)=-i \delta(\eta_1-\eta_2)
\end{equation}
in the integrand and integrate by parts twice.
After integration over $\eta_1$ and $\eta_2$ this leads to the alternative expression
\begin{equation}\label{eq:icont2}
   I_\mathrm{contr}= \int_0^\infty \frac{ \lambda_{3,0}^2(\omega_1^2+2\omega_1 X_1+X_1^2-Y^2) (\omega_1 \omega_2)^\epsilon d\omega_1 d\omega_2}{(\omega_1+X_1+Y)(\omega_2+X_2+Y)(\omega_1+\omega_2+X_1+X_2)}\ .
\end{equation}
The crucial observation is that it is now possible to translate the equality of equations~\eqref{eq:icont1} and~\eqref{eq:icont2} to a differential equation on $I$ by repeatedly applying the integration by parts identity
\begin{equation}\begin{split}
    &\frac{\partial}{\partial X_1} \int_0^\infty \frac{ \omega_1^\alpha \omega_2^\epsilon d\omega_1 d\omega_2}{(\omega_1+X_1+Y)(\omega_2+X_2+Y)(\omega_1+\omega_2+X_1+X_2)}\\
    =& -\alpha \int_0^\infty \frac{ \omega_1^{\alpha-1}\omega_2^\epsilon d\omega_1 d\omega_2}{(\omega_1+X_1+Y)(\omega_2+X_2+Y)(\omega_1+\omega_2+X_1+X_2)}
    \end{split}
\end{equation}
to equation~\eqref{eq:icont2}. This leads to the differential equation
\begin{equation}\label{eq:locality}
    \left(X_1^2-Y^2\right) \frac{\partial^2 I}{\partial X_1^2}+2 X_1(1-\epsilon) \frac{\partial I}{\partial X_1}- \epsilon (1-\epsilon)I=\frac{\partial^2 I_{\mathrm{contr}}}{\partial X_1^2}\, .
\end{equation}
This differential equation relates the single exchange integral with the contracted integral $I_\mathrm{contr}$ that we evaluated in equation~\eqref{eq:icont1}. Therefore, one can loosely think of the differential operator on the left hand-side as contracting the propagator when acting on $I$. Note that a similar derivation can be done for $X_2$, also leading to a second order differential equation for $I$. 

Interestingly, the differential operator on the left hand side only involves $X_1$ and derivatives with respect to $X_1$. As we will see below, $I$ solves a pair of second order differential equations which mixes $X_1$ and $X_2$, as is generally the case. However, due to locality one finds that these differential equation must split, a highly non-trivial property. It turns out that this splitting is deeply connected to the reducibility of the associated GKZ system, a fact which we will explore further in section~\ref{ssec:physics}. Before this, we will need to discuss the GKZ system associated to this integral.

\subsection{GKZ system for the single-exchange integral}
\label{ssec:singexchGKZ}

Having discussed the single-exchange integral, we are now ready to apply the formalism of chapter~\ref{ch:generalGKZ} and study its GKZ system. To do this, it is first necessary to cast it into the form of~\eqref{eq:gengkzint} by promoting the coefficients in the polynomials to parameters $z_I$ and the exponents to complex numbers $\alpha_i$ and $\beta_i$. For the single exchange integral this results in the Euler integral
\begin{equation}\label{eq:exchangegkz}
    I_\Gamma(z;\nu)=\int_\Gamma d^2\omega\frac{\omega_1^{\alpha_1-1} \omega_2^{\alpha_2-1}}{(z_1+z_2 \omega_1)^{\beta_1}(z_3 +z_4 \omega_2)^{\beta_2}(z_5 +z_6 \omega_1 +z_7 \omega_2)^{\beta_3}}\ ,
\end{equation}
and the original integral can be recovered by inserting
\begin{equation}\label{eq:physvars}
    \begin{array}{rl}
z&=(X_1+Y,1,X_2+Y,1,X_1+X_2,1,1)\ ,\\
\nu&=(\beta,\alpha)=(1,1,1,1+\epsilon,1+\epsilon)^T\ ,\\
\Gamma&=\R_+^2\, .
    \end{array}
\end{equation} 
Comparing the form of the integral above with the general form of equation~\eqref{eq:gengkzint} one sees that it consists of the three polynomials
\begin{equation}
    \begin{array}{rl}
        p_1(\omega,z) &= z_1+z_2 \omega_1\ , \\
        p_2(\omega,z) &= z_3+z_4 \omega_2 \ ,\\
        p_3(\omega,z) &= z_5+z_6 \omega_1 + z_7 \omega_2 \, .
    \end{array}
\end{equation}
Continuing with the process described in section~\ref{ch:generalGKZ}, we define the matrices
\begin{equation}
    \begin{array}{rlrlrl}
       \A_1  &\coloneqq  \begin{pmatrix}
           0 & 1 \\
           0 & 0
       \end{pmatrix},&
       \A_2  &\coloneqq \begin{pmatrix}
           0 & 0 \\
           0 & 1
       \end{pmatrix},&
        \A_3  &\coloneqq  \begin{pmatrix}
           0 & 1 & 0\\
           0 & 0 & 1
       \end{pmatrix}
    \end{array}
\end{equation}
from the polynomials $p_j$. These matrices can be combined into the matrix
\begin{equation}\label{eq:exchangeA}
 \A =\left(
\begin{array}{ccccccc}
 1 & 1 & 0 & 0 & 0 & 0 & 0 \\
 0 & 0 & 1 & 1 & 0 & 0 & 0 \\
 0 & 0 & 0 & 0 & 1 & 1 & 1 \\
 0 & 1 & 0 & 0 & 0 & 1 & 0 \\
 0 & 0 & 0 & 1 & 0 & 0 & 1 \\
\end{array}
\right)\, ,
\end{equation}
where the bottom two rows correspond to the matrices $\A_1$, $\A_2$ and $\A_3$. From this matrix we will now obtain the differential operators of the GKZ system.

We first determine the toric operators. Since $\mathrm{dim}(\ker(\A))=2$, there are two linearly independent solutions to
\begin{equation}
    \A u= \A v\, ,
\end{equation}
with $u$ and $v$ in $\N^7$. Here we will consider
\begin{equation}
\begin{array}{cccc}
    u_1= \begin{pmatrix}1\\0\\0\\0\\0\\1\\0\end{pmatrix} ,& v_1= \begin{pmatrix}0\\1\\0\\0\\1\\0\\0\end{pmatrix} ,&u_2= \begin{pmatrix}0\\0\\1\\0\\0\\0\\1\end{pmatrix} ,&v_2= \begin{pmatrix}0\\0\\0\\1\\1\\0\\0\end{pmatrix} ,
\end{array}
\end{equation}
satisfying $\A u_1=\A v_1$ and $\A u_2 =\A v_2$. Using the definition of the toric operators from equation~\eqref{eq:toricopdef}, these result in the toric operators
\begin{equation}\label{eq:exchangetoricop}
    \begin{array}{rl}
         \L_{u_1,v_1}&=\partial_1 \partial_6-\partial_2\partial_5\, ,\\
         \L_{u_2,v_2}&=\partial_3 \partial_7 -\partial_4 \partial_5\, ,
    \end{array}
\end{equation}
and toric equations
\begin{equation}\label{eq:exchangetoric}
    \L_{u_1,v_1}f(z;\nu)=\L_{u_2,v_2}f(z;\nu)=0 \, .
\end{equation}
From these and equation~\eqref{eq:sidef}, the parameters $s_i$ can be obtained, resulting in
\begin{equation}\label{eq:sumvars}
    \begin{split}
        s=\frac{z_1 z_6}{z_2 z_5}\, ,& \quad  t=\frac{z_3 z_7}{z_4z_5}\ ,
    \end{split}
\end{equation}
where we have relabeled $s=s_1$ and $t=s_2$.

Now, the Euler operators are obtained from the matrix $\A$ by calculating $\E=\A\Theta$, and take the form
\begin{equation}\label{eq:exchangeeuler}
    \begin{array}{lll}
       \E_1= \theta_1+\theta_2\, , & \E_2=\theta_3 +\theta_4\, , & \E_3=\theta_5 +\theta_6 +\theta_7\ ,\\
        \E_4=\theta_2+\theta_6 \, , & \E_5=\theta_4+\theta_7 \, . &
    \end{array}
\end{equation}
The Euler equations following from these operators are
\begin{equation}
    (\E_J+\nu_J)f(z;\nu)=0\ ,
\end{equation}
for $1\leq J \leq 5$. These two sets of differential equations describe the complete GKZ system associated to the integral~\eqref{eq:exchangegkz}. 

To obtain the solution basis for this system is somewhat non-trivial. While it is easy to obtain a basis of converging series expansions as described in section~\ref{sec:genGKZ}, these infinite sums are not easily evaluated, even when specializing the parameters $z_I$ and $\nu$ to the physically interesting values from equation~\eqref{eq:physvars}.\footnote{Using some tricks it is actually possible to perform the summations for this particular system, however we consider this system in this chapter because it very nicely illustrates the reduction technique described below.} Alternatively, one can try to make an ansatz like
\begin{equation}\label{eq:generalansatz}
    f(z;\nu)=\frac{z_3^{\nu_5}z_5^{\nu_4}}{z_1^{\nu_1} z_3^{\nu_2} z_4^{\nu_5}z_5^{\nu_3}z_6^{\nu_4}}g(s,t;\nu)
\end{equation}
such as described in equation~\eqref{eq:fpgansatz}, but it is not clear how to solve the resulting coupled system of partial differential equations. However, this GKZ system is reducible which enables us to obtain its solutions individually, without having to solve the entire system at once. In the following section, we will see how reducibility allows us to do this.

\subsection{Determining the reduction operators}\label{ssec:reduction_operators_ex}

We begin our analysis of the single exchange GKZ system by determining the reduction operators, in this section we will use the first method described in section~\ref{ssec:reductionconstructionexplicit}, this is in contrast to chapter~\ref{ch:reductionalgorithm} where we will construct reduction operators directly from faces.
To obtain the different reduction operators, recall that for each $I$, it is necessary to solve
\begin{equation}
    P_I (\nu)\cdot (\E+\nu-a_I)\simeq_I 0\, ,
\end{equation}
where $\simeq_I$ denotes that we apply the toric equivalence relations as well as set $\partial_I$ to zero. This makes it possible to obtain a reduction operator by solving
\begin{equation}\label{eq:pi=qi}
    P_I (\nu)\cdot (\E+\nu-a_I)\simeq Q_I(\nu)\partial_I
\end{equation}
for $Q_I(\nu)$, where the toric relations are now considered without $\partial_I$ set to zero. 

For convenience, let us recall that the defining data of this GKZ system is 
\beq
    \A =\left(
\begin{array}{ccccccc}
 1 & 1 & 0 & 0 & 0 & 0 & 0 \\
 0 & 0 & 1 & 1 & 0 & 0 & 0 \\
 0 & 0 & 0 & 0 & 1 & 1 & 1 \\
 0 & 1 & 0 & 0 & 0 & 1 & 0 \\
 0 & 0 & 0 & 1 & 0 & 0 & 1 \\
\end{array}
\right)\, , \qquad  \nu =\begin{pmatrix}1\\ 1\\ 1\\ 1+\epsilon\\1+\epsilon\end{pmatrix}\ . 
\eeq
Since $\A,\nu$ will be fixed throughout this section, we will omit the dependence on $\nu$ of the reduction operators and the functions, and instead write $f(z)=f(z;\nu)$, $Q_I=Q_I(\nu)$ and $P_I=P_I(\nu)$.

\paragraph{Determining $Q_1$.} Let us use this procedure to obtain $Q_1$. Recall that, for the single exchange integral, the toric operators are given by
\begin{equation}
    \begin{array}{rl}
         \L_{u_1,v_1}&=\partial_1 \partial_6-\partial_2\partial_5\\
         \L_{u_2,v_2}&=\partial_3 \partial_7 -\partial_4 \partial_5\, .
    \end{array}
\end{equation}
Upon setting $\partial_1$ to zero, these reduce to
\begin{equation}\label{eq:toricd1=0}
\begin{array}{rl}
    (\L_{u_1,v_1})|_{\partial_1\rightarrow 0} &=-\partial_2\partial_5\, ,\\
    (\L_{u_2,v_2})|_{\partial_1\rightarrow 0} &= \partial_3 \partial_7 -\partial_4 \partial_5\, .
\end{array}
\end{equation}
Furthermore, the Euler operators are also affected by the replacement $\partial_1 \rightarrow 0$. In particular, inspecting the first Euler operator in equation~\eqref{eq:exchangeeuler}, we see that
\begin{equation}
    (\E_1+(\nu-a_1)_1)|_{\partial_1\rightarrow 0}=(\theta_1+\theta_2))|_{\partial_1\rightarrow 0}= \theta_2\, ,
\end{equation}
where we recall that $\theta_I=z_I \partial_I$ 
and use $\nu-a_1=(0,1,1,1+\epsilon,1+\epsilon)^T$. Combining these observations, we find that
\begin{equation}
    \partial_5 (\E_1+ (\nu-a_1)_1) = z_1 \partial_1 \partial_5+z_2 \partial_2 \partial_5 \simeq_1 0 \ .
\end{equation}
This leads us to consider $P_1=(\partial_5,0,0,0,0)^T$ satisfying
\begin{equation}
    P_1 \cdot (\E+\nu-a_1)=z_1 \partial_1 \partial_5+z_2 \partial_2 \partial_5 \simeq_1 0\, .
\end{equation}
 $P_1$ will give rise to a reduction operator $Q_1$, which can be obtained by solving equation~\eqref{eq:pi=qi}.
 
 To obtain $Q_1$, it is necessary to apply the toric equivalence relations without setting $\partial_1 \rightarrow 0$. This yields
\begin{equation}
    P_1 \cdot (\E+\nu-a_1)\simeq (z_1 \partial_5+z_2 \partial_6)\partial_1\, ,
\end{equation}
where we have used $\partial_1 \partial_6 \simeq \partial_2 \partial_5$. From this expression, we deduce that $Q_1$ has the form
\begin{equation}
    Q_1=z_1 \partial_5+z_2 \partial_6\,.
\end{equation}
Thus we have found the reduction operator at $\nu$ in the direction $1$.

\paragraph{The complete set of reduction operators.} Repeating a similar procedure for the other values of $I$ gives rise to their associated reduction operators. In particular, choosing
\begin{equation}
    \begin{array}{ll}
        P_2 =(\partial_6,0,0,0,0)^T \, ,& P_3 =(0,\partial_5,0,0,0)^T  \, ,\\
        P_4 =(0,\partial_7,0,0,0)^T  \, ,&P_5 =(0,0,\partial_1 \partial_3,0,0)^T \, , \\
        P_6 =(0,0,\partial_2\partial_3,0,0)^T \, , &P_7 =(0,0,\partial_1\partial_4,0,0)^T \, , \\
    \end{array}
\end{equation}
results in the reduction operators
\begin{equation}\label{eq:Qidefs}
\boxed{\rule[-.1cm]{0cm}{.5cm}\quad 
\begin{array}{rl}
    Q_1=Q_2= & z_1 \partial_5+z_2 \partial_6\, , \\
    Q_3=Q_4= & z_3 \partial_5+z_4 \partial_7\, ,\\
    Q_5=Q_6=Q_7=&z_5 \partial_1 \partial_3 + z_6 \partial_2 \partial_3 +z_7 \partial_1 \partial_4\, .
\end{array}\quad}
\end{equation}

We would like to highlight how the reducibility of the GKZ system manifests itself through these reduction operators. First of all, note that each $P_I$ is zero in its last two elements. This is due to the fact that the last two elements of $\nu$ involve the arbitrary complex number $\epsilon$. One can view the operator vectors $P_I$ as determined by the relations between the different operators $\E_J+\nu_J-(a_I)_J$. Having undetermined parameters in $\nu$ then restricts which relations are possible. Secondly, many of the reduction operators are equal. This is an artefact of their associated partial derivatives not being independent, in a precise mathematical sense.\footnote{Precisely, a set of partial derivatives is considered independent if they define a regular sequence on a ring of differential operators, defined in equation~\eqref{eq:SAdef}.} As a result, these partial derivatives, and consequently their respective reduction operators, play a similar role in the GKZ system.

\subsection{Obtaining solutions using reduction operators} \label{ssec:obtaining_solutions}

To obtain the solutions associated with the different combinations of reduction operators, we first make an ansatz similar of the form~\eqref{eq:fpgansatz}, in order to reduce the number of independent variables. Then, we can act on this ansatz with the reduction operators in order to obtain new differential equations. The single exchange GKZ system has four linearly independent solutions and, as we will see, these are annihilated by different combinations of the reduction operators. Here, we will consider the following subsystems:
\begin{equation}\label{eq:subsystems} \boxed{\rule[-.1cm]{0cm}{1cm}\quad  
\begin{array}{lr}
     \text{subsystem $1$, satisfying:}  & Q_1f(z)=Q_3f(z)=0 \, ,\\
     \text{subsystem $2$, satisfying:}  & Q_1f(z)=Q_5f(z)=0 \, ,\\
     \text{subsystem $3$, satisfying:}  & Q_3f(z)=Q_5f(z)=0 \, ,\\
     \text{subsystem $4$, satisfying:}  & Q_5f(z)=0 \, ,
    \end{array}\quad}
\end{equation}
Note that the solutions of subsystem $2$ and $3$ will also be solutions of subsystem~$4$. However, it will be useful for us to consider these subsystems separately.

\paragraph{Solutions of subsystem $1$.} 

Denoting the solutions to the first subsystem by $f_1$, these functions must satisfy
\begin{equation}\label{eq:q1f1=0}
    Q_1f_1(z)=Q_3f_1(z)=0\, .
\end{equation}
To find these solutions, it will be useful to begin with the ansatz
\begin{equation}\label{eq:f1ansatz}
    f_1(z)= \frac{\big(\frac{z_1 z_3}{z_2 z_4}\big)^\epsilon }{z_2 z_4 z_5}g_{1}(s,t)\, ,
\end{equation}
where $s$ and $t$ are defined in~\eqref{eq:sumvars} and $g_1$ is a yet unknown function. This ansatz is of the type found in equation~\eqref{eq:fpgansatz}, where we have a known pre-factor combined with an unknown function of the homogeneous variables.

Inserting the ansatz into 
\begin{equation}\label{eq:qif1=0}
    Q_1 f_1(z)=0
\end{equation}
and using the explicit expression for $Q_1$ given in equation~\eqref{eq:Qidefs}, we find that $g_{1}$ must solve
\begin{equation}
(t-1)\, \frac{\partial g_1(s,t)}{\partial t}+s \, \frac{\partial g_1(s,t)}{\partial s}+g_1(s,t)=0\, .
\end{equation}
This differential equation can be solved for $g_1$, and inserting the solution into equation~\eqref{eq:f1ansatz} results in
\begin{equation}\label{eq:Q1sol}
    f_1(z)=\frac{\big(\frac{z_1 z_3}{z_2 z_4}\big)^{\epsilon } }{z_2 z_4 z_5-z_1 z_4 z_6}\; h_1\left(\frac{z_2 z_3 z_7}{z_1 z_4 z_6-z_2 z_4 z_5}\right)\, ,
\end{equation}
where $h_1$ is yet to be determined. Note that the solutions of both the first and the second subsystem must satisfy equation~\eqref{eq:q1f1=0}. Therefore, solutions to both subsystems must be of the form~\eqref{eq:Q1sol}.

Solutions to the first subsystem will also satisfy
\begin{equation}
    Q_3  f_1(z)=0\, ,
\end{equation}
giving rise to another differential equation. Applying $Q_3$ to the solution in equation~\eqref{eq:Q1sol}, we find that $h_1$ must solve
\begin{equation}
     (u+1) \frac{\partial h_1(u)}{\partial u}+h_1(u)=0\, ,
\end{equation}
where we have defined the variable
\begin{equation}
    u\coloneqq \frac{z_2 z_3 z_7}{z_1 z_4 z_6-z_2 z_4 z_5}\, .
\end{equation}
Solving this differential equation for $h_1$ and inserting the solution into~\eqref{eq:Q1sol}, we find that the solution must be proportional
\begin{equation}\label{eq:f1sol}
 \boxed{\rule[-.1cm]{0cm}{.9cm}\quad     f_1(z)=\frac{ \big(\frac{z_1 z_3}{z_2 z_4}\big)^{\epsilon }}{z_2 z_4 z_5-z_1 z_4 z_6-z_2 z_3 z_7}
 \quad}\,.
\end{equation}
Therefore, the first subsystem has only a single linearly independent solution.

Let us higlight the simplifications that the reduction operators provided here. The full GKZ system consists of five Euler equations and two toric equations, resulting in a system of seven partial differential equations in seven variables. Applying the ansatz of the type~\eqref{eq:generalansatz} already reduces this to only the two toric equations in two variables, however this still results in a system of two coupled second order partial differential equations which is highly nontrivial to solve directly. With the reduction operators though, it was possible to find a solution to this system by solving only two first order differential equations, both of which were easily solved using the general ansatz. Furthermore, we note that, at any convenient point, it is possible to use the toric or Euler equations in order to solve for multiple functions at once. For example, since $f_2$ also satisfies $Q_1f_2=0$, it too takes the form of equation~\eqref{eq:Q1sol}. Inserting this into the toric equations then allows us to find both solutions at once.

\paragraph{Solving the second and third subsystem.} The process for finding the solutions to the second and third subsystem is quite similar to what we have already seen. For example, one can insert the general solution to $Q_1f_2=0$ into $Q_5f_2=0$ and find that the solutions must be proportional to
\begin{equation}\label{eq:f2sol}
 \boxed{\rule[-.1cm]{0cm}{1cm}\quad          f_{2}(z)=\frac{\left(\frac{z_1 \left(z_1 z_6-z_2 z_5\right)}{z_2^2 z_7}\right)^\epsilon}{z_2 z_4 z_5-z_1 z_4 z_6-z_2 z_3 z_7}\,.
\quad}
\end{equation}
Similarly, the equations
\begin{equation}
    Q_3  f_3(z)=Q_5  f_3(z)=0
\end{equation}
have solutions proportional to
\begin{equation}\label{eq:f3sol}
\boxed{\rule[-.1cm]{0cm}{1cm}\quad  
   f_3(z) =\frac{\left(\frac{z_3 \left(z_3 z_7-z_4 z_5\right)}{z_4^2 z_6}\right)^\epsilon}{z_2 z_4 z_5-z_1 z_4 z_6-z_2 z_3 z_7}
   \quad}\,.
\end{equation}
Note that the first three subsystems all consist of only a single linearly independent solution. Now, all that remains is to find the solutions to the last subsystem.

 \paragraph{Ans\"atze from partial solution bases.} 

Finding the solutions to the final subsystem is greatly simplified by using one of a few possible tricks available. One such trick is to use the Euler operators to rewrite the reduction operator $Q_5$ in a more convenient form. An example of this procedure is provided in section~\ref{ssec:qimoving} to obtain $f_1$, but it can be applied here just as well. However, for instructional purposes we want to highlight an additional possible simplification. Namely, that knowing some of the solutions to a set of differential equations often makes it easier to obtain the remaining solutions as well. For ordinary differential equations this technique is known as reduction of order \cite{howell_ordinary_2020} and always leads to simplifications in a very precise manner. However, in practice, one finds that such simplifications also appear often when dealing with partial differential equations.

The main idea is that, knowing a solution $f(z)$ to a differential equation, one can make an ansatz of the form $f(z)g(z)$ and solve for $g(z)$. Since $g(z)$ being a constant must be a solution of this differential equation, the differential equation will only involve derivatives of $g(z)$, while not involving $g(z)$ itself. Defining $u(z)\coloneqq \partial g(z)/ \partial z$ this leads to a differential equation of lower order for $u$. One can then solve this and integrate the solutions to obtain the general solution. As a generalization, for a number of known solutions $f_i$, one can take an ansatz of the form
\begin{equation}
    \sum_i f_i(z) g_i(z)
\end{equation}
where the $g_i$ now are functions to be solved for. The extra degree of freedom from the different $g_i$ can now be used to cancel other terms in the differential equation as well, providing even further simplifications. For partial differential equations these procedures can be a bit more subtle. However, even in these cases Ans\"atze such as described above still often lead to significant simplifications.

\paragraph{General solution of the fourth subsystem.}

To illustrate this, we will use the solutions $f_2$ and $f_3$ in order to obtain simpler differential equations for $f_4$. We begin by making the ansatz
\begin{equation}
    f_4(z)=f_2(z) g_4(s,t)\, ,
\end{equation}
where $f_2$ is given in equation~\eqref{eq:f2sol} and $g_4$ is an unknown function. Inserting this ansatz into $Q_5f_4(z)=0$ implies that $g_4$ must solve
\begin{equation}
   \left( \epsilon\,(2 s-1) +(s-1) s\, \frac{\partial}{\partial s}\right) \frac{\partial g_4(s,t)}{\partial t}=0\, .
\end{equation}
Note that defining $u(s,t)\coloneqq\partial g_4(s,t)/\partial t$ this becomes an ordinary differential equation for $u$. Solving this equation results in
\begin{equation}
    g(s,t)=h_1(s)+\frac{h_2(t)}{(s-s^2)^{\epsilon}}\, ,
\end{equation}
where $h_1$ and $h_2$ are undetermined functions. Inserting this into into the expression for $f_4$ and performing some function redefinitions for $h_1$ and $h_2$, this implies that the general solution to $Q_5f_4(z)=0$ can be written as
\begin{equation}\label{eq:f4h12ansatz}
    f_4(z)=f_2(z) h_1(s)+f_3(z)h_2(t)\, ,
\end{equation}
where $f_3$ is as in equation~\eqref{eq:f3sol}.

\paragraph{Obtaining the remaining linearly independent solution.} 
Using the general solution to $Q_5f_4(z)=0$, we will now obtain the fourth and final basis function, providing us with a full basis of linearly independent solutions to the GKZ system. The advantage of the ansatz~\eqref{eq:f4h12ansatz} is that the resulting differential equations only involve the derivatives of $h_1$ and $h_2$, while not involving $h_1$ and $h_2$ directly. For example, the first toric equation
\begin{equation}
    \L_{u_1,v_1}f_4(z)=0
\end{equation}
reduces to a differential equation of the form
\begin{equation}\label{eq:h1h2eq}
    p_1(s,t) \frac{\partial h_1(s)}{\partial s}+p_2(s,t) \frac{\partial^2 h_1(s)}{\partial s^2}+p_3(s,t) \frac{\partial h_2(t)}{\partial t}=0\, ,
\end{equation}
where the $p_i$ are polynomials in $s,t$ and $\epsilon$. Solving for $\partial h_2/\partial t$ and imposing that
\begin{equation}
    \frac{\partial^2 h_2(t)}{\partial s\partial t }=0
\end{equation}
results in a third order differential equation for $h_1(t)$. Or, since the differential equation does not involve $h_1(t)$ but only its derivatives, a second order differential equation for $\partial h_1(t)/\partial t$. The solution to this differential equation is given by
\begin{equation}
    h_1(t)=\frac{c_1\,_2F_1(1,-2\epsilon,1-\epsilon;1-s)}{(s-s^2)^{\epsilon } }+\frac{c_2 }{\left(s^2-s\right)^{\epsilon }}+c_3\, ,
\end{equation}
where the $c_i$ are undetermined constants and $_2F_1$ is the hypergeometric function. However, inserting this into expression~\eqref{eq:f4h12ansatz} for $f_4$ one finds that the term proportional to $c_2$ can be absorbed into $h_1(s)$, while $c_3$ reproduces a solution proportional to $f_2$. Since we are only interested in finding new solutions to the GKZ system, we will set $c_2=c_3=0$ from now on.

Inserting the solution for $h_2$ into equation~\eqref{eq:h1h2eq} results in the first order inhomogeneous differential equation
\begin{equation}
    \frac{\partial h_2(t)}{\partial t}=\frac{-\epsilon \,c_1}{ (t-t^2)^{\epsilon}}
\end{equation}
for $h_2$, with as its solution
\begin{equation}
    h_2(t)=\frac{c_1\,_2F_1(1,-2\epsilon,1-\epsilon;1-t)}{(t-t^2)^{\epsilon } }+c_4\, ,
\end{equation}
where $c_4$ is a new integration constant. Again, inserting this in to our expression for $f_4$ one finds that the $c_4$ term is proportional to $f_3$ and therefore we will set $c_4=0$ as well. Finally, we will insert the expressions for $h_1$ and $h_2$ into equation~\eqref{eq:f4h12ansatz} resulting in
\begin{equation}\label{eq:f4sol}
\boxed{\rule[-.1cm]{0cm}{1cm}\quad  
\begin{aligned}
    f_4(z)=& \quad\frac{\left(\frac{z_5^2}{z_6 z_7}\right){}^\epsilon \, _2F_1\left(1,-2 \epsilon ;1-\epsilon ;1-\frac{z_1 z_6}{z_2 z_5}\right)}{z_1 z_4 z_6+z_2 z_3 z_7-z_2 z_4 z_5}\\
    &+\frac{\left(\frac{z_5^2}{z_6 z_7}\right){}^\epsilon \, _2F_1\left(1,-2 \epsilon ;1-\epsilon ;1-\frac{z_3 z_7}{z_4 z_5}\right)}{z_1 z_4 z_6+z_2 z_3 z_7-z_2 z_4 z_5}\, ,
\end{aligned}\quad}
\end{equation}
where, since we are only interested in linearly independent solutions, we have set $c_1=1$.

With the four functions~\eqref{eq:f1sol},~\eqref{eq:f2sol},~\eqref{eq:f3sol}, and~\eqref{eq:f4sol} we have a full basis of solutions to the GKZ system at the parameter $\nu$ and, since the original integral in equation~\eqref{eq:singleexchange} is a solution to this system of differential equations, it should be some linear combination of these four functions. We will now provide these coefficients, as well as the explicit form of the solution in terms of the physical variables $X_1$, $X_2$ and $Y$.

\subsection{Solutions in terms of the physical variables}\label{ssec:explicitsols}

Having found the four different functions spanning the solution space of our GKZ system in section~\ref{ssec:obtaining_solutions}, it is now possible to write any convergent integral of the type ~\eqref{eq:exchangegkz} with $\nu=(1,1,1,1+\epsilon,1+\epsilon)^T$ as
\begin{equation}\label{eq:igammagensol}
    I_\Gamma(z;\nu)=c_1(\Gamma;\epsilon)f_1(z)+c_2(\Gamma;\epsilon)f_4(z)c_3(\Gamma;\epsilon)f_{\nu,3}(z)+c_4(\Gamma;\epsilon)f_4(z)\, ,
\end{equation}
where the functions $f_{i}$ are given in equations~\eqref{eq:f1sol},~\eqref{eq:f2sol},~\eqref{eq:f3sol} and~\eqref{eq:f4sol} while the coefficients $c_i(\Gamma;\epsilon)$ can be obtained fixing $\epsilon$ and an integration cycle $\Gamma$, or analytically by considering the integral in specific limits and considering the integral in certain limits. 

The single exchange integral~\eqref{eq:singleexchange} is a special case of the function~\eqref{eq:igammagensol}, obtained by setting the variables $z_I$ to their physical values from equation~\eqref{eq:physvars} and taking the integration cycle $\R^2_+$. With these replacements, the basis functions take the form
\begin{equation}\label{eq:gensols}
\begin{split}
    f_1(z)\vert_{\mathrm{phys}} &=\frac{(X_1+Y)^{\epsilon } (X_2+Y)^{\epsilon }}{2 Y} \, ,\\
     f_2(z)\vert_{\mathrm{phys}} &=\frac{(X_1+Y)^{\epsilon } (X_2-Y)^{\epsilon }}{2 Y}\, , \\
     f_3(z)\vert_{\mathrm{phys}} &=\frac{(X_1-Y)^{\epsilon } (X_2+Y)^{\epsilon }}{2 Y}\, ,\\
     f_4(z)\vert_{\mathrm{phys}} &= \frac{(X_1+X_2)^{2 \epsilon } \left(\, _2F_1\left(1,-2 \epsilon ;1-\epsilon ;\frac{X_1-Y}{X_1+X_2}\right)\right)}{2 Y}\\
     &+ \frac{(X_1+X_2)^{2 \epsilon } \left(\, _2F_1\left(1,-2 \epsilon ;1-\epsilon ;\frac{X_2-Y}{X_1+X_2}\right)\right)}{2 Y}\, ,\\
\end{split}
\end{equation}
where $\vert_\mathrm{phys}$ means that we replace the $z_I$ with their physical values. 

The different coefficients can be obtained by evaluating the functions above, as well as the single exchange integral in certain limits of $X_1$, $X_2$ and $Y$. This results in
\begin{equation}\label{eq:coefssols}
\begin{array}{rl}
     c_1(\R_+^2;\epsilon) &= -\pi ^2 \csc ^2(\pi  \epsilon )\, ,\\
     c_2(\R_+^2;\epsilon) &=0\, ,\\
     c_3(\R_+^2;\epsilon) &=0\, ,\\
     c_4(\R_+^2;\epsilon) &= 2^{-2 \epsilon -1} \sqrt{\pi } \csc (\pi  \epsilon ) \Gamma \left(\frac{1}{2}-\epsilon \right) \Gamma (\epsilon )\, ,
\end{array}
\end{equation}
for the different coefficients. Inserting all of the above into equation~\eqref{eq:igammagensol} we obtain an expression for the single exchange integral. This expression is in agreement with \cite{arkani-hamed_differential_2023} after applying a series of hypergeometric identities. Interestingly, the reduction operators also imply a series of inhomogeneous equations satisfied by the single-exchange integral. In turn, these also lead to boundary conditions using which the coefficients $c_i$ can be obtained as well. We will discuss this in more detail in section~\ref{ssec:physics}.

\subsection{An alternative: relating subsystems using reduction operators}\label{ssec:qimoving}

Before we go on discussing the relation between the reduction operators and locality, we want to highlight an alternative method that one can use to obtain solutions using the reduction operators. This method only works if the reduction operator commutes with the toric operators. However, one finds that this is the case in many examples, including our main example of section \ref{sec:singexreducs}. Furthermore, in chapter~\ref{ch:reductionalgorithm}, we show that this can be extended to all tree-level cosmological correlators.

Recall from section~\ref{ssec:reductionoperators} that, if a reduction operator commutes with the toric operators, it provides a map 
\begin{equation}\label{eq:qimap}
    f(z;\nu)\mapsto Q_If(z;\nu)
\end{equation}
between solutions of the GKZ system at $\nu$ to solutions at some different parameter. And this parameter can be obtained by considering the commutation relations
\begin{equation}
    [\E_J,Q_I(\nu)]=-(q_I(\nu))_J Q_I(\nu)\, ,
\end{equation}
where $q_I$ is an integer vector. 

One reason this map is useful is that the reduction operators satisfy
\begin{equation}
    \partial_I Q_I(\nu)f(z;\nu)=0\, ,
\end{equation}
for all solutions $f$ of the GKZ system at $\nu$. This implies that the image of the map~\eqref{eq:qimap} consists of solutions $\tilde f$ of the GKZ system at $\nu+q_I(\nu)$ satisfying
\begin{equation}
    \partial_I \tilde f(z;\nu+q_I(\nu))=0\, .
\end{equation}
In the notation introduced at the beginning of section \ref{ssec:reductionoperators}, the $\tilde f$ is among the solutions with index $t$. These relations provide the link between the two different kinds of subsystems. 
Therefore, if we can find a solution $\tilde f(z;\nu+q_I(\nu))$ at $\nu+q_I(\nu)$ that is independent of $z_I$, it is possible to obtain new solutions $f$ at $\nu$ by solving the inhomogeneous differential equation
\begin{equation}\label{eq:qif=ft}
    Q_I(\nu)f(z;\nu)=\tilde f(z;\nu+q_I(\nu))\ .
\end{equation}

\paragraph{Solving partial derivative subsystem for $f_1$.}

To put this into practice, let us consider the function $f_1$ from equation~\eqref{eq:f1sol}. Inspecting the different subsystems in equation~\eqref{eq:subsystems}, we find that this is the only solution of the GKZ system satisfying
\begin{equation}
    Q_5f(z)\neq 0\, .
\end{equation}
From the discussion above, we see that this implies that
\begin{equation}\label{eq:q6f1=ft}
    Q_5f_1(z)=\tilde f(z;\nu+q_5)\ ,
\end{equation}
for some solution $\tilde f$ of the GKZ system at $\nu+q_5$. Furthermore, we have that $\tilde f$ must satisfy
\begin{equation}\label{eq:d6ft=0}
    \partial_6 \tilde f(z;\nu+q_5)=0
\end{equation}
by the properties of the reduction operators. Finally, since $Q_5$ satisfies
\begin{equation}
    [\E_J,Q_5]=-(a_1+a_3-a_5)Q_5
\end{equation}
with $a_I$ the column vectors of $\A$, we find that
\begin{equation}
    q_5=a_1+a_3-a_5\,.
\end{equation}
This implies that we should consider solutions of the GKZ system at the parameter
\begin{equation}
    \tilde{\nu}\coloneqq \nu+q_5=(2,2,0,1+\epsilon,1+\epsilon)^T\,.
\end{equation}
In particular, note that $\tilde{\nu}_3=0$.

It is now possible to try to solve for $\tilde f$ and insert it into equation~\eqref{eq:q6f1=ft} to obtain $f_1$. However, there is some further simplification we can do. Since the reduction operators satisfy
\begin{equation}
    Q_5=Q_6=Q_7\, ,
\end{equation}
we have that
\begin{equation}
     Q_5f_1(z)=Q_6 f_1(z)=Q_7f_1(z)=\tilde f(z;\tilde{\nu})\,.
\end{equation}
By the same arguments as before, this implies that $\tilde f$ must not only satisfy equation~\eqref{eq:d6ft=0}, but also
\begin{equation}
    \partial_6 \tilde f(z;\tilde{\nu})=\partial_7 \tilde f(z;\tilde{\nu})=0
\end{equation}
constraining $\tilde f$ further.

Recall from section~\ref{ssec:reducibility} that, when studying solutions of the GKZ system annihilated by partial derivatives, it is possible to interpret these as solutions to a smaller GKZ system. For a set of partial derivatives $\{I_1,\cdots,I_k\}$, this smaller system is obtained by defining the matrix $\A_F$ from the columns $a_I$ of $\A$. Where we take only the columns
\begin{equation}
    I \in \{1,\cdots,N\}\setminus \{I_1,\cdots,I_k\}\,,
\end{equation}
with $N$ being the number of columns of $\A$. The solutions of the original GKZ system annihilated by these partial derivatives will then also be solutions of the GKZ system defined by $\A_F$.

To obtain $\tilde f$, we must therefore construct the matrix with columns
\begin{equation}
    \{1,2,3,4,5,6,7\}\setminus \{5,6,7\}=\{1,2,3,4\}\,.
\end{equation}
In other words, the matrix $\A_F$ consisting of only the first four columns of $\A$. This matrix is given by
\begin{equation}\label{eq:Afmat}
    \A_{F}=\begin{pmatrix}
        1 & 1& 0&0\\
        0&0 &1&1\\
        0&0&0&0\\
        0&1&0&0\\
        0&0&0&1
    \end{pmatrix}\, ,
\end{equation}
and we will now consider the GKZ system it defines. It turns out that this GKZ system is especially simple as it has no toric operators, while the Euler operators are
\begin{equation}
        \begin{array}{lll}
       \tilde \E_1= \theta_1+\theta_2\, , & \tilde \E_2=\theta_3 +\theta_4\, , & \tilde\E_3=0\ ,\\
        \tilde\E_4=\theta_2 \, , & \tilde\E_5=\theta_4\, . &
    \end{array}
\end{equation}
Therefore, $\tilde{f}$ is a solution to the system
\beq\label{EulerF-system}
   (\tilde \E_J + \tilde \nu_J)\tilde f(z;\tilde{\nu}) = 0\ ,
\eeq
with $1\leq J\leq 5$. Note that since $\tilde \E_3=0$ the above implies  $\tilde{\nu}_3\tilde f(z;\tilde{\nu})=0$. Thus, if $ \tilde{\nu}_3\neq 0$,
the system will have no non-zero solutions and the GKZ system is trivial. This exemplifies why a parameter must be in the $\C$-span of $F$ to obtain non-trivial subsystems.

The solution to the differential equations \eqref{EulerF-system} can be found quite easily, and its one-dimensional solution space is spanned by the function 
\begin{equation}
    \tilde{f}(z;\tilde{\nu})=\frac{\big(\frac{z_1 z_3}{z_2 z_4}\big)^{\epsilon}}{z_1 z_2 z_3 z_4}\, .
\end{equation}
Therefore, to obtain $f_1(z)$, we must solve
\begin{equation}\label{eq:q6f=zz}
    Q_5f_1(z)=\frac{\big(\frac{z_1 z_3}{z_2 z_4}\big)^{\epsilon}}{z_1 z_2 z_3 z_4}\, .
\end{equation}
Before we do this, we will showcase a useful way that reduction operators can be rewritten, which greatly simplifies solving equation \eqref{eq:q6f=zz}.

\paragraph{Rewriting reduction operators.} 

As it stands, equation \eqref{eq:q6f=zz} is a bit inconvenient, as it is somewhat involved to solve this differential equation directly. However, we will use this to showcase another useful tool that can be used to simplify such differential equations. The main idea is that, like before, we will consider various differential operators to be equivalent if they act on solutions of the GKZ system in equivalent ways. We have already used this for the toric equations, and the difference here is that we will also apply this reasoning to the Euler equations. 

Consider the Euler equations
\begin{equation}\label{eq:eulerrep}
    \begin{split}
        \E_1+\nu_1&=\theta_1+\theta_2+1\, ,\\
        \E_2+\nu_2&=\theta_3+\theta_4+1\, ,
    \end{split}
\end{equation}
where we recall that $\theta_i=z_i\partial_i$. Note that, by commuting $z_I$ and $\partial_I$, one finds that
\begin{equation}
    \theta_I+1=\partial_I z_I\, ,
\end{equation}
where the derivative acts on everything to its right. Combining this observation with equation~\eqref{eq:eulerrep} and solving for $\partial_4$ and $\partial_2$, one finds that
\begin{equation}\label{eq:dieq}
\begin{array}{rl}
   \partial_4 &\simeq_{\E+\nu} -\partial_3 \frac{z_3}{z_4}\, ,\\
   \partial_2 & \simeq_{\E+\nu} -\partial_1 \frac{z_1}{z_2}\, ,
\end{array}
\end{equation}
where, $\simeq_{\E+\nu}$ means that we consider equivalence relations stemming from the toric equations, as well as the Euler equations. In other words, both sides of equation~\eqref{eq:dieq} act the same on solutions of the GKZ system.

With these replacements, it is possible rewrite the reduction operator $Q_5 (\nu)$ as\footnote{These kind of replacements can also be useful for finding representations of reduction operators that commute with the toric operators, since this property can depend on the representation.}
\begin{equation}\label{eq:tildeq5def}
    Q_5  \simeq_{\E+\nu}  \frac{1}{z_2 z_4}   \partial_1 \partial_3 (z_2 z_4 z_5-z_1 z_4 z_6-z_2 z_3 z_7) \eqqcolon \tilde{Q}_6\, .
\end{equation}
As we will see, this representation simplifies the process of solving for $f_1$ significantly.

\paragraph{inhomogeneous equations from reduction operators.}

Recall that we are interested in solving equation~\eqref{eq:q6f=zz} for $f_1(z)$. By the discussion above, it is possible to replace $Q_5$ with $\tilde{Q}_6$ when acting on $f_1$, leading to the differential equation
\begin{equation}
 \frac{1}{z_2 z_4}   \partial_1 \partial_3 \big((z_2 z_4 z_5-z_1 z_4 z_6-z_2 z_3 z_7) f_1(z) \big)=\frac{\left(\frac{z_1 z_3}{z_2 z_4}\right)^{\epsilon}}{z_1 z_2 z_3 z_4}\, .
\end{equation}
Since the left hand side consists of total derivatives with respect to $\partial_1$ and $\partial_3$, this equation can simply be integrated twice resulting in
\begin{equation}
        f_1(z)=\frac{ \left(\frac{z_1 z_3}{z_2 z_4}\right){}^{\epsilon }}{z_2 z_4 z_5-z_1 z_4 z_6-z_2 z_3 z_7}
\end{equation}
consistent with equation~\eqref{eq:f1sol}. Note that in this integration two integration constants have been ignored. Taking these into account properly would result in including the solutions $Q_5 f=0$ as well. However, since we are only interested in a particular solution, we can simply ignore these. 

Especially when convenient representations of differential operators can be found, solving inhomogeneous differential equations of the type~\eqref{eq:qif=ft} is a way one can easily obtain solutions to the GKZ system. Alternatively, if one reduction operator is much simpler than the others, or if only one exists, this technique makes it possible to only consider differential equations incorporating this single reduction operator. Since solving the inhomogeneous differential equation in full generality includes all functions annihilated by the reduction operator, as well as those that are mapped to the inhomogeneous part.

\subsection{Locality as a consequence of reduction operators} \label{ssec:physics}

As we saw in section~\ref{ssec:singexchangeint}, the single exchange integral satisfies a set of differential equations in $X_1$ and $X_2$ separately, due to the locality of the underlying theory. In this section we show that this property is a direct consequence of the existence of the reduction operators $Q_1 $ and $Q_3 $. The framework of reduction operators also allows us to classify when these kind of equations can exist depending on which components of the parameter $\nu$ are non-integer, providing an alternative perspective to a similar result obtained in \cite{arkani-hamed_differential_2023}. We then speculate about some how this can be generalized. To do this, we first discuss how acting with reduction operators can simplify an integral.

\paragraph{Reduced integrals from reduction operators.}

Recall from section~\ref{ssec:qimoving} that the acting of the reduction operators on a solution maps it to the solution of a smaller GKZ system. Since solutions to GKZ systems can often be described by integrals, it is natural to try to construct the integrals associated to this smaller GKZ system. Note that it is not guaranteed that such integrals provide us with a complete basis of solutions. However, if the parameter is non-resonant for the matrix defining the smaller GKZ system, this is guaranteed.\footnote{Note that a parameter $\nu$ can be simultaneously resonant for $\A$ as well as non-resonant for a submatrix $\A_F$.} 

As an example, let us consider the subsystem considered in section~\ref{ssec:qimoving}, defined from the matrix~\eqref{eq:Afmat}. To construct the associated integrals, one can simply reverse the process described in section~\ref{sec:genGKZ} and identify the polynomials
\begin{equation}
    p_1=z_1+z_2 \omega_1\, ,\qquad p_2=z_3+z_4 \omega_2\,,
\end{equation}
which give rise to the matrix $\A_F$. Note that rows filled with zeroes play no role in this construction. From these polynomials we obtain integrals of the form
\begin{equation}
    \int_\Gamma d^2\omega\frac{(\omega_1\omega_2)^\epsilon}{(z_1 +z_2 \omega_1)^2(z_3 +z_4 \omega_2)^2}
\end{equation}
for the associated GKZ system.

\paragraph{Reduced integrals for the single-exchange integral.} One can apply the same reasoning to the other reduction operators $Q_1 $ and $Q_3 $. With this, one finds the following identities for the single exchange integral:
\begin{subequations}\label{eq:Qactions}
    \begin{align}
        Q_1  I_{\R_+^2}(z;\nu) \vert_{\rm phys}&=-\epsilon \int_{\R_+^2}d\omega_1 d\omega_2\; \frac{(\omega_1 \omega_2)^\epsilon}{(\omega_2+X_2+Y)(\omega_1+\omega_2+X_1+X_2)^2}\, , \label{eq:Q1action}\\
        Q_3  I_{\R_+^2}(z;\nu) \vert_{\rm phys}&=-\epsilon \int_{\R_+^2}d\omega_1 d\omega_2\; \frac{(\omega_1 \omega_2)^\epsilon}{(\omega_1+X_1+Y) (\omega_1+\omega_2+X_1+X_2)^2}\, ,  \label{eq:Q3action}\\
        Q_5  I_{\R_+^2}(z;\nu) \vert_{\rm  phys}&=\epsilon^2 \int_{\R_+^2}d\omega_1 d\omega_2\; \frac{(\omega_1 \omega_2)^\epsilon}{(\omega_1+X_1+Y)^2 (X_2+\omega_2+Y)^2}\, ,  \label{eq:Q5action}\\
        Q_3  Q_1  I_{\R_+^2}(z;\nu) \vert_{\rm phys}&=\epsilon^2 \int_{\R_+^2}d\omega_1 d\omega_2\; \frac{(\omega_1 \omega_2)^\epsilon}{ (\omega_1+\omega_2+X_1+X_2)^3}\, ,  \label{eq:Q13action}
    \end{align} 
\end{subequations}
where $I_{\R_+^2}^2(z;\nu)$ is as defined in equation~\eqref{eq:exchangegkz} and we have specialized the coordinates $z_I$ to their physical values after applying the differential operators. The pre-factors of $-\epsilon$ and $\epsilon^2$ do not follow immediately from this reasoning but have instead been obtained by explicit calculation. It turns out that the right-hand sides of these equations have some interesting physical interpretations.

The different integrals above can be interpreted diagrammatically, due to the fact that the bulk-to-bulk propagator $G(Y,\eta_1,\eta_2)$ has three terms. The first two terms both come with Heaviside step function enforcing either $\eta_1\leq \eta_2$ or $\eta_1 \geq \eta_2$, we will call these the left-ordered or right-ordered parts respectively. The third term does not come with a Heaviside step function and therefore we will call it the non-ordered part. This lack of time ordering implies that the two vertices disconnect and the resulting integral is just a product of the two vertices. Interestingly, if one considers these terms separately and performs the integrations over $\eta_1$ and $\eta_2$, one recovers the right-hand sides of equations~\eqref{eq:Q1action},~\eqref{eq:Q3action} and~\eqref{eq:Q5action} respectively, up to a twist in the powers of the different polynomials. This twist can be realized by acting with derivatives of $X_1$ or $X_2$. The integral in equation~\eqref{eq:Q13action} also has a diagrammatic interpretation. This integral simply corresponds to collapsing the propagator to a point, leading to a single vertex as shown in (d) of figure~\ref{fig:Qdiagrams}.

\begin{figure}[ht]
\centering
    \begin{subfigure}{0.49\textwidth}
    \centering
    \begin{tikzpicture}
        \begin{feynman}
            \vertex (topleft);
            \vertex [right=.75 cm of topleft,boundarydot] (topcenter1) {};
            \vertex [right=1.5 cm of topcenter1,boundarydot] (topcenter2) {};
            \vertex [right=.75 cm of topcenter2,boundarydot] (topcenter);
            \vertex [below=1.5 cm of topcenter] (midcenter) ;
            \vertex[left=1.5 cm of midcenter,bulkdot] (leftdiagram) {};
            \vertex[right=1.5 cm of midcenter,bulkdot] (rightdiagram){};
            \vertex [right=.75 cm of topcenter,boundarydot] (topcenter3) {};
            \vertex [right=1.5 cm of topcenter3,boundarydot] (topcenter4) {};
            \vertex [right =.75 cm of topcenter4] (topright);
    
            \diagram*  {
              (topleft) --[very thick,blue] (topcenter1) --[very thick,blue] (topcenter2) --[very thick,blue] (topcenter3) --[very thick,blue] (topcenter4) --[very thick,blue] (topright);
              (leftdiagram) --[thick,opacity=1] (topcenter1);
              (leftdiagram) --[thick,opacity=1]  (topcenter2);
              (rightdiagram) --[thick,opacity=1] (topcenter3);
              (rightdiagram) --[thick,opacity=1] (topcenter4);
              (rightdiagram) --[dashed,very thick,green,with arrow=.5] (leftdiagram);
            };
            \vertex[below=.5 cm of leftdiagram] (X1) {\textbf{$X_1$}};
            \vertex[below=.5 cm of rightdiagram] (X2) {\textbf{$X_2$}};
            \vertex[above=0cm of midcenter] (Y) {\textbf{\color{green} $Y$}};
        \end{feynman}
    \end{tikzpicture}
    \caption{}
    \vspace*{5mm}
    \end{subfigure}
    \hfill
    \begin{subfigure}{0.49\textwidth}
    \centering
    \begin{tikzpicture}
        \begin{feynman}
            \vertex (topleft);
            \vertex [right=.75 cm of topleft,boundarydot] (topcenter1) {};
            \vertex [right=1.5 cm of topcenter1,boundarydot] (topcenter2) {};
            \vertex [right=.75 cm of topcenter2,boundarydot] (topcenter);
            \vertex [below=1.5 cm of topcenter] (midcenter) ;
            \vertex[left=1.5 cm of midcenter,bulkdot] (leftdiagram) {};
            \vertex[right=1.5 cm of midcenter,bulkdot] (rightdiagram){};
            \vertex [right=.75 cm of topcenter,boundarydot] (topcenter3) {};
            \vertex [right=1.5 cm of topcenter3,boundarydot] (topcenter4) {};
            \vertex [right =.75 cm of topcenter4] (topright);
    
            \diagram*  {
              (topleft) --[very thick,blue] (topcenter1) --[very thick,blue] (topcenter2) --[very thick,blue] (topcenter3) --[very thick,blue] (topcenter4) --[very thick,blue] (topright);
              (leftdiagram) --[thick,opacity=1] (topcenter1);
              (leftdiagram) --[thick,opacity=1]  (topcenter2);
              (rightdiagram) --[thick,opacity=1] (topcenter3);
              (rightdiagram) --[thick,opacity=1] (topcenter4);
              (leftdiagram) --[dashed,very thick,green,with arrow=.5] (rightdiagram);;
            };
            \vertex[below=.5 cm of leftdiagram] (X1) {\textbf{$X_1$}};
            \vertex[below=.5 cm of rightdiagram] (X2) {\textbf{$X_2$}};
            \vertex[above=0cm of midcenter] (Y) {\textbf{\color{green} $Y$}};
        \end{feynman}
    \end{tikzpicture}
    \caption{}
    \vspace*{5mm}
    \end{subfigure}
    
    \begin{subfigure}{0.49\textwidth}
    \centering
    \begin{tikzpicture}
        \begin{feynman}
            \vertex (topleft);
            \vertex [right=.75 cm of topleft,boundarydot] (topcenter1) {};
            \vertex [right=1.5 cm of topcenter1,boundarydot] (topcenter2) {};
            \vertex [right=.75 cm of topcenter2,boundarydot] (topcenter);
            \vertex [below=1.5 cm of topcenter] (midcenter) ;
            \vertex[left=1.5 cm of midcenter,bulkdot] (leftdiagram) {};
            \vertex[right=1.5 cm of midcenter,bulkdot] (rightdiagram){};
            \vertex [right=.75 cm of topcenter,boundarydot] (topcenter3) {};
            \vertex [right=1.5 cm of topcenter3,boundarydot] (topcenter4) {};
            \vertex [right =.75 cm of topcenter4] (topright);
    
            \diagram*  {
              (topleft) --[very thick,blue] (topcenter1) --[very thick,blue] (topcenter2) --[very thick,blue] (topcenter3) --[very thick,blue] (topcenter4) --[very thick,blue] (topright);
              (leftdiagram) --[thick,opacity=1] (topcenter1);
              (leftdiagram) --[thick,opacity=1]  (topcenter2);
              (rightdiagram) --[thick,opacity=1] (topcenter3);
              (rightdiagram) --[thick,opacity=1] (topcenter4);
            };
            \vertex[below=.5 cm of leftdiagram] (X1) {\textbf{$X_1 +$\textbf{\color{green} $Y$}}};
            \vertex[below=.5 cm of rightdiagram] (X2) {\textbf{$X_2 +$\textbf{\color{green} $Y$}}};
        \end{feynman}
    \end{tikzpicture}
    \caption{}\label{eq:singleexchangecut}
    \end{subfigure}
    \hfill
    \begin{subfigure}{0.49\textwidth}
    \centering
    \begin{tikzpicture}
        \begin{feynman}
            \vertex (topleft);
            \vertex [right=.75 cm of topleft,boundarydot] (topcenter1) {};
            \vertex [right=1.5 cm of topcenter1,boundarydot] (topcenter2) {};
            \vertex [right=.75 cm of topcenter2,boundarydot] (topcenter);
            \vertex [below=1.44 cm of topcenter,bulkdot] (midcenter) {};
            \vertex [right=.75 cm of topcenter,boundarydot] (topcenter3) {};
            \vertex [right=1.5 cm of topcenter3,boundarydot] (topcenter4) {};
            \vertex [right =.75 cm of topcenter4] (topright);
    
            \diagram*  {
              (topleft) --[very thick,blue] (topcenter1) --[very thick,blue] (topcenter2) --[very thick,blue] (topcenter3) --[very thick,blue] (topcenter4) --[very thick,blue] (topright);
              (midcenter) --[thick,opacity=1] (topcenter1);
              (midcenter) --[thick,opacity=1]  (topcenter2);
              (midcenter) --[thick,opacity=1] (topcenter3);
              (midcenter) --[thick,opacity=1] (topcenter4);;
            };
            \vertex[below=0.5cm of midcenter] (Xs) {\textbf{$X_1+X_2$}};
        \end{feynman}
    \end{tikzpicture}
    \caption{}
    \end{subfigure}
\caption{The diagrammatical interpretation of the integrals in equation~\eqref{eq:Qactions}. In particular, we have that (a) corresponds to the left-ordered part of the propagator, (b) to the right-ordered part and (c) to the non-time-ordered part. Finally, (d) corresponds to the collapsed propagator.} \label{fig:Qdiagrams}
\end{figure}
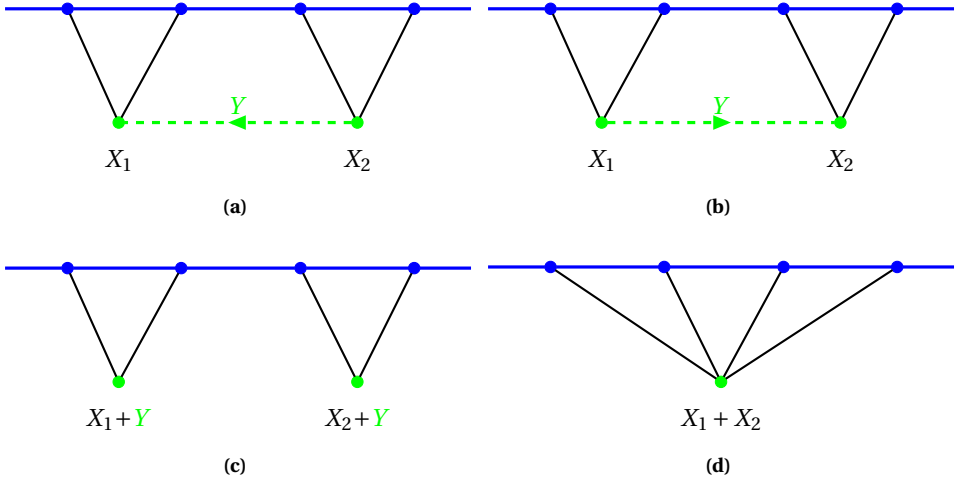

\paragraph{Locality from reduction operators.} The collapsed propagator is particularly interesting since this also appeared in the discussion of locality in section~\ref{ssec:singexchangeint}. Here, collapsing the propagator implied a second order inhomogeneous differential equation for the single exchange integral. Furthermore this differential equation was quite special as it depended on only one of $X_1$ or $X_2$. As we will see, equation~\eqref{eq:Q13action} is exactly equation~\eqref{eq:locality}, only rewritten in terms of $Q_1 $ and $Q_3 $ instead of the derivatives of $X_1$. In particular, the right hand side of equation~\eqref{eq:locality} is exactly the integral appearing in~\eqref{eq:Q13action}, where the twists are realized by the partial derivatives of $X_1$. Therefore it is possible to rewrite
\begin{equation}\label{eq:Q13nu=ict}
    Q_3 Q_1 I_\epsilon \vert_{\mathrm{phys}}=\frac{\partial^2 I_{\mathrm{contr}}}{\partial X_1^2}\, ,
\end{equation}
where $I_{\mathrm{contr}}$ is the contracted integral evaluating to~\eqref{eq:icont1}.

To recover the local differential equation, it is necessary to first rewrite $Q_1 $ and $Q_3 $ in terms of $X_1$ and $X_2$. Following this, we will demonstrate that their product can be reformulated to depend solely on either $X_1$ or $X_2$, thereby recovering equation~\eqref{eq:locality}.

\paragraph{Reduction operators in terms of the physical variables.} 

To find the reduction operators in terms of $X_1$ and $X_2$, it is possible to use the relations between the $z_I$ and the physical variables specified in equation~\eqref{eq:physvars}. A simple application of the chain rule results in
\begin{equation}
        \frac{\partial}{\partial X_1}= \partial_1 +\partial_5  \, ,\quad
        \frac{\partial}{\partial X_2} = \partial_3 +\partial_5  \, ,\quad
         \frac{\partial}{\partial Y} = \partial_1 +\partial_3\, .
\end{equation}
However, these will not all be independent, because the Euler equations~\eqref{eq:exchangeeuler} give additional relations between the partial derivatives $\partial_I$. This makes it possible to eliminate derivatives of $Y$ by replacing
\begin{equation}
    Y \frac{\partial}{\partial Y}\simeq_{\E+\nu} 2\epsilon -1 -X_1 \frac{\partial}{\partial X_1}-X_2\frac{\partial}{\partial X_2}\, ,
\end{equation}
where, as before, $\simeq_{\E+\nu}$ denotes that equality holds when considering the equivalence relations stemming from both the toric and the Euler equations.

With these replacements, it is possible to obtain the action of the reduction operators in terms of $X_1$ and $X_2$. $Q_1 $ and $Q_3 $ then take the form
\begin{equation}\label{eq:Q13phys}
\begin{array}{rl}
     Q_1 \big\vert_{\mathrm{phys}}&\simeq_{\E+\nu} (X_1+Y)\frac{\partial}{\partial X_1} -\epsilon \eqqcolon Q_{1,X} \, , \\
      Q_3 \big\vert_{\mathrm{phys}}&\simeq_{\E+\nu} (X_2+Y)\frac{\partial}{\partial X_2} -\epsilon \eqqcolon Q_{3,X}\, .
\end{array}
\end{equation}
Therefore, equation~\eqref{eq:Q13nu=ict} can be rewritten as
\begin{equation}\label{eq:q13XI}
\left((X_2+Y)\frac{\partial}{\partial X_2} -\epsilon\right)\left((X_1+Y)\frac{\partial}{\partial X_1} -\epsilon\right)I=\frac{\partial^2 I_{\mathrm{contr}}}{\partial X_1^2}    \, ,
\end{equation}
where $I$ is the original single exchange integral~\eqref{eq:singleexchange}. Note that since the single exchange integral only depends on $X_1$ and $X_2$ through the combination $X_1+X_2$ it is already possible to rewrite this into a differential equation only involving one parameter, and this approach will lead to the locality differential equation~\eqref{eq:locality}. However, it turns out that this property is also encoded in the reduction operators themselves, due to a particular property satisfied by all reduction operators.

\paragraph{Reduction of variables and locality.} The key insight is that all reduction operators have the property that
\begin{equation}
    \partial_I Q_I (\nu)f(z;\nu)=0\ ,
\end{equation}
for all solutions $f$ at a parameter $\nu$. This makes it possible to eliminate derivatives of $\partial_I$ when acting on $Q_I (\nu)f$. In particular, it is possible to rewrite 
\begin{equation}
    \partial_1 Q_{1,X} \simeq_{\E+\nu} \frac{1}{2Y}\left(2\epsilon-1 +(Y-X_1)\frac{\partial}{\partial X_1}-(Y+X_2)\frac{\partial}{\partial X_2}\right)Q_{1,X}\, ,
\end{equation}
which can be used to eliminate the $X_2$ derivative in equation~\eqref{eq:q13XI}. After this replacement, equation~\eqref{eq:q13XI} can be rewritten as
\begin{equation}
    \left((X_1-Y)\frac{\partial}{\partial X_1}+\epsilon-1\right)\left((X_1+Y)\frac{\partial}{\partial X_1} -\epsilon\right) I=\frac{\partial^2 I_{\mathrm{contr}}}{\partial X_1^2} \, ,
\end{equation}
which, when expanded, recovers equation~\eqref{eq:locality}. Note that, because $Q_{1,X}$ and $Q_{3,X}$ commute, this procedure also makes it possible to obtain a differential equation involving only $X_2$.

\paragraph{Reduction operators and boundary conditions.}

Interestingly, the reduction operators and the resulting inhomogeneous differential equations can also be used to obtain boundary conditions for the single-exchange integral. In particular, consider $Q_1I_{\R^2_+}(z;\nu)\vert_{\rm phys}$ as in equation~\eqref{eq:Q1action} and let us write $I_{\rm phys}=I_{\R^2_+}(z;\nu)\vert_{\rm phys}$ to lighten the notation somewhat. Then, we want to obtain boundary conditions for $I_{\rm phys}$, such that we can write it in terms of the most general solution of the GKZ system, given by
\begin{equation}\label{eq:gensolphys}
    I_{\rm phys}=c_1(\R^2;\epsilon)f_1\vert_{\rm phys}+c_2(\R^2;\epsilon)f_2\vert_{\rm phys}+c_3(\R^2;\epsilon)f_3\vert_{\rm phys}+c_4(\R^2;\epsilon)f_4\vert_{\rm phys}\,.
\end{equation}
 Here the functions $f_i\vert_{\rm phys}$ are as written in equation~\eqref{eq:gensols} and the coefficients $c_i$ are what we are trying to solve for. 
 
 The first approach to obtaining these coefficients is that one can simply act on this general solution with a reduction operator. Using the inhomogeneous differential equation from~\eqref{eq:Q1action}, this will fix some of the coefficients $c_i$. In particular, acting with $Q_{1,X}$ on $I_{\rm phys}$ as above we find
 \begin{equation}
 \begin{split}
     Q_{1,X}I_{\rm phys} = &\epsilon \,c_3(\R_+^2;\epsilon)(X_1-Y)^{\epsilon-1}(X_2+Y)^\epsilon\\
     & +\epsilon\,c_4(\R_+^2;\epsilon)\frac{(X_1+X_2)^{2\epsilon}}{X_1-Y}\left(\, _2F_1\left(1,-2 \epsilon ;1-\epsilon ;\frac{X_2-Y}{X_1+X_2}\right)-1\right)\,,
\end{split}
 \end{equation}
where $_2F_1$ is again the hypergeometric function. Note that this equation no longer depends on $c_1$ and $c_2$ as $f_1\vert_{\rm phys}$ and $f_2\vert_{\rm phys}$ are annihilated by $Q_{1,X}$. Now, we must impose that this equation is equal to the integral on the right-hand side of equation~\eqref{eq:Q1action}. Evaluating that integral and solving for the coefficients $c_i$ we find
\begin{equation}
    c_3(\R_+^2;\epsilon)=0\, , \quad c_4(\R_+^2;\epsilon)= 2^{-2 \epsilon -1} \sqrt{\pi } \csc (\pi  \epsilon ) \Gamma \left(\frac{1}{2}-\epsilon \right) \Gamma (\epsilon )
\end{equation}
in accordance with equation~\eqref{eq:coefssols}.

A second approach to solving for the coefficients in equation~\eqref{eq:gensolphys} comes from the observation that, for $\epsilon>0$, all the solutions are regular in the limit $X_1+Y\rightarrow 0$\,.\footnote{In fact, this approach also works for general $\epsilon$ as both sides must obey the same scaling behavior in $X_1+Y$.} Therefore, using the explicit expression of $Q_{1,X}$ from equation~\eqref{eq:Q13phys}, we find that 
\begin{equation}
    Q_{1,X}I_{\rm phys}\vert_{X_1+Y\rightarrow 0}=\left((X_1+Y)\frac{\partial I_{\rm phys}}{\partial X_1}-\epsilon I_{\rm phys}\right)\bigg\vert_{X_1+Y\rightarrow 0}=-\epsilon I_{\rm phys}\vert_{X_1+Y\rightarrow 0}
\end{equation}
Using this observation and using the inhomogeneous equation for $Q_{1,X}$, we find that
\begin{equation}
    -\epsilon I_{\rm phys}\vert_{X1\rightarrow -Y} =  -\epsilon \int_{\R_+^2}d\omega_1 d\omega_2\; \frac{(\omega_1 \omega_2)^\epsilon}{(\omega_2+X_2+Y)(\omega_1+\omega_2+X_2-Y)^2}\,.
\end{equation}
Evaluating the integral on the right-hand side and taking the limit of the functions $f_i$, we would obtain the same coefficients as before.

\paragraph{Locality and twists.} As we have seen, the existence of these local differential equations requires the existence of both $Q_1 $ and $Q_3 $. Since the existence of these operators depends on the reducibility of the GKZ system this provides us with a way of classifying when these local differential equations exist. In fact, the reducibility of the GKZ system is fully encoded in the twist parameter $\nu$. It follows from the discussion in chapter~\ref{ch:generalGKZ} that, if there exists a resonant face $F$ with $1 \not\in F$, then the reduction operator $Q_1 $ exists for suitable $\nu$.\footnote{Here, suitable means that $\nu$ is not in the $\C$-span of $F$, while $\nu-a_1$ is in the $\C$-span of $F$. Note that, if this is not the case, it is possible to parameter shift the integral by applying certain differential operators, as described in section~\ref{ssec:reductionoperators}.} Careful analysis then shows that this is the case only if $\nu_1$ is integer. Where we recall that the parameter $\nu$ encodes the twists of the integral
\begin{equation}
     I(X;\nu)\coloneqq \int_{R_+^2}\frac{\omega_1^{\nu_4-1} \omega_2^{\nu_5-1}\;d\omega_1 d\omega_2}{(\omega_1+X_1+Y)^{\nu_1}(\omega_2+X_2+Y)^{\nu_2}(\omega_1+\omega_2+X_1+X_2)^{\nu_3}}\, .
\end{equation}
A similar analysis for $Q_3 $ then implies that $\nu_2$ must also be integer. 

This partially recovers a result from \cite{arkani-hamed_differential_2023} where the locality of the single-exchange integral was also studied. However, there it was found that $\nu_3$ must also be integer, implying that none of the polynomials in the numeral can be twisted. From the perspective of the reduction operators, this implies that $Q_5 $ exists and that the single exchange integral satisfies an equation of the type~\eqref{eq:Q5action}. Interestingly, the diagram corresponding to this equation is the ``cut" diagram of figure~\ref{eq:singleexchangecut}. Therefore, it is possible to obtain differential operators for the single exchange integral that, diagrammatically, correspond to either edge contraction or edge cutting, as shown in figure~\ref{fig:Qedgereduction}. Furthermore, locality is equivalent to both of these reductions being possible.

\begin{figure}
\centering
    \begin{tikzpicture}
        \begin{feynman}
            \vertex (topleft);
            \vertex [right=.5 cm of topleft,boundarydot] (topcenter1) {};
            \vertex [right=0.667 cm of topcenter1,boundarydot] (topcenter2) {};
            \vertex [right=.5 cm of topcenter2,boundarydot] (topcenter);
            \vertex [below=0.667 cm of topcenter] (midcenter) ;
            \vertex[left=0.667 cm of midcenter,bulkdot] (leftdiagram) {};
            \vertex[right=0.667 cm of midcenter,bulkdot] (rightdiagram){};
            \vertex [right=.5 cm of topcenter,boundarydot] (topcenter3) {};
            \vertex [right=0.667 cm of topcenter3,boundarydot] (topcenter4) {};
            \vertex [right =.5 cm of topcenter4] (topright);

            \vertex [right=3 cm of topright] (basepoint);

            \vertex [above=1.333 cm of basepoint] (topleftcut);
            \vertex [right=.5 cm of topleftcut,boundarydot] (topcenter1cut) {};
            \vertex [right=0.667 cm of topcenter1cut,boundarydot] (topcenter2cut) {};
            \vertex [right=.5 cm of topcenter2cut,boundarydot] (topcentercut);
            \vertex [below=0.667 cm of topcentercut] (midcentercut) ;
            \vertex[left=0.667 cm of midcentercut,bulkdot] (leftdiagramcut) {};
            \vertex[right=0.667 cm of midcentercut,bulkdot] (rightdiagramcut){};
            \vertex [right=.5 cm of topcentercut,boundarydot] (topcenter3cut) {};
            \vertex [right=0.667 cm of topcenter3cut,boundarydot] (topcenter4cut) {};
            \vertex [right =.5 cm of topcenter4cut] (toprightcut);

            \vertex [below=1cm of basepoint] (topleftcont);
            \vertex [right=.5 cm of topleftcont,boundarydot] (topcenter1cont) {};
            \vertex [right=0.667 cm of topcenter1cont,boundarydot] (topcenter2cont) {};
            \vertex [right=.5 cm of topcenter2cont,boundarydot] (topcentercont);
            \vertex [below=.667 cm of topcentercont,bulkdot] (midcentercont) {};
            \vertex [right=.5 cm of topcentercont,boundarydot] (topcenter3cont) {};
            \vertex [right=0.667 cm of topcenter3cont,boundarydot] (topcenter4cont) {};
            \vertex [right =.5 cm of topcenter4cont] (toprightcont);

            \vertex[right=.5cm of topright] (arrowanchor);
            \vertex[right=2cm of arrowanchor] (arrowanchor2);
            \vertex[above=.2cm of arrowanchor] (arrowbasetop);
            \vertex[above=.75cm of arrowanchor2] (arrowheadtop);
            \vertex[below=.2cm of arrowanchor] (arrowbasebot);
            \vertex[below=.75 cm of arrowanchor2] (arrowheadbot);            
            
            \diagram*  {
              (topleft) --[very thick,blue] (topcenter1) --[very thick,blue] (topcenter2) --[very thick,blue] (topcenter3) --[very thick,blue] (topcenter4) --[very thick,blue] (topright);
              (leftdiagram) --[thick,opacity=1] (topcenter1);
              (leftdiagram) --[thick,opacity=1]  (topcenter2);
              (rightdiagram) --[thick,opacity=1] (topcenter3);
              (rightdiagram) --[thick,opacity=1] (topcenter4);
              (leftdiagram) --[green,very thick] (rightdiagram);

              (topleftcut) --[very thick,blue] (topcenter1cut) --[very thick,blue] (topcenter2cut) --[very thick,blue] (topcenter3cut) --[very thick,blue] (topcenter4cut) --[very thick,blue] (toprightcut);
              (leftdiagramcut) --[thick,opacity=1] (topcenter1cut);
              (leftdiagramcut) --[thick,opacity=1]  (topcenter2cut);
              (rightdiagramcut) --[thick,opacity=1] (topcenter3cut);
              (rightdiagramcut) --[thick,opacity=1] (topcenter4cut);

              (topleftcont) --[very thick,blue] (topcenter1cont) --[very thick,blue] (topcenter2cont) --[very thick,blue] (topcenter3cont) --[very thick,blue] (topcenter4cont) --[very thick,blue] (toprightcont);
              (midcentercont) --[thick,opacity=1] (topcenter1cont);
              (midcentercont) --[thick,opacity=1]  (topcenter2cont);
              (midcentercont) --[thick,opacity=1] (topcenter3cont);
              (midcentercont) --[thick,opacity=1] (topcenter4cont);

              (arrowbasetop) --[edge label=$Q_5 $, thick,red,with arrow =1]  (arrowheadtop);
              (arrowbasebot) --[edge label'=$Q_3 Q_1$, thick,red,with arrow =1]  (arrowheadbot);
            };
            \vertex[below=.5 cm of leftdiagram] (X1) {\textbf{$X_1$}};
            \vertex[below=.5 cm of rightdiagram] (X2) {\textbf{$X_2$}};
            \vertex[above=0cm of midcenter] (Y) {\textbf{\color{green} $Y$}};

            \vertex[below=.5 cm of leftdiagramcut] (X1cut) {\textbf{$X_1 +$\textbf{\color{green} $Y$}}};
            \vertex[below=.5 cm of rightdiagramcut] (X2cut) {\textbf{$X_2 +$\textbf{\color{green} $Y$}}};

            \vertex[below=.5cm of midcentercont] (x1x2cont) {\textbf{$X_1+X_2$}};
        \end{feynman}
    \end{tikzpicture}
    \caption{Acting with the reduction operators makes it possible to either contract or remove an edge from the diagram.}\label{fig:Qedgereduction}
\end{figure}
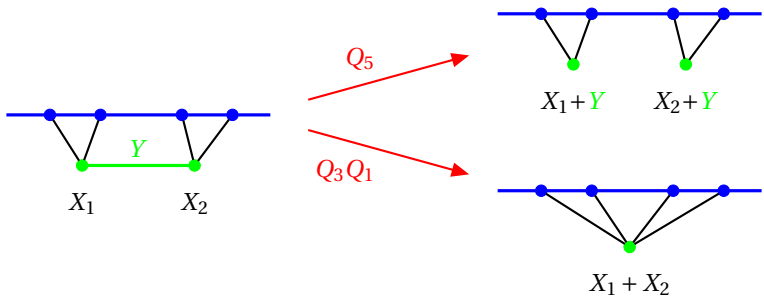

\paragraph{Generalizations.}

One is naturally led to wonder if this story can be generalized, especially since the reduction operators can be obtained for any reducible GKZ system.
As we will see in the following chapter,  this is the case. Here we will carry out an analagous analysis for any tree-level correlator within the toy model we consider. As we will see, here too, the reduction operators can be interpreted as realizing contraction and cut relations between diagrams. And, again, we can use these to obtain inhomogenous differential equations.

\fi

\if\PrintChFour1

\newpage

\stepcounter{thumbcounter}
\setcounter{colorcounter}{4}
\chapter{Recursive Reductions for Cosmological Correlators}
\label{ch:reductionalgorithm}

Having discussed general GKZ systems and cosmological correlators, as well as studied the single-exchange integral in detail, we are now ready for a general study of tree-level cosmological correlators.
In particular, we will construct GKZ systems for such correlators in general, and obtain explicitly their reduction operators.
Here, we will see one key difference between general correlators and the single-exchange integral. Namely, for general diagrams, the GKZ system of the resulting integral has many more independent variables. Thus it is necessary to perform restrictions, a non-trivial feat for GKZ systems~\cite{chestnov_restrictions_2023}. However, we find that special combinations of reduction operators will only involve the physical variables, avoiding this problem.

Besides this, mostly technical, additional difficulty, we will see many similarities to our study of the single-exchange integral. As in section~\ref{ssec:physics}, the reduction operators will admit various physical interpretations, being related to diagrammatical contractions or cuts.
Interestingly, we will also find that the reduction operators give rise to a large number of algebraic relations between different functions, allowing us to greatly reduce the number of independent functions one needs to obtain a particular correlator.

In section~\ref{sec:gencosmGKZ} we begin by describing the GKZ system and reduction operators for general tree-level cosmological correlators. Afterwards, we proceed in section~\ref{sec:physics} by describing how the reduction operators can be interpreted physically, giving rise to various cut and contraction relations. Then, we will construct a closed first-order differential system satisfied by the cosmological correlators in section~\ref{sec:differentialchain}. Finally, we will use additional reduction operators to obtain a large number of algebraic relations satisfied by the solutions to this system in section~\ref{sec:complexity} resulting in drastically improved algorithm for obtaining cosmological correlators.

\section{GKZ system and reduction operators}\label{sec:gencosmGKZ}

In this section, we will construct the GKZ system associated to the correlators of chapter~\ref{ch:cosmology} explicitly. Using this we will obtain a large number of reduction operators for each such tree-level correlators, which we will use throughout the rest of the chapter.

\subsection{Integrals from graph tubings}\label{sec:tubings}

We begin by discussing a more graphical method of obtaining the cosmological correlators from chapter~\ref{ch:cosmology}, introduced in \cite{arkani-hamed_differential_2023}.
This method uses the so-called \textit{tubings} of a diagram, which capture much of the structure of the cosmological correlators and will set the stage for the use of the theory of GKZ systems. Note that different definitions of tubings for cosmological correlators exist in the literature (see e.g. \cite{glew_amplitubes_2025}). 
Here, we adopt the definition given in~\cite[Sec 2]{arkani-hamed_differential_2023}.

\paragraph{Graph tubings and index sets.} Given a Feynman graph with the external propagators removed, a \textit{tube} is defined as a subset of adjacent vertices. Diagrammatically, this is denoted by encircling the corresponding vertices. For example, the single-exchange diagram from figure~\ref{fig:singexchange} has the following three tubes:
\begin{equation}\label{eq:singexchtubes}
\includegraphics[valign=c]{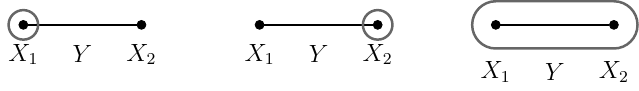}
\end{equation}
These tubes are particularly useful, since they are in one-to-one correspondence with the singularities of the flat space wavefunction coefficient. A \textit{tubing} of a graph is a collection of one or more non-intersecting tubes, and a \textit{complete tubing} is a tubing to which no more tube can be added without violating the non-intersecting condition.

Tubings can be represented in various ways. In this chapter, we will regularly switch between the purely graphical representation used in \cite{arkani-hamed_differential_2023} and the representation of tubes in terms of index sets, where each tube corresponds to the index set of the vertices it contains. For example, the tubings above corresponds to the sets $\{1\}$, $\{2\}$ and $\{1,2\}$ respectively. Then, tubings can be represented simply as sets of tubes, or in other words, sets of index sets. In this representation, the tubing containing all of the tubes in~\eqref{eq:singexchtubes} is denoted by $\{\{1\},\{2\},\{1,2\}\}$.

This representation will have a number of advantages for us. For example, one can now easily sum over the vertices $v$ in a tubing. However, the major reason for introducing tubings as index sets is that the subset structure of the tubes now becomes manifest. In a tubing, a tube $T$ may be graphically fully contained in another tube $T'$, and this will correspond directly to containment of the subsets $T\subseteq T'$. This allows us to conveniently consider all tubes $T'$ contained in a tube $T$, or conversely, all tubes $T'$ contained in $T$. Such collections of tubes will play an important role throughout this chapter, but in particular when obtaining the reduction operators in section~\ref{cosmic_red_op}. Furthermore, we can consider the \textit{successor} or \textit{precursor} of a tube $T$, defined  as the minimal tube containing $T$ and the maximal tube contained in $T$, respectively. These will play an important role in section~\ref{sec:differentialchain}.

\paragraph{Integrals from tubings.}

As shown in~\cite{arkani-hamed_differential_2023}, the tubing of a graph can be used to obtain the associated wavefunction coefficient. To be specific, we define for every tube $T$ a polynomial $p_T$ by setting
\begin{equation} \label{special_pol_cc}
    p_T(X,Y,x)=\sum_{v\in T} (X_v+x_v) +\sum_e Y_e  \,.
\end{equation}
In this expression the first sum is over all the vertices enclosed by the tube, while the second sum includes all edges that cross the tube. The variable $x_v$ will play the role of an integration variable. To every graph tubing $\cT$, we now associate an integral
\begin{equation}\label{eq:tubingintegral_tX}
    I_\mathcal{T}(X,Y;\alpha)=\int_{\R_+^{\Nv}}d^{\Nv} x \, \frac{\prod_{v=1}^{\Nv} x_v^{ \av-1}}{\prod_{T\in \mathcal{T}}p_T(X,Y,x)}\,,
\end{equation}
where the index $v$ runs over the $N_{\rm v}$ vertices in a diagram, and the $\alpha_v$ are variable weights associated to each vertex specified in \eqref{alpha_nu-form}. These integrals will be the key object of interest in describing the structure of cosmological correlators. 
The cosmological correlator associated to a graph $G$ is then recovered as  \cite{arkani-hamed_differential_2023}
\begin{equation}\label{eq:psiG}
    \psi_G(X,Y;\epsilon) =  \sum_{\cT\, \text{complete}} I_\cT(X,Y;\alpha) \,,
\end{equation}
where the sum is over all complete tubings $\cT$ of $G$. 

\paragraph{Convenient variables and permutations.}

Moving forward, we will consider the integrals above in a slightly different perspective. Instead of considering the variables $X_v$ and $Y_e$, we will combine these into variables $z^{(T)}$ for each tube $T$ in the tubing. In particular, we will define
\begin{equation}
    z^{(T)}=\sum_{v\in T} X_v+\sum_e Y_e\,,
\end{equation}
where, as in equation~\eqref{special_pol_cc}, the index $v$ runs over all the vertices enclosed by the tube, while the sum over $e$ is over the edges that cross the tube. These variables are particularly convenient as the polynomials $p_T$ can be written as 
\begin{equation}
    p_T(z,x)=z^{(T)}+\sum_{v\in T} x_v\,.
\end{equation}
Furthermore, the $z^{(T)}$ will map more naturally to GKZ variables defined in the following section.

This change of variables also has another interesting consequence. Rewriting the integral in equation~\eqref{eq:tubingintegral_tX} in terms of these new variables, we see that the only data necessary to define it is combinatorial, namely the data of which tube encircles which vertex. All other diagrammatical data can be re-instated by replacing the $z^{(T)}$ with their definitions in terms of $X_v$ and $Y_e$, as well as choosing the particular values the $\alpha_v$ take. A corollary to this is that many different tubings and diagrams may take the same form after making this replacement. 

For example,  the double-exchange correlator has two complete tubings, given by
\begin{equation}\label{eq:doubleexchangetubings}
\includegraphics[valign=c]{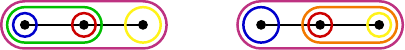}
\end{equation}
In principle, these will lead to different integrals $I_\cT$. However, denoting $T_{\rm b}$, $T_{\rm y}$, $T_{\rm g}$ and $T_{\rm o}$ for the blue, yellow, green and orange tubes respectively, one can make the replacements
\begin{equation}
    z^{(T_{\rm b})}\longleftrightarrow z^{(T_{\rm y})}\,, \quad z^{(T_{\rm g})} \longrightarrow z^{(T_{\rm o})}\,,
\end{equation}
and relate the integral of the left tubing to the right one. We will greatly extend this reasoning in section~\ref{sec:complexity}, where we show that many such relations exist and obtain a minimal set of integrals needed to express the actual correlators. Note that the described abstraction requires keeping track of all relations. However, this is more than compensated for by the significant reduction in the number of integrals to compute, which is an immense advantage in practice.

\subsection{GKZ systems for cosmological correlators} \label{GKZ_for_cosmo}

To start, we will construct the toric and Euler operators for a general correlator explicitly. This allows us in section~\ref{cosmic_red_op} to obtain reduction operators for general cosmological correlators.

\paragraph{Cosmological correlators as GKZ integrals.} 
First, we have to cast cosmological correlators in the form of a general GKZ integral. Recall that, to each such a correlator, we could study its complete tubings $\mathcal{T}$, and for each complete tubing obtain an integral
\begin{equation}
    I_\mathcal{T}(X,Y;\alpha)=\int_{\R_+^{\Nv}}d^{\Nv} x \, \frac{\prod_{v=1}^{\Nv} x_v^{ \av-1}}{\prod_{T\in \mathcal{T}}p_T(X,Y,x)}\,.
\end{equation}
Notice that this is exactly of the form of a general GKZ integral, except that the polynomials $p_T$ do not have arbitrary coefficients $z_{j,m}$. However, one can simply lift the polynomials to functions of $z$ by defining
\begin{equation}\label{eq:ptz_t}
    p_T(z,x)=z^{(T)}+\sum_{v\in T} z^{(T)}_vx_v\, ,
\end{equation}
where we promoted the coefficients of $p_T(X,Y)$ in equation~\eqref{special_pol_cc} to variables $( z^{(T)},z^{(T)}_v)$. 
Let us stress that the $x$-independent term in \eqref{eq:ptz_t} is parametrized by the variable $z^{(T)}$ without an index. This direction 
is special, since the polynomials in the physical variables are recovered 
when setting 
\beq \label{physical_slice}
z^{(T)}\big|_{\rm phys} = \sum_{v\in T} X_v + \sum_e Y_e\ , \qquad z^{(T)}_{v}\big|_{\rm phys}=1\ , 
\eeq
where the second sum is over all edges that cross the tube $T$ as in section~\ref{sec:tubings}. 
We will also refer to this identification as the restriction to the physical slice. The GKZ integral associated to the complete tubing $\mathcal{T}$ is then given by
\begin{equation}\label{eq:tubingintegral_t}
    I_\mathcal{T}(z;\alpha)=\int_{\R_+^N}d^{\Nv} x \, \frac{\prod_{v=1}^{\Nv} x_v^{\alpha_v-1}}{\prod_{T\in \mathcal{T}}p_T(z,x)}\,.
\end{equation}
This integral will then define the GKZ system of differential equations for us.

As an example, let us again consider the single-exchange diagram, which, using tubings, represented as
\begin{equation}
\includegraphics[valign=c]{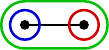}
\end{equation}
where we will denote the blue, red and green tubes as $T_{\rm b}$, $T_{\rm r}$ and $T_{\rm g }$ respectively. We will label the vertices by $v=1,2$ such that $T_{\rm b}=\{1\}$, $T_{\rm r}=\{2\}$ and $T_{\rm g }=\{1,2\}$. From these tubes, applying equation~\eqref{eq:ptz_t} results in the polynomials 
\begin{equation}\label{eq:singexchangepolys}
\begin{array}{rl}
    p_{T_{\rm b}}&=z^{(T_{\rm b})}+z^{(T_{\rm b})}_1 x_1\, ,\\
    p_{T_{\rm r}}&=z^{(T_{\rm r})}+z^{(T_{\rm r})}_2 x_2\, ,\\
    p_{T_{\rm g }}&=z^{(T_{\rm g })}+z^{(T_{\rm g })}_1 x_1+z^{(T_{\rm g })}_2 x_2\, 
\end{array}
\end{equation}
which can be inserted into equation~\eqref{eq:tubingintegral_t} to obtain the GKZ integral for the single-exchange diagram. Note the similarities to equation~\eqref{eq:exchangegkz}.

\paragraph{GKZ data for cosmological correlators.} 

We are now in the position to obtain the GKZ data for a general cosmological correlator. Recall that it consists of the matrix $\A$ and the parameter $\nu$. To obtain the matrix $\A$, recall that one first constructs a matrix $\A_T$ for each polynomial. For each tube $T$, these matrices are obtained by taking the exponents in $x_v$ for each term of $p_T$ and combining these as the column vectors of the matrix $\A_T$. One then combines these matrices into the matrix $\A$ as in \eqref{eq:Adef} by identifying 
\beq \label{eq:Ajorder}
  \A_j = \A_{T_j}\ , \qquad j=1,\ldots,|\mathcal{T}|\, ,
\eeq
where $|\mathcal{T}|$ is the number of tubes  $T_j \in \mathcal{T}$. The parameter $\nu$ can be read off immediately by comparing
\eqref{eq:tubingintegral_t} and \eqref{eq:gengkzint}, resulting in
\begin{equation} \label{nu_cos}
    \nu=(\underbrace{1,\cdots,1}_{\vert \mathcal{T}\vert \text{ times}},\alpha_1,\cdots,\alpha_{\Nv})^{\rm T}\,,
\end{equation}
where we recall that $\Nv$ is the number of integration variables $x_v$, and the $\alpha_v$ are given in \eqref{alpha_nu-form} and depend on $\epsilon$ and the order of the interaction. Note that in this equation, $\rm T$ denotes the transpose and does not refer to a tube. It is interesting to note that much of the following general discussion does not depend on the precise value of $\alpha_v$.

Returning to the example of the single-exchange integral, we can simply read off the exponents of each term in equation~\eqref{eq:singexchangepolys} to obtain the matrices
\begin{equation}
    \begin{array}{rlrlrl}
       \A_{T_{\rm b}}  &=  \begin{pmatrix}
           0 & 1 \\
           0 & 0
       \end{pmatrix},&
       \A_{T_{\rm r}}  &= \begin{pmatrix}
           0 & 0 \\
           0 & 1
       \end{pmatrix},&
        \A_{T_{\rm g}} &=  \begin{pmatrix}
           0 & 1 & 0\\
           0 & 0 & 1
       \end{pmatrix}
    \end{array}
\end{equation}
for each tube. These matrices can be combined into the matrix $\A$, and this matrix together with the parameter $\nu$ then defines the GKZ system of the single-exchange integral. In particular, these are given by
\begin{equation}\label{eq:singexchangeA}
    \A =\left(
\begin{array}{ccccccc}
 1 & 1 & 0 & 0 & 0 & 0 & 0 \\
 0 & 0 & 1 & 1 & 0 & 0 & 0 \\
 0 & 0 & 0 & 0 & 1 & 1 & 1 \\
 0 & 1 & 0 & 0 & 0 & 1 & 0 \\
 0 & 0 & 0 & 1 & 0 & 0 & 1 \\
\end{array}
\right)\, , \qquad  \nu =\begin{pmatrix}1\\ 1\\ 1\\ \alpha_1\\ \alpha_2\end{pmatrix}\ ,
\end{equation}
where the bottom two rows correspond to the matrices $\A_T$ and we recall that $\alpha_1$ and $\alpha_2$ are the twists of the integration variables $x_1$ and $x_2$ respectively. Again, we can compare this to our constructions in section~\ref{ssec:singexchGKZ}. Note that now, we needed to fix an ordering of the tubes. Here, we have chosen 
\begin{equation}
    T_1=T_{\rm b}\, ,\quad T_2=T_{\rm r}\, ,\quad T_3 = T_{\rm g}
\end{equation}
although clearly, the chosen ordering is arbitrary.

\paragraph{Structure in the GKZ data.} 
Let us study the general structure of the matrices $\A$ which we construct for cosmological correlators. Every column vector arises from a particular term in a polynomial $p_T$ for some $T$, and each term corresponds to either a vertex or the constant term. In fact, for every tube $T$ there 
is a set of vectors 
\beq \label{aT-split}
   a^{(T)},\ a^{(T)}_v  \ \ \text{with}\ \ v \in T\ ,
\eeq
and combining these vectors for every tube $T$ we obtain the matrix $\A$. We will collectively denote these column vectors $a^{(T)}_m$, where $m=v$ if the column vector arises from a term in $p_T$ with a vertex, while the index $m$ is removed when it arises from the constant term. Note that these are associated with the coordinates $z^{(T)},z^{(T)}_v$, 
in accordance with \eqref{p-form} and~\eqref{eq:ptz_t}. Labeling 
the tubes as $T_j$ with $j = 1,...,|\mathcal{T}|$ as above, we thus split $\mathcal{A}$ as
\beq \label{cAinvectors}
   \mathcal{A} = \Big( a^{(T_1)} \ \underbrace{a^{(T_1)}_{v_1}}_{v_1 \in T_1} \ |\ a^{(T_2)} \ \underbrace{a^{(T_2)}_{v_2}}_{v_2 \in T_2} \ | \ldots \Big)  \ .
\eeq
To not clutter the notation, we will mostly use the notation \eqref{aT-split}, where it is understood that the index $v$ is associated to the tube $T$.

Comparing \eqref{p-form} and~\eqref{eq:ptz_t} we can now read off the column vectors $a_m^{(T)}$ for any tree-level cosmological correlator. Since the polynomials $p_T$ are all linear $x_v$, these vectors will only consist of ones and zeros. We see that its components split into two parts. First, we have the components of $a_m^{(T)}$ that are introduced when combining the matrices $\mathcal{A}_T$ together, which will consist of the first $\vert \mathcal{T} \vert$ entries. Then, for $1\leq j \leq \vert \mathcal{T}\vert$ we find that the $j$-th entry of $a_m^{(T)}$ is $1$ if $T=T_j$ and $0$ otherwise, where $T_j$ refers to the ordering of tubes we choose when constructing $\A$ in equation~\eqref{eq:Ajorder} or \eqref{cAinvectors}.
The remaining rows admit a similar structure, but now accounts for which vertices appear in the tube $T$. For $a^{(T)}$ this part is zero. However, for the other column vectors $a^{(T)}_v$, $v \in T$, we find that the $\vert \mathcal{T}\vert +v'$-entry of $a^{(T)}_v$ is $1$ if $v=v'$ and zero otherwise. To conclude, we can write the column vectors of 
$\A$ as
\begin{equation} \label{aT-def}
    a^{(T)}=\left(\begin{array}{c} e^{(T)}\\
    \mathbf{0}\end{array}\right) \ , \qquad a^{(T)}_v=\left(\begin{array}{c} e^{(T)} \\ e_v\end{array}\right)\ ,
\end{equation}
where $e^{(T)}$ is a $\vert \mathcal{T}\vert$-dimensional unit vector in the direction associated to $T$, $e_v$ is a $\Nv$-dimensional unit vector in the $v$-th direction, and $\mathbf{0}$ is the $\Nv$-dimensional zero vector. Note that the precise form of these vectors depends on the ordering of $T_j$ and $x_v$ that we have chosen.

To return to our example of the single-exchange integral, we see that \eqref{aT-def}
implies that $\mathcal{A}$ takes the form 
\begin{align}
   \mathcal{A} =& \left(\begin{array}{cc|cc|ccc} a^{(T_{\rm b})} & a^{(T_{\rm b})}_1 & a^{(T_{\rm r})} & a^{(T_{\rm r})}_2 & a^{(T_{\rm g})}& a^{(T_{\rm g})}_1 & a^{(T_{\rm g})}_2 \end{array} \right) \\
   =&  \left(\begin{array}{cc|cc|ccc} e^{(T_{\rm b})} & e^{(T_{\rm b})} & e^{(T_{\rm r})} & e^{(T_{\rm r})} & e^{(T_{\rm g})} & e^{(T_{\rm g})} & e^{(T_{\rm g})} \\ 
   \mathbf{0} &  e_1 &  \mathbf{0} &  e_2 &  \mathbf{0} &  e_1 & e_2
   \end{array} \right) \ . \nonumber
\end{align}
Clearly, upon inserting the unit vectors, we recover the matrix $\mathcal{A}$ given in \eqref{eq:singexchangeA}.

\paragraph{GKZ systems for cosmological correlators.}
We are now in the position to determine the toric and Euler operators associated to $\A,\nu$. Recall that the toric operators arose from vectors $u$ and $v$ in $\N^N$ satisfying $\A u=\A v$. Equivalently, these arise from the relations between the column vectors over the integers. In particular, note that from equation~\eqref{aT-def} it follows that, for any $T$ and any $v$ and $v'$ in $T$, we have
\begin{equation}\label{eq:ATdif}
    a^{(T)}_v-a^{(T)}_{v'}=(0,e_v-e_{v'})\,.
\end{equation}
Note that the right hand side no longer depends on $T$. Therefore, for any two tubes $T$ and $T'$ and $v$, $v'$ contained on both tubes, we have
\begin{equation}
a_v^{(T)}+a_{v'}^{(T')}=a^{(T)}_{v'}+a^{(T)}_v\,.
\end{equation}
A similar story holds for $a^{(T)}$ and $a^{(T)}_v$, resulting in a relation of the form
\begin{equation}   a_v^{(T)}+a^{(T')}=a^{(T)}+a^{(T)}_v\,.
\end{equation}
Both of these relations will give rise to toric operators. In particular, the above implies that for any tubes $T$, $T'$, and vertices $v$, $v'$ we have that
\begin{equation}\label{eq:cosmotoric}
    \begin{array}{rcr}
        v,v'\in T\cap T' &\implies &  \partial^{(T)}_{v} \partial^{(T')}_{v'}-\partial^{(T)}_{v'}\partial^{(T')}_{v} \simeq 0\, ,  \\
        v\in T\cap T' &\implies &  \partial^{(T)}_{v} \partial^{(T')}-\partial^{(T)}\partial^{(T')}_{v} \simeq 0\, ,
    \end{array}
\end{equation}
where we recall that $\simeq$ denotes that equality holds modulo the toric equivalence relations. Here, we have also introduced the notation 
\beq
\partial^{(T)} \equiv \frac{\partial}{\partial z^{(T)}}\ , \qquad \partial^{(T)}_v \equiv \frac{\partial}{\partial z^{(T)}_v}\ ,  
\eeq
for the partial derivatives with respect to $z$-variables. Note that a particular case of these relations arises when a tube is $T$ is completely contained in another tube $T'$. Then, it follows that there are toric operators such as the one above for every $v$ and $v'$ contained in $T$. These relations will be crucial in obtaining the reduction operators. 

Having found the toric operators, we can now turn our attention to the Euler operators. As we have seen before, the rows of $\A$ can be split into rows corresponding to the tubes and rows corresponding to the vertices. Since each row gives rise to an Euler operator, this implies that these can be split in a similar manner. In particular we obtain an operator $\E^{(T)}$ for each tube and an operator $\E_v$ for each vertex. This gives rise to $\vert \mathcal{T}\vert+N_{\rm v}$ operators which take the form
\begin{equation} \label{Euler_cosm}
    \E^{(T)}=\theta^{(T)}+\sum_{v\in T}\theta^{(T)}_v\,,\qquad  \E_v=\sum_{\{ T:v\in T \} }\theta^{(T)}_v  \, ,
\end{equation}
where $\theta^{(T)}\equiv z^{(T)}\partial^{(T)},\ \theta^{(T)}_v\equiv z^{(T)}_v\partial^{(T)}_{v}$. Let us stress that the sum in $\E_v$ is over all tubes that contain the vertex $v$. 

\paragraph{The GKZ system for the single exchange integral.} 

For completeness, let us return to the example of the single exchange integral and show that we recover the system of section~\ref{ssec:singexchGKZ}. From its tubing
\begin{equation}
\includegraphics[valign=c]{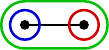}
\end{equation}
we find that the first vertex is enclosed by both the blue and the green tubes, while the second vertex is enclosed by the red and green tubes. Therefore, the GKZ system has toric relations of the form
\begin{equation}
\begin{array}{r}
    \partial^{(T_{\rm b})}_{1} \partial^{(T_{\rm g })}-\partial^{(T_{\rm b})}\partial^{(T_{\rm g })}_{1} \simeq 0\, ,\\
   \partial^{(T_{\rm r})}_{2} \partial^{(T_{\rm g })}-\partial^{(T_{\rm r})}\partial^{(T_{\rm g })}_{2} \simeq 0\,,
\end{array}
\end{equation}
and in fact, these are the only toric relations of this GKZ system. 
The Euler operators come in two parts. First we have the the operators from the tubes, which are given by
\begin{equation}
\begin{array}{rl}
    \E_{T_{\rm b}}&=\theta^{(T_{\rm b})}+\theta^{(T_{\rm b})}_1\, , \\
    \E_{T_{\rm r}}&=\theta^{(T_{\rm r})}+\theta^{(T_{\rm r})}_2\, , \\
    \E_{T_{\rm g }}&=\theta^{(T_{\rm g })}+\theta^{(T_{\rm g })}_1+\theta^{(T_{\rm g })}_2\, .
\end{array}
\end{equation}
Secondly, the Euler operators from the vertices are
\begin{equation}
    \E_1=\theta^{(T_{\rm b})}_1+\theta^{(T_{\rm g })}_1\, , \quad \E_2=\theta^{(T_{\rm r})}_2+\theta^{(T_{\rm g })}_2\,.
\end{equation}
Together with the toric operators above, these completely describe the GKZ system for the single exchange integral.

\subsection{Reduction operators for cosmic GKZ systems} \label{cosmic_red_op}

In this section we determine the reduction operators for 
GKZ systems associated to cosmological correlators. This 
connects the general discussion of chapter~\ref{ch:generalGKZ} 
with the GKZ systems introduced for cosmological correlators above.

We begin by showing that every tube corresponds to a resonant face. Recall that the first $\vert \mathcal{T}\vert$ rows of the matrix $\A$ were associated to the tubes in $\mathcal{T}$. From this, we can obtain a linear functional $L_{F_{T_j}}$ projecting a vector on its $j$-th coordinate. These linear functionals will  satisfy
\beq
\begin{array}{rl}
  L_{F_{T_j}}(a^{(T)}_m) =0 & \text{ for } T\neq T_j \, , \\
  L_{F_{T_j}}(a^{(T)}_m)>0 & \text{ for } T=T_j \, .
\end{array}
\eeq
Therefore, these define a face of $\A$ containing all columns of $\A$ except those arising from the face $T_j$. In particular, this face will correspond to the integral $I_{\mathcal{T}\setminus\{T_j\}}$.

From the above, we see that any tube $T$ will define a face of $\A$. It turns out that these faces are all resonant as well. To see this, we will consider two cases: the case where $T$ is the maximal tube, as well as the case where $T$ is not the maximal tube. Let us consider the latter first. Observe from equation~\eqref{aT-def} that
\begin{equation}
    a^{(T_{\rm max})}_v-a^{(T_{\rm max})}=(\bold{0},e_v)\,,
\end{equation}
where we recall that $\bold{0}$ is a $\vert \mathcal{T}\vert$-dimensional vector of zeroes and $e_v$ is the $N_v$-dimensional unit vector in the $v$-th direction. Inspecting the explicit form of $\nu$ provided in equation~\eqref{nu_cos}, this implies that it is possible to write
\begin{equation}
    \nu = \sum_{v} \alpha_v (a^{(T_{\rm max})}_v-a^{(T_{\rm max})}) + \sum_{T \in \mathcal{T}} a^{(T)}\,,
\end{equation} 
where we recall that the $\alpha_v$ are the twists of the different vertices which will be complex in general. Now note that if $T\neq T_{\rm max}$, the face $F_{T}$ will contain the columns associated to $T_{\rm max}$. Therefore, we see that any non-maximal face is resonant. A similar story holds for the maximal face itself. However, here we must choose any collection of tubes in $\mathcal{T}\setminus \{T_{\rm max}\}$ covering every vertex and proceed along the same lines. If such a covering is not possible, the maximal face will not be resonant. As we will see when constructing the higher-order operators below, in this case we will also obtain no reduction operator for the maximal face. Note that for a complete tubing this is never the case. 

To obtain the actual reduction operators we will proceed as laid out in section~\ref{cosmic_red_op}. We begin by constructing the operator $\E_{F^{(T)}}$ for every tube $T$. However, since the linear functional here is simply a projection, we find that 
\begin{equation}\label{eq:tuberedbase}
    \E_{F^{(T)}}=\E^{(T)}=z^{(T)}\partial^{(T)}+\sum_{v\in T} z^{(T)}_v \partial^{(T)}_{v}\,,
\end{equation}
where $\E^{(T)}$ is the Euler operator of the GKZ system that is associated to $T$, as given in equation~\eqref{Euler_cosm}. Now, recall that a reduction operator is obtained by fixing both a face $F$ and an index $I$ not contained in $F$. For the faces we consider, any column associated to $T$ will be sufficient. Therefore, we will proceed using the column $a^{(T)} $. This implies that we must find $u$ such that
\begin{equation}\label{eq:duETQ}
     \pd_1^{u_1}\cdots \pd_N^{u_N} \E^{(T)} \simeq  Q^{(F_T)}_u \pd^{(T)} \, ,
\end{equation}
although in what follows, we will index $Q$ in different ways.

As we will see, the reduction operators we find fall in to two classes, the first-order operators and the higher-order operators. The first-order operators come with some problems though, as in general it will not be possible to write these solely in terms of the physical variables $z^{(T)}$ and their derivatives. However, we will show that a special combination of the first-order operators can be written in terms of the physical variables only, resulting in a first-order operator for each tube. The higher-order operators do not have this problem, and we will rewrite these directly in terms of the physical variables. Note that we will diverge somewhat from the discussion in section~\ref{cosmic_red_op} and solve equation~\eqref{eq:duETQ} directly, without having to solve equation~\eqref{eq:reducopineq} iteratively for $u$.

\paragraph{First-order reduction operators from a contained tube.}

Let us consider tubes $T$ and $T'$ such that $T$ is fully contained in $T'$. Recall from equation~\eqref{eq:cosmotoric} that this implies that, for every $v$ in $T$ there is a toric relation the form
\begin{equation}\label{eq:firstordertoric}
    \partial^{(T)}_{v} \partial^{(T')} -\partial^{(T)} \partial^{(T')}_{v} \simeq 0\,.
\end{equation}
It then follows that
\begin{equation}\label{eq:pit'0ET}
\begin{array}{rl}
    \partial^{(T')}  \E^{(T)}&=z^{(T)} \partial^{(T)} \partial^{(T')} +\sum_{v\in T} z^{(T)}_v \partial^{(T)}_{v}\partial^{(T')} \\
    &\simeq z^{(T)} \partial^{(T)} \partial^{(T')} +\sum_{v\in T}z^{(T)}_v \partial^{(T')}_v\partial^{(T)} \,,
\end{array}
\end{equation}
where we have inserted equations~\eqref{eq:tuberedbase} and~\eqref{eq:firstordertoric}.
In this equation, $\partial^{(T)}$ can be factored out implying that we have obtained a relation of the form in equation~\eqref{eq:duETQ} and can read off the reduction operator. Writing this reduction operator as $Q^{(T)}_{T'}$, we find that
\begin{equation} \label{QTT'}  
    Q^{(T)}_{T'}=z^{(T)} \partial^{(T')}+\sum_{v\in T} z^{(T)}_v \partial^{(T')}_{v}\ . 
\end{equation}
Thus, we have found a first-order reduction operator whenever a tube $T$ is contained in another tube $T'$. Note that this implies that every non-maximal tube has at least one first-order reduction operator associated to it while the maximal tube has none. 

\paragraph{Physical restriction of first-order reduction operators.}

To use the reduction operators \eqref{QTT'}, we first have to deal with a fundamental challenge that arises when using GKZ systems. In the process of defining the GKZ system we had to introduce many additional parameters $z^{(T)}_v$ that are not present in the physical integral which is evaluated on the slice \eqref{physical_slice}. As mentioned in the general discussion of section~\ref{sec:genGKZ}, the Euler operators impose restrictions on the variables, naturally leading to a choice of homogeneous variables. This can be used to eliminate partial derivatives with respect to some of the  $z^{(T)}_v$. However, the constraint \eqref{physical_slice} is more severe and it turns out to be impossible to write a general reduction operator $Q^{(T)}_{T'}$ only in terms of the physical variables. 
To circumvent this problem, we propose to introduce new operators 
\begin{equation}\label{eq:QTdef_t}
    Q^{(T)}\coloneqq \sum_{T'\supsetneq T} Q^{(T)}_{T'}\Big|_{\rm phys}\ .
\end{equation}
Here the sum is over all tubes $T'$ which contain $T$, excluding $T$ itself and $\vert_{\rm phys}$ means that we restrict to the slice \eqref{physical_slice} and act on solutions of the GKZ system. 
We show in appendix~\ref{red_op_phys} that, using the Euler operators \eqref{Euler_cosm}, $Q^{(T)}$ can be written as
\begin{equation}\label{eq:QTresult}
     \boxed{\rule[-.5cm]{0cm}{1.3cm} \quad 
     Q^{(T)}=z^{(T)}\sum_{ T'\supsetneq T}\partial^{(T')}+\sum_{T'\subseteq T}(\theta^{(T')}+\nu^{(T')})-\sum_{v\in T}\alpha_v\ ,\quad }\,
\end{equation}
which only involves the physical derivatives $\partial^{(T)}$.

Note that in the construction above it was crucial that $T$ was not a maximal tube. We will now show that, by simply inserting the maximal tube $T_{\rm max}$ into equation~\eqref{eq:QTresult}, we obtain an operator $Q^{(T_{\rm max})}$ satisfying
\begin{equation}\label{eq:QTmaxaction}
    Q^{(T_{\rm max})}\simeq_{\E+\nu}0\,,
\end{equation}
where we indicated that it annihilates the integral $I_{\mathcal{T}}$ due to the Euler equations. Technically, this operator is not a reduction operator of the GKZ system, but we will treat as such due to the property \eqref{eq:QTmaxaction}. To check this identity, we insert $T_{\rm max}$ into \eqref{eq:QTresult}
to find 
\begin{align} \label{QTmax}
   Q^{(T_{\rm max})} &= \sum_{T\in \mathcal{T}}\big(\theta^{(T)}+ \nu^{(T)}\big) -\sum_{v\in T_{\rm max}} \alpha_v \\
   &= \sum_{T\in \mathcal{T}} \big(\mathcal{E}^{(T)}+ \nu^{(T)}\big) -\sum_{v\in T_{\rm max}} \big(\mathcal{E}_v + \alpha_v\big)\ , \nonumber 
\end{align}
where we inserted the definitions 
of $\mathcal{E}^{(T)}$ and $\mathcal{E}_v$ given in \eqref{Euler_cosm} to obtain the second line. We now see that the expression on the second line is a sum of the Euler operators and therefore annihilates solutions to the GKZ system.

\paragraph{Higher-order reduction operators.} Having established how a reduction operator for a tube $T$ can be obtained by considering the tubes $T'$ that contain $T$, we now show that there are also reduction operators corresponding to the tubes $T'$ \textit{contained in} $T$. To be precise, we will consider a partition of $T$, i.e.~a collection of tubes $S_\alpha$ contained in $T$ such that every vertex in $T$ is in exactly one of the $S_\alpha$. The tube $T$ can then be recovered as the disjoint union

\begin{equation} \label{T-partition}
    T=\bigsqcup_{\alpha=1}^n S_\alpha\ .
\end{equation}
Furthermore, we can collect the $S_\alpha$ into a set $\pi$ as
\begin{equation}\label{eq:pidef}
    \pi=\{\,S_\alpha\, \vert \, 1\leq \alpha \leq n\,\}\,,
\end{equation}
which we will also refer to as the partition. Note that every partition $\pi$ is also a tubing, in fact it is a minimal tubing containing each vertex in $T$. We will show that, from every partition, we can obtain a new reduction operator. Furthermore, this reduction operator can be written in terms of only the physical derivatives, provided we restrict ourselves to the physical slice.

As for the first-order reduction operators, we will start by considering derivatives acting on the Euler operators \eqref{Euler_cosm}. We first note that, using similar arguments as before, there are toric operators of the form
\begin{equation}
    \partial^{(T)}_{v} \partial^{(S)} -\partial^{(T)} \partial^{(S)}_{v} \simeq 0
\end{equation}
for every $v$ in $S$ and $S$ in $\pi$. Therefore, we find that
\begin{equation}\label{eq:highereulerpart}
    \partial^{(S)} \sum_{v\in S} \theta^{(T)}_v\simeq \sum_{v\in S} z^{(T)}_v \partial^{(S)}_v \partial^{(T)} 
\end{equation}
for each $S$ in $\pi$. Now, we can simply use the decomposition of $T$ to write
\begin{equation}\label{eq:ETdecomp}
    \E^{(T)}=\theta^{(T)} +\sum_{S\in \pi}\sum_{v\in S} \theta^{(T)}_v\,.
\end{equation}
Combining equations~\eqref{eq:highereulerpart} and~\eqref{eq:ETdecomp} we obtain
\begin{equation}
    \prod_{S\in \pi} \partial^{(S)} \E^{(T)} \simeq \Bigg( z^{(T)} \prod_{S\in \pi}\partial^{(S)}  +\sum_{S\in \pi} \sum_{v\in S} z^{(T)}_v\partial^{(S)}_{v}\prod_{\substack{S'\in \pi\\\S'\neq S}} \partial^{(S')} \Bigg)\partial^{(T)} \,,
\end{equation}
from which we can immediately read of the reduction operator associated to $F^{(T)}$ and $a^{(T)}$.

However, we now again run into the issue that this operator involves derivatives with respect to the unphysical variables. To fix this, we will use that on the physical slice we have that $z^{(T)}_v=z^{(S)}_v=1$ for all $S$. Therefore, it is possible to rewrite
\begin{equation}\label{eq:highorderdef}
    \sum_{v\in S} z^{(T)}_v\partial^{(S)}_{v}=\sum_{v\in S} \theta_{S,v}+\cdots\simeq_{\E+\nu} -\theta_{S,0}-\nu^{(S)}+\cdots\, ,
\end{equation}
where the dots denote terms that go to zero in the physical limit, and we made use of the Euler operator $\E_{S}$. Therefore, we find a reduction operator associated to the partition $\pi$ which can be written as 
\begin{equation}\label{eq:highred}     
    Q^{(T)}_{\pi}=\Big(z^{(T)} -\sum_{S\in \pi} z^{(S)} \Big)\prod_{S'\in \pi} \partial^{(S')} -\sum_{S\in \pi}\nu^{(S)}\prod_{\substack{S'\in \pi,\\ S'\neq S}} \partial^{(S')} \ .
\end{equation}
Here we stress that this expression holds only on the physical slice \eqref{physical_slice}, while the existence of the operator is guaranteed for any $z_I$. Interestingly, we will always have $\nu^{(S)}=1$ for each $S\in \pi$. If this is the case, equation~\eqref{eq:highred} can be written as
\begin{equation} \label{higher-order-red_final}
 \boxed{\rule[-.5cm]{0cm}{1.3cm} \quad     Q^{(T)}_{\pi}=\Big(\prod_{S'\in \pi} \partial^{(S')} \Big)\Big(z^{(T)} -\sum_{S\in \pi} z^{(S)} \Big)\,,\quad }
\end{equation}
where the derivatives act on everything to their right. We will see that this form has interesting implications on the singularity structure of the integrals.

\paragraph{A simple example.}

To illustrate the discussions above, let us briefly consider a simple example. We will again consider the single-exchange integral with the tubing
\begin{equation}
\includegraphics[valign=c]{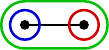}
\end{equation}
and obtain its reduction operators. We begin by obtaining the first-order reduction operator associated to the blue tube $T_{\rm b}$, noting that the reduction operator for $T_{\rm r}$ can be obtained in an almost identical manner. Inspecting equation~\eqref{eq:QTresult} we find that we must consider the tubes contained in $T_{\rm b}$, as well as those that contain it. Here, there are no tubes contained in $T_{\rm b}$. However, it is contained in the green tube $T_{\rm g}$. Thus we find that equation~\eqref{eq:QTresult} reduces to \begin{align} \label{red-op_inexample}
    Q^{(T_{\rm b})}&=z^{(T_{\rm b})}  \partial^{(T_{\rm g})} +z^{(T_{\rm b})} \partial^{(T_{\rm b })}  +\nu^{(T_{\rm b})}-\alpha_1\,\ ,\qquad Q^{(T_{\rm r})} = Q^{(T_{\rm b})}\big|_{T_{\rm b},\alpha_1\rightarrow T_{\rm r},\alpha_2}\ , \nonumber\\
    Q^{(T_{\rm g})}& = \sum_{T\in \{ T_{\rm g},T_{\rm r},T_{\rm b}\}} \big( z^{(T)} \partial^{(T)}  +\nu^{(T)}\big)-\alpha_1-\alpha_2\ . 
\end{align}
Note that it is also possible to write these operators in terms of the physical coordinates $X_v$ and $Y$ as
\begin{align}
     Q^{(T_{\rm b})}&=(X_1+Y)\frac{\partial}{\partial X_1}+\nu^{(T_{\rm b})}-\nu_1\ , \qquad Q^{(T_{\rm r})}=(X_2+Y)\frac{\partial}{\partial X_2}+\nu^{(T_{\rm r})}-\nu_2\ , \nonumber \\
     Q^{(T_{\rm g})}&= X_1\frac{\partial}{\partial X_1}+X_2\frac{\partial}{\partial X_2}+Y\frac{\partial}{\partial Y}+\nu^{(T_{\rm b})}+\nu^{(T_{\rm r})}+\nu^{(T_{\rm g})}-\alpha_1-\alpha_2
\end{align}
as was also found in section~\ref{ssec:physics}. 

For the higher-order reduction operator, we note that $T_{\rm g }$ admits the decomposition $T_{\rm g }=T_{\rm b}\sqcup T_{\rm r}$. Following the notation of equation~\eqref{eq:pidef}, we will denote this partition by
\begin{equation}
\pi=\{T_{\rm b},T_{\rm r}\}=\{ \{1\},\{2\}\}\,.
\end{equation}
From such a decomposition we can obtain a higher-order reduction operator using equation~\eqref{higher-order-red_final} taking the form 
\begin{equation}\label{eq:QTgexplicit}
    Q^{(T_{\rm g })}_\pi= \partial^{(T_{\rm b})} \partial^{(T_{\rm r})}  (z^{(T_{\rm g })} -z^{(T_{\rm b})} -z^{(T_{\rm r})} )\ .
\end{equation}
It is straightforward to write this neatly in terms of $X_v$ and $Y$ as
\begin{equation}\label{eq:singexchangeQhigh}
    Q^{(T_{\rm g })}_\pi= \frac{1}{2}\left( \left(\frac{\partial}{\partial X_1}-\frac{\partial}{\partial X_2}\right)^2 - \frac{\partial^2}{\partial Y^2}\right) Y\,
\end{equation}
where the derivatives act on everything to their right. 

We are now ready to discuss the implications of acting with the reduction operators found in this section on the space of solutions to the cosmic GKZ system.

\section{From reductions to relations, cuts, and contractions}\label{sec:physics}

In this section we discuss how 
the reduction operators derived in section \ref{cosmic_red_op} can be used 
to connect and simplify cosmological correlators. We will first show in section~\ref{ssec:tuberemoval} how the reduction operators remove tubes from a tubing. Subsequently, we will describe in sections~\ref{ssec:contractions} and~\ref{ssec:factors} that their action can be interpreted as either contracting or cutting an edge in the diagram. This leads to relations among integrals associated to different diagrams that are realized via differential operators. Our findings can also be understood diagrammatically via the removal of a tubes, which either 
results in a contraction or a factorization of integrals. 
The resulting relations form the foundation for the algorithm to determine cosmological correlators that we develop in sections~\ref{sec:differentialchain} and \ref{sec:complexity}.

\subsection{Removing tubes using reduction operators}\label{ssec:tuberemoval}

We begin by describing the action of a reduction operator on the integral $I_\mathcal{T}$ and we will see that acting with a reduction operator removes a tube, up to twists in the integrand realized by partial derivatives. To derive this, we will first consider the reduction operators in representations that include derivatives with respect to the unphysical coordinates, as in this form the action of the reduction operator is the simplest. To obtain the action of the physical operators, we note that the unphysical derivatives have been removed using the Euler relations. In a GKZ system, two operators are equivalent modulo an Euler relation if they act equivalently on the integrand of the GKZ integral, modulo a total derivative in one of the integration variables. From this we find that the reduction operators in the physical coordinates must act in the same manner as the unphysical ones.

\paragraph{Tube-removal using $Q^{(T)}$.} We begin by considering the first-order reduction operators. It is useful to first consider the reduction operator $Q^{(T)}_{T'}$ as given in equation~\eqref{QTT'}. Acting on the integrand of~\eqref{eq:tubingintegral_t}, we easily verify the identity
\begin{equation}\label{eq:QTaction}
    Q^{(T)}_{T'}\frac{\prod_{i=1}^n x_v^\epsilon}{\prod_{T\in \mathcal{T}}p_T}=\frac{-p_T}{p_{T'}}\frac{\prod_{i=1}^n x_v^\epsilon}{\prod_{T\in \mathcal{T}}p_T}=\partial^{(T')} \frac{\prod_{i=1}^n x_v^\epsilon}{\prod_{T\in \mathcal{T}\setminus \{T\}}p_T}\,.
\end{equation}
Observe that, in effect, acting with $Q^{(T)}_{T'}$ has removed the tube $T$ from the diagram and replaced it with a derivative in $\partial^{(T')} $. This action generalizes for the operator $Q^{(T)}$ in physical variables up to total derivatives in the integration variables. Because these total derivatives vanish when performing the integrations, the integrals must satisfy \footnote{Note that this equation is consistent with the action~\eqref{eq:QTmaxaction} of the first-order reduction operator $Q^{(T_{\rm max})}$, even though it is, strictly speaking, not a reduction operator of the GKZ system.}
\begin{equation}\label{eq:QTonint}
    \boxed{\rule[-.5cm]{0cm}{1.3cm} \quad   Q^{(T)} I_\mathcal{T}=\bigg( \sum_{T'\supsetneq T} \partial^{(T')}  \bigg)I_{\mathcal{T}\setminus \{T\}}\,, \quad }
\end{equation}
where $Q^{(T)}$ is as in equation~\eqref{eq:QTdef_t}, and the sum is over all tubes $T'$ that strictly contain $T$. We will see later that, in the special case that $T$ is a minimal tube in a complete tubing $\mathcal{T}$, these equations imply contraction identities at the diagrammatical level.

\paragraph{Tube-removal using $Q^{(T)}_{\pi}$.}
Let us now show that there are similar relations for the higher-order reduction operators $ Q^{(T)}_{\pi}$. The procedure to obtain these is  similar to what we did before. However, now we find that the factorization depends on the partition $\pi$ of $T$, where this partition is defined as in equation~\eqref{T-partition}. To be precise, acting with $Q^{(T)}_{\pi}$ on the integral results in
\begin{equation}\label{eq:highordtuberemov}
     \boxed{\rule[-.5cm]{0cm}{1.3cm} \quad  Q^{(T)}_{\pi} I_\mathcal{T}=\left(\prod_{S\in \pi} \partial^{(S)}  \right)I_{\mathcal{T}\setminus \{T\}}\,. \quad}
\end{equation}
We will see that this identity implies an interesting factorization when taking $T$ to be the maximal tube in a tubing.

\subsection{Contractions using reduction operators}\label{ssec:contractions}

In this section, we  further investigate equation \eqref{eq:QTonint} and 
develop an associated diagrammatical interpretation resulting from the action of $Q^{(T)}$. More precisely, we investigate the properties of the integrals with removed tubes, such as $I_{\mathcal{T}\setminus \{T\}}$, and characterize the situations in which they can be represented by 
another integral that arises from a contracted diagram. 

\paragraph{Contractions and tubings.} 
To begin with, let us consider a sub-diagram within a tubing $\mathcal{S}$, with the following properties.  
We consider an edge connecting two vertices $v_1$ and $v_2$. The tubing $\mathcal{S}$
is assumed to contain a tube $T_{\rm g} =\{v_1,v_2\}$ containing both vertices, but both individual vertices are `bare' in the sense that  $\mathcal{S}$ does not contain 
the minimal tubes only encircling $v_1$ and $v_2$, respectively. 
Such a situation can arise, for example, by acting with reduction operators 
$Q^{(T)}$ on a complete tubing $\mathcal{T}$ in such a way that two vertices are bare, as we will discuss after \eqref{two-vertices_circled}. Diagrammatically, we thus consider the following partial tubing
\begin{equation} \label{two-vertices_non-circled}
\includegraphics[valign=c]{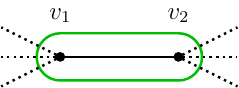}
\end{equation}
The dotted lines denote an arbitrary number of edges that connect to the rest of the diagram. Note that there can also be additional tubes that fully enclose this part of the diagram but these have not been drawn. We denote the integral associated to this tubing by $I_\mathcal{S}$.

We now want to show that $I_\mathcal{S}$ can be computed by evaluating the integral associated to the \textit{contracted diagram}, where the edge is shrunk to a point, and the two vertices $v_1,v_2$ coalesce. 
Diagrammatically, we want to establish an equality 
\begin{equation} \label{simple_two_vertex_conctraction}
\includegraphics[valign=c]{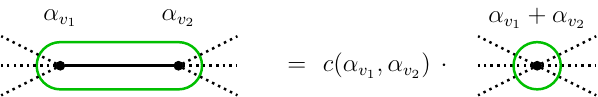}
\end{equation}
where we have displayed the weights associated to each vertex. In this expression $c(\alpha_{v_1},\alpha_{v_2})$ is a universal function of the initial weights of the vertices.
To show this, we first note that since all polynomials $p_T$, $T\in \mathcal{S}$, must enclose both $v_1$ and $v_2$, they can only depend on the combination $ x_{{v_1}}+x_{{v_2}}$. Therefore, changing coordinates to $x_+\equiv x_1+x_2$, $t \equiv x_2/(x_1+x_2)$, we obtain an integral of the form
\begin{equation}\label{int_red_1}
    \int_{\R_+} dx_{v_1}dx_{v_2}\, f(x_+)\, x_{v_1}^{\alpha_{v_1}-1} x_{v_2}^{\alpha_{v_2}-1}= \int_{\R_+} dx_{+} f(x_+) \, x_+^{\alpha_{v_1}+\alpha_{v_2}-1}\int_0^1 dt\, t^{\alpha_2-1}(1-t)^{\alpha_1-1}\,.
\end{equation}
Identifying the integral over $t$ as the Beta function $B(\alpha_1,\alpha_2)$, we can apply this logic to the integral associated to $\mathcal{S}$ to obtain

\begin{equation} \label{contr_integral}
        I_{\mathcal{S}} =B(\alpha_{v_1},\alpha_{v_2}) \cdot \int_{\R_+} d^{N_{\rm v}-2}x dx_+\frac{\prod_{v\neq 1,2} x_v^{\alpha_v-1}}{\prod_{T\in \mathcal{
        S}}p_T} \, x_+^{\alpha_{v_1}+\alpha_{v_2}-1} \,,  
\end{equation}
with the right-hand side being an integral for a diagram with only $N_{\rm v}-1$ vertices and $B$ denoting the Beta function. 

\paragraph{More general contractions.}
It turns out that the above discussion can be generalized further and equally applies to diagrams in which the vertices $v_1,v_2$ are not only connected by an edge, as we considered in \eqref{two-vertices_non-circled}. In fact, a contraction depicted in \eqref{simple_two_vertex_conctraction} generalizes to a tube $T$ that contains any two bare vertices $v_1,v_2$. Repeating the integration steps similar to \eqref{int_red_1} and \eqref{contr_integral} we thus infer the identity
\begin{equation} \label{involved_two_vertex_conctraction}
\includegraphics[valign=c]{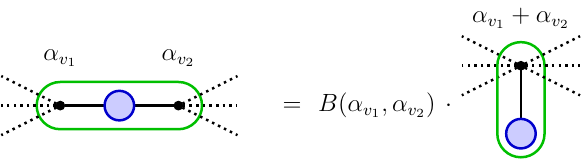}
\end{equation}
where the shaded blue circle denotes an arbitrary sub-tubing. Crucially, this observation applies regardless of the topology of the diagram. This implies that any tubing $\mathcal{T}$ of an $n$-point diagram can be reduced to a tubing of a $\vert \mathcal{T}\vert$-point diagram, relating the functions of higher-point diagrams to the ones of lower-point diagrams. These observations combined allow us to obtain the minimal representation necessary to calculate any $n$-point amplitude, which we will explain in section~\ref{sec:complexity}.

\paragraph{Relation to locality of the theory.} It is interesting to point out that using the reduction operators acting via \eqref{eq:QTonint} and the contractions \eqref{simple_two_vertex_conctraction} leads, at least in the simplest situation, to second-order differential equations that are reminiscent of locality constraints. To see this, we start from a complete tubing $\mathcal{T}$ which contains the two tubes $T_{\rm b}=\{v_1\}$ and $T_{\rm r}=\{v_2\}$ that encircle the individual vertices $v_1,v_2$
as
\begin{equation} \label{two-vertices_circled}
\begin{tikzpicture}
\begin{feynman}
        \vertex [dot](v1) {};

        \vertex[above=.7cm of v1] {$v_1$};
        \vertex [right=2 cm of v1,dot] (v2) {};  
        \vertex[above=.7cm of v2] {$v_2$};        
        \vertex[left=1cm of v1] (lc);
        \vertex[above=.5cm of lc] (lu);
        \vertex[below=.5cm of lc] (ld);

        \vertex[right=1cm of v2] (rc);
        \vertex[above=.5cm of rc] (ru);
        \vertex[below=.5cm of rc] (rd);    
        
        \diagram*  {
          (v1) --[very thick] (v2);
          (v1) --[connect,green!75!black,very thick] (v2);
          
          (v1)--[very thick,dotted] (lu);
          (v1)--[very thick,dotted] (lc);
          (v1)--[very thick,dotted] (ld);
          
          (v2)--[very thick,dotted] (ru);
          (v2)--[very thick,dotted] (rc);
          (v2)--[very thick,dotted] (rd);          
          };
        
        \vertex [above=0 cm of v1,shape=circle,draw=blue!80!black,fill=none,very thick,minimum size=.5cm] (b1) {};
        \vertex [above=0 cm of v2,shape=circle,draw=red!80!black,fill=none,very thick,minimum size=.5cm] (b2) {};
\end{feynman}
\end{tikzpicture}
\end{equation}
The diagram \eqref{two-vertices_non-circled} can be obtained from this tubing $\mathcal{T}$ by the action of the reduction operators $Q^{(T_{\rm b})}$ and  $Q^{(T_{\rm r})}$. In fact, applying \eqref{eq:QTonint} twice, 
we infer the relation 
\begin{equation}\label{eq:Qe1Qe2}
    Q^{(T_{\rm b})} Q^{(T_{\rm r})} I_\mathcal{T}=\bigg( \sum_{T\supsetneq \{v_1\}} \partial^{(T)}  \bigg)\bigg( \sum_{T'\supsetneq \{v_2\}} \partial^{(T')}  \bigg) I_{\mathcal{T}\setminus \{T_{\rm b},T_{\rm r}\}}\,.
\end{equation}
The integral $I_{\mathcal{T}\setminus \{T_{\rm b},T_{\rm r}\}}$ appearing on the right-hand side of this expression, is now associated to a tubing $\mathcal{S} = \mathcal{T}\setminus \{T_{\rm b},T_{\rm r}\}$, that contains a sub-diagram of the type \eqref{two-vertices_non-circled}. 
We now use \eqref{contr_integral} in \eqref{eq:Qe1Qe2}, and 
note that the derivatives on the right-hand side of \eqref{eq:Qe1Qe2} can be replaced by 
partial derivatives $\partial_{X_{v_1}}$ and $\partial_{X_{v_2}}$. Performing a partial integration and dropping boundary terms, we then find that these derivatives merely lead to a modification of the vertex weight $\alpha_{v_1}+\alpha_{v_2}$ to $\alpha_{v_1}+\alpha_{v_2}-2$. The resulting expression can be diagrammatically summarized as
\begin{equation}\label{eq:Qcontraction}
\includegraphics[valign=c]{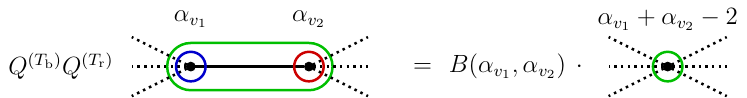}
\end{equation}
Since the reduction operators are first-order operators, we thus determined a second order differential equation relating the integral $I_{\mathcal{T}}$ to 
its contracted version.\footnote{Note that only in de Sitter space, i.e.~for $\epsilon=0$, the right hand side results in the correct weight for the vertex degree as determined from the 
same underlying model. This implies that in this case the integrals are directly related and further simplifications occur.} 

This differential equation is reminiscent of the differential equation obtained 
by using the properties of the propagator $G_e(Y_e,\eta_{v_1},\eta_{v_2})$ in a local quantum field theory. In fact, we can consider a Feynman diagram and replace the propagator $G_e(Y_e,\eta_{v_1},\eta_{v_2})$ by $\delta(\eta_{v_1}-\eta_{v_2})$. Due to the fact that the propagator satisfies the Green's function equation
\begin{equation}
     (\partial_{\eta_{v_1}}^2+Y_e^2)G_e=(\partial_{\eta_{v_2}}^2+Y_e^2)G_e=i\delta(\eta_{v_1}-\eta_{v_2})\ ,
\end{equation}
one can use integration-by-parts relations to equally derive a second-order differential equation relating different diagrams. From this, one finds an equality where on one side a second order differential operator acts on the original integral, while on the other side there is a contracted diagram with one less propagator, similar to~\eqref{eq:Qcontraction}.

\subsection{Cuts and factorizations}\label{ssec:factors}

In section \ref{ssec:contractions} we have considered diagrams that can be contracted due to the absence of minimal tubes encircling individual vertices. We next turn to the case where a maximal tube $T_{\rm max}$ is absent 
and study \textit{factorization identities} and the associated cuts in a diagram. In analogy to section~\ref{ssec:contractions}, we can remove a tube 
by using the reduction operators. In case of a maximal tube $T_{\rm max} \in \mathcal{T}$, however, we will use the higher-order reduction operator $Q^{(T_{\rm max})}_\pi$ and rely on the identity~\eqref{eq:highordtuberemov}. 

\paragraph{Factorization identities.}

While a general integral associated to 
a tubing does not factorize, it is not hard to identify tubings for which the integral splits.
To illustrate this, let us begin with a tubing $\mathcal{S}$ that can be decomposed 
as 
\begin{equation}
    \mathcal{S}=\mathcal{T}_1\sqcup \mathcal{T}_2\ ,
\end{equation}
with the crucial feature that $\mathcal{T}_1$, $\mathcal{T}_2$ are two disjoints tubings that are not inside a bigger tube. Diagrammatically, this can be represented as
\begin{equation}
\includegraphics[valign=c]{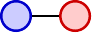}
\end{equation}
where the shaded blue and red circles denote tubings of their associated sub-diagrams.
Recall that for any tube $T$, the polynomial $p_T$ only depends on the integration variables that it encircles. This implies that, for any $T \in \mathcal{T}_1$, $p_T$ cannot depend on the integration variable of any vertex encircled by $\mathcal{T}_2$, and vice versa. This implies that the integral must factorize as 
\begin{equation} \label{max_fac}
    I_{\mathcal{S}}= \left(\int_{\R_+^k}d^kx \, \frac{\prod_{v=1}^k x_v^{\nuv-1}}{\prod_{T\in {\mathcal{T}_1}}p_T}\right)\left(\int_{\R_+^{n-k}}d^{\Nv-k}x \, \frac{\prod_{v=k+1}^{\Nv} x_v^{\nuv-1}}{\prod_{T\in {\mathcal{T}_2}}p_T}\right) = I_{\mathcal{T}_1} I_{\mathcal{T}_2}\,,
\end{equation}
where we have split the vertices such that the first $k$ are encircled by $\mathcal{T}_1$ while the others are encircled by $\mathcal{T}_2$. The resulting integrals 
 $I_{\mathcal{T}_1}$, $I_{\mathcal{T}_2}$ are associated to the blue and red subdiagrams and their respective tubings, leading to the diagrammatical representation
\begin{equation}
\includegraphics[valign=c]{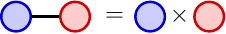}
\end{equation}
We can think of the edge connecting the two sub-diagrams as being cut. For a more general tubing, a similar factorization holds. Assuming a disjoint splitting of a tubing $\mathcal{S}$, we find 
\beq
   \mathcal{S} = \mathcal{T}_1\sqcup \ldots \sqcup \mathcal{T}_n:\qquad I_{\mathcal{S}}=I_{\mathcal{T}_1} \cdot \ldots \cdot I_{\mathcal{T}_n}\ . 
\eeq
In this case, it will result in multiple edges being cut at the same time.

\paragraph{Factorization formulas using reduction operators.}

Having established factorization identities for tubings consisting of disjoint sub-tubings, we next want to show that this situation can always be reached when applying a reduction operator. Let us start with a complete tubing $\mathcal{T}$. For such a tubing, there always is a maximal tube, which then contains two sub-tubings connected by a single edge. Diagrammatically, this can be represented as
\begin{equation}
\includegraphics[valign=c]{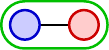}
\end{equation}
where the green tube is the maximal tube. 
From this, we find that it is possible to decompose $\mathcal{T}$ as 
\begin{equation}
    \mathcal{T}=\mathcal{T}_1\sqcup \mathcal{T}_2\sqcup \{T_{\rm max}\}\,,
\end{equation}
where $\mathcal{T}_1$, $\mathcal{T}_2$ are two disjoint tubings of the sub-diagram as above.
We can now use an appropriate reduction operator to remove $T_{\rm max}$. Since the maximal tube is not contained in another tube, it has no first-order reduction operator associated to it and we must consider the higher-order operators $Q^{(T_{\rm max})}_\pi$ associated to partitions of $T_{\rm max}$. There will be multiple of these, and for each such reduction operator it will realize the factorization described above. Here, we will use that $\mathcal{T}_1$ and $\mathcal{T}_2$ both have their own maximal tubes $T_1$ and $T_2$. and that every vertex is enclosed by either of the two. Thus, there is a natural partition
\begin{equation}
    T_{\rm max}=T_{1}\sqcup T_{2}\, , \quad \pi=\{T_1,T_2\}\ .
\end{equation}
Evaluating the higher-order reduction operator~\eqref{higher-order-red_final} with this partition, we find 
\begin{equation}\label{eq:QTmax}
    Q^{(T_{\rm max})}_\pi=\partial^{(T_1)} \partial^{(T_2)} \big(z^{(T_{\rm max})}  -z^{(T_1)} -z^{(T_2)}  \big)\ .
\end{equation}
Inserted in \eqref{eq:highordtuberemov}, this operator will then satisfy
\begin{equation} \label{apply_red_op}
    Q^{(T_{\rm max})}_\pi I_\mathcal{T}= \big(\partial^{(T_1)}  I_{\mathcal{T}_1}\big) \big(\partial^{(T_2)}  I_{\mathcal{T}_2}\big)\ .
\end{equation}
Note that there are many other reduction operator that might realize this factorization. For example, we can consider
\begin{equation}
    T_{\rm max}=\bigsqcup_{v=1}^{N_{\rm v}} \{v\}
\end{equation}
in order to obtain a higher-order reduction operator of degree $N_{\rm v}$ realizing the same factorization.

\paragraph{Singularity structure.}
Interestingly, the higher-order reduction operators have implications for the singularity structure of the cosmological correlators. Using \eqref{eq:QTmax} in~\eqref{apply_red_op}, we find
\begin{equation}  \partial^{(T_1)} \partial^{(T_2)} \big(z^{(T_{\rm max})} -z^{(T_1)} -z^{(T_2)} \big) I_\mathcal{T}=\partial^{(T_1)} \partial^{(T_2)} I_{\mathcal{T}_1}I_{\mathcal{T}_2}\, .
\end{equation}
This equation can be integrated directly, implying that $I_\mathcal{T}$ can be written as
\begin{equation} \label{singularity_split}
    I_\mathcal{T}=\frac{I_{\mathcal{T}_1} I_{\mathcal{T}_2}+f_{T_1}+f_{T_2}}{z^{(T_{\rm max})} -z^{(T_1)} -z^{(T_2)} } = - \frac{I_{\mathcal{T}_1} I_{\mathcal{T}_2}+f_{T_1}+f_{T_2}}{2 Y_e}\ ,
\end{equation}
where $f_{T_1}$ and $f_{T_2}$ are functions independent of $z^{(T_2)} $ and $z^{(T_1)} $ respectively.  Here we have rewritten the denominator in terms of the physical variables using \eqref{physical_slice}, where $Y_e$ is the momentum flowing along the edge connecting $\mathcal{T}_1$ and $\mathcal{T}_2$. Furthermore, since $I_{\mathcal{T}_1} I_{\mathcal{T}_2}$ is independent of $z^{(T_{\rm max})} $, the above equations imply that
\begin{equation}
    {\rm Res}_{Y_e\rightarrow 0}(I_{\mathcal{T}}) = I_{\mathcal{T}_1} I_{\mathcal{T}_2} +\ldots \,,
\end{equation}
where $\rm Res$ denotes the residue around $Y_e\rightarrow 0$ and the dots are some unknown terms due to $f_{T_1}$ and $f_{T_2}$. Factorizations as described above seem related to those obtained for amplitudes in both quantum field theory \cite{cachazo_sharpening_2008,arkani-hamed_locality_2018,travaglini_sagex_2022} as well as for cosmological correlators \cite{frellesvig_decomposition_2021,goodhew_cosmological_2021,melville_cosmological_2021,jazayeri_locality_2021,dipietro_analyticity_2022,baumann_linking_2022,albayrak_perturbative_2024}. Note also that similar factorization formulae will hold for the other higher-order reduction operators.

\section{Differential chains from first-order operators}\label{sec:differentialchain}

One of the crucial observations in the results of the previous section is that acting with a reduction operator effectively removes a tube from a tubing. In this section, we will show that this allows us to use the first-order reduction operators to write derivatives of $I_\mathcal{T}$ as a sum of integrals $I_\mathcal{S}$ with $\cS \subseteq \cT$. This will enable us to determine a system of differential equations for the integral $I_\cT$ with a remarkable similarity to the kinematic flow algorithm of~\cite{arkani-hamed_kinematic_2023}. In section~\ref{ssec:chainconstruction} we present an algorithm to construct the general form of this differential chain. We then illustrate the involved steps in an explicit example in section~\ref{ssec:chainexamples}.

\subsection{Algorithmic construction of the differential chains}\label{ssec:chainconstruction}

\paragraph{A system of differential equations.}

We begin the construction of the differential equations by recalling the key insights from section~\ref{sec:physics}. Consider a tubing $\cT$ and a tube $T$, the corresponding integral $I_\cT$, and the first-order reduction operator $Q^{(T)}$. Now, depending on whether $T\in \cT$ or not, there are two possibilities. Either acting with $Q^{(T)}$ removes the tube as in equation~\eqref{eq:QTonint}, or the tube is already removed and we have
\begin{equation}
    \partial^{(T)} I_\cT=0\,,
\end{equation}
since the polynomial involving $z^{(T)} $ has been removed. Because there is a first-order reduction operator $Q^{(T)}$ for every non-maximal tube $T \in \cT$, this implies that there is the following system of equations
\begin{align} \label{eq:ITsystem}
        Q^{(T)}I_\mathcal{T}&=\sum_{T'\supsetneq T} \partial^{(T')}  I_{\mathcal{T}\setminus \{T\}} & \text{ if } T \in \mathcal{T}\,,\\
        \partial^{(T)} I_\mathcal{T}&=0 &\text{ if } T\not\in \mathcal{T} 
\end{align}
for every tubing $\cT$. Here, for the convenience of the reader, we recall that the first-order reduction operator takes the form  
\begin{equation}\label{eq:recallQ}
    Q^{(T)}=z^{(T)}\sum_{ T'\supsetneq T}\partial^{(T')}+\sum_{T'\subseteq T}(\theta^{(T')}+\nu^{(T')})-\sum_{v\in T}\alpha_v\ .
\end{equation}
This system of equations is the starting point for iteratively constructing a solution.

\paragraph{A differential chain.}
An essential property of the system of differential equations in \eqref{eq:ITsystem} is that the right-hand side only involves tubings containing strictly fewer tubes than $\cT$. This means that the equation can be iterated, leading to an expression in terms of increasingly smaller tubings. The only tube that can not be removed in this manner is the maximal tube $T_{\rm max}$, as from equation~\eqref{eq:ITsystem} it follows that
\begin{equation}\label{eq:QTmaxIcT}
    Q^{(T_{\rm max})}I_\cT=0\,.
\end{equation}
Therefore, the tube removal continues until only the maximal tube remains. 

Then, equation~\eqref{eq:QTmaxIcT} implies that associated integral must satisfy
\begin{equation}\label{eq:ITmaxdif}
    \partial^{(T_{\rm max})} I_{\{T_{\rm max}\}}=
    \frac{1}{z^{(T_{\rm max})}} \left(\sum_{v\in T_{\rm max}}\alpha_v -1\right) I_{\{T_{\rm max}\} }\,.
\end{equation}
Note that all of the other derivatives vanish, since $I_{\{T_{\rm max}\}} $ only depends on $z^{(T_{\rm max})}$. This allows us to solve for $I_{\{T_{\rm max}\}}$, which will consist of $z^{(T_{\rm max})}$ raised to a complex power.

From here, we can iteratively add tubes. In particular, let us first add a single tube tube $T$, and use the system~\eqref{eq:ITsystem} combined with equation~\eqref{eq:ITmaxdif} to write partial derivatives acting on $I_{\{T_{\rm max},T\}}$ in terms of the function itself and $I_{\{T_{\rm max}\}}$. Then, adding another tube $T'$, we can write partial derivatives acting on the new integrals in terms of itself and the functions $I_{\{T_{\rm max},T\}}$, $I_{\{T_{\rm max},T'\}}$ and $I_{\{T_{\rm max}\}}$. Continuing in this manner, we obtain a chain of first-order differential equations which starts with $I_{\{T_{\rm max}\}}$ and ends with the desired integral $I_{\cT}$. As a result, for any tubing $\cT$, the derivatives of $I_{\cT}$ can be expressed in the general form 
\begin{equation}\label{eq:ITpartialfin}
    \partial^{(T)}  I_\mathcal{T} = \sum_{\substack{\mathcal{S}\subseteq \mathcal{T},\\ T_{\rm max}\in \mathcal{S}}} r^{(T)}_\mathcal{S} (\mathcal{T}) \, I_{\mathcal{S}} \,
\end{equation}
where the $r^{(T)}_{\cS}(\cT)$ are rational functions of $\alpha_v$ and $z^{(T)}$ and the sum is over all sub-tubes of $\mathcal{T}$ which contain $T_{\rm max}$. 

In the remainder of this subsection, we will aim to make the structure of this differential chain as explicit as possible. We stress that the procedure is fully algorithmic and can be easily implemented computationally. Nevertheless, one needs to introduce some extra notation if one wants to write down closed-form expressions.

\paragraph{Notation for tube structure.}
 In the following, it is necessary to carefully keep track of the structure of tubes and tubings, and for this purpose we introduce the following notation. Given a tubing $\cT$ and two tubes $S,T\in \cT$, we write 
\begin{equation}
    S \prec_\cT T\ ,
\end{equation}
whenever $S\subsetneq T$ and there exists no $T'\in \cT$ such that $S\subsetneq T' \subsetneq T$. The interpretation is that if we sort the tubes in $\cT$ by inclusion, then $S$ is the precursor of $T$, and $T$ is the successor of $S$. Note that a tube may have any number of precursors, but the successor of a tube is unique.\footnote{The ordering $\prec_\cT$ gives the tubes in $\cT$ the structure of an ordered tree. This tree will be rooted if $\cT$ contains the maximal tube $T_{\rm max}$, which will always be the case for us. Additionally, complete tubings are in one-to-one correspondence with full binary trees. This perspective will be useful for the combinatorial analysis performed later.} We denote the successor of a tube $T$ in a tubing $\cT$ by $T^+_\cT$. Finally, we write
\begin{equation}
    S \sim_\cT T\ ,
\end{equation}
whenever $S$ and $T$ have the same successor in $\cT$. 

To illustrate this notation, consider the tubing $\cT$ given by
\begin{equation}
\includegraphics[valign=c]{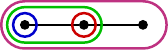}
\end{equation}
where we denote the red, blue, green and magenta tubes by $T_{\rm r}$, $T_{\rm b}$, $T_{\rm g}$ and $T_{\rm m}$ respectively.
Here we have $T_{\rm b} \sim_\cT T_{\rm r}$, since they have the same successor $T_{\rm g}$, which we can express as $(T_{\rm b})_\cT^+=(T_{\rm r})_\cT^+=T_{\rm g}$. Similarly we find  $T_{\rm b}\prec_\cT T_{\rm g} \prec_{\cT} T_{\rm m}$. On the contrary, we see that $T_{\rm m}$ is not a successor of $T_{\rm b}$ or $T_{\rm r}$, i.e.~$T_{\rm{b}} \nprec_{\cT} T_{\rm m}$.

\paragraph{Derivation of the differential chain.}
With this notation, we can derive an explicit form of the differential chain. We begin by observing that, for any $T\in \cT$, there is the following identity:
\begin{equation}
    \sum_{T'\subseteq T}(\theta^{(T')}+\nu^{(T')}) = \theta^{(T)}+\nu^{(T)} +\sum_{S \prec_\cT T}  \sum_{T'\subseteq S}(\theta^{(T')}+\nu^{(T')})\,.
\end{equation}
Similarly, for any $S$ with $S\prec_\cT T$ we have 
\begin{equation}\label{eq:partialSsum}
    \sum_{T'\supsetneq S} \partial^{(T')}=\partial^{(T)}+\sum_{T'\supsetneq T} \partial^{(T')}\,.
\end{equation}
Comparing this with the definition of the first-order reduction operator recalled in \eqref{eq:recallQ} above, it follows that 
\begin{equation}\label{eq:QT-QS}
    Q^{(T)}-\sum_{S \prec_\cT T }Q^{(S)}= \left(z^{(T)}-\sum_{S\prec_\cT T}z^{(S)}\right)\left(\sum_{T'\supseteq T}\partial^{(T')} \right)+\gamma^{(T)}_\mathcal{T}\,,
\end{equation}
where $\gamma^{(T)}_\cT$ is a constant defined by 
\begin{equation}\label{eq:gammadef}
    \gamma^{(T)}_\mathcal{T}= \nu^{(T)}-\sum_{\substack{v\in T,\\ v\not\in S\subsetneq T}} \alpha_v \,.
\end{equation}
Here the sum is over all vertices in $T$ that are not enclosed by any of the sub-tubes of $T$. The relation derived above can be rewritten to 
\begin{equation}\label{eq:partialTinQs1}
\boxed{\rule[-.6cm]{0cm}{1.5cm} \quad
    \sum_{T'\supseteq T}\partial^{(T')}= \frac{    Q^{(T)}-\sum_{S\prec_\cT T }Q^{(S)}-\gamma^{(T)}_\mathcal{T}}{z^{(T)}-\sum_{S\prec_\cT T}z^{(S)}}\,,\quad
    }
\end{equation}
which holds for any tube $T$.\footnote{One might worry that, since these expressions depend heavily on the tubing $\mathcal{T}$, these expressions only hold when acting on $I_\mathcal{T}$. In particular as, when acting on an arbitrary tubing $\mathcal{T}$, these expressions will involve terms of the type $Q^{(T)}I_\mathcal{T}$ with $T$ not in the tubing $\mathcal{T}$. However, one can use the fact that $\partial^{(T)}I_\mathcal{T}=0$ to fix these extractions and one obtains a result  compatible with the above. Note that, for us we will not need such expressions in any case.} This result can straightforwardly be translated to an expression for $\pd^{(T)}I_\cT$ in terms of the reduction operators by noting that
\begin{equation}
    \pd^{(T)} I_\cT = \left(\sum_{T' \supseteq T}\pd^{(T')} - \sum_{T'\supseteq T^+_\cT }\pd^{(T')}   \right) I_\cT
\end{equation}
and using \eqref{eq:partialTinQs1} for the two sums. For brevity we do not display the resulting equation here.

Now, observe that the operator on the left-hand side closely resembles the operator appearing on the right-hand side of equation \eqref{eq:ITsystem}; the difference is that the latter has one extra term. To connect these two equations, note that 
\begin{equation}
    \sum_{T'\supsetneq T}\partial^{(T')}=\sum_{T'\supseteq T^+_\cT}\partial^{(T')} \,.
\end{equation}

Combining this with equations \eqref{eq:ITsystem} and \eqref{eq:partialTinQs1}, we find that the action of a reduction operator $Q^{(T)}$ on $I_\cT$ can be written as  
\begin{equation}\label{eq:QTreduction1}
    Q^{(T)} I_\cT = \frac{  Q^{(T^+_\cT)}-\sum_{S \prec_{\cT \setminus\{T\}} T^+_\cT    }Q^{(S)}-\gamma^{(T^+_\cT) }_{\cT\setminus\{T\}}}{z^{(T^+_\cT)}-\sum_{S \prec_{\cT \setminus\{T\}} T^+_\cT  }z^{(S)}} I_{\cT\setminus \{T\}}\,.
\end{equation}
In this equation the iterative nature of the reduction operator is made manifest; using this equation the expression for $Q^{(T)} I_\cT$ can be recursively reduced until it is a linear combination of integrals $I_\cS$ with $\cS\subseteq \cT$ with rational coefficients. Finally, we note that, as the maximal tube $T_{\rm max}$ has no successor, the right-hand side of~\eqref{eq:QTreduction1} is not well-defined when evaluated for $T_{\rm max}$. Therefore, in this case we must separately replace $Q^{(T_{\rm max})}I_\cT$ by zero.

\paragraph{Compact form of differential chain.}
The equations derived above are explicit, but somewhat complicated. To increase its usability, we now rewrite \eqref{eq:QTreduction1} in a more compact form. We do this by first introducing the matrices 
\begin{align}\label{eq:Mdef}
    M^{T,S}_{\cT} &= \begin{cases}
    1&  \text{if } T \prec_\cT S\,, \\
    -1& \text{if } T\succ_\cT S \text{ or } T\sim_{\cT} S \\
    0  & \text{else}\,.
    \end{cases} 
\end{align}
for each $\cT$. Then, using this notation, we define the functions
\begin{equation}\label{eq:letterM}
    \ell^{(T)}_\cT = \bigg( \sum_{S\in \cT\setminus \{T\}}M^{T,S}_\cT z^{(S)}  \bigg)^{-1} \,.
\end{equation}
as well as the constants
\begin{equation}\label{eq:cdef}
    c^{(T)}_{\cT} = \sum_{S\in \cT\setminus T} M^{T,S}_\cT \sum_{v\in S} \alpha_v-1\,.
\end{equation}
Note that $c^{(T)}_{\cT}=-\gamma^{(T^+_\cT)}_\cT$, with $\gamma$ as in equation~\eqref{eq:gammadef}. With this new notation, equation \eqref{eq:QTreduction1} can be compactly written as
\begin{equation}\label{eq:QTiteration}
\boxed{\rule[-.6cm]{0cm}{1.4cm} \quad 
    Q^{(T)}I_\mathcal{T} = \ell_{\cT}^{(T)} \bigg(\sum_{S\in\cT\setminus \{T\}}M^{T,S}_{\cT} Q^{(S)} +c^{(T)}_{\cT} \bigg)I_{\cT\setminus\{T\}} \,.\quad
}
\end{equation}
Note that, as in equation \eqref{eq:QTreduction1}, this expression does not hold for $Q^{(T_{\rm max})}$, in which case we must impose $Q^{(T_{\rm max})}I_\cT =0$. Furthermore, note that $M^{T,S}_\cT$ is a purely combinatorial object, and can be found algorithmically using the index set representation of the tubes.

In summary, to obtain $Q^{(T)}I_\cT$ one must first apply equation~\eqref{eq:QTiteration}. Then, there will be terms of the form $Q^{(S)}I_{\cT\setminus \{T\}}$ for various $S$ and equation~\eqref{eq:QTiteration} can again be used on these terms to remove yet another tube from the tubing. This procedure can be recursively applied until only the maximal tube remains. As we know the remaining integral satisfies $Q^{(T_{\rm max})}I_{\{T_{\rm max}\}}$, this signals the end of the recursion. Inserting all of this into the original expression for $Q^{(T)}I_\cT$, one is left with an algebraic expression for the action of $Q^{(T)}$ in terms of the other integrals in the chain. Repeating this for each tube in $\cT$, one can then apply equation~\eqref{eq:partialTinQs1} relating the reduction operators with the partial derivative and hence determine the coefficients $r^{(T)}_\cS(\cT)$ in \eqref{eq:ITpartialfin}. As an alternative, we show in appendix~\ref{ap:matrixrep} how the iteration in equation~\eqref{eq:QTiteration} can also be rewritten and solved by interpreting it as a matrix equation on a suitable vector space, resulting in a direct expression of $Q^{(T)} I_\cT$ in terms of the other integrals.

\subsection{An example: the single-exchange diagram  } \label{ssec:chainexamples}

In order to illustrate the construction above, we will again return to the example of the single-exchange integral and construct its differential chain explicitly. Doing this, we will see the iterative nature of this differential chain, motivating the nomenclature of recursive reductions. As the purpose of this section is to illustrate the results above, we will treat this simple example with the general technology, even though directly solving the system~\eqref{eq:ITsystem} would be more efficient in this case.

\paragraph{Functions in the chain.}

To construct the chain, recall that the single-exchange integral arises from the tubing
\begin{equation}\label{eq:singexchagnetubingS5S2}
\includegraphics[valign=c]{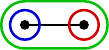}
\end{equation}
and that we have labeled the blue, red and green tubes as $T_{\rm b}$, $T_{\rm r}$ and $T_{\rm g}$ respectively. Furthermore, this GKZ system has three first-order reduction operators, each associated to a tube. As discussed in section~\ref{ssec:chainconstruction}, we can construct the differential chain by studying the action of these reduction operators. In particular, we know that acting with the reduction operators $Q^{(T_{\rm r})}$ and $Q^{(T_{\rm b})}$ will remove the red or blue tube from the tubing, while $Q^{(T_{\rm g})}$ will annihilate the functions in the chain, since $T_{\rm g}$ is the maximal tube. Thus we find that there are four tubings we must consider, organized as 
\begin{equation}\label{eq:singexchdifchain}
\includegraphics[valign=c]{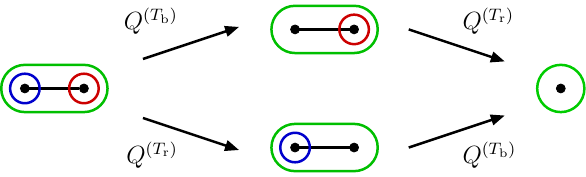}
\end{equation}
where the arrows indicate that the reduction operator acts on the integral associated to the left tubing can be written in terms of derivatives of the right tubing. Note that in the right-most diagram here, we have already used the discussion from section~\ref{ssec:contractions} to contract the edge. Interestingly, the structure of these differential chains is quite similar to the kinematic flow algorithm of \cite{arkani-hamed_differential_2023}.

From the above, we find that we must consider four functions in our differential chain, given by
\beq 
I_\text{\includegraphics[valign=c]{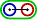}}\,,\quad  
I_\text{\includegraphics[valign=c]{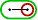}}\,,\quad 
I_\text{\includegraphics[valign=c]{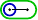}}\,,\quad 
I_\text{\includegraphics[valign=c,scale=.7]{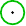}}\ ,
\eeq 
where, in order not to clutter the notation, we have drawn the tubings explicitly in the subscripts. 

\paragraph{General approach.}

The next step is to construct the differential chain. We are now ready to obtain the action of the reduction operators on the functions found above. We will begin with the function $I_\text{\includegraphics[valign=c]{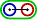}}$ 
and consider the action of $Q^{(T_{\rm b})}$. Recall from section~\ref{ssec:chainconstruction} that, in general, the action of a reduction operator is given in terms of the symbol $M^{T,S}_\cT$, which is determined by the successor structure of $\cT$. Therefore, the first step will be to determine this successor structure for the tubing of interest. Afterwards, we obtain an iterative equation relating the action of $Q^{(T)}$ on $I_\cT$ with reduction operators acting on $I_{\cT \setminus\{T\}}$. Repeating the above procedure we are eventually left with a linear combination of integrals in the differential chain with rational pre-factors. Then, one can use equation \eqref{eq:partialTinQs1} to solve for the partial derivatives in terms of the reduction operators.

\paragraph{The first reduction.}

In general, the successor structure of a diagram can be computed algorithmically using the fact that we can represent tubes as index sets and tubings as sets of tubes. However, given a diagrammatical representation of a tubing, it can also be observed immediately. For us, considering the tubing~\eqref{eq:singexchagnetubingS5S2} we find that the only successor relations are
\begin{equation}
    T_{\rm b} \prec_\text{\includegraphics[valign=c]{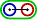}} T_{\rm g}\, , \quad   T_{\rm r} \prec_\text{\includegraphics[valign=c]{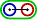}} T_{\rm g}\, , \quad   T_{\rm b} \sim_\text{\includegraphics[valign=c]{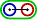}} T_{\rm r}\, .
\end{equation}
From this and the definition of $M^{T,S}_\cT$ in~\eqref{eq:Mdef}, we can immediately read off the non-zero elements of $M^{T,S}_\text{\includegraphics[valign=c]{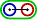}}$, which are given by
\begin{equation}
\begin{array}{rrrrr}
  &  M^{T_{\rm b},T_{\rm b}}_\text{\includegraphics[valign=c]{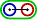}}&=M^{T_{\rm r},T_{\rm r}}_\text{\includegraphics[valign=c]{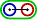}}&= M^{T_{\rm g},T_{\rm g}}_\text{\includegraphics[valign=c]{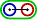}}&=-1\, ,\\[5pt]
       M^{T_{\rm b},T_{\rm r}}_\text{\includegraphics[valign=c]{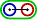}} &=M^{T_{\rm r},T_{\rm b}}_\text{\includegraphics[valign=c]{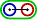}}&= M^{T_{\rm g},T_{\rm b}}_\text{\includegraphics[valign=c]{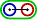}}&= M^{T_{\rm g},T_{\rm r}}_\text{\includegraphics[valign=c]{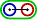}}  &=-1\, ,\\[5pt]
 & &  M^{T_{\rm b},T_{\rm g}}_\text{\includegraphics[valign=c]{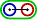}}&= M^{T_{\rm r},T_{\rm g}}_\text{\includegraphics[valign=c]{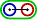}}&=  1 \,.
\end{array}
\end{equation}
The letters $\ell^{(T)}_\text{\includegraphics[valign=c]{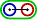}}$, as well as the constants $c^{(T)}_\text{\includegraphics[valign=c]{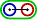}}$ can be readily obtained from these matrices using equations~\eqref{eq:letterM} and~\eqref{eq:cdef}. Note that, as acting with the reduction associated to the maximal tube will always result in zero, we will not need to obtain the letters $ \ell^{(T_{\rm g})}_\cT$ or constants $ c^{(T_{\rm g})}_\cT$ for any of the tubings of the single-exchange diagram. Thus, we find that the remaining letters are given by
\begin{equation}\label{eq:singexchangeletters1}
    \ell^{(T_{\rm b})}_\text{\includegraphics[valign=c]{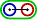}}=\frac{1}{z^{(T_{\rm g})}-z^{(T_{\rm r})}}\,, \quad  \ell^{(T_{\rm r})}_\text{\includegraphics[valign=c]{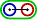}}=\frac{1}{z^{(T_{\rm g})}-z^{(T_{\rm b})}}\,,
\end{equation}
while the constants are given by
\begin{equation}\label{eq:singexchangeconsts}
      c^{(T_{\rm b})}_\text{\includegraphics[valign=c]{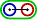}}=\alpha_1-1\,, \quad  c^{(T_{\rm r})}_\text{\includegraphics[valign=c]{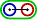}}=\alpha_2-1\,,
\end{equation}
where we recall that $\alpha_1$ and $\alpha_2$ are the twists of the vertices encircled by the blue tube and red tube respectively.

Using the equations above, obtaining the action of the reduction operators on $I_\text{\includegraphics[valign=c]{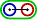}}$ comes from a straightforward application of equation~\eqref{eq:QTiteration}, which yields
\begin{equation}\label{eq:QTbactsingexchange2}
\begin{split}
    Q^{(T_{\rm b})}I_\text{\includegraphics[valign=c]{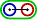}}=&\, \frac{Q^{(T_{\rm g})}-Q^{(T_{\rm r})}+\alpha_1-1}{z^{(T_{\rm g})}-z^{(T_{\rm r})}}I_\text{\includegraphics[valign=c]{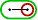}}\,, \\
            Q^{(T_{\rm r})}I_\text{\includegraphics[valign=c]{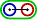}}=& \,\frac{Q^{(T_{\rm g})}-Q^{(T_{\rm b})}+\alpha_2-1}{z^{(T_{\rm g})}-z^{(T_{\rm b})}}I_\text{\includegraphics[valign=c]{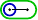}}\,, \\
            Q^{(T_{\rm g})}I_\text{\includegraphics[valign=c]{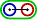}}=& \,0\,,
\end{split}
\end{equation}
where, for the last equality, we have used the fact that acting with the reduction operator of the maximal tube always results in zero.

From equation \eqref{eq:QTbactsingexchange}, the recursive nature of the reduction operators immediately becomes clear. We see that, in order to obtain the action of $Q^{(T_{\rm b})}$ on $I_\text{\includegraphics[valign=c]{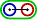}}$, we must now proceed by obtaining the action of the reduction operators on the sub-tubings of~\eqref{eq:singexchagnetubingS5S2}. For general diagrams, this procedure will continue until all tubes but the maximal one are removed.

\paragraph{The second reduction.}
Thus, the next task at hand is to obtain the action of the reduction operators on the sub-tubings of~\eqref{eq:singexchagnetubingS5S2}. Here, we will consider $I_\text{\includegraphics[valign=c]{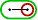}}$ and note that the actions on $I_\text{\includegraphics[valign=c]{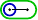}}$ can be obtained by permutations. We must consider two reduction operators now, $Q^{(T_{\rm g})}$ and $Q^{(T_{\rm r})}$. The action of $Q^{(T_{\rm g})}$ must still be zero as $T_{\rm g}$ is the maximal tube. The only successor relation of this diagram is 
\begin{equation}
 T_{\rm r} \prec_\text{\includegraphics[valign=c]{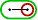}} T_{\rm g}
\end{equation}
resulting in
\begin{equation}
       M^{T_{\rm g},T_{\rm g}}_\text{\includegraphics[valign=c]{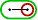}}= M^{T_{\rm r},T_{\rm r}}_\text{\includegraphics[valign=c]{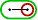}}= M^{T_{\rm g},T_{\rm r}}_\text{\includegraphics[valign=c]{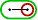}}=-1\, , \quad M^{T_{\rm r},T_{\rm g}}_\text{\includegraphics[valign=c]{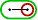}}=1
\end{equation}
for the symbol $M$. Then, proceeding along the same lines as above we obtain
\begin{equation}\label{eq:singexchangelettersandcoefs2}
      \ell^{(T_{\rm r})}_\text{\includegraphics[valign=c]{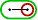}}=\frac{1}{z^{(T_{\rm g})}}\,, \quad       c^{(T_{\rm r})}_\text{\includegraphics[valign=c]{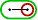}}=\alpha_1+\alpha_2-1\,,
\end{equation}
where again, we note that it is not necessary to obtain the corresponding expressions for $T_{\rm g}$ as it is the maximal tube. Inserting the above into equation~\eqref{eq:QTiteration} we obtain
\begin{equation}\label{eq:QTrInored}
\begin{split}
    Q^{(T_{\rm r})} I_\text{\includegraphics[valign=c]{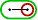}}&=\frac{Q^{(T_{\rm g})}+\alpha_1+\alpha_2-1}{z^{(T_{\rm g})}}I_\text{\includegraphics[valign=c]{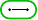}}\,, \\
Q^{(T_{\rm g})}  I_\text{\includegraphics[valign=c]{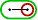}}&= 0\,.
\end{split}
\end{equation}
The corresponding equations can be obtained for $I_\text{\includegraphics[valign=c]{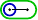}}$ can be obtained by permuting $T_{\rm r}$ with $T_{\rm b}$. We know from section~\ref{ssec:contractions} that $I_\text{\includegraphics[valign=c]{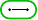}}=B(\alpha_1,\alpha_2)I_\text{\includegraphics[valign=c,scale=.7]{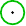}}$, with $B$ the beta-function. This allows us to rewrite the first equation in~\eqref{eq:QTrInored} in terms of $I_\text{\includegraphics[valign=c,scale=.7]{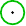}}$. Furthermore, we know that this integral must satisfy
\begin{equation}
    Q^{(T_{\rm g})}  I_\text{\includegraphics[valign=c,scale=.7]{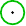}}= 0
\end{equation}
and we find that the iteration terminates here.

Finally, we can simply insert~\eqref{eq:QTrInored} in equation~\eqref{eq:QTbactsingexchange}, combined with the corresponding equations for $T_{\rm b}$, and obtain
\begin{equation}\label{eq:QTbactsingexchange}
\begin{split}
    Q^{(T_{\rm b})}I_\text{\includegraphics[valign=c]{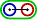}}=&\, \frac{\alpha_1-1}{z^{(T_{\rm g})}-z^{(T_{\rm r})}}I_\text{\includegraphics[valign=c]{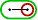}}- \frac{(\alpha_1+\alpha_2-1)B(\alpha_1,\alpha_2)}{z^{(T_{\rm g})}(z^{(T_{\rm g})}-z^{(T_{\rm r})})}I_\text{\includegraphics[valign=c,scale=.7]{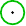}}\,,    \\
            Q^{(T_{\rm r})}I_\text{\includegraphics[valign=c]{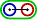}}=& \frac{\alpha_2-1}{z^{(T_{\rm g})}-z^{(T_{\rm b})}}I_\text{\includegraphics[valign=c]{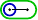}}- \frac{(\alpha_1+\alpha_2-1)B(\alpha_1,\alpha_2)}{z^{(T_{\rm g})}(z^{(T_{\rm g})}-z^{(T_{\rm b})})}I_\text{\includegraphics[valign=c,scale=.7]{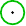}}\,,     \\
            Q^{(T_{\rm g})}I_\text{\includegraphics[valign=c]{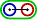}}=& \,0\,,
\end{split}
\end{equation}
Using these expression, we can now obtain the action of the partial derivatives on $I_\text{\includegraphics[valign=c]{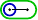}}$.

\paragraph{Partial derivatives.}

Now that we have found the action of all the reduction operators, the next step is to apply equation \eqref{eq:partialTinQs1} to rewrite the partial derivatives in terms of the reduction operators. Considering this equation for all $T$, one can straightforwardly solve for the partial derivatives.  In the following, we will focus on $\partial^{(T_{\rm b})}$, although the process will be similar for the other derivatives in the chain. 
To obtain the partial derivatives for the single exchange integral, let us insert $T=T_{\rm b}$ in equation~\eqref{eq:partialTinQs1} and act with it on $I_\text{\includegraphics[valign=c]{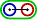}}$. 
In this case, we find
\begin{equation}
\begin{split}
  (\partial^{(T_{\rm b})}+\partial^{(T_{\rm g})})I_\text{\includegraphics[valign=c]{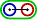}}=&
        \frac{\alpha_1-1}{z^{(T_{\rm b})}}\gamma^{(T)}_\text{\includegraphics[valign=c]{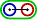}}I_\text{\includegraphics[valign=c]{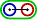}}     +\frac{\alpha_1-1}{z^{(T_{\rm b})}(z^{(T_{\rm g})}-z^{(T_{\rm r})})}I_\text{\includegraphics[valign=c]{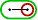}}\\
        &- \frac{(\alpha_1+\alpha_2-1)B(\alpha_1,\alpha_2)}{z^{(T_{\rm b})}z^{(T_{\rm g})}(z^{(T_{\rm g})}-z^{(T_{\rm r})})}I_\text{\includegraphics[valign=c,scale=.7]{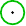}}\,.        
\end{split}
\end{equation}
Again, the corresponding equation for $T_{\rm r}$ can be found in an identical manner.

The final equation we need is obtained by inserting $T_{\rm g}$ in equation~\eqref{eq:partialTinQs1}. The combined system of equations is easily solved for the partial derivatives, giving
\begin{equation}\label{eq:singexchange3}
    \partial^{(T_{\rm b})}   I_\text{\includegraphics[valign=c]{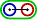}} 
        = 
        r_\text{\includegraphics[valign=c]{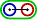}}I_\text{\includegraphics[valign=c]{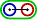}}        +         r_\text{\includegraphics[valign=c]{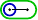}}I_\text{\includegraphics[valign=c]{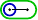}}         +         r_\text{\includegraphics[valign=c]{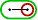}}
        I_\text{\includegraphics[valign=c]{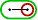}}         +         r_\text{\includegraphics[valign=c,scale=.7]{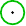}}I_\text{\includegraphics[valign=c,scale=.7]{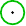}}   
\end{equation}
where the coefficients are given by
\begin{equation}
\begin{array}{ll}
     r_\text{\includegraphics[valign=c]{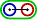}}&= \frac{\alpha_1 -1}{z^{(T_{\rm b})} }+\frac{1}{z^{(T_{\rm g})} -z^{(T_{\rm b})} -z^{(T_{\rm r})} }  \, ,\\
   r_\text{\includegraphics[valign=c]{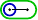}}  &=  \frac{\alpha_2-1}{(z^{(T_{\rm g})} -z^{(T_{\rm b})} )(z^{(T_{\rm g})} -z^{(T_{\rm b})} -z^{(T_{\rm r})} )}\, ,\\
   r_\text{\includegraphics[valign=c]{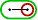}} &= \frac{\alpha_1-1}{z^{(T_{\rm b})} (z^{(T_{\rm g})} -z^{(T_{\rm b})} -z^{(T_{\rm r})} )}\, , \\
r_\text{\includegraphics[valign=c,scale=.7]{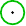}} &= \frac{(\alpha_1+\alpha_2-1)B(\alpha_1,\alpha_2)}{z^{(T_{\rm b})} (z^{(T_{\rm g})} -z^{(T_{\rm b})} )(z^{(T_{\rm g})} -z^{(T_{\rm b})} -z^{(T_{\rm r})} )}\,,
\end{array}
\end{equation}
Using the same methods, similar expressions can be found for $\partial^{(T_{\rm r})}$ and $\partial^{(T_{\rm g})}$, as well as how the derivatives act on other functions in the chain.

\begingroup
   \renewcommand{\drawFM}[1]{ \begin{tikzpicture}[baseline=-0.5ex] \begin{feynman} #1 \end{feynman}\end{tikzpicture}}

\section{Algebraic relations and the recursive reduction algorithm}
\label{sec:complexity}

In the previous section we have shown that to parameterize any tree-level cosmological correlator one can construct a basis of functions that is closed under partial derivatives by merely using the first-order reduction operators. The next natural step is to consider the role of the higher-order reduction operators. In this section we will study these operators in more detail and argue that they imply algebraic relations between various basis functions.

We begin in section~\ref{ssec:higherrelations} by showing explicitly how the higher-order reduction operators  lead to algebraic relations. Afterwards, we will showcase some examples of such relations in section~\ref{ssec:higherrelsexample}. Then, we will explain in section~\ref{ssec:minimalbasis} how these relations help to obtain a more minimal set of basis functions. Furthermore, we will illustrate the reduction in complexity by showing that the full double-exchange correlator can be expressed in terms of only four such functions. We will leave the exact counting of these minimal representation functions to the following chapter, in particular this can be found in section~\ref{ssec:minrepcounting}.

\subsection{Algebraic relations from higher-order operators}\label{ssec:higherrelations}

In this section we will explain how the higher-order reduction operators lead to algebraic relations between different integrals. Concretely, this follows from two observations. Firstly, as we have seen already in section~\ref{sec:physics}, acting with higher-order reduction operators removes tubes, similar to first-order reduction operators. Secondly, we have shown in section~\ref{sec:differentialchain} that acting on an integral $I_\mathcal{T}$ with a differential operator must result in a linear combination of integrals associated to sub-tubings of $\mathcal{T}$ with rational coefficients. Using this, the derivatives of a higher-order reduction operator acting on an integral $I_\mathcal{T}$ can be rewritten in terms of these integrals resulting in a purely algebraic relation between the various basis functions. The exact terms that can appear in these relations will vary, depending on whether the higher-order reduction operator comes from a maximal tube in which case the factorization relations of section~\ref{ssec:factors} become important. Therefore, we will treat the two cases separately, beginning with the non-maximal case.

\paragraph{Algebraic relations from non-maximal tubes.}

Recall that, if a tube $T$ admits a partition $\pi$, it is possible to obtain a higher-order reduction operator $Q^{(T)}_\pi$ using equation~\eqref{eq:highred}. Furthermore, when acting on an integral $I_\mathcal{T}$ with $T$ contained in this tubing, we have seen in section~\ref{ssec:tuberemoval} that the reduction operator will act as
\begin{equation}
    Q^{(T)}_\pi I_\mathcal{T} = \prod_{S\in \pi}\partial^{(S)}  I_{\mathcal{T}\setminus \{T\}}\,.
\end{equation}
Now, using equation~\eqref{eq:ITpartialfin} to iteratively rewrite the derivatives acting on an integral in terms of sub-tubings, it is possible to turn this differential relation in to an algebraic one. Furthermore, this equation can be solved for $I_\mathcal{T}$ resulting in an algebraic relation between $I_\mathcal{T}$ and integrals $I_\mathcal{S}$ for sub-tubings $\mathcal{S}$ of $\mathcal{T}$.

If the tube $T$ is not a maximal tube, this algebraic relation will only involve sub-tubings $\mathcal{S}$ that contain the maximal tube, and therefore are already included in the differential chain constructed in section~\ref{ssec:chainconstruction}. Therefore, we find that not all the functions in the differential chain are algebraically independent and we do not actually need to solve the differential equation for all of these functions. Instead, we can solve the differential equations only for a subset of these functions and obtain the others using the algebraic relations.

In conclusion, we find that if a tubing $\mathcal{T}$ contains \textit{any} non-maximal tube $T$ that admits a partition, there is a relation of the form
\begin{equation}\label{eq:ITalgebraic}     \boxed{\rule[-.5cm]{0cm}{1.3cm} \quad 
    I_\mathcal{T}=\sum_{\substack{\mathcal{S}\subsetneq \mathcal{T},\\ T_{\rm max} \in \mathcal{S}}} \tilde{r}_\mathcal{S}(\cT) I_{\mathcal{S}} \quad}
\end{equation}
where the $\tilde{r}_\mathcal{S}$ are rational functions of $z$ and $\alpha$ and the sum is over all strict sub-tubings of $\mathcal{T}$ that contain $T_{\rm max}$. In other words, the sum is over all sub-tubings of $\mathcal{T}$ that are contained in the differential chain. Note that the functions $\tilde{r}$ can be obtained explicitly using the procedure above. However, we leave a general explicit expression for these coefficients for future work.

\paragraph{Factorization relations.}

We now turn our attention to the case where the maximal tube admits a partition, which will lead to similar algebraic relations to the ones found above. However, as explained in section~\ref{ssec:factors}, removing the maximal tube results in a factorization formula. Thus, acting with the corresponding reduction operators results in
\begin{equation}
    Q^{(T_{\rm max})}_\pi I_\mathcal{T} = \prod_{S\in \pi}\partial^{(S)}  \prod_{\alpha=1}^{k} I_{\mathcal{T}_\alpha}\,,
\end{equation}
where we have labeled the tubings of the different factors as $\mathcal{T}_\alpha$, and denoted the number of such factors by $k$. 

Similar to the approach above, the derivatives on both sides of this equation can be rewritten in terms of functions that belong to a differential chain. However, there is one key difference: the integrals $I_{\cT_\alpha}$ do not include the maximal tube $T_{\rm max}$. As a result, these integrals and their derivatives are not part of the original chain. Instead, they form their own separate differential chains.
This slightly changes the algebraic relations, as now these new functions and their derivatives have to be incorporated as well. Thus we see that the higher-order reduction operators will not allow us to immediately decrease the number of functions we need to solve for. However, one should keep in mind that the diagrams associated to each tubing $I_{\cT_\alpha}$ are much simpler than the original diagram. Therefore, the resulting algebraic relation will still result in an algebraic relation that simplifies $I_\mathcal{T}$. Furthermore, as we will see in section~\ref{ssec:minimalbasis}, these new functions can be written in terms of the same set of minimal representation functions as the ones already part of the chain.

To conclude, given a tubing $\mathcal{T}$ we find that whenever the maximal tube $T_{\rm max}$ admits a partition $\pi$, the integral $I_\mathcal{T}$ will satisfy an algebraic relation of the form
\begin{equation}\label{eq:ITmaxalgebraic}     \boxed{\rule[-.5cm]{0cm}{1.3cm} \quad 
    I_\mathcal{T}=\sum_{\substack{\mathcal{S}\subsetneq \mathcal{T},\\ T_{\rm max} \in \mathcal{S}}} \tilde{r}_\mathcal{S} I_{\mathcal{S}} +\prod_{\alpha=1}^k \bigg(\sum_{\substack{\mathcal{S}\subsetneq \mathcal{T}_\alpha,\\ T_{\rm max,\alpha} \in \mathcal{S}}} \tilde{r}_\mathcal{S} I_{\mathcal{S}}\bigg) \quad}
\end{equation}
where again, $\mathcal{T}_\alpha$ are the different factors appearing after removing $T_{\rm max}$, the first sum is over all sub-tubings of $\mathcal{T}$ that contain $T_{\rm max}$ while the second sum is over all sub-tubings of the factor $\mathcal{T}_\alpha$ containing its respective maximal tube.

\subsection{Some relations for the single- and double-exchange integrals} \label{ssec:higherrelsexample}

To make the above more explicit, we will now showcase how these algebraic relations can be obtained in two examples, in particular one for a maximal tube and one for a non-maximal tube. We will first derive a factorization relation for the single-exchange integral. Afterwards, we will derive an algebraic relation for functions in the differential chain of the double-exchange integral. We have chosen somewhat simple examples here in order to keep the formulas from becoming too involved. However, the same procedures generalize to any tree-level cosmological correlator.

\paragraph{Factorization relation for the single-exchange integral.}

We begin by considering the single-exchange integral again, since here the formulas will be the most simple. As we have seen in equation~\eqref{eq:singexchangeQhigh}, the maximal tube in 
\begin{equation}
\includegraphics[valign=c]{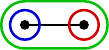}
\end{equation}
admits a partition by the blue and red tubes. Therefore, there is a reduction operator \begin{equation}
    Q^{(T_{\rm g})}_\pi=\partial^{(T_{\rm r})} \partial^{(T_{\rm b})} \big(z^{(T_{\rm g})}  -z^{(T_{\rm r})} -z^{(T_{\rm b})}  \big)\,,
\end{equation}
where have denoted the partition by $\pi$, recall that $T_{\rm r}$, $T_{\rm b}$ and $T_{\rm g}$ are the red, blue and green tubes respectively and note that the derivatives act on everything to their right. As discussed in section~\ref{ssec:factors}, this reduction operator will lead to a differential relation of the form
\begin{equation}\label{eq:singexchangehighfact}
    \partial^{(T_{\rm r})} \partial^{(T_{\rm b})} \big(z^{(T_{\rm g})}  -z^{(T_{\rm r})} -z^{(T_{\rm b})}  \big)
    I_\text{\includegraphics[valign=c]{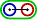}}=\big(\partial^{(T_{\rm b)}}   I_\text{\includegraphics[valign=c,scale=.7]{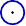}} \big) \big(\partial^{(T_{\rm r})}I_\text{\includegraphics[valign=c,scale=.7]{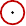}}\big)\,.\end{equation}
in accordance with equation~\eqref{apply_red_op}.

\paragraph{Algebraic relation for the single-exchange integral.}

Now, we will use the differential chain constructed in section~\ref{ssec:chainexamples} to rewrite this differential relation into an algebraic one. We will begin by rewriting the right-hand side. 

In section~\ref{ssec:chainconstruction} we have seen that, if a tubing only consists of a single tube, the only non-zero differential equation it satisfies can be obtained from equation~\eqref{eq:ITmaxdif}. This implies that
\begin{equation}
    \begin{array}{rl}
    \partial^{(T_{\rm b)}}   I_\text{\includegraphics[valign=c,scale=.7]{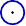}}&  \displaystyle  =\frac{\alpha_1-1}{z^{(T_{\rm b})}} I_\text{\includegraphics[valign=c,scale=.7]{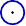}}\, , \\
     \partial^{(T_{\rm r})}I_\text{\includegraphics[valign=c,scale=.7]{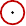}}&  \displaystyle   =\frac{\alpha_2-1}{z^{(T_{\rm r})}} I_\text{\includegraphics[valign=c,scale=.7]{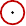}} \, ,
    \end{array}
\end{equation}
where we recall that the vertices are ordered such that the blue tube encircles the first vertex while the red tube encircles the second. Inserting these identities into equation~\eqref{eq:singexchangehighfact} results in
\begin{equation}
     \partial^{(T_{\rm r})} \partial^{(T_{\rm b})} \big(z^{(T_{\rm g})}  -z^{(T_{\rm r})} -z^{(T_{\rm b})}  \big)
    I_\text{\includegraphics[valign=c]{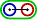}}=\frac{(\alpha_1-1)(\alpha_2-1)}{z^{(T_{\rm b})}z^{(T_{\rm r})}}I_\text{\includegraphics[valign=c,scale=.7]{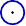}}I_\text{\includegraphics[valign=c,scale=.7]{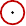}}\ ,
\end{equation}
and we see already that one side of the equation is now purely algebraic.

Using a similar reasoning, we apply the strategy of section~\ref{ssec:chainexamples} to rewrite the derivatives acting on $I_\text{\includegraphics[valign=c]{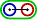}}$ 
in terms of functions in the differential chain. This process, while somewhat tedious, is straightforward and results in
\begin{equation}
\begin{split}
     \partial^{(T_{\rm r})} \partial^{(T_{\rm b})} \big(z^{(T_{\rm g})}  -z^{(T_{\rm r})} -z^{(T_{\rm b})}  \big)
    I_\text{\includegraphics[valign=c]{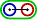}}=&\frac{(\alpha_1-1)(\alpha_2-1)\big(z^{(T_{\rm g})}-z^{(T_{\rm b})}-z^{(T_{\rm r})}\big) I_\text{\includegraphics[valign=c]{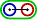}}}{z^{(T_{\rm b})} z^{(T_{\rm r})}}\\
        & + \frac{(\alpha_1-1)(\alpha_2-1)\left(I_\text{\includegraphics[valign=c]{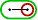}}+ I_\text{\includegraphics[valign=c]{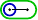}}\right)}{z^{(T_{\rm b})} z^{(T_{\rm r})}}\,.
\end{split}
\end{equation}
Inserting this equation into~\eqref{eq:singexchangehighfact} and solving for $ I_\text{\includegraphics[valign=c]{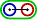}}$, we obtain
\begin{equation}\label{simplified-full-I}
     I_\text{\includegraphics[valign=c]{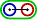}}=\frac{ I_\text{\includegraphics[valign=c,scale=.7]{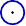}}   I_\text{\includegraphics[valign=c,scale=.7]{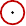}}
-  I_\text{\includegraphics[valign=c]{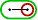}}-I_\text{\includegraphics[valign=c]{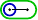}}}{z^{(T_{\rm g})}-z^{(T_{\rm b})}-z^{(T_{\rm r})}}
  \,,
\end{equation}
which is the algebraic relation within the single-exchange chain due to the higher-order reduction operator $Q^{(T_{\rm g})}_\pi$.

Interestingly, equation~\eqref{simplified-full-I} implies a concrete simplification for 
$I_\text{\includegraphics[valign=c]{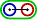}}$
at the functional level. The differential chain~\eqref{eq:singexchdifchain} for the single exchange integral results in a coupled system of second-order differential equations satisfied by 
$I_\text{\includegraphics[valign=c]{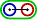}}$. 
In general, one would expect that the solution would be some two-variable generalized hypergeometric function such as an Appell function. However, from the explicit form of the single exchange integral obtained in~\cite{arkani-hamed_differential_2023,grimm_reductions_2025} it follows that it can be written as a sum of single-variable hypergeometric functions, as well as polynomials raised to complex powers. The functions 
$I_\text{\includegraphics[valign=c]{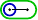}} $ and $I_\text{\includegraphics[valign=c]{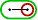}} $ 
take the form of such single-variable hypergeometric functions while 
$I_\text{\includegraphics[valign=c,scale=.7]{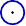}}   I_\text{\includegraphics[valign=c,scale=.7]{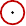}}$ 
can be written in terms of polynomials raised to complex powers. Therefore, equation~\eqref{eq:singexchangehighfact} encodes exactly this simplification. We will see in section~\ref{ssec:minimalbasis} that, for general diagrams, many such functional simplifications will happen. In section~\ref{ssec:recursivereduction}, we will explain how to obtain the minimal set of such functions necessary.

\paragraph{A relation for the double-exchange integral.}

As a second example, let us briefly examine the type of algebraic relations that appear when considering reduction operators that are not associated to a maximal tube. In this case, it is necessary to introduce an example that is slightly more involved than the single exchange integral, namely the double-exchange integral. In particular, we will consider the double-exchange diagram with the tubing
\begin{equation}\label{diag:doubexchforred}
\includegraphics[valign=c]{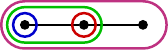}
\end{equation}
and, since the green tube admits a partition, obtain an algebraic relation for $I_\text{\includegraphics[valign=c]{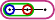}}$.

In order to obtain this relation, let us first provide the functions in the differential chain needed to construct $I_\text{\includegraphics[valign=c]{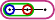}}$. These can be obtained simply by considering the sub-tubings of~\eqref{diag:doubexchforred}, resulting in the differential chain
\begin{equation}
\includegraphics[valign=c]{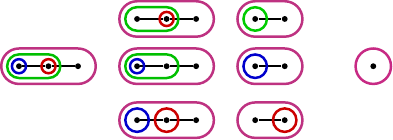}
\end{equation}
where we have not drawn the arrows relating different diagrams in order to avoid clutter and again have contracted any edges using the arguments of section~\ref{ssec:contractions}. We emphasize that the integrals being part of this differential chain implies that a partial derivative acting on any of the integrals
\begin{equation}
\begin{array}{llll}
     I_\text{\includegraphics[valign=c]{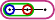}}\, ,& I_\text{\includegraphics[valign=c]{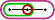}} \,, &
  I_\text{\includegraphics[valign=c]{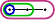}}\,,&
 I_\text{\includegraphics[valign=c]{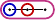}}\, ,\\
I_\text{\includegraphics[valign=c]{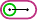}}\,,&
I_\text{\includegraphics[valign=c]{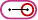}}\,,&
I_\text{\includegraphics[valign=c]{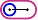}}\,,&
I_\text{\includegraphics[valign=c,scale=.7]{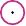}}\, ,     
\end{array}
\end{equation}
can be expressed as a linear combination of the others with rational coefficients.

Since, in the left-most diagram, the green tube admits a partition using the blue and red tubes, there will be a differential relation
\begin{equation}\label{eq:doubleexchangereductionop}
    Q^{(T_{\rm g})}_\pi I_\text{\includegraphics[valign=c]{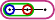}}= \partial^{(T_{\rm r})}  \partial^{(T_b)}  I_\text{\includegraphics[valign=c]{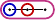}} \,,
\end{equation}
where $Q^{(T_{\rm g})}_\pi$  takes the form\footnote{Note that the expression for $Q_\pi^{(T_{\rm g})}$ is the same here as for the single-exchange integral, for which $T_{\rm g}$ admits the same partition.}
\begin{equation}
   Q^{(T_{\rm g})}_\pi=  \partial^{(T_{\rm r})} \partial^{(T_{\rm b})} \big(z^{(T_{\rm g})}  -z^{(T_{\rm r})} -z^{(T_{\rm b})}  \big)\,.
\end{equation}
Now, again using the fact that derivatives acting on any of these functions can be expressed as other functions in the differential chain, this will lead to an algebraic relation. The process of constructing the differential chain, as well as solving for the partial derivatives is computationally more involved in this case. However, there is no fundamental difficulty and one can proceed along the same lines as above. 

Interestingly, the resulting algebraic relation is remarkably similar to the equation~\eqref{simplified-full-I}, being of the form
\begin{equation}
    I_\text{\includegraphics[valign=c]{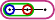}}= \frac{ I_\text{\includegraphics[valign=c]{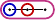}}-I_\text{\includegraphics[valign=c]{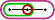}} - 
  I_\text{\includegraphics[valign=c]{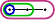}}} {z^{(T_{\rm g})}-z^{(T_{\rm b})}-z^{(T_{\rm r})}}\,.
\end{equation}
Note that in this case, all the functions on the right-hand side of this equation are already part of the differential chain. Therefore, this algebraic relation actually reduces the number of functions that one has to determine for the double-exchange integral.

\subsection{Minimal representation functions}\label{ssec:minimalbasis}

Having discussed the algebraic relations, we observe that 
they often lead us to consider integrals of the factorized diagrams.  
Staying within the differential chain, this would not bring a simplification since the number of functions one has to determine has not decreased. 
This leads us to consider a change of perspective: instead of simply counting how many functions appear in a certain differential chain, we consider the types of functions that can appear. In effect, this implies that we must consider functions equivalent when they merely differ by permuting or shifting inputs. Computationally, one only needs to obtain each such function once, since the permutations and shifts are simple operations. If we implement all such simplifications, we find a minimal set of functions necessary to describe tree-level cosmological correlators, which we call the minimal representation functions.

\paragraph{Permutations.}

During the construction of the differential chains in the examples above, we have already seen many functions appear multiple times with differently permuted inputs. For example, for the single-exchange integral the functions $I_\text{\includegraphics[valign=c]{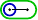}}$ and $I_\text{\includegraphics[valign=c]{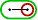}}$ where 
permuting the inputs $z^{(T_{\rm b})}$ and $z^{(T_{\rm r})}$ as well as the twists $\alpha_1$, and $\alpha_2$ results in an equivalence
\begin{equation}
    I_\text{\includegraphics[valign=c]{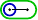}} \vert_{z^{(T_{\rm b})} ,\alpha_1 \rightarrow z^{(T_{\rm r})} ,\alpha_2} = I_\text{\includegraphics[valign=c]{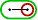}}
\end{equation}
as one can immediately see from the diagrams themselves. However, this procedure similarly works for more complicated diagrams. For example, one can obtain an equivalence of the tubings
\begin{equation}
\includegraphics[valign=c]{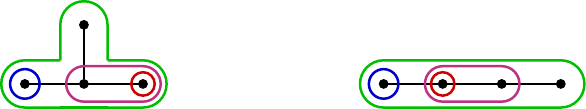}
\end{equation}
now involving four edges, simply by permuting the inputs $z^{(T)} $ and $\alpha_v$. This is a consequence of the fact that, as described in section~\ref{sec:tubings}, the GKZ system is agnostic of any topological properties of the diagram. Instead, the only information that enters the GKZ system is combinatorial, consisting strictly of the vertices contained in each tube.

\paragraph{Permutations for factorizations.}

Let us note that symmetries are also prevalent in the factorization relations for the single-exchange integral. In this case three integrals $I_\text{\includegraphics[valign=c,scale=.7]{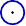}}$, $I_\text{\includegraphics[valign=c,scale=.7]{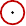}}$, and $I_\text{\includegraphics[valign=c,scale=.7]{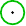}}$  appear, which merely differ by permuting the variables and the twists, with possibly some additional shifts. In fact, this behavior is rather general. Let us consider an integral $I_\cT$ admitting a factorization of the form
\begin{equation}
    Q^{(T_{\rm max})}_\pi I_\cT = \prod_{S\in \pi} \partial^{(S)} \prod_{\alpha=1}^k I_{\cT_\alpha}\,
\end{equation}
where there are $k$ factors $\cT_{\alpha}$. Then there are two options, either a non-maximal tube in $\cT$ also admits a partition, implying that there is an algebraic relation relating $I_\cT$ in terms of functions part of the differential chain, or the factors $I_{\cT_\alpha}$ are permutations of functions already part of the differential chain. This implies that any integral of a tubing with a factorization relation can be fully written in terms of permutations of its sub-tubings.

To see this, we will assume that $I_{\cT}$ does not contain a non-maximal tube that admits a partition. Then, let us choose any of the factors $\cT_{\alpha}$. We will show that there is a tubing $\cS_\alpha$ such that $I_{\cT_\alpha}$ is a permutation of $I_{\cS_\alpha}$ and $\cS_\alpha$ is a sub-tubing of $\cT$. We begin by removing all non-maximal tubes from $\cT$ that are not contained in $\cT_\alpha$, note that removing non-maximal tubes will result in a function that is in the differential chain. Furthermore, we will consider the maximal tube of $\cT_{\alpha}$ and also remove it, we will denote the resulting tubing by $\cS_\alpha$. Note that the maximal tube of $\cT_{\alpha}$ is not the maximal tube of $\cT$, therefore $I_{\cS_\alpha}$ will be contained in the differential chain of $\cT$. Since $I_{\cT_\alpha}$ does not admit a partition, its maximal tube must contain at least one bare edge. Therefore, we can use the contraction identities from section~\ref{ssec:contractions} to contract all edges in $\cS_\alpha$ that are not fully contained in $\cT_\alpha$. The resulting tubing will be a permutation of $\cT_\alpha$ in which only the maximal tubes are permuted.

Let us illustrate the above with an example. We will consider the double-exchange integral with the tubing
\begin{equation}\label{eq:minimalfactor}
\includegraphics[valign=c]{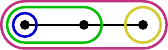}
\end{equation}
The magenta tube admits a partition by the green and yellow tubes, and the resulting algebraic relation will involve the factors
\begin{equation}\label{eq:minimalfactors}
\includegraphics[valign=c]{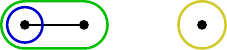}
\end{equation}
which naively should be added to the differential chain separately. However, removing the yellow and green tube from~\eqref{eq:minimalfactor} results in
\begin{equation}
\includegraphics[valign=c]{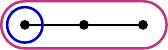}
\end{equation}
which can be contracted to
\begin{equation}
\includegraphics[valign=c]{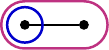}
\end{equation}
that is equivalent to the left factor in~\eqref{eq:minimalfactors} by a permutation of their maximal tubes. Similarly, removing all tubes except for the maximal tube in~\eqref{eq:minimalfactor} would result in the right factor of~\eqref{eq:minimalfactors}.

\paragraph{Minimal representation functions.}

The permutation symmetry above, as well as the algebraic relations found throughout this section, lead us to a natural question: what is the set of function that remains after all redundancy has been removed? The resulting functions, which we dub the minimal representation functions, will have as their defining property that these are the minimal set of functions that must be solved using their differential equations, as there can be no further algebraic or permutative identities. In other words, these functions are the building blocks that all other functions in the differential chain can be constructed from, using the algebraic and permutative relations. 

Interestingly, the minimal representation functions are shared for all tree-level cosmological correlators, independent of any particular tubing or topology in a diagram. As described in section~\ref{sec:tubings}, this is rooted in the fact that the GKZ system is agnostic to this information. To signify that we only care about the functions themselves and are agnostic to the particular tubing or diagram that they arise from, we will denote the minimal representation functions by removing the color from their tubings, as in $   I_\text{\includegraphics[valign=c]{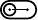}}$ and $I_\text{\includegraphics[valign=c,scale=.7]{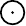}}$.
Note that, in order  to solve the differential equations satisfied by a minimal representation function, it may be necessary to color in these tubes again.

The minimal representation functions also give an intuitive handle on the complexity of the functions that can appear. For example, for the single-exchange integral one shows, see e.g.~\cite{grimm_reduction_2025}, that it consists only of polynomials to complex powers and $_2F_1$ hypergeometric functions. We can motivate the expected complexity by the minimal chain these functions can be contained in. For example, the function $   I_\text{\includegraphics[valign=c]{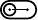}}$ can be minimally contained in the chain
\begin{equation}
  \includegraphics[valign=c]{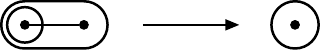}
\end{equation}
Here, the length of the chain will describe the order of the full differential equations, while the different arrows at each layer are related to the number of variables each function depends on. Note that each function also implicitly satisfies the differential equation due to the maximal tube. However, this differential equation will only fix an overall scaling of the variables, the arrows here then denote how many remaining variables the function depends on. From this, we find that the chain above corresponds to a second order differential equation in one variable, giving rise to a $_2F_1$ hypergeometric function.

\paragraph{Tubings for minimal representation functions.}

Even though the minimal representation functions no longer correspond to any particular tubing, they can still be represented by a tubing. Now, it turns out that these tubings must have a few specific properties. In particular, note that, if any tube contains no bare vertices, the tubing admits a partition and can therefore be removed. Conversely, if a tube contains multiple bare vertices, these can be contracted using the methods of section~\ref{ssec:contractions}. This implies that, for a minimal representation function, each tube in the corresponding tubing must contain \textit{exactly} one bare vertex. This property will help us greatly in section~\ref{ssec:minrepcounting}, where we will provide counting formulas for the minimal representation functions. 

Furthermore, we find that these conditions imply that a diagram must contain exactly the same number of vertices as the number of tubes. From this, we find that the minimal representation functions are naturally ordered by the number of vertices. Moreover, since acting with reduction operators removes tubes, we find that a derivative acting on a minimal representation function with $n$ vertices can be expressed in terms of the function itself, as well as minimal representation functions with $n-1$ vertices. Therefore, we find that the differential chain of a minimal representation function with $n$ vertices consists of itself, alongside minimal representation functions that have strictly fewer vertices.

\subsection{The recursive reduction algorithm}\label{ssec:recursivereduction}

In this section, we will summarize the results obtained throughout this chapter in to a single algorithm, the recursive reduction algorithm. We outline the key steps required for performing the reductions, referring to earlier sections for explicit formulas. We then illustrate the reduction process schematically for the double-exchange integral, demonstrating that it can be expressed in terms of just four minimal representation functions.

\paragraph{The recursive reduction algorithm.}

The recursive reduction algorithm is based on the idea that it is beneficial 
to decompose the cosmological correlators into the simplest set of building block functions. While this introduces combinatorial complexity, solving the differential equations for the building block functions will be significantly simpler. 

The algorithm proceeds in the following steps: 

\noindent
\textbf{Step 1:} Considering a cosmological correlator with a fixed number of external momenta, one first has to write down all tree-level diagrams that contribute. Each diagram is initially studied separately. Focusing on a diagram one needs to find all complete tubings $\cT$. The goal is then to construct the minimal representation functions for the sum $\sum_{\cT} I_{\cT}$. 

\noindent
\textbf{Step 2:} Next, one considers a specific tubing $\cT$. To obtain $I_{\cT}$ one constructs the differential chain in which $I_{\cT}$ resides. 
This requires finding all the sub-tubings of $\cT$ and then using the reduction operators as in section~\ref{ssec:chainconstruction}. Note that in this step, we are not required to obtain the different functions in the chain explicitly, only the differential equations they satisfy, which follow by recursively applying equations~\eqref{eq:partialTinQs1} and~\eqref{eq:QTiteration}.

\noindent
\textbf{Step 3:} Once the differential relations have been obtained, one uses the algebraic relations of section~\ref{ssec:higherrelations} to eliminate the integrals associated to any tubing admitting a partition. In addition, one also contracts any edges using the methods of section~\ref{ssec:contractions}, and identifies the remaining functions up to permutations in the variables, as described in section~\ref{ssec:minimalbasis}. The resulting set of functions will be the minimal representation functions.

\noindent
\textbf{Step 4:} It remains to find the minimal representation functions by solving the differential equations that they satisfy. 
This is computationally the most difficult step, as solving such coupled systems of differential equations is a hard problem. 
Note that, as described in section~\ref{ssec:minimalbasis}, acting with a reduction operator on a minimal representation function removes edges. These first-order relations suggest that the minimal representation functions could admit an iterated integral representation.

\noindent
\textbf{Step 5:}
In the final step, we invert all of the algebraic and permutation relations used in step 3, in order to express the integral $I_{\cT}$ in terms of the minimal representation functions. This step will consist of keeping track of a large number of identities between different functions, and will therefore be computationally tedious. However, no fundamental difficulties remain in this step.

To illustrate the algorithm above, we will now partly apply it to the double-exchange integral. Note that we will be somewhat schematic, as keeping track of and displaying such large numbers of identities is quite tedious and not very insightful. Instead, we will mostly focus on step 2 and step 3 to obtain the minimal representation functions. This illustrates the large number of symmetries and identities that can be found in this example.

\paragraph{Minimal representation for the double-exchange integral.}

To obtain the minimal representation of the double-exchange integral, we must first construct its differential chain. We will begin with the tubing
\begin{equation}\label{eq:doubleexchangecomptub}
\includegraphics[valign=c]{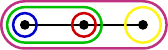}
\end{equation}
and note that the other tubing for the double-exchange integral can be obtained from this one by symmetry. Then, to construct the differential chain, we must find all sub-tubings of this tubing. There are 16 such tubings, given by
\begin{equation}
\includegraphics[valign=c]{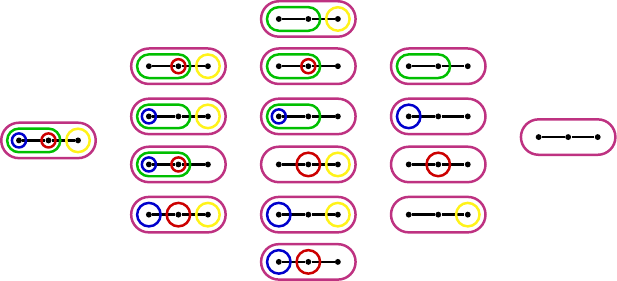}
\end{equation}
where again we have not drawn the arrows relating different diagrams in order to avoid clutter.

Now, we must eliminate all functions which can be algebraically removed using the higher-order reduction operators. This means that any tubing which contains a tube that can be partitioned must be removed. Note, if the reduction operator results in a factorization relation, one should in principle keep track of both of the factors. However, as we have seen in section~\ref{ssec:minimalbasis}, these will lead to the same minimal representation functions. Thus, we will ignore them here. Removing all of these redundant tubings, we are left with the following set:
\begin{equation}
\includegraphics[valign=c]{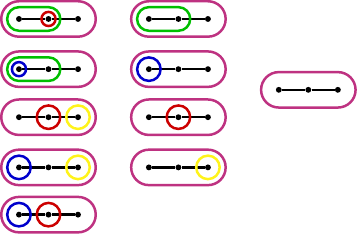}
\end{equation}
Note that the remaining tubes are localized in the right-most columns. This is general behavior as, when a tubing contains more tubes than vertices, it must contain a tube admitting a higher-order reduction operator.

Then, before we identify the different functions up to symmetry, we must contract all possible edges using the techniques of section~\ref{ssec:contractions}. From this, we find that each tubing in the $n$-th column of our differential chain can be contracted to include only diagrams with $n$ edges. In particular, we find
\begin{equation}
\includegraphics[valign=c]{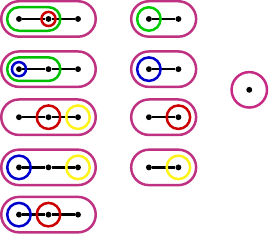}
\end{equation}
Note that this does not decrease the number of tubings, but will greatly increase the number of functions that can be identified up to permutations. Performing this identification is the final step of the algorithm, after which we are left with the four tubings
\begin{equation}
\includegraphics[valign=c]{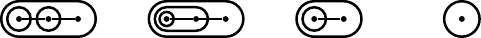}
\end{equation}
where we have again removed the colors of the tubings to represent that, in the actual correlator these will appear with differently permuted variables.

We would like to emphasize how drastic the decrease in necessary functions is after these reductions. Initially, we found that the double-exchange integral can be obtained using a differential chain containing sixteen different tubings, and thus required solving differential equations for sixteen different functions. Furthermore, analyzing the differential chain one would expect a solution of these equations to be some four-variable generalized hypergeometric function. However, applying all the possible simplifications and reductions, we are left with only four different minimal representation functions, using which it must be possible to express the original correlator. Additionally, two of those functions can already be obtained when solving for the single-exchange integral, while the other two are new two-variable generalized hypergeometric functions. These functions are substantially less complex than a generically expected solution of the original system, a direct consequence of the great number of relations present.

\endgroup

\fi

\if\PrintChFive1


\newpage

\stepcounter{thumbcounter}
\setcounter{colorcounter}{5}
\chapter{Complexity Reduction}
\label{ch:complexity}

In this chapter we will turn our attention to some more precise notions of complexity, which allow us to make precise some of the simplifications and reductions in complexity we have seen in the previous chapters. We will see that encoded in the differential equations discussed so far, there is concrete knowledge regarding the topological and computational properties of cosmological correlators. 

Concretely, this means that  the differential equations constructed in chapter~\ref{ch:reductionalgorithm} are special: they have a triangular form and are specified by polynomials. 
The theory of Pfaffian functions is thoroughly developed and we will use the data of the differential equations to give upper bounds for the \textit{topological complexity} and the \textit{computational complexity} of these functions. The former notion captures information such as the number of poles and zeros of these functions while the latter is the running time of an algorithm to check formulas satisfied by the correlators. 

Before we turn our attention to the algorithms from chapter~\ref{ch:reductionalgorithm}, we first discuss another way of obtaining the differential equations for cosmological correlators: the kinematic flow algorithm. Introduced in~\cite{arkani-hamed_differential_2023}, this algorithm is also Pfaffian and thus we can study its Pfaffian complexity. Interestingly, we find that the resulting complexities vastly overestimate known topological properties of cosmological correlators, hinting at the existence of a simpler representation. This was the original motivation for the application of reduciblity to cosmological correlators, as a method of incorporating any redundancies within the algorithms. While this does bring with it a significant reduction in complexity, this also brings to light some of the limits of the Pfaffian framework. In particular, the incorporation of symmetries and permutations remains a challenge.

In section~\ref{sec:pfaffian}, we provide a general overview of Pfaffian functions and their complexity. Afterwards, we analyze the Pfaffian complexity of the kinematic flow algorithm in section~\ref{sec:complexitykinematic}. Finally, we study the recursive reduction algorithm of chapter~\ref{ch:reductionalgorithm} in section~\ref{sec:complexityrecursive}. Here, we also comment on some of the limitations of the Pfaffian framework.

\section{Complexity and Pfaffian systems}\label{sec:pfaffian}

To obtain an explicit and useful notion of the complexity of a function, one natural perspective is to consider the differential equations it satisfies. 
This is the perspective taken for the measure of complexity we will use throughout this chapter: Pfaffian complexity.
Briefly, to show that a function is Pfaffian, and thus admits a Pfaffian complexity, one has to show that it satisfies a specific type of differential equation. These specific differential equations, known together as the \textit{Pfaffian chain}, then determine its complexity. Therefore, we will begin by defining such a chain.

\paragraph{Definition of Pfaffian chains.}
A Pfaffian function is a function which is defined by a triangular system of algebraic differential equations. More precisely, given a domain $U\subseteq \bbR^n$, a Pfaffian chain is a finite sequence of functions $\zeta_1,\ldots,\zeta_r:U\to \bbR$ which satisfies 
\begin{equation} \label{Pfaffian_chain-ODEs}
    \frac{\pd\zeta_i}{\pd x_j} =  P_{ij}(x_1,\ldots,x_n,\zeta_1,\ldots,\zeta_i)  \qquad \text{for all }i,j,
\end{equation}
where each $P_{ij}$ is a polynomial of $n+i$ variables. The triangularity condition, i.e.~the assumption that the derivatives of $\zeta_i$ depends only on $\zeta_1,\ldots,\zeta_i$ and not on $\zeta_{i+1},\ldots,\zeta_{r}$, is essential to ensure that the functions in the chain are sufficiently well-behaved. Given such a chain, a Pfaffian function is a function of the form
\begin{equation}
    f(x_1,\ldots,x_n) = P(x_1,\ldots,x_n,\zeta_1,\ldots,\zeta_r)
\end{equation}
where $P$ is a polynomial in $n+r$ variables. Note that, equivalently, equation~\eqref{Pfaffian_chain-ODEs} can be prhased in terms of the total differential
\begin{equation}\label{eq:pfaffianchaintotaldif}
    d \zeta_i = \sum_{j=1}^n P_{ij}(x_1,\ldots,x_n,\zeta_1,\ldots,\zeta_i) \,d x_j \,.
\end{equation}
We will use both perspectives throughout this chapter.

As an example, consider the function $\zeta(x_1,\ldots,x_n)=x_1^{m_1}\cdots x_n^{m_n}$, which satisfies
\begin{equation}
        \frac{\pd \zeta}{\pd x_j} =  m_j \zeta_j \zeta \,,
\end{equation}
for each $j$, where $\zeta_j$ are the functions $\zeta_j(x_1,\ldots,x_n)=1/x_j$ which satisfy
\begin{equation}\label{eq:expfaff}
        \frac{\pd \zeta_j}{\pd x_k} = -\delta_{jk}\zeta_j^2 .
\end{equation}
In this way, the functions $(\zeta_1,\ldots,\zeta_n,\zeta)$ form a Pfaffian chain. For another example, consider the case where all polynomial $P_{ij}$ are linear in the functions $\zeta_i$. The system from equation~\eqref{eq:pfaffianchaintotaldif} can then be rewritten as
\begin{equation}
    d\mathbf{\zeta} = A \mathbf{\zeta} \, ,
\end{equation}
where $\mathbf{\zeta}$ is a column vector with elements $\zeta_i$ and $A$ is a matrix defined in accordance with the system~\eqref{eq:pfaffianchaintotaldif}. The condition that the original differential chain is Pfaffian then implies that $A$ is an upper-triangular matrix.

\subsection{Pfaffian complexity}

The relevance of Pfaffian functions is that they have finiteness features which can be precisely quantified using their Pfaffian chain description. In particular, the data of the chain can be used to define a notion of complexity for Pfaffian functions. It consists of four numbers, namely the number of variables, $n$; the length of the chain, $r$ (also called the \textit{order}); the degree of the chain, $\alpha$, defined by $\alpha=\max_{i,j}(\deg(P_{ij}))$; and the degree of the Pfaffian function, $\beta$, defined by $\beta=\deg(P)$. These together define the \textit{Pfaffian complexity} of $f$, schematically denoted by
\begin{equation} \label{PfaffC}
    \cC(f) = (n,r,\alpha,\beta).
\end{equation}

The Pfaffian complexity $\mathcal{C}(f)$ can be viewed as giving a measure of how much information is needed to define the function $f$. An essential feature is that it depends on the description of the function. Since a given function may have several different descriptions, its complexity is not uniquely defined. In particular, this feature can be used to compare the complexity of different descriptions, which will provide us with a way of quantifying the effect of the reduction operators on cosmological correlators.

\paragraph{Interpretations of Pfaffian complexity.}

What is the meaning of the Pfaffian complexity introduced in equation \eqref{PfaffC}?  Loosely, it can be viewed as giving a measure of how much information is minimally needed to define the function $f$. It may at first seem peculiar that one needs to specify four numbers to describe this measure of information, but it turns out that a single number is is not rich enough to specify the information content of a function \cite{binyamini_sharply_2022}. 
Crucially, the Pfaffian complexity also depends on the precise representation of the function, and one might be able to reduce the complexity by making clever choices, for example, for the Pfaffian chain. As explained in more detail in \cite{grimm_complexity_2024} this freedom is a key feature of this consistent notion of complexity and pertains beyond the Pfaffian context. Hence, our result \eqref{PfaffC-ex} gives an upper bound on the information content. There are thus two natural questions to ask:
\begin{itemize}
\item[(1)] What characterizes a minimally complex representation of a function $f$? How should one minimize the list of integers $\cC(f)$?
\item[(2)] What is the physical interpretation of the minimal complexity of $f$?
\end{itemize}
Here, we will make some general comments regarding these questions. In the following sections, we will be more specific regarding the complexity of cosmological correlators. 

\paragraph{Topological complexity.}

One of the essential features of Pfaffian complexity is its ability to encode other notions of complexity \cite{vorobjov_complexity_2004}. Specifically, as alluded to in the introduction, Pfaffian complexity encodes topological and computational complexity. Though we will use these terms rather loosely, the precise meaning is that a geometric object constructed from Pfaffian functions obeys complexity bounds which are controlled by the underlying Pfaffian complexity. As a basic example, a Pfaffian function $f$ with complexity $\cC(f)=(1,r,\alpha,\beta)$ has bounds such as
\begin{equation}\label{eq:Bezout}
    \text {number of isolated zeroes} \leq 2^{r(r-1)} \beta \big(\alpha+\beta\big)^r \,,
\end{equation}
which may be interpreted as a generalized B\'ezout bound~\cite{vorobjov_complexity_2004}. Since it counts the number of connected components of the solution set of an equation, it provides a coarse measure of topological complexity. Let us note that this example only scratches the surface of the rich theory of Pfaffian complexity and one can determine the complexity of much more involved sets \cite{vorobjov_complexity_2004}.
These more involved examples and applications of these ideas to physical settings are discussed in \cite{grimm_complexity_2024}.

Consequently, a representation which minimizes complexity could be one which minimizes the bound on one of the derived notions of complexity, such as the topological complexity given above. 
However, one characteristic of all the complexity bounds derived in \cite{binyamini_sharply_2022} is that they grow exponentially in $r$, the order of the chain. Therefore, one natural choice of a minimal representation is one which minimizes $r$.

\paragraph{Computational complexity.}

Another measure of complexity derived from Pfaffian complexity is computational complexity. In general, this is a quantity which measures how complicated it is to algorithmically compute a geometric object built from Pfaffian functions. The algorithms in question are based on \textit{real numbers machines}, which are computational devices capable of performing exact computations on real numbers. If $X$ is a $d$-dimensional set defined by $M$ Pfaffian equalities or inequalities of complexity $(n,r,\alpha,\beta)$, then the computational complexity of $X$ is estimated by \cite{vorobjov_complexity_2004} 
\begin{equation}
   M^{(r+n)^{O(d)}}(\alpha+\beta)^{(r+n)^{O(d^2n)}}. 
\end{equation}
As a special case, one may consider the complexity of computing the zeros of a Pfaffian function, which is estimated by
\begin{equation}\label{eq:zerosofPfaffian}
    (\alpha+\beta)^{(r+n)^{O(d^2n)}}\, 
\end{equation}
Looking forward to the applications for cosmological correlators, this bound may be interpreted as the computational complexity of a cosmological correlator attaining a certain prescribed value.

\paragraph{Connection to sharp o-minimality.}

Before discussing the applications to cosmological correlators, it is worthwhile to discuss another perspective on Pfaffian complexity. The four numbers comprising the Pfaffian complexity are not arbitrary, but in fact part of a larger mathematical program aiming to assign a meaningful notion of complexity to large classes of functions, called sharp o-minimality \cite{binyamini_sharply_2022,binyamini_tameness_2023}. The meaning comes from generalizing the computational and topological properties of algebraic functions. In the algebraic case, i.e.~when the functions of interest are polynomials, these properties can be captured in terms of the maximum degree of the polynomials, $\cD$, and number of variables $\cF$. Crucially, the computational complexity of algorithms performed on these algebraic functions then admit bounds which are polynomial in $\cD$ and exponential in $\cF$ \cite{binyamini_sharply_2022}. 

The aim of sharp o-minimality is to assign a suitable pair $(\F,\D)$ to more general functions, while keeping similar bounds on computational complexity. For the Pfaffian functions, the right generalization turns out to be given by $\F=n+r$, and $\D = \deg P + \sum_{i,j}\deg P_{ij}$ \cite{binyamini_sharply_2022}. These ideas have previously been applied to various physical settings \cite{grimm_complexity_2024,grimm_structure_2024,grimm_complexity_2024a}, where these concepts are explained in more detail. For our purposes, it suffices that the Pfaffian complexity is a measure of the complexity of a function which can be given a computational meaning. 

\section{Complexity of the kinematic flow algorithm}\label{sec:complexitykinematic}

To apply the above formalism to the cosmological correlators of the previous chapters, we will first focus on an existing algorithm for obtaining their differential equations: the kinematic flow algorithm \cite{arkani-hamed_differential_2023,arkani-hamed_kinematic_2023}. This algorithm has been extensively studied in the literature recently \cite{hang_note_2024,baumann_kinematic_2024,he_differential_2025}. Here, we will show that the resulting system of differential equations is Pfaffian and study its complexity. In order to do this, we begin by describing the kinematic flow algorithm itself.

\subsection{The kinematic flow algorithm}

As in chapter~\ref{ch:cosmology}, we will consider the cosmological correlator associated to a single diagram. However, in contrast to most of chapter~\ref{ch:reductionalgorithm}, here we consider the full diagram at once instead of focusing on a single tubing. In particular, we will obtain a differential equation for the wave function $\psi$ arising from a single Feynman graph. Furthermore, this equation will take the form of
\begin{equation}
    dI = AI
\end{equation}
where $I=(\psi,f_{2},\ldots,f_{N_{\rm }})$ is a basis vector of functions and $A$ is an $N_{\rm k}\times N_{\rm k}$ matrix of one-forms. 
The goal of the kinematic flow algorithm is then to obtain the matrix $A$ using a combinatorial procedure. In what follows we will provide a minimalistic review of its steps, referring the reader to \cite{arkani-hamed_differential_2023,arkani-hamed_kinematic_2023} for a more detailed description and derivation.

To prepare the algorithm, one starts with a Feynman diagram and removes the external propagators. Furthermore, one marks the remaining edges of the graph with crosses. On the resulting marked graphs, one considers graph tubings, which are
clusters of adjacent vertices and crosses including at least one vertex. Note that these are \textit{not} the same tubings as those discussed in chapter~\ref{ch:reductionalgorithm}. 
This collision of nomenclature is unfortunate, and in~\cite{arkani-hamed_differential_2023} both are referred to simply as the tubings of a graph. As we will need to refer to both, we will refer to the tubings used for the kinematic flow algorithm as \textit{kinematic tubings} and refer to the previously introduced tubings as \textit{graph tubings} in what follows.

Kinematic tubings with a single component provide a convenient representation for the \textit{letters} of the differential equation, which encode the singularity structure. The letter associated to a kinematic tubing is given by the sum of the vertex energies $X_i$ in the tube and the internal energies $Y_j$ of the edges that enter the tube. If the tube includes the cross on such an edge, the sign of the internal energy is flipped. For example, for the two-vertex the letters may be represented by the following kinematic tubings
\begin{equation}
\vcenter{\hbox{\includegraphics{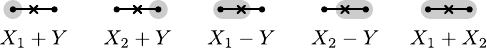} }}
\end{equation}
We will now explain the steps of the algorithm, and afterwards we give an example for the two-vertex chain.
\begin{enumerate}
    \item[(i)] Select a complete kinematic tubing of the graph, for which the differential $d f$ of the corresponding function $f$ is to be computed.
    \item[(ii)] Write down a `kinematic flow tree' according to the following steps:
\begin{enumerate}    \item[1.] Activation. For each component of the kinematic tubing, write down a descendant in which that component is `activated' (indicated by a coloring).
    \item[2.] Growth. For each activated component which contains no crosses, descendants are generated by `growing' the component by including adjacent crosses in all possible combinations.
    \item[3.] Merger. If the tube resulting from the previous step intersects another tube component, the two components merge and the union becomes activated. 
    \item[4.] Absorption. If a cross contained in an activated component is adjacent to another component containing a cross, the other component is `absorbed' whereupon the union becomes activated, generating another descendant.
\end{enumerate}
    \item[(iii)] The expression for $d f$ is read off from the kinematic flow tree as follows: for each graph, include a term $ d \log \Phi$ where $\Phi$ is the letter of the active tube, and multiply this term by the  function associated to the graph minus the functions corresponding to the direct descendant graphs. Finally, multiply all terms by an overall factor $\epsilon$ and the number of vertices included in the tube upon activation.
\end{enumerate}
Applying these steps to all functions $\psi,f_2,\ldots,f_{N_{\rm k}}$ in the basis yields the matrix of one-forms $A$.

The steps of the algorithm are rather abstract, so let us apply it to a simple example, the two-vertex chain. There are four complete kinematic tubings, and hence four functions
\begin{equation}
\vcenter{\hbox{\includegraphics{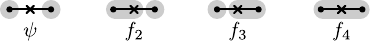} }}
\end{equation}

We start by computing $d \psi$, and write down the required kinematic flow tree below.
\begin{equation}
\vcenter{\hbox{\includegraphics{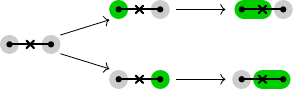} }}
\end{equation}
There are two components, and hence the activation step yields two descendant graphs. The next layer is generated by the growth of the active tubes to include the adjacent crosses. The kinematic flow terminates here, since the conditions for further growth, merger, or absorption are not fulfilled. Using the letters for the two-vertex chain given above, step (iii) now tells us that 
\begin{align}
    d \psi = \epsilon \big[ &(\psi-f_2) d \log(X_1+Y) + f_2 d\log(X_1-Y)\\ + &(\psi-f_3) d \log(X_2+Y) + f_3 d\log(X_2-Y)      \big] .\nonumber
\end{align}
Next, the kinematic flow tree for $f_2$ is given by
\begin{equation}
\vcenter{\hbox{\includegraphics{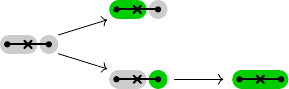} }}
\end{equation}
In the second layer the two components become active. The top channel terminates already at this step. The bottom channel shows the merger step, in which the entire graph is activated. The algorithm then tells us that 
\begin{align}
    d f_2 = \epsilon \big[ f_2 &d \log(X_1-Y) + (f_2-f_4)d \log(X_2+Y) \\ + f_4 &d \log(X_1+X_2)   \big] \nonumber
\end{align}
The equation for $f_3$ follows by symmetry. Finally, the kinematic flow for the final function is simply 
\begin{equation}
\vcenter{\hbox{\includegraphics{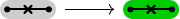} }}
\end{equation}
which gives (noting the factor of $2$ coming from the two vertices included in the tube upon activation)
\begin{equation}
    d f_4 = 2\epsilon f_4 d \log(X_1+X_2).
\end{equation}

\subsection{Pfaffian chains for cosmological correlators}\label{sec:Pfaff}

We will now argue that the basis functions discussed in the previous section form a Pfaffian chain. As a preparation, we need to include the $d\log$ terms of the letters into the chain. For each letter $\Phi_i(Z_I)= \sum_{I} c_i^I Z_I $, we have
\begin{equation}
    d\log \Phi_i = \frac{1}{\Phi_i}\sum_I c_i^Id Z_I.
\end{equation}
The coefficient $\ell_i=1/\Phi_i$ can be realized as a Pfaffian function by means of the differential equation
\begin{equation}
    d \ell_i = -\sum_{I}\ell_i^2 c_i^I d Z_I .
\end{equation}
Thus, the first part of the Pfaffian chain for cosmological correlators consists of the functions $(\ell_1,\ldots,\ell_{N_{\rm L}})$.

To set up a chain for the basis functions in the kinematic flow, we first organize them by the number of \textit{cuts} (line segments between vertices and crosses) appearing in the complete kinematic tubing. Complete kinematic tubings of a graph may then be uniquely described by the placement of these cuts. 

The crucial observation is that, as a consequence of the rules of the algorithm, the differential of every basis function only depends on the function itself and basis functions corresponding to complete kinematic tubings with \textit{strictly fewer} cuts. In particular, the growing of the tubes in step 2.-4.~always implies a reduction in the number of cuts. Meanwhile, there is a single basis function $\zeta_0$ which corresponds to zero cuts, i.e.~the complete kinematic tubing with one component. Taking this as the first basis function in the chain we can iteratively add the functions with an increasing number of cuts, thereby obtaining a Pfaffian chain. To be precise, let us denote the basis functions by $\zeta_{i,j}$, where $i$ is the number of cuts and $j=1,\ldots,n_i$ enumerates the basis functions with $i$ cuts. By the above argument we then have
\begin{equation}
    d \zeta_{i,j} = \sum_{I}  P_I (\ell_1,\ldots,\ell_{N_{\rm L}},\zeta_{i,j},\zeta_{i-1,1},\ldots)d Z_I \, ,
\end{equation}
which precisely fits the defining condition of a Pfaffian chain.

Let us again turn to the two-vertex chain to exemplify our argument. The inverse letters comprise the first five functions of the chain. For instance, one has $\ell_1=1/(X_1+Y)$, for which the required Pfaffian differential equation is given by $d\ell_1 = -\ell_1^2 d X_1 -\ell_1^2 d Y$.
Now we are able to write down the differential equations satisfied by the basis functions $(\psi,f_2,f_3,f_4)\equiv(\psi,\zeta_{1,1},\zeta_{1,2},\zeta_0)$ in terms of a Pfaffian chain, starting with $f_4$, which satisfies
\begin{align}
    d \zeta_0 &= 2\epsilon \ell_5 \zeta_0 (d X_1 +d Y) \,.
\end{align}
Next, we define the basis functions belonging to graphs with one cut, namely $\zeta_{1,1}$ and $\zeta_{1,2}$. Employing the algorithm, one finds that they satisfy the Pfaffian equations
\begin{align}
     d\zeta_{1,1} =& \,\epsilon \ell_2 \zeta_{1,1}(d X_1-d Y) +\epsilon \ell_5 \zeta_0(d X_1 +d X_2) \\
     &+ \epsilon \ell_3(\zeta_{1,1}-\zeta_0)(d X_2 +d Y)\,, \nonumber\\ 
     d\zeta_{1,2} =& \,\epsilon \ell_4 \zeta_{1,2}(d X_2-d Y) +\epsilon \ell_5 \zeta_0(d X_1 +d X_2)\nonumber \\
     &+ \epsilon \ell_3(\zeta_{1,2}-\zeta_0)(d X_1 +d Y)\,. \nonumber
\end{align}
Finally, the wavefunction itself is recovered in the Pfaffian chain as
\begin{align}
d \psi=\,\, &\epsilon \ell_1(\psi-\zeta_{1,1})(d X_1 + d Y) + \ell_2 \zeta_{1,1} (d X_1-d Y)  \nonumber \\
 + &\epsilon \ell_3(\psi-\zeta_{1,2})(d X_2 + d Y) + \ell_4 \zeta_{1,2} (d X_2-d Y) \,. \nonumber \\
\end{align}
We will now use this algorithm to obtain Pfaffian complexities for cosmological correlators.

\subsection{Pfaffian complexity of cosmological correlators}

The results from the previous section allow us to estimate the complexity of a wavefunction coefficient in terms of the number of vertices of the graph, $N_{\rm V}$. The number of variables in the chain, i.e.~the number of kinematic variables, is given by $n=2N_{\rm V}-1$, since there is one variable for each vertex and edge. The order of the chain, $r$, is the component of the Pfaffian complexity which is most strongly dependent on the number of vertices. For a given graph, the Pfaffian chain consists of $N_{\rm L}$ letters and $N_{\rm k}$ basis functions. The number of letters strongly depends on the topology of the graph. For instance, we have
\begin{align} \label{N-bounds}
    N_{\rm L}^{\rm chain}(N_{\rm V})&= 2N_{\rm V}^2 -2N_{\rm V}+1, \\
    N_{\rm L}^{\rm star}(N_{\rm V}) &= 3^{N_{\rm V}-1} +2N_{\rm V}-3.    \nonumber
\end{align}
It is worth noting that for fixed $N_{\rm V}$, the chain graph has the least letters and the star graph has the most letters, i.e.~we always have 
\begin{equation} \label{N_Lbound}
   N_{\rm L}^{\rm chain}(N_{\rm V}) \leq N_{\rm L}\leq N_{\rm L}^{\rm star}(N_V)\ . 
\end{equation}
Note that a star graph with $k$ legs is only present when the theory under consideration contains a $\phi^k$ interaction. In particular, in the notation of section \ref{sec:toymodel} we require $k\leq D$ and $\lambda_k\neq 0$. Meanwhile, counting the number of basis functions one finds that
\begin{equation}
    N_{\rm k} = 4^{N_{\rm V}-1}.
\end{equation}
Therefore, for large graphs, we have $r\sim 4^{N_{\rm V}}$. The degrees of the Pfaffian chain are independent of $N_{\rm V}$, since the system of differential equations is linear in the basis functions and the coefficients are given by the inverse letters $\ell_i$. In particular, we have $\alpha=2$ and $\beta=1$, so the complexity of an $N_{\rm V}$-vertex wavefunction coefficient is 
\begin{equation} \label{PfaffC-ex}
    \cC(\psi) = (2N_{\rm V}-1, 4^{N_{\rm V}-1} + N_{\rm L},2,1).
\end{equation}
The dependence of the complexity as a function of $N_{\rm V}$ is most apparent in the order $r$ of the chain, which shows an exponential growth. Also, note that the order $r$ increases by one for every function added to the Pfaffian chain. Since the steps of the kinematic flow algorithm generate more functions to be added to the chain, we note that \textit{complexity grows along the kinematic flow in a quantifiable way}. The rest of this section is devoted to the interpretation of these observations.

Let us note that the preceding discussion refers to the contribution to the wavefunction coefficient coming from a single Feynman graph viewed as a function of the kinematic variables $Z_I$. In general, the full wavefunction coefficients $\psi_n$ arise from a sum of graphs, and it will be a function of the couplings and kinematic variables $\psi_n(Z_I,\lambda_p)$. However, at tree level the sum over graphs is finite, implying that the tame structure in $Z_I$ is preserved while the coupling dependence is now a polynomial in the $\lambda_p$ with $p\leq n$. The complexity of the full tree level wavefunction coefficient viewed as Pfaffian function in $Z_I$ and $\lambda_p$ can then be obtained through the complexities of the underlying graphs which are determined by the interaction terms in the Lagrangians. It is an exciting task to study possible reductions of this total complexity, particularly due to relations or symmetries in the Lagrangian. 

\paragraph{Topological bounds.}

Now, let us apply the complexity found above with the explicit bounds from section~\ref{sec:pfaffian}. Recall that, for a Pfaffian function with complexity $\cC(1,r,\alpha,\beta)$, it satisfies the bound
\begin{equation}
   \text {number of isolated zeroes} \leq 2^{r(r-1)} \beta \big(\alpha+\beta\big)^r \,.
\end{equation}
Combining this with the actual complexity of wavefunction coefficients from equation~\eqref{PfaffC-ex}, we obtain 
\begin{equation} \label{BezoutCC}
    2^{(4^{N_{\text{V}}-1} +N_{\rm L})(4^{N_{\rm V}-1} +N_{\rm L}-1) }\big(4N_{\rm V}-1\big)^{4^{N_{\rm V}-1} +N_{\rm L}}  \,,
\end{equation}
for an $N_{\rm V}$-vertex graph as a function of the kinematic variables.
Note that the number of letters $N_{\rm L}$ is bounded in terms of $N_{\rm V}$ as in \eqref{N_Lbound} with \eqref{N-bounds} implying  doubly exponential grown in $N_{\rm V}$. 

The expression \eqref{BezoutCC} also gives a bound on the 
poles of $\psi$, since we can equally consider $1/\psi$ in \eqref{eq:Bezout} by minimally increasing the chain. However, from the integral representation of the cosmological correlators one would expect that the number of physical poles is only single-exponentially related to the number of letters and hence grows much slower with $N_{\rm V}$. Hence, we conclude that the representation of the cosmological correlators with the Pfaffian chain found in section \ref{sec:Pfaff} using the kinematic flow algorithm is far from optimal. It is apparent that the fast growth arises from the exponential growth of length of the chain with $N_{\rm V}$. In order to minimize the topological complexity, one would ideally like to find a representation that has a slowly growing chain, with maximally exponentially growing degrees $\alpha,\beta$ of the polynomials defining it. There is also a physical argument for this reduction of complexity. While a general solution to the differential equations may have a complicated singularity structure, many of these are not singularities of the wavefunction coefficient \cite{arkani-hamed_cosmological_2017,armstrong_graviton_2023,salcedo_analytic_2023,lee_amplitudes_2024,fan_cosmological_2024}. Likewise, a further reduction is observed when passing from the wavefunction coefficients to cosmological correlators \cite{chowdhury_subtle_2025}. Therefore, the true topological complexity should be smaller than estimated here.

A guiding principle for finding a simpler representation  arises from the locality of the physical theory. It was shown in \cite{arkani-hamed_differential_2023} that, for some specific examples, locality implies the existence of simpler sets of differential equations for the cosmological correlators. From these one can then find a shorter Pfaffian chain than initially expected. In fact, these are exactly the additional relations between basis functions we have seen in chapter~\ref{ch:reductionalgorithm}. We will explain the resulting simplifications in more detail in section~\ref{sec:complexityrecursive}.

\paragraph{Computational complexity and the emergence of time.}

Recall that Pfaffian complexity also allows us to bound the computational complexity of functions. In particular, we saw in equation~\eqref{eq:zerosofPfaffian} that calculating the zeros of a Pfaffian function has a computational complexity estimated by
\begin{equation}
    (\alpha+\beta)^{(r+n)^{O(d^2n)}}\, .
\end{equation}
Inserting the complexity of the wavefunction coefficients we obtain
\begin{equation}
     3^{(4^{N_{\rm V}}+ N_{\rm L}+ 2N_{\rm V}-1)^{O(d^2(N_{\rm L}-1))}}\,,
\end{equation}
where $d$ is the dimension of the locus on which $\psi$ attains a certain value $\psi_0$. 

The authors of \cite{arkani-hamed_differential_2023,arkani-hamed_kinematic_2023} argue that the kinematic flow algorithm is a boundary avatar of the cosmological time evolution in the bulk. In this static description, time arises as an emergent concept. The discussion above tells us that there is a quantifiable growth of computational complexity along the kinematic flow. It is therefore tempting to speculate on the connection between time and complexity, and the idea that complexity may provide an emergent description of time. In physics there are many hints that time and complexity are related, for instance through entropy, or more recently the idea of holographic complexity and time evolution in the bulk \cite{susskind_addendum_2014,carmi_time_2017,susskind_entanglement_2021}. The notion of complexity used here is somewhat different than the one used in these works, and thereby provides a complementary perspective. Building further upon the idea of emergent time through kinematic flow, our findings provide a possible step in giving a quantitative connection between complexity and time, through the computational complexity of wavefunction coefficients. It is a compelling idea whether the aforementioned notion of computational complexity has a physical interpretation in which the universe acts as a computational device. We emphasize that these are speculative comments, and leave a further exploration of these ideas open to future work.  

\section{Complexity of recursive reduction}\label{sec:complexityrecursive}

Having discussed the complexity of the kinematic flow algorithm, we now turn our attention to the complexity of the recursive reduction algorithm of chapter~\ref{ch:reductionalgorithm}. As we've seen above, analyzing the kinematic flow algorithm in the Pfaffian framework we obtain vast overestimation the complexities, most clearly illustrated by the bounds on the number of poles of the wavefunction coefficients.
As we've discussed, such an overestimation can be a sign that there exists some other, more optimal algorithm. As we've seen in chapter~\ref{ch:reductionalgorithm} that one natural method of incorporating such simplifications is by studying the reduciblity of the differential equations. This avenue of inquiry eventually lead us to the construction of the recursive reduction algorithm of chapter~\ref{ch:reductionalgorithm}. Here, we will investigate the resulting complexity from a Pfaffian perspective and describe some problems that arise when trying to directly compare the two algorithms. Afterwards, to obtain an alternative concrete measure of the complexity of the two algorithms, we perform a counting of the different distinct functions needed.

\subsection{Pfaffian complexity}

The differential chain structure found in section~\ref{sec:differentialchain} bears a striking resemblance to the Pfaffian chains reviewed above, and indeed we will show below that it is possible to write this representation of cosmological correlators in the form of a Pfaffian chain. This will lead to an alternative measure of the complexity of cosmological correlators, although we will see that comparing the two directly comes with a few problems.. However, with having a Pfaffian chain in terms of first-order reduction operators will allow us to explicitly study the reduction of complexity implemented by the higher-order reduction operators in section \ref{sec:complexity}. 

\paragraph{Pfaffian chain for cosmological correlators.}

Let us now discuss how to set up a Pfaffian chain for the cosmological correlators, based on equation \eqref{eq:QTiteration}. To begin, let us recall that given a graph tube $T$ in a graph tubing $\mathcal{T}$,\footnote{Recall the difference between the graph tubings used in the recursive reduction algorithm with the kinematic tubings used in the kinematic flow algorithm. In this section we will exclusively discuss the graph tubings.} the action of a reduction operator was described using three objects: a rational function $\ell^{(T)}_\cT$, the matrices $M^{T,S}_\cT$ for various other graph tubes $S$ and the constant $c^{(T)}_\cT$. We begin by focusing on the rational functions $\ell^{(T)}_\cT$, which we will refer to as letters. 

In analogy to equation \eqref{eq:expfaff}, the required Pfaffian differential equations take the simple form
\begin{equation}\label{eq:PCletters}
\pd^{(S)} \ell^{(T)}_\cT  = -M^{S,T}_\cT\left(\ell_\cT^{(T)}\right)^2  \,. 
\end{equation}
The Pfaffian complexity will then depend on the number $n_{\rm L}$ of letters which we need to specify. Recalling the definition of $\ell^{(T)}_\cT$ in terms of the matrices $M^{T,S}_\cT$, we see that the number of letters $N_{\rm L}$ is determined by the number of pairs $(T,\cT)$ with distinct precursors and successors. In general this is a complicated counting problem which depends on the topology of the Feynman graph and the chosen complete graph tubing. It is bounded by the number of pairs $(T,\cT)$, which grows exponentially in the number of vertices. 

The recursive nature of the differential chain found in section~\ref{sec:differentialchain}, expressed in the form of equation \eqref{eq:ITpartialfin} guarantees that it is a Pfaffian chain. In this chain we need a differential equation for every function $I_\cS$ with $T_{\rm max}\in\cS\subseteq\cT$, and there are $2^{|\cT|-1}-1=2^{N_{\rm v}-1}-1$ such functions; this determines the order $r$ of the Pfaffian chain. The degree $\alpha$ depends on the number of iterations of the recursion equation \eqref{eq:QTiteration} that are required. In turn, the number of recursions depends on the depth of the graph tubing $\cT$, i.e.~the length of the longest ascending chain of graph tubes in $\cT$. For a complete graph tubing, the depth is always equal to the number of vertices $N_{\rm v}$. Finally, the degree $\beta$ is equal to $1$, since the function of interest $I_\cT$ is already part of the chain. From these observations we deduce that the Pfaffian complexity of $I_\cT$ is bounded by
\begin{equation}\label{eq:recurscomplexity}
    \cC(I_\cT) =  (N_{\rm v}, N_{\rm L}+2^{N_{\rm v}-1}-1 , N_{\rm v} ,1) \,.
\end{equation}
The precise growth depends on the topology of the graph and the chosen graph tubing $\cT$ through the number $N_{\rm L}$. 

Note that direct comparison with the complexity of the kinematic flow algorithm in equation~\eqref{PfaffC-ex} would seem to suggest that the complexity of the first-order part of the recursive reduction is already lower than that of the kinematic flow algorithm. However, we should emphasize that the complexity in equation~\eqref{eq:recurscomplexity} is only that of the function associated to a \textit{single} graph tubing. The full wavefunction coefficient associated to a graph, which is what is considered in equation~\eqref{PfaffC-ex}, will consist of many of such tubings. 
This makes it somewhat difficult to compare the two algorithms directly, as they  are measuring the complexity of slightly different objects. Therefore, we will instead focus on how the additional algebraic relations of section~\ref{sec:complexity} does lead to an explicit complexity reduction within the recursive reduction algorithm. We do expect that it is possible to modify the recursive reduction algorithm to incorporate these other tubings, but leave that to future work.

\paragraph{Pfaffian perspective on complexity reduction.}

The construction of the Pfaffian chain above only consisted of a part of the recursive reduction algorithm, the second and arguably most important part consists of leveraging the reduction operators to obtain a large amount of algebraic and permutative identities. Now, we will comment on the implications of these relations from a Pfaffian perspective.

From this perspective, the clearest results follow when an algebraic relation is contained fully already within the Pfaffian chain. Recall from section~\ref{sec:complexity} that this is the case whenever we use a higher-order reduction operator associated to a non-maximal tube, e.g.\ when one does not need to use a factorization relation. 
Whenever there is such a relation, one can algebraically eliminate one of the functions in the Pfaffian chain, resulting in a new chain where the order $r$ is reduced by one. Since there are many such relations, the order $r$ will reduce significantly.

However, there are two aspects of the reduction which are not captured by the Pfaffian framework. Firstly, the Pfaffian chain structure demands that we separately define all the letters $\ell^{(T)}_\cT$ by the differential equation \eqref{eq:PCletters}. The algebraic relations in the recursive reduction algorithm do not lead to a clear reduction in the number of letters needed for the minimal representation functions, so the order $r$ of the Pfaffian chain will still have a contribution $n_{\rm L}$ which grows exponentially in the number of vertices.

Secondly, part of the reductions in the recursive reduction algorithm require permutations among the variables in the integrals. This type of symmetry is however not detected by Pfaffian complexity, since it assumes a fixed ordering on the variables. For example, consider a Pfaffian chain containing a function $f(z_1,z_2)$. The function $g(z_1,z_2)=f(z_2,z_1)$, obtained by swapping the two variables, cannot be obtained as a Pfaffian function without adding it to the Pfaffian chain separately and increasing the complexity. An example of this arises from the factorization relations, where new functions must be introduced which will always be permutations of functions already in the Pfaffian chain.
This is closely related to the challenge in establishing a connection between complexity and symmetry, as pointed out in \cite{grimm_complexity_2024a}. We believe that this calls for a complexity framework in which symmetries of this form are more naturally detected, and we leave this as a future direction of research. 

\subsection{Function counting versus Pfaffian complexity.}\label{ssec:minrepcounting}

Having analyzed the Pfaffian complexity of the recursive reduction algorithm, we now turn our attention to a more direct comparison between the kinematic flow and recursive reduction algorithms. In particular we will count the number of new functions needed to obtain an $N_v$-vertex tree-level diagram, assuming that the $N_v-1$ diagram has already been solved. From this perspective, the power of the recursive reduction becomes much  more apparent, as this can fully incorporate all the permutative and algebraic identities.

\paragraph{Counting minimal representation functions.}

Recall that, when performing the recursive reduction algorithm, the remaining tubings are characterized by having exactly one bare vertex, no tubes which admit a partition, and all tubings related by permutations removed. See for example table~\ref{tab:minrep}, where the minimal representation functions for $N_{\rm v}=1,2,3,4$ vertices are displayed.

\begingroup 
\def\arraystretch{1.5}
\arrayrulecolor{arnoblue}
\begin{table}[h]
    \centering
    \begin{tabular}{
        c
        !{\color{arnoblue}\vrule width 1.0pt}
        l
    }
\rowcolor{arnoblue-faded}
    $N_{\rm v}$ & Minimal representation functions\\
\specialrule{1.0pt}{0pt}{0pt}
        $1$ & $
\begin{tikzpicture}[baseline=-0.5ex]
    \begin{feynman}
        \vertex [circle,fill,inner sep=1pt](v51) {};
        \diagram*  {
};
    	\vertex [above=0 cm of v51,shape=circle,draw=black,fill=none,very thick,minimum size=.6 cm] (bv51) {};
    \end{feynman}
\end{tikzpicture}$ \\
         $2$ & $
\begin{tikzpicture}[baseline=-0.5ex]
    \begin{feynman}
        \vertex [circle,fill,inner sep=1pt](v411) {};
        \vertex [right=.5cm of v411,circle,fill,inner sep=1pt] (v412) {};            
        \diagram*  {
	     (v411) --[very thick] (v412);
          (v411) --[connect=3mm,black,very thick] (v412);        
};
        \vertex [above=0 cm of v411,shape=circle,draw=black,fill=none,very thick,scale=1] (bv411) {};
    \end{feynman}
\end{tikzpicture}$ \\
         $3$ & $
\begin{tikzpicture}[baseline=-0.5ex]
    \begin{feynman}
        \vertex [circle,fill,inner sep=1pt](v211) {};
        \vertex [right=.5cm of v211,circle,fill,inner sep=1pt] (v212) {};    \vertex [right=.5cm of v212,circle,fill,inner sep=1pt] (v213) {}; 
        \vertex[right=1.5 cm of v213] (s2); 
        
        \vertex [above=0cm of s2,circle,fill,inner sep=1pt](v311) {};
        \vertex [right=.5cm of v311,circle,fill,inner sep=1pt] (v312) {};    \vertex [right=.5cm of v312,circle,fill,inner sep=1pt] (v313) {};         
        \diagram*  {
          (v211) --[very thick] (v212) --[very thick] (v213);
          (v211) --[connect=3mm,black,very thick] (v213);     
        
          (v311) --[very thick] (v312) --[very thick] (v313);
          (v311) --[connect=2mm,black,very thick] (v312);
          (v311) --[connect=3mm,black,very thick] (v313);           
};
        \vertex [above=0 cm of v211,shape=circle,draw=black,fill=none,very thick,scale=1] (bv211) {};
        
        \vertex [above=0 cm of v212,shape=circle,draw=black,fill=none,very thick,scale=1] (bv212) {};
        
        \vertex [above=0 cm of v311,shape=circle,draw=black,fill=none,very thick,scale=.6] (bv311) {};
    \end{feynman}
\end{tikzpicture}$\\
         $4$ & $
\begin{tikzpicture}[baseline=-0.5ex]
    \begin{feynman}
        \vertex [circle,fill,inner sep=1pt](v11) {};
        \vertex [right=.5cm of v11,circle,fill,inner sep=1pt](v12) {};
        \vertex [right=.5cm of v12,circle,fill,inner sep=1pt](v13) {};
        \vertex [right=.5cm of v13,circle,fill,inner sep=1pt](v14) {};
        
        \vertex [right=1cm of v14, circle,fill,inner sep=1pt](v21) {};
        \vertex [right=.5cm of v21,circle,fill,inner sep=1pt](v22) {};
        \vertex [right=.5cm of v22,circle,fill,inner sep=1pt](v23) {};
        \vertex [right=.5cm of v23,circle,fill,inner sep=1pt](v24) {};  

        \vertex [right=1cm of v24, circle,fill,inner sep=1pt](v31) {};
        \vertex [right=.5cm of v31,circle,fill,inner sep=1pt](v32) {};
        \vertex [right=.5cm of v32,circle,fill,inner sep=1pt](v33) {};
        \vertex [right=.5cm of v33,circle,fill,inner sep=1pt](v34) {};     

        \vertex [right=1cm of v34, circle,fill,inner sep=1pt](v41) {};
        \vertex [right=.5cm of v41,circle,fill,inner sep=1pt](v42) {};
        \vertex [right=.5cm of v42,circle,fill,inner sep=1pt](v43) {};
        \vertex [right=.5cm of v43,circle,fill,inner sep=1pt](v44) {};     
        
        \diagram*  {
          (v11) --[very thick] (v12) --[very thick] (v13)--[very thick] (v14);
          (v11) --[connect=3mm,black,very thick] (v14);    

          (v21) --[very thick] (v22) --[very thick] (v23)--[very thick] (v24);
          (v21) --[connect=3mm,black,very thick] (v24);  
          (v22) --[connect= 2mm,black,very thick] (v23);

          (v31) --[very thick] (v32) --[very thick] (v33)--[very thick] (v34);
          (v31) --[connect=3mm,black,very thick] (v34);
          (v31) --[connect=2mm,black,very thick] (v33);  

          (v41) --[very thick] (v42) --[very thick] (v43)--[very thick] (v44);
          (v41) --[connect=3mm,black,very thick] (v44);
          (v41) --[connect=2.3mm,black,very thick] (v43);
          (v41) --[connect=1.6mm,black,very thick] (v42);
          
};
        \vertex [above=0 cm of v11,shape=circle,draw=black,fill=none,very thick,scale=1] (bv11) {};
       \vertex [above=0 cm of v12,shape=circle,draw=black,fill=none,very thick,scale=1] (bv12) {};
        \vertex [above=0 cm of v13,shape=circle,draw=black,fill=none,very thick,scale=1] (bv13) {};

        \vertex [above=0 cm of v21,shape=circle,draw=black,fill=none,very thick,scale=1] (bv21) {};
       \vertex [above=0 cm of v22,shape=circle,draw=black,fill=none,very thick,scale=.6] (bv22) {};

        \vertex [above=0 cm of v31,shape=circle,draw=black,fill=none,very thick,scale=.6        ] (bv31) {};
       \vertex [above=0 cm of v32,shape=circle,draw=black,fill=none,very thick,scale=.6] (bv32) {};

        \vertex [above=0 cm of v41,shape=circle,draw=black,fill=none,very thick,scale=.45        ] (bv41) {};       
    \end{feynman}
\end{tikzpicture}$
    \end{tabular}
    \caption{The minimal representation functions for all diagrams up to four vertices.}
 \label{tab:minrep}
\end{table}
\endgroup

To obtain an expression for the number of new minimal representations needed $N_{\rm m}$, we derive a recursion relation as follows. We start with a chain of $N_{\rm v}$ vertices, and encircle all vertices by the maximal tube. 
Since this maximal tube must have exactly one bare vertex, which by permutation symmetry can be taken to be the right-most vertex, the remaining $N_{\rm v}-1$ vertices must be encircled by adding more tubes. In order to count in how many ways this can be done, we note that the counting receives contributions from all possible ways of partitioning the $(N_{\rm v}-1)$-chain into smaller chains. For these smaller chains, the same counting problem holds. This observation implies the following recursion relation:
\begin{equation}
    N_{\rm m}(N_{\rm v}) = \sum_{\pi\in P(N_{\rm v}-1)} \prod_{j\in \pi} N_{\rm m}(j) \,.
\end{equation}
Here $P(N_{\rm v}-1)$ denotes the set of integer partitions of $N_{\rm v}-1$. Note that this formula denotes the number of minimal representation functions with \textit{exactly} $N_{\rm v}$ vertices, and therefore this counting does not include the functions with fewer edges. To incorporate these one would simply take the sum of $N_{\rm m}(n)$ from $n=1$ to $N_{\rm v}$.

To clarify the meaning of this formula, consider for example $n=5$. Then sum then runs over all integer partitions of $4$, which are given by 
\begin{equation}
    \{4\},\,\{3,1\},\,\{2,2\},\,\{2,1,1\},\,\{1,1,1,1\} \,.
\end{equation}
The number of minimal representation functions of the 5-chain is then given by
\begin{align*}
    N_{\rm m}(5)  =& N_{\rm m}(4) + N_{\rm m}(3)N_{\rm m}(1) + N_{\rm m}(2)N_{\rm m}(2) \\
    & + N_{\rm m}(2)N_{\rm m}(1)N_{\rm m}(1) + N_{\rm m}(1)N_{\rm m}(1)N_{\rm m}(1)N_{\rm m}(1)  \\
    =& \,9.
\end{align*}
The sequence $N_{\rm m}(N_{\rm v})$ coincides with the one documented in \cite{OEIS}, and no closed-form expression is known.

For comparison, let us consider the number of functions $N_{\rm f}(N_{\rm v})$ needed to express a cosmological correlator $\psi$ in terms of the differential chain from section~\ref{sec:differentialchain}, i.e.~without the implementation of the higher-order reduction operators. Recall from equation~\eqref{eq:psiG} that, for a given graph, $\psi$ is given by a sum of the form
\begin{equation}
    \psi = \sum_{\cT\, \text{complete} } I_\cT\,,
\end{equation}
where this sum is over all complete tubings of the graph.
For each term $I_\cT$, we have to solve the differential chain from the previous section. However, many of the functions in the various chains will overlap, since a tubing $\cS$ can be a sub-tubing of several distinct complete tubings $\cT$. Effectively, this means that we need to solve for  $I_\cS$ for \textit{every} tubing $\cS$ containing the maximal tube. In other words, the number $N_{\rm f}(N_{\rm v})$ is given by the number of such tubings. 

This counting depends on the topology of the graph, so for concreteness let us consider a chain of $N_{\rm v}$ vertices. In this case, the counting problem is equivalent to the number of ways in which a list of $N_{\rm v}$ items can be grouped into nested sublists, which is discussed in \cite{OEIS2}. The first few values of this sequence are compared to the number of minimal representation functions in table \ref{tab:growths}.

\begin{table}[h]
    \centering
\arrayrulecolor{arnoblue}
\begin{tabular}{
    c
    !{\color{arnoblue}\vrule width 1.0pt}
    r r r r r r r r r
}
\rowcolor{arnoblue-faded}
$N_{\rm v}$ & 1 & 2 & 3 & 4 & 5 & 6 & 7 & 8 & 9   \\
\specialrule{1.0pt}{0pt}{0pt}
$N_{\rm m} $ & 1 & 1 & 2 & 4 & 9 & 20 & 49 & 117 & 297   \\ 
\specialrule{.5pt}{0pt}{0pt}
 $N_{\rm f}$ & 1 & 4 & 24 & 176 & 1440 & 12,608 & 115,584 & 1,095,424 & 10,646,016    \\ 
\specialrule{.5pt}{0pt}{0pt}
 $N_{\rm k}$ & 1 & 4 & 16 &  64 &256 & 1024 & 4096 & 16,384 & 65,536    \\
    \end{tabular}
    \caption{Comparison of the number of new functions needed to compute the contribution to a cosmological correlator from a graph of $N_{\rm v}$ vertices, in the minimal representation ($N_{\rm m}$), the differential chain ($N_{\rm f}$), and the kinematic flow algorithm ($N_{\rm k}$).}
    \label{tab:growths}
\end{table}

In table~\ref{tab:growths}, we also include the number of functions needed for a chain of $N_{\rm v}$ vertices in the kinematic flow algorithm of \cite{arkani-hamed_differential_2023,arkani-hamed_kinematic_2023}, which is given by $N_{\rm k}(N_{\rm v})=4^{N_{\rm v}-1}$. The table shows that, compared to the differential chain and the kinematic flow representations, the recursive reduction algorithm achieves a significant reduction in complexity. 

\fi

\if\PrintConclusion1


\newpage

\stepcounter{thumbcounter}
\setcounter{colorcounter}{6}
%
%
%


\chapter[Summary and Outlook]{Summary and Outlook}\label{ch:conclusion}


In the final chapter of this thesis we will summarize its main results and outline potential directions for future research, some of which are already under investigation at the time of writing.

\section{Summary}

In chapter~\ref{ch:generalGKZ}, we have begun this thesis with the mathematical underpinning necessary for the following chapters. We began with a general review on GKZ system, explaining how a certain class of integrals gives rise to differential equations. Furthermore, we discussed how the resulting differential equations could be fully encoded in a single matrix and parameter vector, the so-called GKZ data. Then, we discussed some well-known properties of the resulting system of differential equations. In particular, we discussed two common methods of obtaining the solutions of these differential equations, in terms of convergent series expansions or a useful ansatz.

Afterwards, we took a more formal perspective with a discussion on D-modules and their reducibility. D-modules are modules of the ring of polynomial differential operators and provide a useful formal framework for studying systems of differential equations. In this setting, solutions of the differential equations roughly correspond to homomorphisms from the D-module itself to the space of holomorphic functions. 

For a D-module reducibility is defined as the existence of non-trivial submodules. Crucially, we explained that this property can be translated usefully to the solutions of the differential equations. From this perspective, reducibility corresponds to the existence of differential operators that annihilate some, but not all, of the solutions. These additional differential operators, that we dubbed reduction operators, can be used to obtain partial solution bases for the full system of differential equations. This allows us to solve a system of differential equations using a ``divide and conquer'' approach, where it is first divided up into simpler subsystem. These simpler subsystems can be solved separately and, in some cases, combining several of them results in a full solution basis for the original system.

Using this general perspective, we then turned our attention back to GKZ systems. Here, there is already mathematical literature studying the reducibility using which reducibility can be completely classified from the GKZ data. Our main contribution here was using the proofs in~\cite{schulze_resonance_2012} to obtain the various submodules explicitly. Interestingly, while the proofs and constructions themselves were phrased in quite abstract terms, mostly being stated in terms of exact sequences on so-called Euler-Koszul homologies, we were able to frame it completely in terms of differential operators acting on partial solution bases. This made it possible for these abstract results to be used when solving the actual systems of differential equations, without needing the full technical formalism.

Our attention then turned to these differential operators themselves, the reduction operators. We constructed various algorithms that can be used to obtain the reduction operators explicitly, as well as classified their properties. In particular, we want to highlight the various counting formulae of section~\ref{sec:reductionoperators} which were previously unpublished.

With the general framework established we then turned towards cosmological correlators in chapter~\ref{ch:cosmology}. We introduced the toy model, introduced originally in~\cite{arkani-hamed_cosmological_2017}, that we studied in the remainder of the thesis. The rest of chapter~\ref{ch:cosmology} was devoted to studying the single-exchange integral, a simple example of a cosmological correlator. For this example, we provided its GKZ system and, as it is reducible, constructed the reduction operators. Afterwards, we used those to obtain the partial solution bases, combining these to form a full solution basis and obtain the correlator itself.

For the single-exchange integral, we had only used the reduction operators to obtain additional homogenous differential equations that allowed us to obtain the partial solution bases. However, in the process we also found that it satisfied certain inhomogeneous differential equations, which led to contraction and cut identities on a diagrammatical level. In chapter~\ref{ch:reductionalgorithm}, we extended this observation greatly.

There, we continued studying the correlators of the same toy model. In particular, we used the result of~\cite{arkani-hamed_differential_2023} that tree-level diagrams of this theory can be decomposed using so-called tubings. Each tubing corresponds to an integral and the correlator of the diagram consists of a sum over these tubings. The tubings were a convenient perspective for us, as the allowed us to construct their GKZ systems for any tubing of an arbitrary diagram. Furthermore, we were also able to find the reduction operators for such a tubing in full generality.

Here, a possible roadblock appeared, namely that the construction of a GKZ system usually involves the introduction of additional variables that play no role in the physical correlator. The differential system then involves partial derivatives with respect to these superfluous variables, usually leading to an unnecessary increase in complexity. For the cosmological correlators, the reduction operators provided us with a convenient way around this problem, as we were able to show that it was possible to combine reduction operators such that only the physical variables appeared.

Having found these physical combinations of reduction operators, we then studied their structure. Using these operators, we constructed a closed system of first-order non-homogenous differential equations. The resulting first-order system could in principle be solved for the tubing integral, and possesses similarities to other known differential systems for cosmological correlators. The key finding was that the reduction operators actually implied a large redundancy in this first-order system. To be precise, the functions that appeared in this system were related to each other using algebraic relations, which were a direct consequence of the existence of various reduction operators.

Besides the algebraic relations within the first-order system, there also exist algebraic relations relating the systems of different cosmological correlators. In particular, the reduction operators could be used to relate the system of one diagram to the systems of diagrams where various edges were contracted or cut. Here, we saw that these basis functions in these various systems were actually very similar to the ones already in the differential chain, just with permuted inputs.

The existence of these algebraic and permutative relations naturally lead to the following question: once all possible relations of these kinds are used, what functions will be left? We called such a minimal set the minimal representation functions and showed that they are universal for all cosmological correlators, regardless of topology or tubing of the diagram. Furthermore, we saw that the number of such functions was surprisingly small, especially when compared to other known algorithms.

The minimal representation functions naturally lead to the introduction of our recursive reduction algorithm. Here, solving a cosmological correlator broadly happens in two phases. First, all of the needed first-order systems are written down and, using \textit{only} algebraic and permutative relations, these are all expressed in terms of the minimal representation functions. Secondly, the minimal representation functions must be found. As these functions represent the minimal building blocks that the full correlators are build out of, these can only be solved directly using the differential equations they satisfy. Once solved, these functions can be inserted into an expression for the full correlator.
Crucially, this algorithm first leverages as much symmetry as possible, as solving algebraic identities or permuting function inputs is a much simpler task than solving differential equations.

Finally, we studied the complexity of the various algorithms in chapter~\ref{ch:complexity}. The goal here was to make the intuition of complexity reduction and simplicity precise, using the framework of Pfaffian functions. A Pfaffian function is defined using a first-order system of differential equations having a particular structure. Furthermore, it allows for explicit quantification of its complexity, with topological and computational consequences. This made it a natural framework for analyzing cosmological correlators, as we are already considering them through their differential equations.

We analyzed the complexity of two different algorithms, one is the kinematic flow algorithm of~\cite{arkani-hamed_differential_2023}, the other the first-order system we introduced in chapter~\ref{ch:reductionalgorithm}.
We were able to cast both of these differential systems in terms of a so-called Pfaffian chain, implying that their complexity could be analyzed using the Pfaffian framework.

Crucially, in this framework the complexity of the cosmological correlators depends on their explicit representation. Thus, if the explicit bounds we find are much too high, this could be an indication that some other simpler representation exists.
For the kinematic flow algorithm, we observed that this is the case and the topological bounds resulting from this analysis were great overestimation of what was expected from physical grounds. 

This overestimation of complexity was one of the original motivations for the recursive reduction algorithm of chapter~\ref{ch:reductionalgorithm}: introducing a framework in which such additional simplifications and relations are automatically incorporated. Interestingly, while the recursive reduction algorithm succeeded in incorporating these simplifications, it did so in a way that is not completely visible from a Pfaffian perspective. While many of the algebraic relations obtained for the first-order system lead directly to a reduction in Pfaffian complexity, the permutation relations do not. Consequently, the Pfaffian framework still overestimates and one wonders if there is a more suitable measure of complexity that does incorporate these relations.

\section{Outlook}

With the results found above, there are a number of possible future directions, some of which we are already currently pursuing. One source of future directions is that the results on GKZ systems in chapter~\ref{ch:generalGKZ} are quite broadly applicable. GKZ systems appear throughout physics, with examples in Feynman diagrams~\cite{nasrollahpoursamami_periods_2016,vanhove_feynman_2018,delacruz_feynman_2019,klemm_lloop_2020,klausen_hypergeometric_2020,feng_feynman_2022,ananthanarayan_feyngkz_2023,tellander_cohenmacaulay_2023,chestnov_restrictions_2023,caloro_ahypergeometric_2023,levkovich-maslyuk_yangian_2024}, string theory~\cite{hosono_gkzapp_1996,hosono_gkzCY_1996,hosono_maximal_1997,stienstra_gkz_2005,li_picardfuchs_2012,aspinwall_mirror_2017,li_superpotentials_2022}, and many other settings.

In many cases, these GKZ systems are reducible. For example, in geometric settings, such as string theory compactifications, this is always the case. Furthermore, it has been shown that the period integrals of algebraic varieties are solutions to GKZ systems that often have a larger solution space than expected from geometry~\cite{hosono_mirror_1995,hosono_gkzCY_1996,hosono_gkzapp_1996}. 
Interestingly, it turns out that the resulting set of differential equations factorizes, and the identification of suitable factors results in the correct set of differential equations -- the Picard-Fuchs equations \cite{hosono_mirror_1995,hosono_gkzCY_1996}. This factorization is a consequence of the reducibility of the underlying GKZ system and the tools introduced in this thesis can be readily applied to finding such geometric differential operators. For example, for the quintic hypersurface in $\P^4$, one easily obtains the actual Picard-Fuchs equations from the GKZ system by finding an appropriate reduction operator. One can expect that further reductions arise if the underlying variety has special additional features, such as a fibration structure, and it would be interesting to explore this further.

Similarly, we expect that the reduction techniques are generally useful in the study of Feynman integrals. Considering them in their Feynman or Lee-Pomeransky representation \cite{weinzierl_feynman_2022}, one shows that these integrals can be interpreted as GKZ systems and the this system often is reducible. Furthermore, there are known reduction formulas for the polynomials in these integrals from a graph theoretic perspective \cite{bogner_feynman_2010}. Therefore, it seems likely that possible reductions for the polynomials in the integral will lead to reductions for the underlying GKZ system. If this is possible it could greatly simplify solving certain Feynman integrals and hopefully allow for systematic studies of classes of Feynman integrals, similar to those performed in \cite{bonisch_analytic_2021}. We are currently exploring this and will discuss this in future work.

A different interesting future direction is to further explore reducibility for cosmological correlators. For the tree-level correlators studied in this thesis, there are interesting aspects about the interplay of first-order and higher-order operators that might simplify our discussion further. In particular, we expect that using the higher-order operators earlier in the reduction could be beneficial when focusing on the singularity structure of the correlator discussed in section~\ref{ssec:factors}. The next natural step is then to investigate  loop-level cosmological correlators, which can also be represented using tubings \cite{baumann_kinematic_2024}. Also at loop-level the main task will be to understand the space of reduction operators and which relations they impose on the amplitudes. We expect that much of our strategy carries over to these cases and it would be desirable to go through the construction in a follow-up investigation. Eventually, one can aim at finding a full-fledged recursion for the complete amplitude. Another natural extension is the application of reduction operators to cosmological correlators where the propagators have arbitrary masses, or more generally, the examination of other phenomenological models replacing the conformally coupled scalar action \eqref{model_action}. 

Finally, we expect there are improvements to be made regarding the explicit complexities of chapter~\ref{ch:complexity}. Here, the Pfaffian framework was used to provide a measure of complexity, but that these bounds are rather weak and that the inclusion of the algebraic relations and the simplifications due to permutations and shifts into this construction is very challenging.
There are two issues that we believe hinder us to present better estimates of the full complexity. Firstly, we know that the Pfaffian complexity is very sensitive to adding new functions, since the bounds also have to hold even for the worst-case solutions to a given Pfaffian system. However, a crucial part of the Pfaffian chain are the letters \eqref{eq:PCletters}. These are actually rather simple functions, but we did not succeed to find a simple representation to incorporate them. Secondly, we are not aware of a refined Pfaffian framework that incorporates symmetries and provides stronger bounds. We believe that both issues should be addressed in the future. Eventually, we hope to fully compare the complexity of algorithms. The algorithm giving the best bounds on the number of poles, which matches our physical exceptions, would then have the most minimal representation.

\fi


\if\PrintAppendices1
\cleardoublepage


\appendix
\stepcounter{thumbcounter}
\renewcommand{\isAppendix}{1}

%
%
%


\chapter{Scientific Appendices}\label{ch:appendices}

\section{First-order reduction operators in physical variables} \label{red_op_phys}

In this appendix we will show that the operator $Q^{(T)}$ from equation~\eqref{eq:QTdef_t} can be rewritten using only the physical coordinates. To show this, we first note that, when restricting to the physical slice $z^{(T)}_v\rightarrow1$, we have
\begin{equation}
    z^{(T)}_v\partial^{(T')}_{v}=\theta^{(T')}_{v}+\ldots
\end{equation}
for tubes $T$, $T'$ and all vertices $v$ in $T\cap T' $. Here, the $\ldots$ denote that this equality holds up to terms which go to zero under the restriction and we recall that $\theta^{(T)}_v=z^{(T)}_v\partial^{(T)}_v$. Then, considering the definition~\eqref{eq:QTdef_t} of $Q^{(T)}$ and inserting equation~\eqref{QTT'} for the various reduction operators, we find
\begin{equation}\label{eq:physQTderiv}
 \sum_{T'\supsetneq T}Q_{T,T'}=z^{(T)}  \sum_{T'\supsetneq T}\partial^{(T')} +\sum_{v\in T} \sum_{T'\supsetneq T} \theta^{(T')}_{v}+\ldots\,.
\end{equation}
Note that the first term on the right-hand side is already in terms of the physical variables only. Thus, we will now use the Euler operators of the GKZ system to rewrite the second term in this expression.

Recall that the Euler operators of the GKZ system associated to cosmological correlators are given by \eqref{Euler_cosm}. From these expressions, we construct the following useful combination of Euler operators
\begin{equation}\label{eq:physredeuler}
    \sum_{T'\subseteq T}\E_{T'}-\sum_{v\in T}\E_i=\sum_{T'\subseteq T} \theta^{(T')} +\sum_{T'\subseteq T}\sum_{v\in T'}\theta^{(T')}_{v}-\sum_{v\in T}\sum_{\{T':v\in T'\}}\theta^{(T')}_{v}\,,
\end{equation}
where the sum $\sum_{\{T':v\in T'\}}$ is over all tubes $T'$ containing $v$. Notice that, by the non-crossing condition, we have that $v\in T$ and $v\in T'$ if and only if either (1) $v\in T'$ and $T'\subseteq T$, or (2) $v\in T$ and $T'\supsetneq T$. Therefore, we find that
\begin{equation}
\sum_{v\in T}\sum_{\{T':v\in T'\}}\theta^{(T')}_{v}=\sum_{T'\subseteq T}\sum_{v\in T'}\theta^{(T')}_{v}+\sum_{T'\supsetneq T}\sum_{v\in T}\theta^{(T')}_{v}\,.
\end{equation}
Inserting this into equation~\eqref{eq:physredeuler}, we can solve for $\sum_{T'\supsetneq T}\sum_{v\in T}\theta^{(T')}_{v}$ and obtain
\begin{equation}
     \sum_{T'\supsetneq T}\sum_{v\in T}\theta^{(T')}_{v}=\sum_{v\in T}\E_i- \sum_{T'\subseteq T}\E_{T'}+\sum_{T'\subseteq T} \theta^{(T')} 
\end{equation}
When acting on solutions of the GKZ system, an Euler operator $\E_J$ may be replaced with $\nu_J$. Therefore, we have the equality
\begin{equation}
    \sum_{v\in T}\sum_{T'\supsetneq T}\theta^{(T')}_{v}\simeq_{\E+\nu}\sum_{T'\subseteq T}(\theta^{(T')} +\nu^{(T')})-\sum_{v\in T}\nu_i\,,
\end{equation}
where we recall that $\simeq_{\E+\nu}$ means that this equality holds only when acting on solutions of the GKZ system at the parameter $\nu$.

Finally, we insert this equation into equation~\eqref{eq:physQTderiv} and obtain
\begin{equation}
     Q^{(T)}\simeq_{\E+\nu}\sum_{T'\supsetneq T}Q_{T,T'}=z^{(T)}  \sum_{T'\supsetneq T}\partial^{(T')} +\sum_{T'\subseteq T}(\theta^{(T')} +\nu^{(T')})-\sum_{v\in T}\nu_i+\ldots\,.
\end{equation}
Recall that the $\ldots$ terms go to zero in the physical limit. Thus we find that, when acting on GKZ systems and in the physical limit, the combination of reduction operators will act as 
\begin{equation}
    \sum_{T'\supsetneq T}Q_{T,T'}\vert_{\rm phys}=z^{(T)}  \sum_{T'\supsetneq T}\partial^{(T')} +\sum_{T'\subseteq T}(\theta^{(T')} +\nu^{(T')})-\sum_{v\in T}\nu_i\,,
\end{equation}
which we identify as being $Q^{(T)}$ as stated in~\eqref{eq:QTresult}.

\section{Matrix form of the first-order system}\label{ap:matrixrep}

This appendix is an addition to chapter~\ref{ch:reductionalgorithm} devoted to solve equation~\eqref{eq:QTiteration} as a matrix equation, thereby solving the iterative equation for $Q^{(T)}I_\cT$ in terms of linear combinations of integrals $I_\cS$ with rational coefficients. Thus, we must first fix a tubing $\cT$, such that we can obtain $Q^{(T)}I_\cT$ for each $T\in \cT$. Then, we must identify a suitable vector space, which enables us to keep track of both a tubing as well as a tube contained in this tubing. This leads us to construct a basis of vectors $e_{(S,\cS)}$, where $S$ is a tube contained in $\cS$, and $\cS$ is a subset of $\cT$. Furthermore, it will be useful to take linear combinations of such pairs, which we will denote as $c \cdot e_{(S,\cS)}$ where $c$ is some matrix of coefficients. Note that, in the end, we will take $c$ to be rational functions in the $z$. Using this notation, we obtain the vector space of all such formal combinations as
\begin{equation}
    \mathbf{V}\coloneqq \big\{ \, \sum_{(S,\cS)} c_{(S,\cS)} e_{(S,\cS)}\; : \;S\in \cS ,\,\cS \subseteq \cT\, \big\}\,.
\end{equation}
This vector space will be the key in solving for $Q^{(T)}I_\cT$.

To relate this vector space to the actual integrals we are trying to solve for, we must first define a mapping between the two. Thus, we begin by defining the integral mapping $\mathrm{int}$, which sends each basis element $\mathbf{V}$ to an integral and extending linearly. In other words, we have
\begin{equation}
  \mathrm{int}(e_{(S,\cS)})=I_{\cS}
\end{equation}
for the basis elements.  Note that this function does not take into account the tube $S$ in $e_{(S,\cS)}$.

Now, we will define a variety of operators on $\mathbf{V}$ such that we can rewrite~\eqref{eq:QTiteration} in terms of matrices acting on $\mathbf{V}$. We begin by defining the operator $\mathbf{Q}$ implicitly, using equation~\eqref{eq:QTiteration}. In particular, we will use that, as we know that equation~\eqref{eq:QTiteration} can be solved iteratively, each combination $Q^{(S)}I_{\cS}$ must be a linear combinations of integrals $I_{\cS'}$. Therefore, there must exist an operator $\mathbf{Q}$ such that
\begin{equation}
   \mathrm{int}(\mathbf{Q} \cdot e_{(S,\cS)})=Q^{(S)} I_\mathcal{S}
\end{equation}
for each combination $(S,\cS)$.
Note that, since the operator $\mathrm{int}$ does not take into account the tube $S$, there is some ambiguity in this definition of $\mathbf{Q}$. However, since eventually we will always apply $\mathrm{int}$ to the equations which we obtain, this ambiguity will not be important for us. Therefore, we are free to fix $\mathbf{Q}$ such that its image lies in the set
\begin{equation}
    \emptyset \times \mathcal{PT}=\big\{ \,(\emptyset,\cS)\;\vert\; \cS\subseteq \cT\,\}\,
\end{equation}
where $\mathcal{PT}$ is the power set of $\cT$.

It turns out that equation~\eqref{eq:QTiteration} implies that $\mathbf{Q}$ must satisfy an analogous matrix equation. In particular, we will define three operators $\mathbf{A}$, $\mathbf{L}$ and $\mathbf{G}$ on the vector space $\mathbf{V}$, and solve for $\mathbf{Q}$ in terms of these operators. We begin with a matrix $\mathbf{A}$, which is defined as 
\begin{equation}
    \mathbf{A}_{(S,\cS),(S',\cS')}=\left\{ \begin{array}{rll}
    1 & \text{if } S=S' &\text{and } \cS=\cS'\, ,\\
    -1 & \text{if } S\prec S' &\text{and } \cS=\cS'\, , \\
    0 & \text{otherwise}\, ,&
\end{array} \right.
\end{equation}
where we recall from section~\ref{ssec:chainconstruction} that $S\prec S'$ implies that $S$ is a maximal tube contained strictly in $S'$.  Additionally, we will define the matrix $\mathbf{L}$ as
\begin{equation}
    \mathbf{L}_{(S,\cS),(S' ,\cS')}=\left\{ \begin{array}{ll}
    \ell^{(S')}_{\cS'} & \text{if } S\succ S' \text{ and } \cS=\mathcal{S}'\setminus\{T'\}\\
    0 & \text{otherwise}
\end{array} \right.\, ,
\end{equation}
where $\ell^{(S')}_{\cS'}$ are the letters from \eqref{eq:letterM}.\footnote{One can also define $\mathbf{L}$ in terms of $1/p^{(S')}_{\cS'}$, with $p^{(S')}_{\cS'}$ the denominator in equation~\eqref{eq:QT-QS}.} Observe that, while the actions of $\mathbf{L}$ and $\mathbf{A}$ on the whole of $\mathbf{V}$ is rather involved, its action can be obtained element-wise quite easily. Finally we will define the operator $\mathbf{G}$ acting as
\begin{equation}
   \mathbf{G}_{(S,\cS),{(S',\cS')}}=\left\{ \begin{array}{ll}
    \gamma^{(S')}_{\cS'} & \text{if } S=\emptyset \text{ and } \cS=\cS'\\
    0 & \text{otherwise}
\end{array} \right.\, ,
\end{equation}
with $\gamma^{(S')}_\mathcal{S'}$ as in equation~\eqref{eq:gammadef}.

With all of this notation, equation~\eqref{eq:QTiteration} can be written as an equation for $\mathbf{Q}$, which is given by
\begin{equation}
    \mathbf{Q}=\left(\mathbf{Q}\mathbf{A}-\mathbf{G}\right)\mathbf{L}\,.
\end{equation}
This can simply be solved for $\mathbf{Q}$, from which we find that
\begin{equation}
    \mathbf{Q}=-\mathbf{G} \mathbf{L}\left(1-\mathbf{A}\mathbf{L}\right)^{-1}\,.
\end{equation}
Note that, as $\mathbf{A}\mathbf{L}$ is nil-potent, this can also be written as 
\begin{equation}
    \mathbf{Q}=-\mathbf{G} \mathbf{L} \sum_{i=0}^k \left(\mathbf{A}\mathbf{L}\right)^i\,.
\end{equation}
where $k$ is the nil-potent degree of $\mathbf{A}\mathbf{L}$. 

Then, equation~\eqref{eq:QT-QS} can be written as
\begin{equation}
   Q^{(T)}I_\mathcal{T} = -\sum_{i=0}^k\sum_{(S,\cS)}\left( \mathbf{G} \mathbf{L} (\mathbf{A}\mathbf{L})^i\right)_{(S,\cS),(T,\cT)}I_{\mathcal{T'}}\,.
\end{equation}
Note that the image of $\mathbf{G}$ is contained in $\emptyset \times \mathcal{PT}$, with $\mathcal{PT}$ the set of all possible tubings, confirming that the image of $\mathbf{Q}$ is as well. Furthermore, recall that in the construction of $\mathbf{V}$, we only considered tubings $\cS$ that are subsets of $\cT$. Therefore, the equation above can be written as
\begin{equation}
           Q^{(T)}I_\mathcal{T} = -\sum_{i=0}^k\sum_{\cS\subseteq \cT}\left( \mathbf{G} \mathbf{L} (\mathbf{A}\mathbf{L})^i\right)_{(\emptyset,\cS),(T,\cT)}I_{\mathcal{S}}\,,
\end{equation}
and we have found that, using the matrices $\mathbf{G}$, $\mathbf{L}$ and $\mathbf{A}$, one can solve the iterative equation for $Q^{(T)}I_\cT$ in terms of the integrals $I_\cS$.


\cleardoublepage
\fi



\thumbfalse

\if\PrintSamenvatting1


\chapter*{Nederlandse Samenvatting}
\addcontentsline{toc}{chapter}{Nederlandse Samenvatting}
\addcontentsline{chaptoc}{chapter}{Nederlandse Samenvatting}
\setheader{Nederlandse Samenvatting}

{\selectlanguage{dutch}

In dit proefschrift hebben we gekeken naar kosmologische correlators, door de lens van differentiaalvergelijkingen en zogenoemde ``GKZ'' systemen. We zijn in hoofdstuk~\ref{ch:generalGKZ} begonnen met een algemene studie van dit soort systemen, en hebben uitgelegd hoe men voor een brede klasse aan integralen de differentiaalvergelijkingen van het GKZ systeem kan vinden. Verder hebben we behandeld hoe deze vergelijkingen volledig gespecificeerd worden door een enkele matrix en een parameter vector, de zogenoemde GKZ data. Daarna hebben we enkele welbekende eigenschappen van deze systemen behandeld. In het bijzonder twee gebruikelijke methoden om oplossingen voor deze systemen te verkrijgen, doormiddel van convergente machtreeksen ofwel een geschikt ansatz.

Hierna hebben we een formeler perspectief genomen, met een discussie van D-modules en diens reduceerbaarheid. D-modules zijn modules in de ring van polynomiale differentiaaloperatoren. In deze context corresponderen oplossingen van de differentiaalvergelijking ruwweg met homomorfismen van de D-module naar de ruimte van holomorfe functies.

Voor een D-module wordt reduceerbaarheid gedefinieerd als het bestaan van niet-triviale submodules. Cruciaal is dat deze eigenschap vertaald kan worden naar de oplossingen van de differentiaalvergelijking. Vanuit dit perspectief correspondeert reduceerbaarheid met het bestaan van differentiaaloperatoren, die wij reductie operatoren hebben genoemd, die gebruikt kunnen worden om parti\"ele oplossingsbasissen voor het volledige stelsel te kunnen construeren. Dit maakt een oplossingsstrategie volgens een ``verdeel en heers'' principe mogelijk, waar een system eerst wordt opgedeeld in eenvoudiger subsystemen, die afzonderlijk kunnen worden opgelost. In sommige gevallen, zoals degene beschreven in dit proefschrift, kan dan een volledige oplossingsbasis voor het oorspronkelijke systeem worden verkregen door de verschillende deeloplossingen te combineren.

Gebruik makende van dit algemene perspectief keerden wij terug naar de GKZ systemen. In deze context is er al wiskundige literatuur waarin de reduceerbaarheid van GKZ systemen wordt bestudeerd en volledig geclassificeerd wordt met behulp van de GKZ data. Onze belangrijkste bijdrage hier was, gebruik makende van de bewijzen in~\cite{schulze_resonance_2012}, het expliciet construeren van de verschillende subsystemen. 
Hoewel de oorspronkelijke bewijzen in abstracte termen waren geformuleerd, voornamelijk gebruik makende van exacte sequenties op zogeheten Euler-Koszul homologie\"en, hebben wij deze resultaten volledig kunnen herformuleren in termen van differentiaaloperatoren en parti\"ele oplossingsbasissen. Hierdoor was het mogelijk om de abstracte resultaten op concrete toepassingen te kunnen gebruiken, zonder dat daarbij het volledige technische formalisme nodig was.

Onze aandacht richtte zich hierna op de reductie operatoren. We hebben algoritmen ontwikkeld om deze operatoren expliciet te construeren, en hun eigenschappen weten te classificeren. Met name willen we hier wijzen op de telresultaten die te vinden zijn in de sectie~\ref{sec:reductionoperators}, die voorheen nog niet in de literatuur verschenen zijn.

Met dit raamwerk zijn wij in hoofdstuk~\ref{ch:cosmology} overgegaan naar kosmologische correlatoren. Hier hebben we het speelgoedmodel ge\"introduceerd, oorspronkelijk afkomstig uit~\cite{arkani-hamed_cosmological_2017}, dat we de in de rest van dit proefschrift hebben bestudeerd. Het resterende deel van hoofdstuk~\ref{ch:cosmology} was gewijd aan de studie van een specifieke kosmologische correlator in dit model: de enkele-uitwisselingsintegraal. Voor deze correlator hebben wij het bijbehorende GKZ systeem gegeven en, aangezien dit systeem reduceerbaar bleek, de reductie operatoren geconstrueerd. Met behulp daarvan hebben we parti\"ele oplossingsbassisen verkregen, die gecombineerd konden worden tot een volledige oplossingsbasis voor het systeem en dus tot de correlator zelf.

Voor de enkele-uitwisselingsintegraal hebben we de reductie operatoren uitsluitend gebruikt om, met behulp van homogene differentiaalvergelijkingen, parti\"ele oplossingsbasissen te vinden. In dit proces ontdekten wij echter dat deze integraal ook bepaalde inhomogene differentiaalvergelijkingen vervulde, hetgeen leidde tot contractie- en sij-identiteiten op diagramniveau. In hoofdstuk~\ref{ch:reductionalgorithm} hebben we deze observatie aanzienlijk uitgebreid.

In hoofdstuk~\ref{ch:reductionalgorithm} zijn we verder gegaan met het bestuderen van de kosmologische correlatoren van hetzelfde speelgoedmodel. In het bijzonder maakte wij gebruik van het resultaat uit~\cite{arkani-hamed_differential_2023} dat boomdiagrammen in deze theorie kunnen worden ontbonden met behulp van zogenaamde buizenstelsel. Elke buizenstelsel correspondeert met een integraal en de correlator zelf is en som over verschillende buizenstelsels. Deze stelsels bleken een bijzonder geschikt perspectief, aangezien zij ons in staat stelden om GKZ systemen voor willekeurige buizenstelsels te construeren en de reductie operatoren in volle algemeenheid te bepalen.

Een mogelijk obstakel bij de constructie van GKZ systemen is dat doorgaans de introductie van extra variabelen, die geen rol spelen in de fysische correlator, vereist is.
Het resulterende differentiaalstelsel bevat dan parti\"ele afgeleide naar deze overbodige variabelen, wat doorgaans leidt tot een nodeloze toenamen van complexiteit.
Voor kosmologische correlatoren boden de reductie operatoren hier echter een elegante oplossing, aangezien wij konden aantonen dat deze operatoren gecombineerd konden worden op zodanige wijze dat uitsluitend de fysische variabelen overbleven.

Na de constructie van deze fysische combinaties van de reductie operatoren hebben wij hun structuur nader onderzocht. Met behulp van deze operatoren construeerden wij een gesloten systeem van eerste-orde inhomogene differentiaalvergelijkingen. Het resulterende stelsel kan in principe worden opgelost voor de buizenstelselintegraal en vertoont gelijkenissen met andere bekende differentiaalstelsels voor kosmologische correlatoren. 
De cruciale bevinding was dat de reductie-operatoren een aanzienlijke overtolligheid in dit stelsel impliceerden. Om precies te zijn, de oplossingen van dit differentiaalsysteem zijn gerelateerd aan elkaar met algebra\"ische relaties, als directe consequentie van het bestaan van bepaalde reductie operatoren.

Behalve de relaties binnen het systeem van een bepaalde correlator, leiden de reductie operatoren ook tot algebra\"ische relaties tussen de systemen van verschillende kosmologische correlators. Om precies te zijn leiden de reductie operatoren tot relaties tussen een diagram en verscheidene andere diagrammen waar lijnen gecontracteerd of doorgesneden zijn.
Hierbij bleek dat de basisfuncties in deze verschillend systemen in hoge mate overeenkomen, slechts met permutaties van de invoer.

Het bestaan van deze algebra\"ische en permutatieve relaties leidde tot de vraag: welke functies blijven over nadat al zulk mogelijke relaties zijn toegepast? Wij hebben de overgebleven functies minimaalrepresentatiefuncties genoemd en aangetoond dat deze universeel zijn voor alle kosmologische correlatoren, onafhankelijk van de topologie of buizenstelsel van het diagram.
Bovendien bleek het aantal van dergelijke functies verassend klein, zeker in vergelijking met andere bekende algoritmen.

De minimaalrepresentatiefuncties vormden de basis voor ons recursieve reductie algoritme. In dit algoritme verloopt het oplossen van een kosmologische correlator ruwweg in twee fases. 
Eerst worden alle benodigde eerste-ordestelsels opgesteld en, met uitsluitend gebruik van algebraïsche en permutatieve relaties, uitgedrukt in termen van de minimaalrepresentatiefuncties. Vervolgens dienen deze functies zelf te worden bepaald. Omdat zij de elementaire bouwstenen van de correlatoren vormen, kunnen zij enkel direct worden opgelost aan de hand van de differentiaalvergelijkingen die zij vervullen. Eenmaal gevonden kunnen deze functies worden ingevoegd in een uitdrukking voor de volledige correlator. Cruciaal hierbij is dat het algoritme zo veel mogelijk symmetrie benut: het oplossen van algebraïsche identiteiten of permuteren van functiebasissen is immers aanzienlijk eenvoudiger dan het oplossen van differentiaalvergelijkingen.

Ten slotte hebben wij in hoofdstuk~\ref{ch:complexity} de complexiteit van de diverse algoritmen bestudeerd. Het doel was om de intuïtie van complexiteitsreductie en eenvoud te kwantitatief te maken met behulp van het raamwerk van Pfaffiaanse functies. Een Pfaffiaanse functie wordt gedefinieerd via een stelsel eerste-orde differentiaalvergelijkingen met een specifieke structuur, en laat bovendien een expliciete kwantificatie van complexiteit toe, met zowel topologische als computationele implicaties. Dit maakte het een natuurlijk kader voor de analyse van kosmologische correlatoren, aangezien we deze reeds via differentiaalvergelijkingen hebben bestudeerd.

Wij analyseerden de complexiteit van twee algoritmen: enerzijds het kinematische flow-algoritme uit~\cite{arkani-hamed_differential_2023}, anderzijds het eerste-ordestelsel dat wij in hoofdstuk~\ref{ch:reductionalgorithm} introduceerden. Beide konden in termen van een zogenoemde Pfaffiaanse keten worden gegoten, zodat hun complexiteit binnen dit kader onderzocht kon worden.

Essentieel is dat de complexiteit van de kosmologische correlatoren in dit kader afhankelijk is van hun expliciete representatie. Wanneer de gevonden bovengrenzen aanzienlijk te hoog zijn, kan dit dus erop wijzen dat een eenvoudigere representatie bestaat. Voor het kinematische flow-algoritme stelden wij inderdaad vast dat dit het geval is: de topologische bovengrenzen die hieruit volgden waren een sterke overschatting van wat op fysische gronden verwacht werd.

Deze overschatting van de complexiteit vormde een van de oorspronkelijke motivaties voor het recursieve reductie-algoritme uit hoofdstuk~\ref{ch:reductionalgorithm}: het introduceren van een kader waarin dergelijke bijkomende vereenvoudigingen en relaties automatisch worden verwerkt. Interessant genoeg slaagde het recursieve reductie-algoritme hierin, maar op een wijze die niet volledig zichtbaar is binnen het Pfaffiaanse raamwerk. Terwijl veel van de algebra\"ische relaties uit het eerste-ordestelsel direct tot een reductie in Pfaffiaanse complexiteit leiden, geldt dit niet voor de permutatieve relaties. Als gevolg daarvan blijft het Pfaffiaanse kader een overschatting geven, en rijst de vraag of er een meer geschikt complexiteitsbegrip bestaat dat ook deze relaties meeneemt.

}

\cleardoublepage

\chapter*{Popular Science Summaries}
\addcontentsline{toc}{chapter}{Popular Science Summaries}
\addcontentsline{chaptoc}{chapter}{Popular Science Summaries}
\setheader{Popular Science Summaries}

\section*{English summary}

\addcontentsline{toc}{section}{English Summary}
\addcontentsline{chaptoc}{section}{English Summary}

One of the biggest mysteries in physics is how the universe looked and behaved during its earliest moments. Mathematically, physicists study this using so-called \textit{cosmological correlators}. These correlators contain all information about the early universe. However, it is also notoriously difficult to obtain these correlators and extract this physical information. 

My dissertation develops new mathematical tools that make extracting this information more manageable. The key idea here is called \textit{reducibility}. Simply put, reducibility is the observation that, instead of trying to solve a complex problem all at once, it is often easier to break it up into more manageable chunks and solve these chunks separately. 

For cosmological correlators, we have identified exactly how solving them can be split up like this. We find the simplest building block they consist of, and then construct the correlators themselves from these building blocks. We call these building blocks \textit{minimal representation functions} and using them, can greatly simplify solving cosmological correlators. Therefore, these techniques can allow us to obtain a deeper understanding of the early universe.

We have focused here on cosmological correlators, but it bears mentioning that the underlying mathematics, using so-called differential equations and their reductions, is more broadly applicable. In the future, we hope that these techniques will find applications in many other parts of physics.

\section*{Nederlandse Samenvatting}

\addcontentsline{toc}{section}{Nederlandse Samenvatting}
\addcontentsline{chaptoc}{section}{Nederlandse Samenvatting}

E\'en van de grootste mysteries in de natuurkunde is hoe het universum eruitzag en zich gedroeg in zijn allereerste momenten. Wiskundig bestuderen natuurkundigen dit met behulp van zogeheten \textit{kosmologische correlatoren}. Deze correlatoren bevatten alle informatie over het vroege universum. Het is echter ontzettend moeilijk om deze correlatoren te berekenen en de fysieke informatie eruit te halen.

Mijn proefschrift ontwikkelt nieuwe wiskundige hulpmiddelen die dit proces makkelijker maken. Het belangrijkste idee hier is \textit{reduceerbaarheid}. Simpel gezegd houdt dit in dat het vaak makkelijker is een complex probleem op te lossen door het op te delen in kleinere, beter hanteerbare stukken, en die dan afzonderlijk proberen aan te pakken.

Voor kosmologische correlatoren hebben we precies uitgewerkt hoe dit opdelen kan gebeuren. We hebben de eenvoudigste bouwstenen ge\"identificeerd waaruit ze bestaan, en de correlatoren vervolgens weer opgebouwd uit deze bouwstenen. We noemen deze bouwstenen de \textit{minimaalrepresentatiefuncties} en met hun hulp kunnen we het berekenen van kosmologische correlatoren aanzienlijk vereenvoudigen. Hierom kunnen deze technieken ons helpen om een dieper inzicht te krijgen in het vroege universum.

We hebben ons hier gefocust op kosmologische correlatoren, maar het is belangrijk te vermelden dat de onderliggende wiskunde, gebaseerd op zegenoemde differentiaalvergelijkingen en hun reducties, veel breder toepasbaar is. In de toekomst hopen we dat deze technieken ook in andere delen van de natuurkunde gebruikt zullen worden.
\fi


\if\PrintAcknowledgements1

\chapter*{Acknowledgements}

\addcontentsline{toc}{chapter}{Acknowledgements}
\addcontentsline{chaptoc}{chapter}{Acknowledgements}
\setheader{Acknowledgements}

Now that the scientific content of the PhD is done, it is time to spend some time thanking the many people who guided me along the way, helped me when I needed it, or in any other way made life the last four year just that much better. 

I will begin with thanking my supervisor, Thomas Grimm. Throughout the years I think we've built a great collaboration and I thoroughly enjoy working together. I want to thank you for the many great pieces of advice and guidance you gave me, both about physics and about my future.

Secondly, I want to thank some people in the department. Let me start with Mick, who is one of my paranymphs. We've collaborated quite a bit during my PhD and I really enjoyed these collaborations. We've had many discussions in 7.12 and, even if they were about some counting problem we were stuck on for months, I look back on them very fondly. 

There are many people within the department who made my time here so enjoyable, either during lunches, borrels or in any other way. There are Georgios, Guoen and Nico, with whom I started the PhD and spent many weeks traveling around Europe during Solvay and afterwards, a time I enjoyed immensely. Jeroen, who guided me during my master's thesis and at times even managed to get me to go the gym in the morning. Also Hossein and Valentina, I feel so lucky to have had such great office mates during my PhD and whether it was when we installed our amazing seasonal decorations, or when we sang our hearts out at the karaoke bar, you never failed to make me laugh and smile.

There are many other people who made my time during lunches or borrels at the department so much more enjoyable. For example, I want to thank the many people in group: Casey, C\'esar, Claire, David, Edwan, Guim, Javi, Maaike, Pedro, Ra\'ul, Shradha, Stefan, Thorsten, Tuna and Wilke. As well as some past members such as Erik, Damian, Filippo, Jeroen, Lorenz and Stefano. I also want to thank Umut, whose presence is missed dearly.

Outside of our group I also want to thank the people who were in the PhD Councils during my stay there, both in the institute and for the graduate school. I really enjoyed working together and it is exciting to see that some of the projects are (finally!) reaching their conclusions. 
Additionally, the people in the borrel committee (which at least officially, I was a part of), as well as the people who went to the borrels. So many times where you've made me stay way longer than I intended just by being too fun and giving me Aperol Spritz.

Now I want to thank some of the people outside of the university. Depending on who these are for, some of them will be in Dutch.

Ik wil natuurlijk graag Hidde bedanken, ook mijn paranymph. Je hebt me altijd geholpen wanneer en hoe ik dat nodig had en me gruwelijk hard laten lachen. Je steun betekent ontzettend veel voor me en ik ben ontzettend dankbaar voor onze vriendschap. 

Ashley, Bj\"orn, Oscar and Robert, also deserve special mention. I always love the many late night discussions about whatever physics (or philosophy) topics we would come up with. I also greatly appreciate being able to rant together about whatever thing we were stressed about or stuck on that moment.

Ook mijn ex-huisgenoten, Dylan, Lies, Klaas en Pam. Jullie gezelschap was, zeker in coronatijd, onmisbaar en ik ben blij dat ik zulke vrienden eraan heb overgehouden.
Lies ook nog bedankt voor het nakijken van welke tekst ik ook naar je toe stuurde, inclusief delen van dit proefschrift.

There is an uncountable number of other friends that made the last four years so enjoyable, and all of you I also want to thank. These are too many to name explicitly, but you know who you are. This also includes the many great people I've met on conferences throughout the years, who made going to these conferences always incredibly enjoyable.

Ook wil ik graag mijn familie, Frits, Marian, Johan, Liza, Ramon, Moniek, Daniel en Lotte, bedanken omdat jullie support ook onmisbaar is geweest. Allen ook dank voor het begrip als ik even een periode weer druk was en dus van de aardbodem verdween. Ik waardeer ook ontzettend dat jullie oprecht ge\"interesseerd waren in mijn onderzoek en wat ik nou deed, zelfs al was het zo abstract. Frits specifiek ook nog bedankt voor het ontwerpen van de kaft, uitnodiging en schutbladen van dit proefschrift. Ik ben niet altijd een makkelijke klant voor je geweest maar ik waardeer ontzettend dat je dit voor mij hebt willen doen.

Tot slot wil ik  mijn vriendin, Kari, bedanken. 
Tijdens mijn PhD ben je me ontzettend tot steun geweest. Als ik gestresst kon je altijd dingen relativeren en ik waardeer het ontzettend dat we onze ervaringen met het PhD leven met elkaar konden delen. Ik weet zeker dat de PhD, en dit proefschrift, niet hetzelfde was geweest zonder jou. Als kers op de taart zijn we ook nog gaan samenwonen vorig jaar en ik geniet daar elke dag ontzettend van. Ik houd van jou.

\fi

\if\PrintCV1
\cleardoublepage

\chapter*{About the Author}

\addcontentsline{toc}{chapter}{About the Author}
\addcontentsline{chaptoc}{chapter}{About the Author}
\setheader{About the Author}

Arno Hoefnagels was born on 21 January, 1997 in Amersfoort, the Netherlands, where he also grew up and obtained his secondary school diploma from Het Vathorst College in 2014. In the same year he started a bachelor in analytical chemistry at the Hogeschool Utrecht, from which he graduated with honours in 2018.

Afterwards, he transitioned to physics with a pre-master and master in theoretical physics at Utrecht University, the latter of which he completed with the distinction cum laude. During his master's thesis, he studied conformal quantum mechanics with applications in Hodge theory under the supervision of Thomas Grimm. 

He continued with a PhD at Utrecht University, again under the supervision of Thomas Grimm, albeit with a slightly different topic. During his PhD, he studied the mathematics underlying Feynman diagrams and cosmological correlators, with a focus on applications in the latter setting. The research culminated in this PhD thesis, which he will defend on October 8, 2025.
\fi
\if\PrintBackCover1
\fi

\addcontentsline{toc}{chapter}{References}
\addcontentsline{chaptoc}{chapter}{References}

\printbibliography[title={\color{thumb\arabic{thumbcounter}}References}]


\end{document}